\documentstyle[aps,epsfig,isolatin1,psfig]{revtex}

\newcommand{\widtheps}{10.5cm}
\newcommand{\widthepstwo}{11cm}
\newcommand{\widthepsthree}{11.5cm}
\newcommand{\widthepsfour}{11.2cm}
\newcommand{\widthepsxmgr}{8cm}
\newcommand{\widthepsxmgrtwo}{9cm}
\newcommand{\widthepsxmgrthree}{10cm}

\newcommand{\espace}{\vspace{0.5cm}}

\newcommand{\boite}[1]{\mbox{\scriptsize #1}}
\newcommand{\dslash}{\partial\!\!\!/}
\newcommand{\sslash}[1]{{#1}\!\!\!\!\!/\ }
\newcommand{\barray}{\begin{equation}\begin{array}{l}}
\newcommand{\earray}{\end{array}\end{equation}}
\newcommand{\barraystar}{\begin{displaymath}\begin{array}{l}}
\newcommand{\earraystar}{\end{array}\end{displaymath}}
\newcommand{\qbar}{\bar{q}}
\newcommand{\Gu}[1]{\Gamma^{#1}} 
\newcommand{\Gd}[1]{\Gamma_{#1}} 
\newcommand{\gd}[1]{\gamma_{#1}}  
\newcommand{\gu}[1]{\gamma^{#1}}  
\newcommand{\ld}[1]{\lambda_{#1}}  
\newcommand{\lu}[1]{\lambda^{#1}}  
\newcommand{\one}{1\!\!1}         
\newcommand{\p}{\partial}          
\newcommand{\f}[2]{\frac{#1}{#2}}
\newcommand{\tr}{{\rm tr}}
\newcommand{\be}{\begin{equation}}
\newcommand{\ee}{\end{equation}}
\newcommand{\beqn}{\begin{eqnarray}}
\newcommand{\eeqn}{\end{eqnarray}}
\newcommand{\Tr}{{\rm Tr}}
\newcommand{\tg}{\rm tg}

\newcommand{\fpi}{\mbox{$f_{\pi}$}}
\newcommand{\mpi}{\mbox{$m_{\pi}$}}
\newcommand{\gpiqq}{\mbox{$g_{\pi\bar{q}q}$}}
\newcommand{\diag}{$\mbox{diag}$}
\newcommand{\s}{\sigma}
\newcommand{\st}{\sigma^T}
\newcommand{\Sig}{\Sigma}
\newcommand{\Sigt}{\Sigma^T}
\newcommand{\G}{\Gamma}

\newcommand{\ginv}{g^{-1}}
\newcommand{\gdeux}{g^2}
\newcommand{\ginvdeux}{g^{-2}}
\newcommand{\Vinv}{V^{-1}}
\newcommand{\qq}{q^2}
\newcommand{\micarre}{m_i^2}
\newcommand{\mjcarre}{m_j^2}
\newcommand{\souligne}[1]{\vspace{0.0cm}\par
\noindent\underline{{#1}}\par
\vspace{0.0cm}\par}

\title{The 3-flavor A-scaled Nambu--Jona-Lasinio model}

\author{Bruno Van den Bossche}
\address{Physique Nucl\'eaire Th\'eorique\\
Institut de Physique B5, Universit\'e de Li\`ege Sart Tilman\\
4000 Li\`ege, Belgium}

\begin{document}

\everymath={\displaystyle}

\maketitle

\begin{abstract}
  We review the so called A-scaled Nambu--Jona-Lasinio (NJL) model both in the vacuum and at finite temperature and density in the case of 3 flavors ($u,d,s$). Starting from QCD and integrating out the gluons, we show how to reintroduce some of their effects in order to take into account the trace (or {\it scale}) anomaly of QCD. The axial $U_A(1)$ anomaly is also present in the model in order to discuss the $\eta'$ particle. In the present work we treat the pseudoscalar and scalar sectors, pointing out  the defects related to the omission of the vector particles. We use our model to study the restoration of  chiral symmetry as a function of temperature, the thermodynamical functions and the equation of state, and we survey  the mixing between the  scalar isoscalar particles and the glueball. 
\end{abstract}

\thispagestyle{empty}
\tableofcontents

\vspace{3cm}

Pictures 4--7, 9--17, 20--23 are reprinted from 

\vspace{.25cm}

\begin{itemize}
\item Nuclear Physics {A582} , M. Jaminon, B. Van den Bossche, "{$SU(3)$} scaled effective {L}agrangians for a hot and
strange system", 517--567, Copyright 1995
\item Nuclear Physics {A592}, D. Blaschke, M. Jaminon, Yu. L. Kalinovsky, P. Petrow, S. Schmidt, B. Van den Bossche, "Anomalous pion decay in effective {QCD} at finite temperature", 561--580, Copyright 1995
\item Nuclear Physics {A598}, J. Cugnon, M. Jaminon, B. Van den Bossche, "Phase transition and thermodynamics of a hot and dense system in a scaled {NJL} model", 515--538, Copyright 1996
\end{itemize} 

\vspace{.25cm}

with permission from Elsevier Science.

\newpage

\section{Introduction} 
\label{introduction}
Quantum chromodynamics (QCD), the theory of strong interaction, is well suited to perform calculations at high energy-momentum. The non-linearity in the gluon sector (three- and four-gluon vertices) together with the high value of the coupling constant make it difficult, if not impossible, to apply standard calculational techniques for the description of the hadronic sector. One has then to choose between two strategies: either to put QCD on a lattice or to search for simplified models believed to  mimic QCD in a given range of energy-momentum. Needless to say, due to the structure of QCD (high number of degrees of freedom (d.o.f.)) and problems related to the treatment of fermions on the lattice (fermion doubling, fermion determinant), lattice gauge theory is restricted to a  very small number of space and time slices.
Although improved discrete actions can lead to the hope that these small lattices already give stable outputs, it can be advantageous to turn to the second strategy. It is not, of course, solving QCD but it can shed some light on the way the true theory behaves. \par

It is not our purpose here to describe the different models which exist. We shall mainly focus on an extension of the Nambu Jona-Lasinio (NJL) model, while keeping in mind that it is a simplification of the Global Color Model (GCM). The NJL model was invented by analogy with superconductivity. The physical phenomenon underlying the BCS theory is the creation of an energy gap, resulting from  interactions of electrons with phonons, near the Fermi surface in the spectrum of one-fermion excitations. However, the gap is not sufficient to realize superconductivity whose essence is related to the appearance of the gap through the  dynamical breaking of $U(1)$ gauge symmetry.\par

In the absence of a current quark mass term $m$, QCD possesses an additional symmetry named the chiral symmetry which is related to the fact that the two inequivalent representations of the Lorentz group for spinors ($(1/2,0)$ and $(0,1/2)$) transform independently (if $m=0$). As soon as this symmetry is dynamically broken, one might expect that the fermions will acquire (dynamical) masses. For a review of  the dynamical breaking of chiral symmetry in quantum field theory, see \cite{miransky93}.

\subsection{Symmetries}
\label{symmetries}
In this section, we discuss the symmetries connected to the QCD Lagrangian which are of concern for the construction of effective theories modeling it.

The basic theorem for the  examination of symmetries is  N\oe ther theorem. It states that
{\em for any invariance of the action under a continuous transformation of the fields, there exists a classical charge $Q$ which is time-independent and is associated with a conserved current, $\p_{\mu}J^{\mu}=0$}.\par 

A deviation from this theorem implies that either the symmetry is only an approximate one (for example the chiral symmetry for light quark masses) or that quantum corrections have broken it: we enter the world of anomalies. Moreover, there still exists the possibility that the symmetry is spontaneously broken (this is the case of the chiral symmetry).

Although gauge invariance is of prime importance in QCD, because it is the basis to justify the QCD Lagrangian, it is of little interest in this work since we  shall work within the frame of the NJL model which is not locally gauge invariant. (It is however globally color invariant.) The relevant symmetries  in the present context are shown below, where we indicate the corresponding transformations of the fields, and the status of the conservation\footnote{In addition to these symmetries, QCD is also invariant under Lorentz transformations, $P$, $C$, $T$ and $G-$parity, which are also symmetries of the NJL action.}. We restrict ourselves to   three flavors of quarks $u,d,s$ and we take the convention to denote by $m$ the current quark mass matrix $\diag(m_u,m_d,m_s)$, and by $q$ the vector representing quarks in the flavor space $q=\diag(q_u,q_d,q_s)$. 

\begin{itemize}
\item {\bf Global gauge symmetry}: as stated above, NJL is not locally gauge invariant but only globally invariant. The color enters only through the number $N_c$ of each quark flavor. In the GCM model, the important features of local gauge invariance of the color group $SU(3)_c$ such as confinement and asymptotic freedom can be parametrized in the form of an effective two-point gluon propagator (see below). However, this is not the case for the NJL model where an unphysical meson to quark-antiquark pair threshold  is a direct consequence of the lack of confinement. 
\goodbreak
\item  {\bf Scale symmetry}: 
\[
\left\{
\begin{array}{l}
x_{\mu}\rightarrow \lambda^{-1}x_{\mu}, \\
A_{\mu}^a\rightarrow \lambda A_{\mu}^a, \\
q\rightarrow \lambda^{3/2}q,
\end{array}
\right.
\]
is exact in the limit of  vanishing current quark mass $m=0$. However, this symmetry is broken by quantum effects (see below).
\item {\bf Vector $U(1)_V$ symmetry or Baryonic number conservation}:  $q\rightarrow \exp(i\alpha)q$.
\item {\bf Quark number conservation}: $q_i\rightarrow
\exp(i\alpha_i)q_i$ ($i=u,d,s$): each flavor has its own conserved  number.
\item {\bf Axial $U(1)_A$ symmetry}: $q\rightarrow\exp(i\gd{5}\alpha)q$ is exact in the limit of  vanishing current quark mass $m=0$. This symmetry is broken by quantum effects, which explains why it is not seen in the spectrum of physical states.
\item {\bf Isospin symmetry}: \[
q\equiv\left(\begin{array}{c}u\\
d\end{array}\right)\rightarrow\exp(i\tau^a
\alpha_a)q\ ,\ \ a=1,...,3.
\] Contrary to axial  symmetry, isospin symmetry is exact in the limit where light quark masses are equal $m_u = m_d$ ($\tau_a$ are the Pauli matrices).
\item {\bf Vector $SU(3)_V$ symmetry}: \[
q\equiv\left(\begin{array}{c}u\\
d\\ s\end{array}\right)\rightarrow\exp(i\f{\lambda_a}{2}
\alpha_V^a)q\ ,\ \ a=1,...,8.
\] The isospin symmetry can be generalized to the three-flavor case and is exact if $m_u=m_d=m_s$ ($\ld{a}$ are the $SU(3)$ flavor Gell-Mann matrices).
This approximate symmetry explains why the hadrons are ordered into multiplets.
\item {\bf Axial $SU(2)_A$ and $SU(3)_A$ symmetries}: \[
q\equiv\left(\begin{array}{c}u\\
d\\ s\end{array}\right)\rightarrow\exp(i\gd{5}\f{\lambda_a}{2}
\alpha_A^a)q\ ,\ \ a=1,...,8 \ \ \ (SU(2)_A: \ld{a}\rightarrow \tau_a).
\] The quark flavors can also be mixed in the axial case. This symmetry  is exact as long as $m_u=m_d=m_s=0$.
This approximate symmetry is not seen in  the spectrum. Axial transformations alter the parity that is associated with a state. A manifestation of $SU(2,3)_A$ in nature would require that each isospin (or $SU(3)_V$) multiplet be accompanied by a mirror multiplet of opposite parity. In the same way, since we do not observe opposite parity partners to all hadrons, the $U(1)_A$ symmetry cannot be realized directly by QCD. While the axial $SU(3)_A$ symmetry is realized in the Goldstone mode through the dynamical breaking of chiral symmetry, $U(1)_A$ is never realized, being completely broken by quantum effects (the anomaly).
\item {\bf Chiral symmetry}:
vector and axial $SU(3)$ symmetries can be combined to realize transformations on the left and right parts of the quarks ($q_{R\atop L}\equiv\f{\displaystyle 1\mp\gd{5}}{\displaystyle2}q$):
\[
SU(3)_V\otimes SU(3)_A\leftrightarrow SU(3)_L\otimes SU(3)_R,
\]
where 
\[
SU(3)_L\Rightarrow q_L\rightarrow\exp(i\frac{\lambda_a}{2}\alpha_L^a)q_L, \mbox{    }
SU(3)_R\Rightarrow q_R\rightarrow\exp(i\frac{\lambda_a}{2}\alpha_R^a)q_R.
\]
Under chiral symmetry, left-handed  and right-handed quarks transform independently. This symmetry is broken by the quark mass matrix $m$. Apart from this explicit breaking, chiral symmetry is also spontaneously broken down to $SU(3)_V$.
\end{itemize}

\subsubsection*{Spontaneous symmetry breaking and anomalies}
As mentioned above, a symmetry can be manifested in several ways.
\begin{itemize}
\item It may remain exact. This is the case of the electromagnetic gauge symmetry (except for superconductivity) and the color gauge symmetry of QCD\footnote{See however the works \cite{rajagopal97,rajagopal98,rajagopal98b} and \cite{shuryak97} where the authors, taking care of both the quark and diquark d.o.f., show that the global color symmetry might be dynamically broken by the formation of a $<qq>$ condensate of quark Cooper pairs.}. 

\item It may be explicitly broken. This is the case of the isospin symmetry if $m_u\ne m_d$ or the axial $SU(2)$ or $SU(3)$ symmetry if $m\ne 0$.

\item It may be hidden. It is an invariance of the action but not of the ground state: the symmetry is not seen in the spectrum of physical states. Two types of mechanisms \cite{donoghue92} are possible.
  \begin{enumerate}
  \item  The symmetry can be spontaneously broken. It is the case of the $SU(2)_L$ symmetry in electroweak interactions. The breaking requires the presence of scalar fields (Higgs fields in electroweak interactions) which give rise to vacuum expectation values. The order parameter associated with the breaking is based on an elementary operator. In the case of the Higgs mechanism, the order parameter is the vacuum average of the scalar field.
  \item The symmetry can be dynamically broken: it does not require any scalar field. This is the case of the chiral $SU(2,3)_L\otimes SU(2,3)_R$ symmetry in QCD, or the $U(1)$ electromagnetic gauge symmetry in superconductivity. The order parameter is given by the vacuum average of a composite (and not elementary) operator. In the case of dynamical chiral symmetry breaking, the order parameter is the quark condensate\footnote{Other order parameters are possible. The quark condensate is the simplest one. However, it should be stressed that although $<0|\bar{q}q|0>\ne0$ implies dynamical chiral symmetry breaking, the reverse is not true.} $<0|\bar{q}q|0>$.
\end{enumerate}

In the following we shall use the term ``spontaneous symmetry breaking'' to describe both cases of hidden symmetry, making the distinction when appropriate.
To the notion of hidden symmetry can be attached the Goldstone theorem: {\em if a theory has a continuous symmetry of the Lagrangian which is not a symmetry of the vacuum\footnote{This is the Goldstone mode of the symmetry, by opposition to the  Wigner mode. Goldstone mode can only occur in the case of an infinite number of d.o.f.}, there must exist one or more massless bosons} (Goldstone bosons).  This means that  spontaneous or dynamical breaking of a continuous symmetry will entail massless particles in the spectrum. 
The Goldstone theorem, associated to the dynamical breaking of chiral symmetry, explains the small pseudoscalar nonet mass ($\pi_0, \pi_{\pm},K_0,\bar{K}_0,\eta,\eta'$), except for the $\eta'$. In fact, in the limit of vanishing current quark mass, 
both pions, kaons and $\eta$ have a vanishing mass in accordance with the Goldstone theorem. Their nonzero value is just the reflection of the explicit breaking of  chiral symmetry by current quark mass. The fact that the strange quark mass is heavier than the up and down quark mass implies a greater corresponding explicit breaking and thus a heavier mass for the kaons (and $\eta'$ without the axial anomaly, see section~\ref{bs}) compared to the pions. 
The breaking of chiral symmetry can be summarized by
\[
SU(3)_L\otimes SU(3)_R\Rightarrow SU(3)_V.
\]
\item The symmetry may have an anomaly. An anomaly is a symmetry of the action (classical level) which can be broken by quantum effects: the N\oe ther theorem is no longer valid. In QCD, this is for example the case of the axial $U(1)_A$ and scale symmetries: even if  the quark masses are identically equal to zero, the corresponding current does not vanish. 
For the axial $U(1)$ symmetry, we get ($F, \tilde{F}$ are the $SU(3)$ field strength tensor and its dual, respectively, $\beta_{\mbox{\scriptsize
QCD}}$ is the Callan-Symanzik $\beta$-function of QCD and $\gamma_m$ is the mass anomalous dimension) 
\be
\partial^{\mu}(\bar{q}\gd{\mu}\gd{5}\f{\lu{0}}{2}q)
=2i\bar{q}\gd{5}m^0\f{\lu{0}}{2}q+\sqrt{\f{3}{2}}\f{g^2}{32\pi^2}
F^a_{\mu\nu}\tilde{F}^{\mu\nu}_a,
\label{anomalieaxialeforte}
\ee
while the  scale anomaly reflects itself in the form
\be
\p_{\mu}J^{\mu}=\theta^{\mu}_{\mu}=(1+\gamma_m)
\sum_{i=1}^{N_F}\bar{q}_im_iq_i +\f{\beta_{\mbox{\scriptsize
QCD}}}{2g}F_{\mu\nu}^aF_a^{\mu\nu}.
\label{anomalietrace}
\ee
An other example of anomaly\footnote{The reference \cite{donoghue92} is a good introduction on anomalies in the standard model. Its chapter VII-3 deals with purely hadronic processes through the study of the Wess-Zumino-Witten anomaly action.} concerns the decay of neutral pions into two photons $\pi_0\rightarrow\gamma\gamma$. The isovector axial current $J_{5\mu}^{(3)}\equiv\bar{u}\gd{\mu}\gd{5}u-\bar{d}\gd{\mu}\gd{5}d$ is not conserved in the quantum world: its divergence leads to a term $F\tilde{F}$, where $F$ is now the electromagnetic field strength tensor. The strong axial anomaly (\ref{anomalieaxialeforte}) is believed to give its high mass to the $\eta'$ particle compared to the other members of the pseudoscalar nonet.
\end{itemize}

With this small introduction on symmetries, dynamical breaking and anomalies, we are now in a position to understand the basis of the NJL model which is described below as a limiting case of the global color model.

\subsection{From QCD to GCM}
\label{qcdtogcm}
Our purpose is to discuss the NJL model as an approximation of  QCD and its extensions.  Our starting point is  the global color model (GCM) of which NJL is a particular case.\par
 
QCD is the theory of quarks and gluons. However, gluons are often neglected in favor of an effective interaction between quarks. This is the case for non-relativistic models such as the quark model as well as for relativistic models such as the NJL model. For models incorporating  the dynamical breaking of chiral symmetry, one argues that the interaction between quarks and antiquarks, coming from complicated processes of gluon exchange,  is attractive and  leads to an effective interaction between quarks. This interaction is responsible for  quark-antiquark condensation in the vacuum, when the interaction exceeds a critical strength. Although still a symmetry of the Lagrangian (if $m=0$), the condensation is responsible for the fact that the vacuum is no longer symmetric under $SU(3)_A$ transformations. This leads to massless Goldstone bosons as well as a dressing of quarks by $\bar{q}q$ pairs.

 The shape of the effective interaction is important to model QCD and its choice is a competition between mathematical simplicity (still retaining some important aspects of QCD such as  dynamical symmetry breaking and, more generally, the relevant global symmetries) and conceptually more appealing approaches, at the price of loosing simplicity. While NJL belongs to the former, the latter corresponds to the more general GCM-type model.

In this section, we summarize the main steps which enable  going from QCD to GCM. 

The QCD \cite{cheli84} Lagrangian is given by:
\begin{equation}
{\cal L}_{QCD}=\bar{q}(i\gamma^{\mu}\partial_{\mu}-m)q
-\frac{1}{4}(F^a_{\mu\nu})^2+g\bar{q}\gamma^{\mu}A_{\mu}q,
\label{Lqcd}
\end{equation}
where $q$ is the quark field in flavor and color spaces (in the fundamental representation) and $g$ is the coupling constant. The quantity $A_{\mu}$ is a shortened notation for the eight gluon fields $A_{\mu}=A_{\mu}^a\lambda^a/2$ in the adjoint representation, where
$\lambda^a$ ($a=1...8$) acts in the color space. The second term is the Yang-Mills Lagrangian, constructed from the gluon field strength:
\begin{equation}
F^a_{\mu\nu}=\partial_{\mu}A^a_{\nu}-\partial_{\nu}A^a_{\mu}+gf^{abc}
A^b_{\mu}A^c_{\nu}.
\label{fieldstr}
\end{equation}

The last term of eq. \ref{fieldstr} gives an interaction term between gluons (three- and four-gluon couplings) and is due to the  non-Abelian structure of the theory
(the totally antisymmetrical coefficients $f^{abc}$ are the structure constants of  $SU(3)$). The last term of the Lagrangian~(\ref{Lqcd}) gives the coupling between the gluon fields
$A_{\mu}^a$ and the color quark currents
\be
\label{courantfort}
j^{\mu}_a=\bar{q}\f{\lambda^a}{2}\gamma^{\mu}q.
\ee
In  quantum theory, a given  Lagrangian has to be supplied with
a quantification procedure which is chosen here to be the Feynman prescription: solving the quantum theory is equivalent to  search for the vacuum-vacuum transition amplitude defined through the  following functional integral in Minkowski space:
\begin{equation}
{\cal
Z}_{QCD}=\int{\cal D}q{\cal D}\bar{q}\int{\cal D}A_{\mu}^a\exp
\left(i\int d^4x{\cal L}_{QCD}\right).
\label{Zqcd}
\end{equation}
The integration over gauge fields has to be carried out only over  gauge inequivalent orbits, which is performed by fixing the gauge and using the Faddeev-Popov trick \cite{cheli84}. In the following, this is supposed to be included in the gluon measure ${\cal D}A_{\mu}^a$. Several procedures have been investigated to 
manage the functional integral (\ref{Zqcd}). For example, Alkofer and Reinhardt
\cite{reinhardt91,reinhardt92,reinhardt95} use the field strength method to obtain an effective action for an auxiliary field, describing the non-perturbative vacuum already at tree level. Making a quadratic expansion around this auxiliary field, these authors are able to obtain a color current-current interaction, which is local in the limit of small momenta, while being identical to  one-gluon exchange at high momenta.

Another, more usual, approach is possible and described for example in \cite{roberts87,cahill92}. We shall however follow the notation of \cite{reinhardt95}. The integration over the gauge fields in (\ref{Zqcd}) is not Gaussian. It cannot be evaluated exactly\footnote{\protect In the field strength formalism, Alkofer and Reinhardt
\cite{reinhardt92} get a Gaussian form for the auxiliary fields they introduce. This can be integrated exactly, leading to an effective functional for the auxiliary fields.}$^{,}$\footnote{\protect An exact calculation can also be performed for the approximations $QCD_2$ and Abelian $QCD_4$
\cite{ebekas91,reinhardt94}.}. Equation (\ref{Zqcd}) can be rewritten in the form 
\begin{equation}
{\cal
Z}_{QCD}=\int{\cal D}q{\cal D}\bar{q}\exp
\left(i\int
d^4x\bar{q}(i\gamma^{\mu}\partial_{\mu}-m^0)q+iW[j]\right),
\label{Zqcd2}
\end{equation}
where
\begin{equation}
iW[j]=\ln\int{\cal D}A_{\mu}^a\exp\left(-\frac{i}{4}\int
d^4x(F_{\mu\nu}^a)^2+ig\int d^4xA_{\mu}^aj^{\mu}_a\right).
\label{Zqcd3}
\end{equation}
The idea is to expand~(\ref{Zqcd3}) in powers of quark currents 
\begin{eqnarray}
iW[j]&=&iW[0]+g\int W^{(1)}(x_1)^a_{\mu}j^{\mu}_a(x_1)
d^4x_1\nonumber\\
&&\mbox{}+\frac{g^2}{2!}\int
W^{(2)}(x_1,x_2)^{a_1,a_2}_{\mu_1,\mu_2}
j^{\mu_1}_{a_1}(x_1)j^{\mu_2}_{a_2}(x_2)d^4x_1d^4x_2+\ldots\nonumber\\
&&\mbox{}+\frac{g^n}{n!}\int W^{(n)}(x_1\ldots x_n)^
{a_1\ldots a_n}_{\mu_1\ldots\mu_n} j^{\mu_1}_{a_1}(x_1)\ldots
j^{\mu_n}_{a_n}(x_n)d^4x_1\ldots d^4x_n\nonumber\\
&&\mbox{}+\ldots
\label{gammaj}
\end{eqnarray} 
The coefficients $W^{(n)}(x_1\ldots x_n)^{a_1\ldots a_n}_{\mu_1\ldots\mu_n}$ are the  connected  $n$-point  functions of the gluons without quark-loop contributions. The second term is
\begin{equation}
  \label{propagator}
  i^{-2}W^{(2)}(x_1,x_2)^{a_1a_2}_{\mu_1\mu_2}=
\left<A_{\mu_1}^{a_1}(x_1)A_{\mu_2}^{a_2}(x_2)\right>-
\left<A_{\mu_1}^{a_1}(x_1)\right>\left<A_{\mu_2}^{a_2}(x_2)\right>,
\end{equation}
which readily shows that $W^{(2)}(x_1,x_2)$ is the gluon correlation function\footnote{We have defined 
\[
\left<O\right>=\frac{\int{\cal
D}A_{\mu}^aO\exp\left(-i/4\int d^4x(F_{\mu\nu}^a)^2\right)}
{\int{\cal
D}A_{\mu}^a\exp\left(-i/4\int
d^4x(F_{\mu\nu}^a)^2\right)}.
\]}.
The GCM model consists in keeping only $W^{(2)}$, modeling the effect of the suppression of the higher $n$-point Green functions ($n\ge3$) through the use of a particular shape for $W^{(2)}$ \cite{roberts87,roberts85,roberts88,roberts89}. This shape can be chosen to reproduce important features of QCD such as asymptotic freedom and confinement \cite{roberts92,roberts92b,roberts92c}. However it can be shown that the inclusion of non-canonical quark-gauge boson vertices is still necessary.

Each function $W^{(n)}$   is separately
invariant under  Lorentz and global gauge color transformations.
However, only the total sum~(\ref{gammaj}) 
is  invariant  under local gauge transformations: its truncation to a given order is then gauge dependent.\par

The  model  is finally described by the Lagrangian
\begin{equation}
{\cal L}_{GCM}=\bar{q}(i\gamma^{\mu}\partial_{\mu}-m)q+\f{g^2}{2}j^{\mu}_a(x)W^{(2)}(x,y)_{\mu\nu}^{ab}j^{\nu}_b(y),
\label{LGCM}
\end{equation}
where $W^{(2)}(x,y)_{\mu\nu}^{ab}$ has  the interpretation of an effective gluon propagator.

This model is very appealing in the sense that, while mathematically tractable, it 
\begin{itemize}
\item gives the correct high momentum limit (through the asymptotic freedom form of the propagator);
\item leads to quark confinement (although giving no insight on the mechanism of color confinement, choosing a form of the gluon propagator so that the quark propagator has no pole on the real axis is a sufficient -- albeit not necessary -- condition to ensure the absence of propagation of colored free quarks \cite{roberts92b});
\item is renormalizable (the non-locality of the interaction makes a natural cut-off on  high momenta implied in loops and makes the theory renormalizable. We shall however not discuss the renormalization procedure further on in this paper).
\end{itemize}

However, the model has the drawbacks associated to its qualities, apart from the lack of local color invariance: 
\begin{itemize}
\item the first limitation is of mathematical nature: although tractable, the model is quite heavy to handle;
\item a second  limitation is of conceptual nature. To ensure the convergence of calculations -- see the $i$ factor in the exponential of (\ref{Zqcd}) --  the model has to be rewritten in Euclidean space\footnote{Note also that results from lattice QCD and from studies through Schwinger-Dyson equations are obtained in Euclidean space.}, which has none of the problems due to the indefinite norm inherent to Minkowski space. However, since the transcription of non-perturbative equations from one space to the other is not equivalent (except for simple models) to the analytic continuation of the solutions, the field theory  has to be considered as defined in Euclidean space. This means that a solution of eq.~(\ref{LGCM}) written in Minkowski space is not a solution of eq.~(\ref{LGCM}) written in Euclidean space. Indeed, the essential singularity at infinity which is encountered when the quark propagator is an entire function implies that the Wick rotation {\em cannot} be used to justify a change of space (Euclidean $\rightarrow$ Minkowski) by just transcripting the form of the equations \cite{tandy97}; 
\item one should not forget that there is no reason to expect that the effective interaction  should have the form of a gluon exchange interaction. The ``derivation'' we have given is at best an attractive intellectual game (which, however, gives a beautiful frame to the model).
\end{itemize}

\subsection{Fierz identities, Feynman-like gauge, covariant gauge}
\label{fierztransfo}
As we saw in the previous section, removing the full set of gluon Green functions except for the 2-point one implies that {\em i}) we loose the local gauge invariance of the theory and {\em ii}) we have to model the 2-point Green function. The modeling depends not only on space-time but also on internal indices. Without loss of generality, the gluon propagator can be written\footnote{Translational invariance implies $D(x,y)=D(x-y)$. To derive general relations, we shall however keep the first form up to the end of the derivation of the Fierz identities.}

\begin{equation}
  \label{gluonprop}
 D_{\mu\nu}^{ab}(x,y)\equiv g^2W^{(2)}(x,y)_{\mu\nu}^{ab}=\delta_{ab}D_{\mu\nu}(x,y). 
\end{equation}

In this section,  the summation over repeated indices is implicit. We work in Minkowski space and the notational conventions are from Itzykson and Zuber \cite{itzykson85}, see appendix \ref{conventions}. In particular, we have
the metric $g^{\mu\nu}=\mbox{diag}(1,-1,-1,-1)$, $\gamma^5=\gamma_5=
i\gamma^0\gamma^1\gamma^2\gamma^3$, and the totally antisymmetric Levi-Civita tensor
is given by (with $\epsilon_{\mu\nu\rho\sigma}=-\epsilon^{\mu\nu\rho\sigma}$)
\beqn
\epsilon^{\mu\nu\rho\sigma}=\left\{
\begin{array}{ll}
+1 & \mbox{if $\{\mu,\nu,\rho,\sigma\}$ is an even permutation of 
$\{0,1,2,3\}$,}\\
-1 & \mbox{if it is an odd permutation,}\\
0 & \mbox{otherwise.}
\end{array}
\right.
\eeqn 

Fierz transformations are useful identities allowing to rewrite the color-color interaction  (\ref{LGCM}) in terms of physically observable quantum numbers. This is a standard procedure when dealing with mesons and for a $\gu{\mu}\otimes\gu{\mu}$ structure such as in  the Feynman gauge. This procedure has also been recently applied  to the case of diquarks, taking them as building blocks for baryons. However, there has been little effort to derive these identities in other gauges. This is not  important in the Schwinger-Dyson (SD) and Bethe-Salpeter approach, where Fierz identities are not needed. (For a review of the SD approach  and its application to hadronic physics, see \cite{roberts94}.)  However,  the choice of the gauge is important in the GCM model, where one introduces the physical fields in terms of the corresponding quantum numbers as early as in the GCM action coming in the path integral formalism. In fact, this choice leads the appearance and the coupling between these physical d.o.f. For example, there is no tensor particle in a Feynman-like gauge. (We denote by Feynman-like gauge \cite{tandy97} a gauge whose Dirac structure is the same as the  Feynman gauge for the perturbative gluon propagator: $D_{\mu\nu}(x,y)=g_{\mu\nu}D(x,y)$. It is clear that, in non-perturbative studies, the Feynman-like gauge is not a conceptually  good choice, since it does not yield  the Slavnov-Taylor identity $k_{\mu}k_{\nu}D^{\mu\nu}=k_{\mu}k_{\nu}D^{\mu\nu}_0$ ($D^{\mu\nu}_0$ being the free gluon propagator) which is valid in arbitrary covariant gauges.) However this ``could be'' gauge
is often used in GCM-type studies because it simplifies considerably the use of Fierz identities. Although the NJL model is also based on the Feynman-like gauge, we  treat below the case of an arbitrary covariant gauge.

In the following, we shall first focus on the color, then the flavor Fierz transformations (both for mesons and diquarks) before turning to the spin structure. We shall then summarize our results.

\subsubsection{Fierz identities in color space}
\label{fierztransfocolor}
\subsubsection*{$\qbar q$ sector}

As long as one is not interested in the diquark sector, one can use the standard identity
\begin{equation}
\left(\frac{\lambda^c}{2}\right)_{\alpha\beta}
\left(\frac{\lambda^c}{2}\right)_{\delta\gamma}=
\frac{1}{2}(1-\frac{1}{N_c^2})
\delta_{\alpha\gamma}\delta_{\delta\beta}-\frac{1}{N_c}
\left(\frac{\lambda^c}{2}\right)_{\alpha\gamma}
\left(\frac{\lambda^c}{2}\right)_{\delta\beta},
\label{fierzcouleur}
\end{equation}
whose derivation starts from the obvious decomposition of a Hermitian matrix
$A=A_0\one+A_c\lambda^c$. Taking the traces $\tr A$ and $\tr \lu{c}A$ 
allows  to obtain $A_0$ and $A_c$:
\begin{equation}
A=\f{1}{N_c}\tr (A)\one+\f{1}{2}\tr (A\lambda^c)\lambda^c.
\end{equation}
We then have
\begin{equation}
A_{\alpha\beta}=\f{1}{N_c}A_{\delta\delta}
\delta_{\alpha\beta}+\f{1}{2}(A_{\gamma\delta}\lambda^c_{\delta\gamma})
\lambda^c_{\alpha\beta},
\label{eqfierz5}
\end{equation} 
or
\begin{equation}
A_{\alpha\beta}\equiv
A_{\gamma\delta}\delta_{\gamma\alpha}\delta_{\delta\beta}
=A_{\gamma\delta}(\f{1}{N_c}\delta_{\alpha\beta}\delta_{\gamma\delta}+
\f{1}{2}\lambda^c_{\delta\gamma}\lambda^c_{\alpha\beta}),
\end{equation} 
i.e.
\begin{equation}
\f{1}{2}\lambda^c_{\delta\gamma}\lambda^c_{\alpha\beta}=
\delta_{\gamma\alpha}\delta_{\delta\beta}
-\f{1}{N_c}\delta_{\alpha\beta}\delta_{\gamma\delta}.
\label{fierzusuel}
\end{equation} 
Permuting the $\beta$ and $\delta$ indices and injecting the factor $\delta_{\alpha\beta}\delta_{\gamma\delta}$ obtained in this way in (\ref{fierzusuel})  leads trivially to eq.~(\ref{fierzcouleur}).\par

We note that the sign of the color octet term in the r.h.s. of (\ref{fierzcouleur}) is the opposite of the sign of the color singlet. If channels are attractive for the singlet, they will be repulsive for the octet. This is what happens within the NJL model (for the GCM model, see below) where it is usual to neglect this octet: at the tree level it cannot contribute to give a colorless object while, anyway, the colored object would be unbound.

\subsubsection*{$qq$ and $\qbar\qbar$ sectors}

It has been shown by Cahill and collaborators \cite{cahill92,cahill89,cahill89b}
that an alternative bosonization to meson modes $\qbar q$ is possible. It amounts to rewrite the interaction to build $qq$ and $\qbar\qbar$ diquark modes\footnote{These Fierz transformations leading to diquarks have also been used in the NJL model. See for example works by  Reinhardt \cite{reinhardt90} and Weise \cite{weise92}.} (which, when combined with a third quark, can give rise to baryons and anti-baryons).
Although not conceptually necessary in the NJL model, the situation is different within the GCM model: since it is based on a model gluon propagator which makes the quark interaction non-local, bosonization  implies the introduction of bilocal fields. With the Fierz identities of the previous section, this would amount to the presence of  bilocal fields describing singlet and octet states. Physical mesonic states are obtained through the expansion of these bilocal fields into local ones. However, since the gluon exchange is repulsive between color octet states, the corresponding local expansion cannot be performed. We are then left to another, alternative, bosonization: now, the bilocal fields describe $\qbar q$ singlet $\one_c$ color states, as in the previous section,  and $qq$ antitriplets $\bar{3}_c$ color states, i.e. diquark states. This result is important because the gluon exchange between  quarks in the $\bar{3}_c$ states is attractive. An expansion of bilocal diquark fields into local ones is then possible. These $qq$ $\bar{3}_c$ states play an important role in the baryonic structure: a baryon is a color singlet formed of three quarks. Two of the quarks have then to be in a $\bar{3}_c$ state. This comes from the fact that, since the quarks are in the fundamental representation $3_c$ of the color group $SU(3)$, two quarks can only form antitriplet $\bar{3}_c$ or sextet $6_c$ states\footnote{For a recent review of group theory with applications to nuclear physics, see \cite{stancu96}.} ($3\otimes3=\bar{3}\oplus6$). The third quark can only be added to the antitriplet and not to the sextet ($3\otimes\bar{3}=1\oplus8$ as for the mesons, while\footnote{Moreover, Cahill et al. have shown that in  GCM type of models -- current-current interaction -- Fierz transformations do not yield color sextet diquark states.} $3\otimes6=10\oplus8$).\par

An easy way to make diquarks appearing in the formalism is to add and subtract
$\delta_{\alpha\gamma}\delta_{\delta\beta}$ in the r.h.s. of~(\ref{fierzusuel}), leading to
\be
\left(\frac{\lambda^c}{2}\right)_{\alpha\beta}
\left(\frac{\lambda^c}{2}\right)_{\delta\gamma}=
\frac{1}{2}(1-\frac{1}{N_c})
\delta_{\alpha\gamma}\delta_{\delta\beta}+\frac{1}{2N_c}
\epsilon_{\rho\alpha\delta}
\epsilon_{\rho\gamma\beta},
\label{fierzdiquark}
\ee 
where we used  $(\epsilon^{\rho})_{\alpha\delta}
(\epsilon^{\rho})_{\gamma\beta}$ =
$\delta_{\alpha\gamma}\delta_{\delta\beta}-
\delta_{\alpha\beta}\delta_{\gamma\delta}$. ($\epsilon_{\rho\alpha\beta}$ is a possible representation of $\bar{3}$.) 

The singlet color part $\delta_{\alpha\gamma}\delta_{\delta\beta}$ 
of~(\ref{fierzdiquark}) will be associated to flavor and Dirac Fierz identities connected to $\qbar q$ modes while the color octet part $\epsilon_{\rho\alpha\delta}\epsilon_{\rho\gamma\beta}$ will be associated to flavor and Dirac Fierz identities connected to $qq$ and $\qbar \qbar$ modes (see below).

\subsubsection{Fierz identities in flavor space}
\label{fierzflavor}
\subsubsection*{$\qbar q$ sector}

We have to go from singlet flavor d.o.f. to any flavor d.o.f. We just rewrite eq. (\ref{fierzusuel}) in flavor space, with  the last term on the r.h.s. written as $(1/2)\lambda^0_{\alpha\beta}\lambda^0_{\delta\gamma}$:
\be
\delta_{ij}\delta_{kl}=\sum_e(G^e)_{il}(G^e)_{kj},
\label{fierzsaveur}
\ee
where (with $F$ meaning flavor space)
\be
G^e=(\f{1}{\sqrt{3}}\one\equiv \f{(\lu{0})_F}{\sqrt{2}},\f{(\lu{a})_F}{\sqrt{2}}), a=1,...,8.
\label{fierzsaveur2}
\ee
We have ordered $G^e$ in the way singlet + octet.

\subsubsection*{$qq$ and $\qbar\qbar$ sectors}

We  reorder~(\ref{fierzsaveur},\ref{fierzsaveur2}) to correspond to diquarks in the $\bar{3}_F$ (antisymmetric) and $6_F$ (symmetric) representations:
\be
\delta_{ij}\delta_{kl}=\sum_e(G^e)_{ik}(G^e)_{lj},
\label{fierzsaveurdiq}
\ee
where
\be
G^e\equiv (G^e_A,G^e_S)=(\f{(\lu{2})_F}{\sqrt{2}},\f{(\lu{5})_F}{\sqrt{2}},
\f{(\lu{7})_F}{\sqrt{2}};\f{1}{\sqrt{3}}\one\equiv\f{(\lu{0})_F}{\sqrt{2}},\f{(\lu{1})_F}{\sqrt{2}},
\f{(\lu{3})_F}{\sqrt{2}},\f{(\lu{4})_F}{\sqrt{2}},
\f{(\lu{6})_F}{\sqrt{2}},\f{(\lu{8})_F}{\sqrt{2}}).
\ee

\subsubsection{Fierz identities in Dirac space}
\label{fierztransfodirac}
Since we do not yet specify the chosen gauge, we work explicitly with $D_{\mu\nu}(x,y)$ instead of the more familiar choice corresponding to the Feynman-like gauge $g_{\mu\nu}D(x,y)$. 

Any 4 x 4 matrix can be expanded (we recall that we use the notation of\cite{itzykson85}) on the basis $\Gamma^{\alpha}$ with

\vspace{0.3cm}
\begin{center}
\begin{tabular}{|ccccc|}
\hline
scalar & vector & tensor & axial & pseudoscalar\\
$\Gamma^S$ & $\Gamma^V$ & $\Gamma^T$ & $\Gamma^A$ & $\Gamma^P$\\
\hline
\hline
$\one$ & $\gamma^{\mu}$ & $\sigma^{\mu\nu}\equiv
i/2[\gamma^{\mu},\gamma^{\nu}]$ 
& $\gamma^5\gamma^{\mu}$ & $i\gamma^5$\\
\hline
\end{tabular}
\end{center}

\vspace{0.3cm}
With these matrices, the forms $\bar{q}\Gamma^{\alpha}q$ are hermitian.
If we define $\Gamma_{\alpha}\equiv(\Gamma^{\alpha})^{-1}$, we then have
\be
\Tr(\Gamma^{\alpha}\Gamma_{\beta})=4\delta^{\alpha}_{\beta},\ \ \ 
1\le\alpha,\beta\le16.
\ee
The matrices $\Gamma_{\alpha}$ are obtained through their definition:
\vspace{0.3cm}
\begin{center}
\begin{tabular}{|ccccc|}
\hline
scalar & vector & tensor & axial  & pseudoscalar\\
$\Gamma_S$ & $\Gamma_V$ & $\Gamma_T$ & $\Gamma_A$ & $\Gamma_P$\\
\hline
\hline
$\one$ & $\gamma_{\mu}$ & $\sigma_{\mu\nu}$ & $-\gamma^5\gamma_{\mu}$ &
 $-i\gamma^5$\\
\hline
\end{tabular}
\end{center}

\vspace{0.3cm}
Since any 4 x 4 matrix $X$ has the expansion
\be
X=x_{\alpha}\Gamma^{\alpha}=\frac{1}{4}\Gamma^{\alpha}\tr(X\Gamma_{\alpha})=
\frac{1}{4}\Gamma_{\alpha}\tr(X\Gamma^{\alpha}),
\label{xongamma}
\ee
one can then show that
\be
\delta_{ab}\delta_{cd}=\frac{1}{4}\Gu{\alpha}_{ac}\Gd{\alpha,db}.
\label{deltadelta}
\ee
\subsubsection*{Four-quark interaction}

From  $b(4,2;3,1)=\bar{q}(4)\Gu{\alpha}q(2)\bar{q}(3)\Gd{\alpha'}
q(1)$, we deduce (taking properly into account the Fermi field
 anticommutation)
\beqn
b(4,2;3,1)&=&-\bar{q}_{a_4}(4)\bar{q}_{a_3}(3)(\Gu{\alpha})_{a_4a_2}
(\Gd{\alpha'})_{a_3a_1}q_{a_2}(2)q_{a_1}(1)\\
&=&-\bar{q}_{a_4}(4)\bar{q}_{a_3}(\Gu{\alpha})_{a_4a_2}(\Gd{\alpha'})_{a_3a_1}
q_{a_2'}(2)q_{a_1'}(1)\delta_{a_2'a_2}\delta_{a_1'a_1}.
\eeqn  
With $\delta_{a_2'a_2}=\delta_{a_2a_2'}$ and
eq. (\ref{deltadelta}), we have
\beqn
b(4,2;3,1)&=&-\frac{1}{4}\bar{q}_{a_4}(4)\bar{q}_{a_3}(3)
(\Gu{\alpha})_{a_4a_2}(\Gu{\beta})_{a_2a_1'}
(\Gd{\alpha'})_{a_3a_1}(\Gd{\beta})_{a_1a_2'}q_{a_2'}(2)q_{a_1'}(1)\\
&=&-\frac{1}{4}\bar{q}(4)\Gu{\alpha}\Gu{\beta}q(1)
\bar{q}(3)\Gd{\alpha'}\Gd{\beta}q(2).
\eeqn 

This is the basic formula in the mesonic sector. 

\subsubsection*{Color quark current-current interaction}
The complete demonstration of Fierz identities in Dirac space is given in appendix \ref{fierzappendix} (where we explain the notation and physical contents related to $G^e_{A,S}$). Here, we give  only the main result, that we mix with sections \ref{fierztransfocolor} and \ref{fierzflavor} to obtain the complete Fierz-transformed interaction part of the Lagrangian:
\beqn
\lefteqn{
\hspace{-1cm}
\bar{q}(x){\f{\lambda_a}{2}}\gamma^{\mu}q(x)D_{\mu\nu}(x,y)
\bar{q}(y){\f{\lambda_a}{2}}q(y)=-\frac{1}{4}D_{\mu\nu}(x,y)\frac{1}{3}}
\nonumber\\
&&\times\Bigg\{
\bigg[
g^{\mu\nu}\sum_{a,e}\Big(\qbar(x)G^eK'^aq(y)\Big)
\Big(\qbar(y)G^eK'_aq(x)\Big)\nonumber\\
&&\mbox{}-2\Big(\qbar(x)G^ei\gu{\mu}q(y)\Big) \Big(\qbar(y)G^ei\gu{\nu}q(x)\Big)
-2\Big(\qbar(x)G^ei\gu{5}\gu{\mu}q(y)\Big)
\Big(\qbar(y)G^ei\gu{5}\gu{\nu}q(x)\Big)\nonumber\\
&&\mbox{}-4\Big(\qbar(x)G^e\frac{{\sigma^{\nu}}_{\mu'}}{\sqrt{2}}q(y)\Big)
\Big(\qbar(y)G^e\frac{\sigma^{\mu\mu'}}{\sqrt{2}}q(x)\Big)
\bigg]
\nonumber\\
&&\mbox{}+\bigg[
g^{\mu\nu}\sum_{a,e}\Big(\qbar(x)G^e_{A,S}\frac{i\epsilon^{\rho}}{\sqrt{2}}K'^aC\qbar^T(y)\Big)
\Big(q^T(y)CG^e_{A,S}\frac{i\epsilon^{\rho}}{\sqrt{2}}K'_aq(x)\Big)\nonumber\\
&&\mbox{}-2\Big(\qbar(x)G^e_Si\gu{\mu}\frac{i\epsilon^{\rho}}{\sqrt{2}}C\qbar^T(y)\Big) \Big(q^T(y)CG^e_S\frac{i\epsilon^{\rho}}{\sqrt{2}}i\gu{\nu}q(x)\Big)\nonumber\\
&&-2\Big(\qbar(x)G^e_A\frac{i\epsilon^{\rho}}{\sqrt{2}}i\gu{5}\gu{\mu}C\qbar^T(y)\Big)
\Big(q^T(y)CG^e_A\frac{i\epsilon^{\rho}}{\sqrt{2}}i\gu{5}\gu{\nu}q(x)\Big)\nonumber\\
&&\mbox{}-4\Big(\qbar(x)G^e_S\frac{i\epsilon^{\rho}}{\sqrt{2}}\frac{{\sigma^{\nu}}_{\mu'}}{\sqrt{2}}C\qbar^T(y)\Big)
\Big(q^T(y)CG^e_S\frac{i\epsilon^{\rho}}{\sqrt{2}}\frac{\sigma^{\mu\mu'}}{\sqrt{2}}q(x)\Big)
\bigg]
\Bigg\}\\
\nonumber
\eeqn
with $K'^{a}=(1,i\gu{5},i\gu{\mu'},i\gu{5}\gu{\mu'},\sigma^{\mu'\nu'}/\sqrt{2})$.

We can introduce meson (diquark) fields in the usual way. Because of the part of the propagator proportional to $k_{\mu}k_{\nu}$ (see below), we have tensor mesons as well as non-diagonal couplings between vector mesons (diquarks), between axial mesons (diquarks), and between tensor mesons (diquarks).  For example, we get, for the meson part,
\barray
\hspace{-1cm}
-\frac{1}{4}D_{\mu\nu}(x,y)\frac{1}{3}
\Bigg\{
g^{\mu\nu}\phi^{a}(x,y)\phi_a(y,x)-2
\phi^{\mu}(x,y)\phi^{\nu}(y,x)
-2
\phi_5^{\mu}(x,y)\phi_5^{\nu}(y,x)\\
-4
\phi_{\sigma}^{\mu\mu'}(x,y)\phi_{\sigma,\mu'}^{\nu}(y,x)
\Bigg\}
\label{bosonizationmeson}
\earray
or, separating each contribution,
\barray
\hspace{-1cm}
-\frac{1}{4}D_{\mu\nu}(x,y)\frac{1}{3}
\Bigg\{
g^{\mu\nu}\phi(x,y)\phi(y,x)+g^{\mu\nu}\phi_5(x,y)\phi_5(y,x)
+\left(g^{\mu\nu}\phi^{\alpha}(x,y)\phi_{\alpha}(y,x)-2\phi^{\mu}(x,y)\phi^{\nu}(y,x)\right)\\
+\left(g^{\mu\nu}\phi^{\alpha}_5(x,y)\phi_{5,\alpha}(y,x)-2\phi_5^{\mu}(x,y)\phi_5^{\nu}(y,x)\right)\\
+\left(g^{\mu\nu}\phi^{\alpha\beta}_{\sigma}(x,y)\phi_{\sigma,\alpha\beta}(y,x)
-4
\phi_{\sigma}^{\mu\mu'}(x,y)\phi_{\sigma,\mu'}^{\nu}(y,x)
\right)
\Bigg\},
\earray
where we have defined $\phi^a(x,y)\equiv (\phi(x,y),\phi_5(x,y),\phi^{\mu}(x,y),\phi_5^{\mu}(x,y),\phi_{\sigma}^{\mu\nu}(x,y))$ with
\beqn
\phi(x,y)&=&\qbar(x)G^eq(y),\\
\phi_5(x,y)&=&\qbar(x)G^ei\gu{5}q(y),\\
\phi^{\mu}(x,y)&=&\qbar(x)G^ei\gu{\mu}q(y),\\
\phi_5^{\mu}(x,y)&=&\qbar(x)G^ei\gu{5}\gu{\mu}q(y),\\
\phi_{\sigma}^{\mu\nu}(x,y)&=&\qbar(x)G^e\frac{\sigma^{\mu\nu}}{\sqrt{2}}q(y).
\eeqn

\subsubsection{Conclusions}
\label{fierzconclusion}
\begin{itemize}
\item From the previous section, it is clear that Fierz identities are a very useful tool to rewrite the interaction in terms of d.o.f. having the good low energy mesonic quantum numbers. It is also clear that, as soon as the most general form of the gluon effective propagator is retained, all quantum numbers are allowed (scalar, pseudoscalar, vector, axial-vector, tensor\footnote{This has also been seen in the works by Shrauner \cite{shrauner77} and Roberts and Cahill \cite{roberts86,roberts87b}.}). The above derivation has also shown an extra interaction, compared to the Feynman-like gauge, among each family (vector-vector, ...). Indeed, in the Feynman-like gauge, the gluon propagator takes the simple form $D_{\mu\nu}(x,y)=g_{\mu\nu}D(x,y)$ and, leaving aside diquark contributions,  (\ref{bosonizationmeson}) is  replaced by
\be
-\frac{1}{4}D(x,y)\frac{1}{3}
\left(
4(\phi^2+\phi_5^2)+2(\phi_{\mu}\phi^{\mu}+\phi_5^{\mu}\phi_{5,\mu})
\right).
\label{feynmanlike}
\ee

One  effectively sees that there is no longer tensor term and that the interaction inside each species is simpler. In a covariant gauge where the gluon propagator writes (because of the translational invariance, the propagator is only a function of $(x-y)$, i.e. of one variable $k$ in Fourier space)
\be
D_{\mu\nu}(k)=\left(
g_{\mu\nu}-\f{k_{\mu}k_{\nu}}{k^2}
\right)
\f{d(k^2)}{k^2}
+\xi\f{k_{\mu}k_{\nu}}{k^2},
\ee 
it is clear that, together with the $g_{\mu\nu}$ term $g_{\mu\nu}{d(k^2)}/{k^2}$
which gives rise to the usual Feynman-like gauge result, there is an additional term depending both on the gauge parameter $\xi$ and the particular shape of $d(k^2)$,
leading to interaction terms of the form $k^2\phi^a\phi_a$, $(k.\phi)^2$, ...

Note also from eq.~(\ref{feynmanlike}) that we recover the usual factor $1/2$ when comparing the strength vector-pseudovector versus scalar-pseudoscalar.
 
\item The NJL model can be seen as a limiting case of the GCM class where the interaction is {\em i}) written in the Feynman-like gauge and {\em ii}) is a contact interaction: $D_{\mu\nu}(x,y)\propto g_{\mu\nu}\delta^{(4)}(x-y)$. The interacting part of the Lagrangian is then
\be
{\cal
L}_{int}\propto \f{1}{3}[(\bar{q}\Lambda_{\alpha}q)(\bar{q}\Lambda^{\alpha}q)
+(\bar{q}\Gamma_{\alpha}C\bar{q}^T)(q^TC\Gamma^{\alpha}q)]
={\cal L}_{int}^{\bar{q}q}+{\cal L}_{int}^{qq},
\label{interqq}
\ee
where we have defined (with $K^a=(1,i\gd{5},i\gu{\mu}/\sqrt{2},i\gu{\mu}\gd{5}\sqrt{2}$))
\beqn
\Lambda^{\alpha}&\equiv&\one_c\otimes G^e\otimes K^a,
\nonumber\\
\Gamma^{\alpha}&\equiv&\f{1}{\sqrt{2}}
\epsilon^{\rho}\otimes G^e_{A,S}\otimes K^a.
\eeqn
It is  useful to note that, in order to reproduce the pseudoscalar and vector spectra, different coupling constants  have to be chosen as free parameters (i.e. the factor $1/2$, found\footnote{This factor is not constrained by the symmetries of the model, so that choosing different coupling constants still respect them.} when comparing pseudoscalar and vector strength, is not suitable). 

\item When $N_c \ne 3$, the factor 1/3 in the interacting Lagrangian~(\ref{interqq}) is replaced by $(N_c-1)/(2N_c)$ for ${\cal L}_{int}^{\bar{q}q}$ and $1/N_c$ for ${\cal L}_{int}^{qq}$. When $N_c\rightarrow\infty$, ${\cal
L}_{int}^{qq}$ is reduced by a factor $1/N_c$ as compared to the mesonic part ${\cal L}_{int}^{\bar{q}q}$. We are then left with a pure mesonic interaction, in agreement with model independent large $N_c$ analyses, so that the baryons have to be exclusively described as solitonic solutions of the  Lagrangian (see for example the works by Reinhardt and collaborators
 \cite{reinhardt95,reinhardt95b}, and the works by the Bochum group 
\cite{goeke93,goeke94,goeke93c}). In the last reference, the interested reader will find results concerning mesons, baryons, and links with  chiral perturbation theory and the Skyrme model.

\item Apart from solitonic solutions, baryons can also be obtained from (\ref{interqq}), treating them as a 
quark-diquark system.   When a third quark is added to the diquark,
one has to consider states formed by the product
$(3\otimes\bar{3})_c$
$\otimes$ $(3\otimes\bar{3})_F$, which leads to
($\one_c\otimes\one_F$)
$\oplus$
($\one_c\otimes8_F$) $\oplus$ ($8_c\otimes\one_F$) $\oplus$
($8_c\otimes8_F$). Calculations performed by Cahill \cite{cahill92}
have shown that for the ($\one_c\otimes\one_F$) and ($8_c\otimes
8_F$) states, quark rearrangement is repulsive, while it is attractive for the remaining terms. The state
($8_c\otimes\one_F$) has then to be suppressed by hand. This defect, present both in NJL and GCM studies, can be traced back to the lack of local color gauge invariance.
\end{itemize}

\subsection{The 3-flavor scaled NJL models and anomalies}
\label{3flavors}

We showed in the previous section how to rewrite the color current-current interaction as a function of bilinear forms having the physical quantum numbers seen in the spectrum. This procedure amounted to use  Fierz identities in color, flavor and Dirac (spin) spaces. We also mentioned that the NJL model can be viewed as a limiting case of the GCM model, where the interaction is local (4-quark point interaction) and a Feynman-like gauge is used: eq.~(\ref{LGCM}) together with Fierz identities limited to mesonic scalar and pseudoscalar modes leads to 
\be
{\cal
L}_{NJL}=\bar{q}(i\dslash-m)q+G_S\sum_{i=0}^8[(\bar{q}
\f{(\lu{i})_F}{2}q)^2+(\bar{q}i\gd{5}\f{(\lu{i})_F}{2}q)^2].
\label{LNJL}
\ee

The NJL model has already been extensively studied by several groups of which we can only quote a few. Our purpose being here to describe its scaled version, the reader is referred to the following literature\footnote{The number of published papers in this field of research is enormous and still growing. We apologize  if important contributions are omitted here. Since our main purpose is to make an introduction to the scaled NJL model, we mainly mention review papers and school proceedings.} (and references therein) to more general studies about the model. The most recent work has been done by Ripka \cite{ripka97} in a book which contains an in-depth analysis of regularization procedures and symmetry conserving approximations;
Klevansky \cite{klevansky92} and Hatsuda and Kunihiro \cite{hatsuda94} give a general introduction to NJL, both in the vacuum and at finite temperature and density; Alkofer, Reinhardt and Weigel \cite{reinhardt95b}  mainly applied the model to discuss baryons as chiral solitons, as  also done by Goeke and collaborators \cite{goeke94} who  discuss, using heat kernel and gradient expansion techniques, the link between NJL and fully bosonized approaches of the Skyrme type on the one hand and the chiral sigma model on the other hand; 
Bijnens \cite{bijnens96} discusses chiral perturbation theory within NJL; Alkofer, Ebert, Reinhardt and Volkov \cite{reinhardt95,reinhardt94} hadronize the NJL model (both mesons and baryons\footnote{In this way, baryons are no more pure solitonic solutions but are also described by a three valence quark Faddeev equation. See also \cite{reinhardt90,ball90b,huang94} for this approach and \cite{buck92,ishii93,ishii93b,buck95,krewald95} for numerical results.}) through the use of the alternative Fierz identities leading to diquarks; Vogl and Weise \cite{weise91b} and Weise \cite{weise92} review several of the above mentioned topics: bosonization and hadronization, finite density and temperature effects.

\par

It is clear that the Lagrangian (\ref{LNJL}) describes the interaction between flavor carrying colorless objects. In the chiral limit where the current quark mass matrix $m$ is identically zero, it is invariant under  $SU(3)_V\otimes SU(3)_A\otimes U(1)_V\otimes U(1)_A$ (see notation in section~\ref{symmetries}). However, the $U(1)_A$ invariance is broken by quantum effects (strong axial anomaly) which cannot be taken into account since the true QCD Lagrangian has been replaced by an effective form. The Lagrangian~(\ref{LNJL}) is used at the tree level. If we want this order to implement the full quantum aspects of QCD, it is necessary to supplement it with a term which, while still invariant under the true symmetries of QCD, has to break the axial $U(1)_A$ invariance.
This anomaly can be related to the formation of instantons \cite{veneziano79,witten79b,thooft86} and yields anomalous contributions to the $\eta$ and $\eta'$ masses. The $U(1)_A$ symmetry is explicitly broken by instanton induced interactions which have the form of a 't Hooft determinant. The phenomenological term which has to be added to~(\ref{LNJL}) can thus take the form of a 't Hooft determinant. Other forms can be chosen (see \cite{veneziano79,witten79b,witten80,veneziano80,schechter80,schechter80b}). For example, working on the bosonized (see section~\ref{AandBmodels}) version of~(\ref{LNJL}), one can show that the Schechter~\cite{schechter80b} proposition\footnote{$\pi_a$ and $\sigma_a$ will be introduced in section~\ref{AandBmodels}. They are related to pseudoscalar and scalar d.o.f., respectively.}
\be
{\cal L}_{U(1)_A}\propto\left(
\tr_F\ln(\sigma_a\lu{a}+i\pi_a\lu{a})-\tr_F\ln(\sigma_a\lu{a}-i\pi_a\lu{a})
\right)^2, a=0,...,8,
\label{Lschechter}
\ee
is also compatible with the symmetries of QCD, while breaking the $U(1)_A$ one. This Lagrangian is interesting, being simpler\footnote{This form does not produce any flavor mixing, while the 't Hooft determinant implies maximum flavor mixing.} than the 't Hooft determinant, while containing it as a limiting case. In particular, (\ref{Lschechter}) does not modify the gap equations. Up to  second order expansion into the mesonic fields, it yields extra mass terms for the  $\eta$ and $\eta'$ particles. We shall use here an even simpler approach which consists in just adding a mass term for the $\eta_0$ particle (i.e. the $\pi_0$ in the above notation). In the chiral limit, all the pseudoscalar particles would then be massless (in agreement with the Goldstone theorem applied to the dynamical chiral symmetry breaking mechanism) except for the $\eta'$ ($\eta_0$ in this limit), see section~\ref{bs}. This procedure is related (although not equivalent) to the approach of Frère and collaborators \cite{frere89,frere96}: in these works, there is a coupling between the $\eta_0$ and $\eta_8$ through the quark mass difference {\em and} the anomaly whose net effect is to modify the mixing angle from its ideal value. It can be shown that, in the mean field approximation, our approach only explicitly breaks (apart for  quark masses)  the $U(1)_A$ symmetry, then being a possible parameterization of the axial anomaly. The expression for the divergence of the axial current  is given in (\ref{anomalieaxialeforte}) where the last term is the anomaly. It has the quantum number $J^{PC}=0^{-+}$ of  pseudoscalar isoscalars. It is then tempting to identify it as an interpolating field for the pseudoscalar glueball and to incorporate it into the formalism so that the $\eta$ and $\eta'$ have to be searched from a three state ($\eta$, $\eta'$, glueball) diagonalization procedure\footnote{This is similar to the approach of Shore and  Veneziano \cite{veneziano92}. See also the works of Nekrasov \cite{nekrasov92,nekrasov94}}. We shall however restrict ourselves to the $\eta_0$-mass procedure.

So far we  only discussed  the axial anomaly. As mentioned in section \ref{symmetries}, another symmetry at the classical level of the QCD action is broken by quantum effects:  scale symmetry. In fact, this symmetry is broken in three ways: explicitly, through the quark masses which introduce a dimensional parameter in the theory (see the first term on the r.h.s. of eq. (\ref{anomalietrace})); dynamically as for  chiral symmetry down to vector symmetry; because of the anomaly, see the last term  on the r.h.s. of eq. (\ref{anomalietrace}). As for the axial anomaly, the effect of the trace anomaly has to be added by hand. In that way, QCD quantum effects are present to the lowest order of the effective theory we are interested in.

Several steps are necessary in order to construct the modified NJL model in the perspective of the symmetries and anomalies as described above. In view of the way of treating the axial anomaly (mass term for the $\eta_0$), it is better to work on the bosonized version of the model. The main points to bosonize the theory are sketched below.

\subsubsection{Bosonization}

For a review of hadronization techniques, the reader is referred to \cite{kleinert76,kleinert78}.

We start from 
\be
\label{eq3.1}
{\cal
Z}_{QCD}\approx
\int{\cal D}q{\cal
D}\bar{q}\exp
\left(i\int d^4x{\cal L}_{NJL}\right)\equiv{\cal Z}_{NJL},
\ee
with ${\cal L}_{NJL}$ given in (\ref{LNJL}).
Unity is introduced in~(\ref{eq3.1}), through
auxiliary fields $\phi^a$ and $\varphi^a$
\be
1=\int{\cal
D}\phi^a\delta(\phi^a-\bar{q}\Gamma^aq)=\int{\cal
D}\phi^a{\cal D}\varphi^a\exp\left(i\int
d^4x\varphi^a(\phi^a-\bar{q}\Gamma^aq)\right),
\label{auxiliaire}
\ee
where $\phi^a=(\sigma^a,\pi^a)$ and
$\Gamma^a=(\lu{a}/2,i\gd{5}\lu{a}/2)$, $a=0,...,8$. (The first component is relative to scalars, the second component to pseudoscalars; to simplify the notation, we have removed the flavor index $F$.)  Eq. (\ref{auxiliaire}) is -- apart form unimportant constant factors -- the functional equivalent of a possible definition of the Dirac distribution. From
$f(x)\delta(x-x_0)=f(x_0)\delta(x-x_0)$, we get
\be
{\cal Z}_{NJL}=\int{\cal D}q{\cal D}
\bar{q}{\cal D}\phi^a{\cal
D}\varphi^a\exp\left(i\int
d^4x[\bar{q}(i\dslash-m^0)q+G_S(\phi^a)^2
+\phi^a\varphi^a-\varphi^a\bar{q}\Gamma^aq]\right).
\ee
With the identity
\be
G_S(\phi^a)^2+\phi^a\varphi^a=G_S(\phi^a+\f{1}{2G_S}\varphi^a)^2
-\f{1}{4G_S}(\varphi^a)^2=G_S(\phi'^a)^2-\f{1}{4G_S}(\varphi^a)^2,
\ee
the $\phi'^a$ field integration is Gaussian and can be performed exactly. (The Jacobian of the 
transformation $\phi^a\rightarrow\phi'^a$ is unity.) Since its value
contributes only  a  constant term it can be omitted. The effective functional takes the form
\be
{\cal Z}_{NJL}=\int{\cal D}q{\cal D}
\bar{q}{\cal
D}\varphi^a\exp\left(i\int
d^4x[\bar{q}(i\dslash-m^0-\varphi^a\Gamma^a)q-
\f{1}{4G_S}(\varphi^a)^2]\right).
\label{Znjl}
\ee
Then, the quark Gaussian integration has to be performed (for Grassman variables $q$ and $\bar{q}$), so that the model  describes only mesonic d.o.f.\par

We already mentioned the fact that the GCM model (and studies performed with SD equations) is best expressed in the Euclidean space. This is also the case here, although not for the same reason: since we are interested in finite temperature effects, we found that the Matsubara (or imaginary time) formalism is well suited for our purpose so that the Euclidean space will be used from now on.
The
functionals (\ref{Zqcd}) and (\ref{Znjl}) become
partition functions describing  the thermodynamics of the systems.
Formally, we go first to Euclidean metric, then making the substitution\footnote{In the NJL model, this substitution does not cause any of the problems mentioned at the end of section~\ref{qcdtogcm}.}
$t=-i\tau$ with real $\tau$.  We have
\beqn
x^{\mu}=x_{\mu}=(\tau,\vec{x}),\nonumber\\
x^2=\tau^2+\vec{x}^2,\nonumber\\
\gu{\mu}=\gd{\mu}=(\gd{0}=i\beta,\vec{\gamma}),
\eeqn
where $\beta$ is
the Dirac matrix\footnote{There should be no confusion with the inverse temperature.}
${\rm diag}(1,1,-1,-1)$ -- see appendix~\ref{conventions} for our conventions --.\par

We define
\cite{jamvdb94,jamvdb95} 
\be
\Tr O=\tr\int d^4x \left<x|O|x\right>\ ,\ \ \ \ \int d^4x=
\int_0^{\beta}d\tau\int_{\Omega}d^3x,
\ee 
where $\tr$ is the trace w.r.t. internal d.o.f.
(Dirac, color, flavor), $\beta=1/T$ is the inverse 
temperature and
$\Omega$ is the volume of the system. We get
\be
{\cal Z}_{NJL}=\int{\cal
D}\varphi^a\exp\left(-[-\Tr\ln(-i\dslash+m^0+\varphi^a\Gamma^a)+
\int d^4x\f{1}{4G_S}(\varphi^a)^2]\right), 
\label{detfermionique}
\ee
where we have defined the NJL effective action in the Euclidean space
\be
I_{eff}=-\Tr\ln(-i\dslash+m^0+\varphi^a\Gamma^a)+
\int d^4x\f{1}{4G_S}(\varphi^a)^2.
\label{Ieffnjl}
\ee
\par

Following the standard procedure
 \cite{jamvdb94,jamvdb95,fetwal71}, density effects are considered through the introduction of a chemical potential for each flavor (Lagrange multiplier for the quark number conservation). In our formalism, the equivalent relation to
$H-\mu N$, where $H$ is the Hamiltonian of the system and $\mu$ the Lagrange multiplier associated to a conserved  quantity $N$,
consists in subtracting  the
quantity\footnote{$\beta$ is the Dirac matrix and $\mu$ a diagonal
matrix in flavor space.}
$\bar{q}\mu\beta q$ from the Lagrangian.   In Euclidean metric, this is equivalent to the introduction of a vector
 $W_{\nu}=(-i\mu,0,0,0)$, 
where
$\mu$ is a notation for   $\diag(\mu_u,\mu_d,\mu_s)$. Each $\mu_i$ ($i=u,d,s$)  is a chemical potential associated to the corresponding flavor. When all chemical potentials are equal\footnote{Note that different chemical potentials break the invariance under axial transformations.},  or in the case of a single flavor,
the usual result
$\mu\bar{q}\beta q$ is recovered.\par

With these prescriptions, the effective action describing a system of quarks at finite temperature and density is 
\be
I_{eff}=-\Tr\ln(-i\dslash+m^0+\varphi^a\Gamma^a-\sslash{W})+
\int d^4x\f{1}{4G_S}(\varphi^a)^2,
\ee
\par
\espace
which still has to be modified in order to implement the QCD axial and trace anomalies.

We already mentioned in the beginning of this section that a mass term for the $\eta_0$ (or $\pi_0$ in our notation) is well suited and particularly easy to implement. We have thus
\be
I_{eff}=-\Tr\ln(-i\dslash+m^0+\varphi_a\Gamma_a-\sslash{W})+
\int d^4x\f{a^2}{2}\varphi_a^2+\int d^4x\f{a^2}{2}\xi\pi_0^2
\label{chap314}
\ee
where, to make the notation similar to the original papers
\cite{jamvdb94,jamvdb95,jamvdb94b}, we have introduced
the coupling constant $a^2=2/(G_S)$ and have redefined
$\Gamma^a=(\lu{a},i\gd{5}\lu{a})$, $a=0,...,8$.\par

The scale anomaly requires several steps to be included in our formalism. First of all, we have to make the NJL action scale invariant. This comes from the fact that without anomaly, the QCD action is scale invariant except for the quark  mass term. The mass term in NJL has the same scale dimension as in QCD so that this term has not to be modified. However, the coupling constant $a^2$ in (\ref{chap314}) has a dimension and has then to be modified.
The second step is connected to the anomaly itself. It is generally modelized by a scalar dilaton field. This has been used to modify different models of QCD at low energy. For example the Syracuse group \cite{schechter86,schechter88} studies in this way scalar mesons and chiral solitons.
More recently, this dilaton field has been used 
to render non-linear
$\sigma$ models scale invariant  for examining the effect of density on meson masses
\cite{brown91} and the deconfining phase transition\cite{ellis90,ellis90b}. The authors of \cite{ellispj94,ellispj96,ellispj97,ellispj98}  used a modified linear sigma model to look at the  mesons and nucleons at finite temperature and density. Cohen et al. \cite{cohen92}
studied the behavior of quark and gluon condensates 
w.r.t. density in a model independent way (at low density) and within NJL. Kusaka and Weise \cite{weise94} study some hadronic properties, mainly as a function of density, and indicate conditions which have to be satisfied in order to get a scaling of these properties. They show in that work that the universal scaling as proposed in \cite{brown91} can only occur for small glueball mass (compared to the mass of the chiral scalar mesons). They show, and it has been confirmed in \cite{cohen92}, that the low-density behavior of hadronic properties is insensitive to the specific form of the interaction. We shall comment further on this point later.
The condition that the glueball potential must satisfy to reproduce the anomaly is described in
\cite{reinhardt94,schechter86}. In the works
\cite{ellis90,ellis90b}, the dilaton field is an order parameter associated
 with gluon confinement. We shall adopt this point of view, mainly to facilitate the language, while keeping in mind that it is questionable \cite{roberts92b}.
\par

At the  NJL model level, the dilaton field is put by hand in the interaction term in order to restore the scale invariance.
Let  $\chi$ be the dilaton field, of scale dimension 1. The 
interaction term
\be
\int d^4x\f{a^2}{2}\chi^2\varphi_a^2+\int
d^4x\f{a^2}{2}\xi\chi^2\pi_0^2,
\label{intinvechelle}
\ee
where the coupling constant $a^2$ has been replaced by 
$a^2\chi^2$,  of scale
dimension~2, is then scale invariant\footnote{The new coupling constant $a^2$ 
 has now no dimension.}. With this procedure,
the NJL model is scale invariant\footnote{In the field strength method described in \cite{reinhardt95,reinhardt94}, the first term of eq. (\ref{intinvechelle}) can be derived in the NJL limit, i.e. the scale dimension is correct and in relation to the gluon condensate. It is then legitimate to argue that it is the elimination of the gluons from the theory which gives rise to the effective interaction (\ref{intinvechelle}).} except for a mass term of scale dimension 3, as in QCD.\par

One drawback of the NJL model is its lack of renormalizability, which leads to a subtlety in the scaled model: the regularization parameter
$\Lambda$, chosen is this work as a 4-dimensional cut-off, cannot be eliminated from the theory. In this way, it is a fundamental parameter of the model.
This cut-off $\Lambda$ enters  the game when one wishes to evaluate the fermionic determinant~(\ref{chap314}) and breaks its scale invariance.
The work of Ripka and Jaminon
\cite{jamrip92} is of prime importance in the sense that  the scale invariance of the fermionic determinant is kept
by 
introducing the dilaton field in the regularization factor: $\Lambda\rightarrow\Lambda\chi$. The effective action is then
\be
I_{eff}=-\Tr_{\Lambda\chi}\ln(-i\dslash+m^0+\varphi_a\Gamma_a-\sslash{W})+
\int d^4x\f{a^2}{2}\chi^2\varphi_a^2+\int
d^4x\f{a^2}{2}\xi\chi^2\pi_0^2,
\ee
where we have explicitly indicated by $\Tr_{\Lambda\chi}$ the fact that
the regularization parameter depends on the dilaton.\par

Up to this point, all we have done is to make NJL scale invariant. We still have to consider the anomaly. It can be introduced by adding  the Lagrangian
\cite{schechter86,brown91,ellis90}
\be
{\cal L}_{\chi}
={1 \over 2}{\left({{\partial
}_{\mu }\chi }\right)}^{2}+V_{\chi},
\label{Lagchi}
\ee
with
\be
V_{\chi}={{b}^{2} \over 16}\left({{\chi }^{4}
\ln({{\chi }^{4} \over {\chi }_{G}^{4}})-({\chi }^{4}-{\chi
}_{G}^{4})}\right).
\label{eq3.21}
\ee
The trace of the energy-impulsion tensor associated
to the Lagrangian~(\ref{Lagchi}) is
\be
\theta^{\mu}_{\mu}\equiv\f{\beta_{\mbox{\scriptsize
QCD}}}{2g}F_{\mu\nu}^aF^{\mu\nu}_a=\chi\f{\p V_{\chi}}{\p\chi}-4V_{\chi}=   
 \f{b^2}{4}\chi^4,
\label{eq3.22}
\ee
where the first equality comes from~(\ref{anomalietrace}). (In the last equality, we omitted the subtraction of
$(b^2/4)\chi_G^4$, coming from the last term of~(\ref{eq3.21}), included there  only to make the potential equaling zero at its minimum $\chi=\chi_G$.)

\subsubsection{A and B scaled models}
\label{AandBmodels}

Finally,  the bosonized,  scale invariant (except for the quark mass term) NJL action is 
\beqn
\lefteqn{I_{eff}=-\Tr_{\Lambda\chi}\ln(-i\dslash+m^0+\varphi_a\Gamma_a
-\sslash{W})+\int d^4x\f{a^2}{2}\chi^2\varphi_a^2}\nonumber\\
&&\mbox{}+\int
d^4x\bigg\{
\f{a^2}{2}\xi\chi^2\pi_0^2+{1 \over 2}{\left({{\partial
}_{\mu }\chi }\right)}^{2}+{{b}^{2} \over 16}\left({{\chi }^{4}
\ln({{\chi }^{4} \over {\chi }_{G}^{4}})-({\chi }^{4}-{\chi
}_{G}^{4})}\right)
\bigg\},
\label{modeleA}
\eeqn
\par
taking into account axial and scale anomalies.
This action\footnote{Without the $\xi$-term, the model described by this action has also been studied by Kusaka, Volkov and Weise \cite{volkov93}.}, called A-scaling model, is the simplest one verifying the above mentioned conditions of symmetries and anomalies. In the works
\cite{jamvdb94,jamvdb95,jamvdb94b,jamrip92,jamrip93} several models have been studied, differing each other by the way the quadratic part $\varphi_a^2$ is made scale invariant.
For example, the B-scaled model
respects all the symmetries of QCD, and in this sense is as valid as the A-scaling model (\ref{modeleA}).
The 3-flavor extension \cite{jamvdb95} of the B-scaling model 
\cite{jamrip92} writes
\beqn
\lefteqn{I_{eff}=-\Tr_{\Lambda\chi}\ln(-i\dslash+m^0+\varphi_a\Gamma_a
-\sslash{W})+\int
d^4x\bigg\{
\f{a^2}{2}(\varphi_a^2+\xi\pi_0^2-\eta\chi^2)^2}\hspace{2cm}
\nonumber\\  
&&\mbox{}\,+{1\over 2}{\left({{\partial }_{\mu }\chi
}\right)}^{2}+{{b}^{2}
\over 16}\left({{\chi }^{4}
\ln({{\chi }^{4} \over {\chi }_{G}^{4}})-({\chi }^{4}-{\chi
}_{G}^{4})}\right)
\bigg\}.
\label{modeleB}
\eeqn
\par
The justification of this model was to try to reproduce the scaling behavior of the hadronic properties as proposed in \cite{brown91}. However as was noted in \cite{weise94} (in the NJL model) and in \cite{cohen92} (model independent study),  universal scaling is not expected to occur. This can be traced back to the high value ($M_{GL}\sim1500$ MeV) of the glueball mass. In the following we shall restrict ourselves to the A-scaling models. For results and comments upon the B-scaling version, the reader is referred to \cite{jamvdb95}.

\subsubsection{Gap equations (or Schwinger-Dyson equations)}
\label{dsgap}
From now on we shall use  the model (\ref{modeleA}). We work in the spirit of the effective model: the tree or classical approximation is used throughout the study, which corresponds to the lowest order in a $1/N_c$ expansion. (For a study at next order in $N_c$, see \cite{klevansky94b,klevansky94c,lemmer94,blaschke95c,schmidt95,zhuang95,ripka96}.) We work in the saddle-point approximation which consists in evaluating the effective action at the point in $\sigma_a,\pi_a$ space where its first derivative vanishes identically. More generally, the saddle-point approximation consists in Taylor expanding the effective action, putting the first derivative equal to zero and evaluating the higher order terms in the standard way. For example, the second order is Gaussian and can be  evaluated exactly. Note however that, to get the effective potential, one should reactualize the expanding point at each order \cite{itzykson75}.  We shall work with the lowest order. This means that we expand (\ref{modeleA}) around its stationary points
\beqn
\left.
\f{\delta I}{\delta\sigma_a}
\right|_{\sigma_a^s,\pi_a^s,\chi_s}=0,\label{gapsigma}\\
\left.
\f{\delta I}{\delta\pi_a}
\right|_{\sigma_a^s,\pi_a^s,\chi_s}=0,\label{gappi}\\
\left.
\f{\delta I}{\delta\chi}
\right|_{\sigma_a^s,\pi_a^s,\chi_s}=0.\label{gapchi}
\eeqn
These equations lead to a parity invariant vacuum $\pi_a^s=0$. The first one is called the gap equation (by analogy with superconductivity) or the Schwinger-Dyson equation for the quark self-energy. The last equation, which is not present in usual NJL models, is a consequence of the introduction of the dilaton to modelize  the trace anomaly. It can also be shown that, out of the nine equations summarized by (\ref{gapsigma}), only $a=0,3,8$ leads $\sigma_a^s\ne0$. Moreover, if the isospin limit $m_u=m_d$ is considered, then we also have $\sigma_3^s=0$. In the following, we shall work in this isospin limit so that we have to manage three coupled equations for $\sigma_0^s,\sigma_8^s,\chi_s$.

Working at tree level means that the field stationary values  will be used in order to
\begin{itemize}
\item determine the thermodynamical properties of the system (see section \ref{sectionthermodynamics});
\item determine some static meson properties (see section \ref{bs});
\item study some mesonic processes such as the neutral pion decay into  two photons (see section \ref{sectionanomalouspiondecay}) or scalar isoscalar meson decay into two photons, including the scalar glueball (see section \ref{sectionscalardecay}).
\end{itemize}
The last two  mentioned points require  introducing  the electromagnetic field in the formalism and expanding the action up to third order 
or, equivalently,  working with triangle Feynman diagrams (see sections \ref{sectionanomalouspiondecay} and \ref{sectionscalardecay}).

The two ($\times 9$) gap equations (\ref{gapsigma},\ref{gappi})  lead to (the calculations are well documented in \cite{jamvdb95,vdb96})
\beqn
\pi_a^s&=&0, a=0,...,8,\\
\sigma_a^s&=&0, a=1,...,7 \mbox{  ($\sigma_3^s=0$ in the isospin limit)},\\
{a}^{2}\chi_s^2M_u\left({1-\frac{{m}_{u}}{{M}_{u}}}\right)
&=&\ 8{N}_{c}M_u{g}_{{M}_{u},\beta ,{\mu }_{u}},
\label{gapup}\\
{a}^{2}\chi_s^2M_s\left({1-\frac{{m}_{s}}{{M}_{s}}}\right)
&=&\ 8{N}_{c}M_s{g}_{{M}_{s},\beta ,{\mu }_{s}},
\label{gapstrange}
\eeqn
%
with
\be
{g}_{{M}_{i},\beta ,{\mu }_{i}}=\frac{1}{\beta \Omega }
\sum_k\frac{1}{k^{*2}_i+{M_i}^2},\ \ \ \,i=u,s,
\label{eq2125}
\ee
and $k_i^*=(k_0-i\mu_i,\vec{k}), k_0=(2n+1)\pi\beta$, the odd Matsubara frequencies being related to the anti-commutating nature of the fermions. The constituent quark masses $M_u$ and $M_s$ are related to $\sigma_0^s,\sigma_8^s$ by 
\beqn
M_u&=&\f{1}{\sqrt{3}}(\sigma_8^s+m_3)+m_1+\sqrt{\f{2}{3}}\sigma_0^s,
\label{eqMu}\\
M_s&=&-\f{2}{\sqrt{3}}(\sigma_8^s+m_3)+m_1+\sqrt{\f{2}{3}}\sigma_0^s.
\label{eqMs}
\eeqn
As already mentioned, the NJL model is not renormalizable, so that the regularization parameter cannot be removed. Several procedures are available \cite{ripka97}. Here we shall focus on a 4-dimensional regularization scheme. One of its advantages is that it does not break the Lorentz invariance\footnote{Regularizations with a 3-momentum cut-off $\lambda$ break it. A meson, described as a $\bar{q}q$ excitation, cannot have a center of mass momentum $|\vec{k}|$ exceeding $2\lambda$.  This is unphysical and leads to distortions of $\pi$-mesons thermal distributions. See however \cite{blaschke95} for a way out of this difficulty.} of the theory at zero temperature and density. Another one is that it gives the correct (since the temperature and density dependent part is not regularized) high temperature limit\footnote{For a 3-momentum $\lambda$ cut-off, the temperature part decreases as $\lambda/T$ as $T$ is increased, see, e.g., \cite{klevansky94c}.} corresponding to a  plasma of free massive quarks.
The regularization procedure chosen here is the following:

Let $A(\beta,\mu)$ be a temperature
$T=1/\beta$ and a chemical potential $\mu$ dependent quantity.
 We can always write
\be
A(\beta,\mu)=A(\beta,\mu)-A(\infty,0)+A(\infty,0).
\label{regA}
\ee
All the terms are diverging quantities. However, the subtraction between the first two gives a finite result which does not have to be regularized\footnote{The subtraction in (\ref{regA}) also leads to diverging results when higher order loops are taken into account.} (in fact, it cannot be regularized with a 4-dimensional cut-off because of the heat bath as a preferred referential).  The last term is a zero temperature and density quantity which is Lorentz invariant and can be regularized with a 4-dimensional cut-off (in Euclidean space).
We write
\be
A^F(\beta,\mu)=A(\beta,\mu)-A(\infty,0)
\label{eqFermi}
\ee
as  the Fermi part of the quantity $A$, while
\be
A^D(\infty,0)=A(\infty,0)
\label{eqDirac}
\ee
is the Dirac part.

This regularization procedure will be applied to every diverging quantity we shall encounter in the following. Note however that the pressure (and hence the action) has to be  specified with respect to a reference point. We choose it to be the non-perturbative vacuum.

With these prescriptions, (\ref{eq2125}) can be evaluated and gives
\be
{g}_{{M}_{i},\beta ,{\mu }_{i}}={g}_{{M}_{i},\beta,{\mu}_{i}} -{g}_{{M}_{i},\infty,0}+{g}_{{M}_{i},\infty,0}
\ee
with
\beqn
{g}_{{M}_{i},\infty,0}&=&\int_{0}^{\infty
}\!\!\f{{d}^{4}k}{(2\pi)^4}\frac{1}{{k}^{2}+{{M}_{i}}^{2}}\nonumber\\
&\Rightarrow&
\int_{0}^{{\left({\Lambda
\chi_s}\right)}^{2}}\!\!\f{d{k}^{2}}{16\pi^2}\frac{{k}^{2}}{{k}^{2}+{{M}_{i}}^{2}}=
\frac{1}{16{\pi}^{2}}\left[{{\left({\Lambda \chi_s}\right)}^{2}-
{{M}_{i}}^{2}\ln\left({1+
\frac{{\left({\Lambda
\chi_s}\right)}^{2}}{{{M}_{i}}^{2}}}\right)}\right]
\label{eqB1}
\eeqn
\goodbreak
and
\beqn
{g}_{{M}_{i},\beta,{\mu}_{i}} -{g}_{{M}_{i},\infty,0}&=&\frac{
1}{\beta} \sum_{n=-\infty}^{\infty} \int_{
0}^{\infty}\frac{d^3 k}{{\left({2\pi}\right)}^{3}}
\frac{1}{
\left[
(2n+1)\pi/\beta-i\mu_i
\right]^2+{E}_{i}^{2}
}\nonumber\\
&&\qquad\mbox{}-\int_{-\infty}^{\infty
}\frac{d{k}_{0}}{{\left({2\pi}\right)}^{}}\int_{0}^{\infty
}\frac{{d}^{3}k}{{\left({2\pi}\right)}^{3}}\frac{1}{{{k}_{0}}^{2}+
{E}_{i}^{2}}\nonumber\\
&=&-\frac{1}{4{\pi}^{2}}
\int_{0}^{\infty}dk\frac{{k}^{2}}{{E}_{i}}
({n}_{i+}+{n}_{i-}),
\label{eqB2}
\eeqn
where we have defined the single-particle energies $E_i=(k^2+{M_i}^2)^{1/2}$ and 
\be
n_{i\pm }=\frac{1}{1+\exp\left[{\beta
\left({{E}_{i}\pm{\mu}_{i}}\right)}\right]}.
\label{eqB3}
\ee
The gap equations are intimately related to the quark condensates since they are given by
\beqn
\left\langle{\bar{u}u}\right\rangle&\equiv&{\frac{1}{\beta\Omega}}
{\frac{{\partial I}_{eff}^{c}}{{\partial m}_{u}}}
\left({{M}_{u},{M}_{s},{\chi}_{s}}\right)=-{\frac{1}{2}}{a}^{2}
{\chi}_{s}^{2}\left({{M}_{u}-{m}_{u}}\right)=
-4N_cM_ug_{M_u,\beta,\mu_{u}},\hspace{0.25cm}
\label{eq13}\\
\left\langle{\bar{s}s}\right\rangle&\equiv&{\frac{1}{\beta\Omega}}
{\frac{{\partial I}_{eff}^{c}}{{\partial m}_{s}}}
\left({{M}_{u},{M}_{s},{\chi}_{s}}\right)=-{\frac{1}{2}} {a}^{2}{\chi
}_{s}^{2}\left({{M}_{s}-{m}_{s}}\right)=
-4N_cM_sg_{M_s,\beta,\mu_{s}}.\hspace{0.25cm}
\label{eq14}
\eeqn 
Note that these condensates are not independent, contrarily to the pure NJL model: they are related explicitly (see $\chi_s$ in (\ref{eq13},\ref{eq14})) and implicitly by the gluon condensate (implicitly because the value of $M_u$ and $M_s$ is extracted from the gap equations (\ref{gapup},\ref{gapstrange}) which are coupled to the equation for the $\chi$ field -- see below).

These quark condensates show that a quark condensate can be considered  as a good candidate for the order parameter of chiral symmetry breaking. When this symmetry is explicitly broken, the quarks acquire   constituent quark masses $M_i>m_i$, solutions of the gap equations. Because of the explicit breaking of the chiral symmetry by the current quark mass $m_u,m_s$, it is better to set them to zero to study its dynamical breaking. We  then see that the up quark gap equation (\ref{gapup}) (and similarly for the strange quark) admits two solutions: one with $M_u=m_u=0$, the other one with $M_u\ne0$. When the solution $M_u\ne0$ is energetically favored, the chiral symmetry is dynamically broken and the quark condensate does not vanish. The solution $M_u=0$ can be reached at high temperature or densities: the quark condensate vanishes and the chiral symmetry is restored.
If the chiral transition is not discontinuous (what we shall call second order in the following), the critical exponents are well known since we work at the mean field level. It can be shown that there is a correspondence between the quark condensate and the magnetization, and between the current quark mass and the magnetic field. This means that when $m_u=0$, the shape of the quark condensate near $T_c$ is like $(T_c-T)^{1/2}$ (where $T_c$ is the critical temperature at which  the quark condensate vanishes) while there is a long tail extending from $T_c$ if $m_u\ne0$.

However, the phase structure of our extended NJL model is  richer than in the usual NJL model. Technically this is due to the fact that the gap equations are coupled to each other and with the equation for the $\chi$ field (\ref{gapchi}), which reads:

\beqn
\lefteqn{
\exp\left\{{-\frac{{a}^{2}}{2{\chi }_{0}^{2}}
\left[{{{M}_{u}^{0}}^{2}{\left({1-
\frac{{m}_{u}}{{M}_{u}^{0}}}\right)}^{2}+
\frac{{{M}_{s}^{0}}^{2}}{2}{\left({1-
\frac{{m}_{s}}{{M}_{s}^{0}}}\right)}^{2}}\right]}\right\}}
\nonumber\\ &&\mbox{}\times\frac{{\left[{{\left({\Lambda {\chi }_{
0}}\right)}^{2}+{{M}_{u}^{0}}^{2}}\right]}^{\frac{{2N}_{c}{\Lambda
}^{4}}{8{\pi }^{2}}}}{{\chi }_{0}^{{b}^{2}/2}} {\left[{{\left({\Lambda
{\chi}_{0}}\right)}^{2}
+{{M}_{s}^{0}}^{2}}\right]}^{\frac{{N}_{c}{\Lambda }^{4}}{8{\pi }^{2}}}
 \nonumber\\
&&\mbox{}=\exp\left\{{-\frac{{a}^{2}}{2{\chi_s}^{2}}
\left[{{M}_{u}^{2}{\left({1-
\frac{{m}_{u}}{{M}_{u}}}\right)}^{2}+\frac{{M}_{s}^{2}}{2}{
\left({1-\frac{{m}_{s}}{{M}_{s}}}\right)}^{2}}\right]}
\right\}\nonumber\\
&&\mbox{}\times
\frac{{\left[{{\left({\Lambda {\chi }_{
s}}\right)}^{2}+{{M}_{u}}^{2}}\right]}^{\frac{{2N}_{c}{\Lambda
}^{4}}{8{\pi }^{2}}}}{{\chi_s}^{{b}^{2}/2}} {\left[{{\left({\Lambda
{\chi}_{s}}\right)}^{2}
+{{M}_{s}}^{2}}\right]}^{\frac{{N}_{c}{\Lambda }^{4}}{8{\pi
}^{2}}},
\label{gapchi2}
\eeqn
where  $\chi_s$  has been eliminated in favor of $\chi_0$, the vacuum expectation value\footnote{This is related to the above mentioned fact that  the pressure is only defined w.r.t. a reference point. This point is chosen to be the vacuum at zero temperature and chemical potential.} of $\chi$ at $T=\mu=0$.\par

We show in section \ref{resultsCondensates} (see also section~\ref{sectionparameters}) that the coupling between (\ref{gapup},\ref{gapstrange},\ref{gapchi2}) enables also to get first order transitions, depending upon the free parameters of the model.

As a conclusion to this section, it is useful to note that
\begin{itemize}
\item in field theory, gap equations are called Schwinger-Dyson equations for the self-energy.
Defining
\be
\Sigma=m^0-\mu\beta+\varphi_a^s\Gamma_a,
\ee
one can effectively show that (\ref{gapsigma})
is equivalent to
\be
\Sigma=m^0-\mu\beta-\f{1}{(a\chi_s)^2}\Gamma_a\bigg(\Tr\Big[
\f{1}{i\dslash-\Sigma}\Gamma_a\Big]\bigg);
\label{gapSD}
\ee
\item gap equations, written in the form~(\ref{gapup}) and~(\ref{gapstrange}) or~(\ref{gapSD}), show clearly that the coupling constant can be redefined by absorbing  the color factor $N_c$.
This suggests that, in the limit  $N_c\rightarrow\infty$, the effective coupling constant is
\be
\tilde{a}^2=\f{a^2}{N_c}.
\label{atilde}
\ee
In term of the strong coupling constant $g$, we find
\be
\lim_{N_c\rightarrow\infty}g^2N_c=\mbox{cst.}=\tilde{g}^2
\label{gtilde}
\ee
which is in agreement with the high $N_c$ analysis \cite{witten79};
\item high $N_c$ arguments can be used to justify the saddle-point approximation used to derive the gap equations. If we admit (\ref{atilde}), we have, leaving aside the dilaton contribution,
\beqn
I_{eff}
&=&-N_c\bigg\{\Tr'_{\Lambda}\ln(-i\dslash+m^0+\varphi_a\Gamma_a
-\sslash{W})-\int d^4x\f{\tilde{a}^2}{2}\varphi_a^2\bigg\},
\label{modeleAmodifie}
\eeqn
where $\Tr'$ indicates that the color trace has to be omitted, allowing a factor $N_c$ to factorize. Besides justifying the saddle-point approximation, this indicates that $N_c$ is a relevant factor to organize the higher order correction series\cite{klevansky94b,klevansky94c,blaschke95c,schmidt95,zhuang95,ripka96}. However, to recover the low energy theorems, a particular treatment has to be applied to these higher orders \cite{lemmer94}, related to the symmetry conserving approximation program \cite{ripka97,ripka96}.
\end{itemize}

\subsubsection{Equations for meson masses (or Bethe-Salpeter equations)}
\label{bs}
As far as we are interested in  meson masses, wave function normalization and coupling constants, we have to expand the action up to  second order in the fields. Expanding up to second order means that the operator sandwiched between meson fields can be interpreted as a meson propagator. This can be cast in the form of a Bethe-Salpeter equation which, in the NJL model case, is tractable because of the inherent simplicity of the model: both the kernel and the vertices are known, so that we only have to search for the eigenvalues which are the masses  entering the meson propagators. (For a discussion of the Bethe-Salpeter formalism in the GCM and NJL models, see \cite{tandy97}.)

Expanding the action (\ref{modeleA}) up to second order gives\footnote{The quantities $P$ and $\Gamma$ are given in  appendix~\ref{PandGamma}.}

\be
{I}_{eff}^{2}  ={I}_{s}^{2}+
{I}_{ps}^{2}
\ee
with
\be
 {I}_{s}^{2}=\frac{1}{2\beta \Omega} 
\sum_{q}\left({\tilde{\sigma}}_{0q}
,{\tilde{\sigma}}_{8q},{\tilde{\chi}}_q,\vec{a}_{0q},
{K_0^*}_{aq}\right)
\left(\begin{array}{ccccc} {\Gamma}^{00}
&{\Gamma}^{08}&{\Gamma}^{0\chi}&0&0\\
{\Gamma}^{80}&{\Gamma}^{88}&{\Gamma}^{8\chi}&0&0\\
{\Gamma}^{\chi 0}&{\Gamma}^{\chi
 8}&{\Gamma}^{\chi \chi}&0&0\\
0&0&0&{\Gamma}^{a_0a_0}&0\\ 
0&0&0&0&{\Gamma}^{K_0^*K_0^*}
\end{array}\right)
\left(\begin{array}{c}
{\tilde{\sigma}}_{0-q}\\
{\tilde{\sigma}}_{8-q}\\
{\tilde{\chi}}_{-q}\\
\vec{a}_{0-q}\\
{K_0^*}_{{a-q}}
\end{array}\right)
\label{actionscalaire}
\ee
and
\be
{I}_{ps}^{2}=\frac{1}{2\beta \Omega} 
\sum_{q}\left(
{\tilde{\pi}}_{0q}
,{\tilde{\pi}}_{8q},{\vec{\pi}}_{q},{K}_{aq}
\right) 
\left(
\begin{array}{cccc}
{P}^{00}&{P}^{08}&0&0\\
{P}^{80}&{P}^{88}&0&0\\
0&0&{P}^{\pi \pi}&0\\
0&0&0&{P}^{KK}
\end{array}
\right)
\left(
\begin{array}{c}
{\tilde{\pi}}_{0-q}\\
{\tilde{\pi}}_{8-q}\\
{\vec{\pi}}_{-q}\\
{K}_{a-q}
\end{array}
\right),
\label{eq2135}
\ee
which shows that, in the pseudoscalar sector, $\pi_0$ and $\pi_8$, corresponding to $\eta_0$ and $\eta_8$, are coupled in agreement with the fact they have the same quantum numbers (pseudoscalar isoscalar $J^{PC}=0^{-+}$) while, owing to the glueball, the scalar sector has three coupled particles\footnote{Three coupled particles can also be described in the pseudoscalar channel by allowing the presence of the pseudoscalar glueball $J^{PC}=0^{-+}$. This is for example the approach of the composite operator formalism of Nekrasov \cite{nekrasov92,nekrasov94}. See also the works of Schechter \cite{schechter80,schechter80b}.} (scalar isoscalar $J^{PC}=0^{++}$).

The meson masses  can be obtained from (\ref{actionscalaire},\ref{eq2135}). Only pseudoscalar pions and kaons, and scalar $K^*$ and $a_0$ can be directly extracted. Masses for $\eta,\eta'$ and the three scalar isoscalars have to be obtained from a suitable diagonalization. In the isospin limit and for symmetric matter, these masses are given by ($\Delta\mu\equiv\mu_u-\mu_s$)
\beqn
{m}_{\pi}^{2}(q)&=&\frac{{a}^{2}{\chi}_{s}^{2}}{{Z}_{\pi}(\beta,\mu)}
\frac{{m}_{u}}{{M}_{u}},
\label{massepion}\\
({m}_{K^{\pm}}(q)\pm\Delta\mu)^2&=&\left(
{M}_{u}-
{M}_{s}\right)^2+\frac{{a}^{2}{\chi}_{s}^{2}}{2{Z}_{K^{\pm}}(\beta,\mu)}
\left(
\frac{{m}_{u}}{{M}_{u}}+
\frac{{m}_{s}}{{M}_{s}}
\right),
\label{massekaon}
\eeqn
for the pion-kaon sector ($K_0$ has the same mass -- except for isospin breaking corrections that we do not consider here -- as $K^+$ while $\bar{K}_0$ has the same mass as $K^-$), while the $\eta-\eta'$ sector is given by
\beqn
\lefteqn{
{m}_{\eta'
\choose
\eta}^{2}(q) =
{a}^{2}{\chi}_{s}^{2}
\left[
{\frac{{m}_{u}}{{M}_{u}}+
\frac{{m}_{s}}{{M}_{s}}+\xi \pm
\frac{1}{3\cos2{\theta}_{q}}
\left(
{\frac{{m}_{u}}{{M}_{u}}-\frac{{m}_{s}}{{M}_{s}}+3\xi}
\right)}
\right]}\hspace{3cm}\nonumber\\
&&\mbox{}\times
{\left(
{{Z}_{\pi}^{u} +{Z}_{\pi}^{s}\pm
\frac{\left(
{{Z}_{\pi}^{u}-{Z}_{\pi}^{s}}
\right)}
{3\cos2{\theta}_{q}}}
\right)}^{-1},
\label{eq2150}
\eeqn
where $\theta_q$ is the solution of
\beqn
\tg2{\theta}_{q}&=&\frac{2{P}^{08}}{{P}^{00}-{P}^{88}}\nonumber\\
&=&2\sqrt{2}
\left\{
1+\frac{3\xi
{\left(
{a{\chi}_{s}}
\right)}^{2}}
{\left[
{{q}^{2}\left({{Z}_{\pi
}^{u}-{Z}_{\pi}^{s}}
\right)+{\left({a{\chi
}^{s}}
\right)}^{2}
\left(
{\frac{{m}_{u}}{{M}_{u}}-
\frac{{m}_{s}}{{M}_{s}}}
\right)}
\right]}
\right\}^{-1}.
\label{eq2146}
\eeqn
In these relations, the functions $Z$ are the field normalizations. Defining
\be
{F}_{\mu \beta}({M}_{i},{M}_{j})=\frac{1}{\beta \Omega}
\sum_{k}
\frac{1}{
\left(
{{k}_{i}^{*2}+{{M}_{i}}^{2}}
\right)
\left[({{k}_{j}^{*}-q})^2+
{{M}_{j}}^{2}\right]}\ \ i,j=u,s,
\label{eq2139}
\ee
these normalizations are
\beqn
{Z}_{\pi}(\beta,\mu)&=&4{N}_{c}{F}_{\mu\beta}({M}_{u},{M}_{u}),
\label{zpion}\\
{Z}_{K^\pm}(\beta ,\mu)&=&
4{N}_{c}\left(
  \begin{array}{c}
{F}_{\mu\beta}({M}_{u},{M}_{s})\\
{F}_{\mu\beta}({M}_{s},{M}_{u}) 
 \end{array}
\right),
\label{zkaon}\\
{Z}_{\pi}^{u} \equiv  {Z}_{\pi}(\beta,\mu)&=&4{N}_{c}{F}_{\mu \beta}
({M}_{u},{M}_{u}),
\label{eq2147}\\
{Z}_{\pi}^{s} & =&4{N}_{c}{F}_{\mu
\beta}({M}_{s},{M}_{s}).
\label{eq2148}
\eeqn
\par

For the scalar sector, the masses are (there is the same correspondence between the charged and the neutral $K_0^*$ mass as for the kaon) 
\beqn
m_{a_0}^2(q)&=&4M_u^2+\f{a^2\chi_s^2}{Z_{\pi}(\beta,\mu)}\f{m_u}{M_u}\\
(m_{K_0^{*,\pm}}(q)\pm\Delta\mu)^2&=&\left(M_u+M_s\right)^2+\f{a^2\chi_s^2}{2Z_{K^{\pm}}(\beta,\mu)}
\left(
\f{m_u}{M_u}+\f{m_s}{M_s}
\right).
\eeqn
 The solution for the isoscalar sector is not illuminating (the masses are solutions of a cubic equation) so that we do not write it here. A numerical evaluation of these masses will be given in section~\ref{sectionhybrids}.

These formulae can be used to study  the mass variation as a function of temperature and density (chemical potential) and are the starting point for studying meson propagation in hot and dense matter.

Formulae (\ref{massepion}--\ref{eq2150}) are interesting because they readily show the Goldstone nature of the pseudoscalar mesons, and the role played by $\xi$ as a modelization of the strong axial $U(1)_A$ anomaly. Indeed, working in the limit $\mu_u=\mu_d=\mu_s$, we note that
when $\xi=0$, we get
\beqn
m_{\eta}^2 &=&\f{{a}^{2}{\chi}_{s}^{2}}{Z^u_{\pi}}\f{m_u}{M_u},
\label{uneta}\\
m_{\eta'}^2  &=&\f{{a}^{2}{\chi}_{s}^{2}}{Z^s_{\pi}}\f{m_s}{M_s}.
\label{unetaprime}
\eeqn
Moreover, in the vector symmetry limit $m_u=m_d=m_s$, there is a complete mass degeneracy
\be
m_{\pi}^2=m_K^2=m_{\eta}^2=m_{\eta'}^2=
\f{{a}^{2}{\chi}_{s}^{2}}{Z^u_{\pi}}\f{m_u}{M_u}.
\ee
In the chiral limit all these mesons have a vanishing mass, as it should (Goldstone theorem).

If the strong axial $U(1)_A$ anomaly is taken into account, we have (in the chiral limit)
\beqn
m_{\eta'}^2 &=&\f{{a}^{2}{\chi}_{s}^{2}}{Z^u_{\pi}}\xi,
\label{eqmasseetaprime}\\
m_{\eta}^2  &=&0.
\label{eqmasseeta}
\eeqn
The particle $\eta$ is still a Goldstone boson, but the $\eta'$ has lost this property
 (in a world with three colors $N_c$ = 3). Large $N_c$ studies \cite{donoghue92,witten79b} indicate that the axial symmetry is restored when  $N_c\rightarrow\infty$. This implies that the $\xi$ parameter behaves like $1/N_c$. This can can be seen by writing
\be
m_{\eta'}^2=\f{N_c{\tilde{a}}^{2}{\chi}_{s}^{2}}{4N_cF_{\mu\beta}(M_u,M_u)}
\xi,
\ee
where we have used eqs. (\ref{atilde}) and (\ref{zpion}). It is clear that a vanishing of $m_{\eta'}$ at large $N_c$  requires $\xi=\tilde{\xi}/N_c$.

\begin{figure}[hbt]
\vbox{\begin{center}\psfig{file=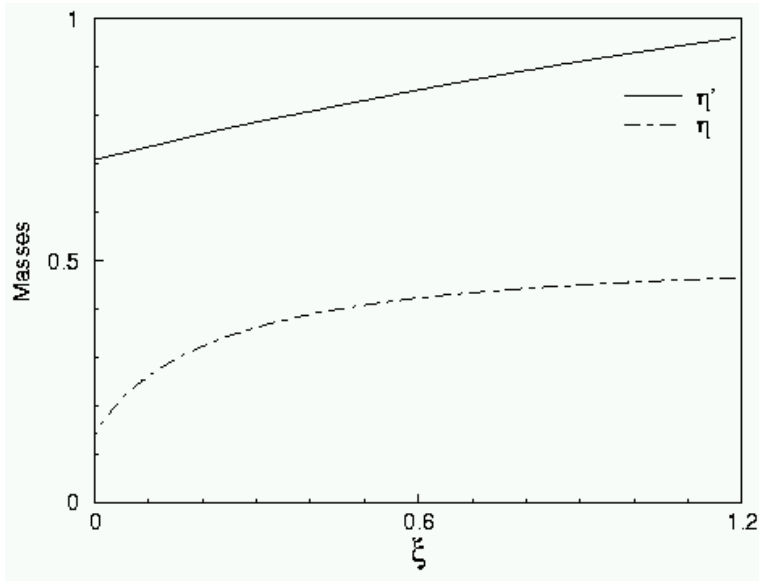,width=\widthepsxmgrthree}\end{center}}
\caption{\em $\eta,\eta'$ mass variation as a function of the $\xi$ axial anomaly  parameter.
\label{etaetaprime.eps}}
\end{figure}

In figure~\ref{etaetaprime.eps}, we show the mass variation of $\eta$ and $\eta'$  as a function of the $\xi$ parameter (for $M_u^0=400$ MeV, $\chi_0=350$ MeV). It is clear that, without the axial anomaly, the $\eta$ is degenerate with the pion while the $\eta'$ mass is identical to that of $\pi$ with the replacements $m_u\rightarrow m_s$, $M_u\rightarrow M_s$.

\subsection{Parameters}
\label{sectionparameters}
The main aim of the model is, starting from zero temperature $T$  and density $\rho$, to study the evolution of the system with respect to these thermodynamical external parameters. The model has intrinsic parameters that we can fix at $T=\rho=0$ so that the evolution is a prediction. In the isospin limit, the 3-flavor version of the model has seven parameters: the current quark masses $m_u, m_s$, the coupling constants $a^2,b^2$, the vacuum gluon condensate $\chi_0$, the cut-off $\Lambda$ and the $\eta_0$ mass parameter $\xi$ modelizing the strong axial anomaly. The gap equation~(\ref{gapup}) in the vacuum can be used to eliminate the coupling constant $a^2$ in term of the constituent up quark mass $M_u^0$. We choose $M_u^0$ and $\chi_0$ as free parameters; the five remaining parameters are adjusted to reproduce, in the vacuum, the pion mass\footnote{\label{mpifootnote} It is well known (see for example \cite{lemmer95}) that the mass difference between neutral and charged pions is of electromagnetic nature. Isospin effects account at most for a five percent difference. This justifies neglecting these corrections. In order to improve our calculations, it is necessary to look at these electromagnetic corrections rather than at the isospin breaking ones. Note also that we use, as  always  in the literature, the value $m_{\pi}=139$ MeV for the pion mass in the vacuum, even if we are interested only in neutral pions. This will have to be corrected for effects which are strongly dependent on $\mpi$. See for example section \ref{sectionanomalouspiondecay}.} $m_{\pi}=139$ MeV, the kaon mass $M_K=495$ MeV (same remark as in footnote \ref{mpifootnote}), the $\eta'$ mass $m_{\eta'}=958$ MeV, the glueball mass\footnote{There is some debate in the literature concerning the glueball mass. It is now believed that the scalar glueball mass is   1500 MeV   \cite{amsler95} or 1710 \cite{sexton95} (see also \cite{close97}). As we are not interested in the scalar sector, the exact value is not important for the analysis. To make contact with our previous works we shall keep the value 1300 MeV until section~\ref{sectionmixing}.} $m_{GB}=1300$ MeV and the on-shell pion decay constant. These parameters are determined in the following way: once $M_u^0$ and $\chi_0$ are given, the cut-off is obtained  from the approximate\footnote{We have neglected a term $q^2\partial Z\partial q^2$ evaluated at $q^2=-m_{\pi}^2$.} equation for the pion decay constant
\be
f_{\pi}\approx M_u^0Z_{\pi}^{1/2}(\infty,0)_{q^2=-m_{\pi}^2}.
\label{fpiapprox}
\ee
The pion mass (\ref{massepion}) fixes the product $a^2m_u$ and $a^2$ is obtained from the gap equation~(\ref{gapup}) in the vacuum. Since we are not interested in  the scalar sector, the parameter $b^2$ is then obtained from the glueball mass in an equivalent sigma model: $m_{GL}^2\approx \chi_0^2(b^2-3N_c\Lambda^4/2\pi^2)$ (see  section~\ref{sectionmixing} for its exact determination from the mass value of one of the $f_0$ hybrids). $M_s$ and $m_s$ are obtained from equations (\ref{gapstrange},\ref{massekaon}) and, finally, $\xi$ is fitted to reproduce the $\eta'$ mass.

The values of the parameters, together with the quark condensates,  are given in table~\ref{table1}, adapted from \cite{jamvdb95}, for four sets of input ($M_u^0,\chi_0$). The first two sets have been studied in \cite{jamvdb94,jamrip92}. The choice $\chi_0=350$ MeV is estimated from QCD sum rules applied to charmonium data \cite{svz79,narisson89}. The coupling of the quark and gluon condensates is then weak and the dilaton field is frozen to its vacuum value\footnote{With such a value of the gluon condensate, the model is almost a pure NJL.} (we can say that, with this value of the gluon condensate, QCD does not scale, see \cite{cohen92}). Lower values such as $\chi_0=125$ MeV (which gives a bag constant $B'^{1/4}\approx 200$ MeV), have to be used to enhance the coupling. Low values of the gluon condensate are required to obtain a chiral phase transition at  relatively low temperature ($T\approx 140$ MeV, as suggested by lattice calculations \cite{laermann96}). This justifies the use of the last two sets of parameters of table~\ref{table1}, which were first studied in \cite{jamvdb95,jamvdb94b}. 

\begin{minipage}[c]{\textwidth}
\begin{table}
\caption{\em Parameters of the A-scaling model ($m_{GL}=1.3$ {\rm GeV})}
\begin{tabular}{lr@{}lr@{}lr@{}lr@{}l}
$\chi_0$ (MeV) \phantom{123456789}  & \phantom{1234} 350&&
\phantom{1234} 125&& \phantom{1234} 80&&\phantom{1234} 90&\\
$M_u^0$ (MeV)        & 450&   & 600&   & 300&   & 800&\\
$m_{GL}$ (GeV)       & 1&.3   & 1&.3   & 1&.3   & 1&.3\\
$m_u$ (MeV)          & 9&.1   & 8&.7   & 7&.7   & 7&.7\\
$m_s$ (MeV)          & 205&   & 196&   & 186&   & 177&\\
$M_s^0$ (MeV)        & 645&   & 761&   & 539&   & 929&\\
$\Lambda$            & 2&.06  & 5&.85  &10&.13  & 8&.59\\
$-<\bar{u}u>^{1/3}_0$  & 207&.9 & 211&.4 & 219&.8 &220&.2\\ 
$-<\bar{s}s>^{1/3}_0$  & 207&.8 & 208&.2 & 234&.1 & 216&.4\\
$a^2$                & 0&.333   & 2&.044 & 11&.359 & 3&.327\\
$b^2$                & 22&.05 & 642&.6 & 5062&  & 2691&\\
$\xi$                &   1&.09   & 0&.639 &   1&.3   & 0&.385
\end{tabular}
\label{table1}
\end{table}
\end{minipage}

\begin{figure}[bt]
\vbox{\begin{center}\psfig{file=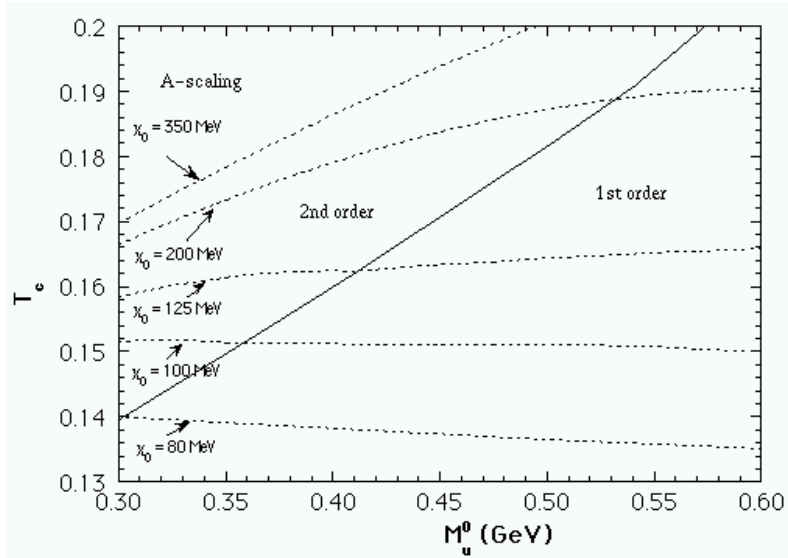,width=\widtheps}\end{center}}
\caption{\em Critical temperature as a function of the parameters $M_u^0,\chi_0$.
\label{figureZeit1.eps}}
\end{figure}

Note that a special treatment has to be reserved to  $\xi$ for the two sets of parameters $M_u=450$ MeV and $M_u=300$ MeV because of the threshold problem due to the lack of confinement of the NJL model (the meson can unphysically decay into two free quarks). We explain this problem, and the way to still extract results, in  section~\ref{sectionanomalouspiondecay}.
In figure~\ref{figureZeit1.eps} we show the variation of the critical\footnote{By critical temperature and/or density we mean the temperature and/or density associated with the restoration of the chiral symmetry.} temperature as a function of the two free parameters in the two degenerate flavor version of the model. This picture is instructive since it shows that 

\begin{itemize}
\item  the scaled model is able to reproduce a low critical temperature ($\approx 140$ MeV) contrarily to a pure NJL model ($T\approx 190$ MeV) -- see figures\footnote{In figures \ref{figuresu31a.eps} and \ref{figuresu31b.eps}, we have kept a finite value for the current quark masses. This explains the long tail near $T_c$. The $T_c$ value that we give is always the true value, obtained in the chiral limit so that the tail does not contribute (see the difference between figure \ref{figuresu31b.eps} and figure \ref{figurethermo4a.eps}). This can readily be seen on the gap equation which gives, for a second order phase transition,
\be
T_c=\sqrt{\f{3}{2}}\f{\Lambda\chi_c}{\pi}
\left(
1-\f{8\pi^2a^2}{4N_c\Lambda^2}
\right)^{1/2}.
\label{eqtempcritique}
\ee
The second term under the square root is due to the chiral limit which then reduces  the critical temperature. This equation also shows clearly that the scaled model allows a reduction of the critical temperature compared to a pure NJL model, in the ratio $\chi_c/\chi_0$.}~\ref{figuresu31a.eps} and \ref{figuresu31b.eps} of section \ref{resultsCondensates} --;
\item the model allows for first order transitions (this is not possible with the pure NJL model which only gives rise to second order transitions\footnote{To keep things clear, we adopt the somewhat unconventional approach of calling a second order transition a transition which is continuous although, strictly speaking, we should have made the distinction between true second order and crossover.} when $\mu_u=\mu_s=0$). Physically this comes from the fact that quark and gluon condensates are not independent. Mathematically, this comes from the fact that we have to deal with coupled equations (\ref{gapup},\ref{gapstrange},\ref{gapchi2}). An example of first order transition is given in figure~\ref{figuresu32a.eps}, section \ref{resultsCondensates}. It is clear on this picture that quark and gluon condensates are strongly coupled: the chiral phase transition is abrupt and felt by the gluon condensate;
\item in scaled models, a transition temperature as low as $T\approx 140$ MeV can only be attained if the vacuum gluon condensate is small. Then, there is the problem that the chiral symmetry can be restored at densities close to, or even lower than, the normal nuclear matter density. However, this defect can be related to the absence of vector mesons in our calculations. It is in fact well known that these vector mesons make the vacuum stiffer against the restoration of  chiral symmetry with respect to density \cite{jamrip93,jamrip91,jamrip92bis}. We have however to check if these mesons do not change also too drastically the critical temperature.
\end{itemize}

\subsection{Drawbacks of the models}
\label{drawbacks}

Effective Lagrangians are very attractive: they are quite simple while giving opportunities to have insights in an energy-momentum regime where lattice QCD is the only way to examine the true theory. There is a price for this  simplicity. Physically,  effective interactions are built in order to reproduce some effects at a given range of energy-momentum. This implies that their validity is restricted to this range. Even if calculations can be performed at another scale, there is nothing which enables  to relate these calculations to the true theory. Mathematically, there are also several drawbacks which have to be handled if we want the model to be meaningful. In the NJL model, these drawbacks are  the lack of renormalizability and the lack of confinement.
\begin{itemize}
\item The lack of renormalizability comes from the point-like nature of the quark-quark interaction (see (\ref{LNJL})). The model has to be regularized and this regularization cannot be removed by a renormalization procedure. Several regularization procedures are available (the references we indicate below are examples of application of these regularization procedures to the NJL model) such as Pauli-Villars \cite{goeke93c,lemmer94,lemmer95}, proper time \cite{goeke93c}, three dimensional cut-off \cite{klevansky94c,blaschke94,klevansky94}, four dimensional cut-off \cite{weise91b,vdb96}, regularizators which depend only on the orthogonal component to the 4-momentum \cite{blaschke95}, ... An in-depth analysis of regularization procedures and more references to the corresponding literature is given in \cite{ripka97}. Because there is no renormalization, it is clear that the regularization procedure is part of the model: to specify it requires to give both the Lagrangian and the regularizator. The hope is that observables are not too  sensitive to a particular choice of the regularizator. For logarithmic divergences, this low sensitivity is quite understandable. However,  this is a nontrivial task for quadratic divergences. However, fitting the regularizator to reproduce an observable such as the pion decay constant makes the results less sensitive to the particular choice.\par

\item The lack of confinement can be traced back to the fact that the quark propagator has a pole. A quark propagator without pole (with a $p^2$-dependent mass) is not a necessary condition to get a confined quark but it is sufficient \cite{roberts92b} since it ensures that the quark propagator cannot have any Lehmann representation. (Note, however, that the confined quark does not mean the absence of trouble to get meson properties. When calculating on-shell meson masses, we are looking for poles for time-like momenta $q^2<0$ (in Euclidean space). The propagators have then to be continued analytically, which can introduce unphysical poles: the continuation is a rather hard task whose difficulty is related to the confining property. To quote Ripka \cite{ripka97} ``{\em Our ignorance as how to continue propagators in the complex plane reflects our ignorance of the confining mechanism}''.) The lack of confinement of the NJL model means  that quark loops for physical meson processes will generate a $\bar{q}q$ production threshold at $s=4M^2$. Celenza et al. \cite{shakin95,shakin96} still use the  NJL quark propagator (constant quark mass). However they get loops without threshold by modifying the vertices in such a way that they cancel unphysical poles. Physically, these vertices are related to the confining part of potential models (this supposes that the confining part of a vertex can be separated from the full one, i.e. that it is possible to decouple the discussion of confinement from that of chiral-symmetry breaking). 
As a first step, we prefer to treat the NJL model as it is, trying to find a prescription to  still give a meaning to it, even above threshold. We shall explain the prescription we use in section~\ref{sectionanomalouspiondecay}. Simply speaking, we can say that we use a phenomenological interaction where the confinement is introduced by throwing away, by hand, the unphysical part. This is justified as long as the unphysical decay width of the meson into free quarks is small compared to the mass.
Since lattice calculations tend to show that the chiral phase transition and the confining transition are coincident \cite{laermann96}, we shall use the shortcut of speaking about the confined phase for the phase where the chiral symmetry is spontaneously, dynamically, broken, while we shall identify the unconfined phase with the phase where the chiral symmetry is restored ($M_u=m_u, M_s=m_s$).\par

\item Our model is an extension of the NJL model which takes into account the QCD trace or scale anomaly. Within the NJL model it is introduced in two steps  \cite{jamrip92,jamrip93}: first a scalar $\chi$ field is used to make the effective Lagrangian scale invariant (this includes also the treatment of the regularizator) and, second, an effective term is added which mimics the scale anomaly at  tree level \cite{schechter80b,schechter86}. The vacuum is driven to a nonzero value of $\chi_0$ and provides a mass for the dilaton excitations. The $\chi$ field can be related to the trace of the energy-momentum tensor (\ref{eq3.22}).
One drawback of this approach is that it is assumed that the use of a single dilaton field is enough to saturate the anomaly, i.e. that the anomaly equation can be used to define one interpolating field for the gluball. Although working for the pion (PCAC), this idea of the interpolating field is however questionable for such a massive state as the glueball ($M_{GB}\approx 1.5$ GeV). The defect of our approach can be seen by looking at the particle data  book \cite{pdb96} where it is clear that the region near 1.5 GeV is populated by several scalar resonances. However, we take here  the drastic assumption that all the scalars can be  described with only three resonances\footnote{In view of this, all the observed scalars near 1.5 GeV are  manifestations of the same particle: the $f_0(1500)$.}: the $f_0(1370)$, $f_0(1500)$ and $f_{J=0}(1710)$. The robustness of this approximation has to be tested phenomenologically within the model. This will be performed when studying the scalar sector in section~\ref{sectionmixing}.\par

Without the anomaly, the scale invariance would be broken in two ways: {\em i}) explicitly, by the current quark masses and {\em ii}) spontaneously, leading to a Goldstone boson (massless dilaton field). However, the quantum breaking of the axial symmetry  completely destroys this situation (massive dilaton field) and it can be argued that QCD does not scale \cite{birse94}. This is also shown in the model-independent study \cite{cohen92} and within the extended NJL model \cite{jamvdb95}. In \cite{weise94} it is explicitly shown  in the extended NJL model that, to get a scaling of  hadron masses and decay constants as suggested by   Brown and Rho \cite{brown91}, the dilaton mass has to be much smaller that the lightest ($\bar{q}q$) scalar meson.  Since this is not the case, the Brown-Rho scaling should not be expected in the data. Note that the model independent estimate made in \cite{cohen92} concerns matter up to normal densities. It does not prevent the dilaton field to play a role at higher densities. In fact, our studies \cite{jamvdb95,vdb96} have shown that the gluon condensate only plays a role  near the phase transition. This means that, although QCD does not scale (at least in the chirally broken phase), the gluon condensate can be relevant for  phase transition studies. In our formalism, this can be traced back to the fact that the gap equations for the quark condensates are coupled with the gluon condensate (eqs. (\ref{gapup},\ref{gapstrange},\ref{gapchi2})). 
It is this coupling which allows our model to describe second order phase transitions as well as first order transitions. This also allows to get low critical temperature and densities. However, the most interesting cases (departure from the pure NJL case) show up only when $\chi_0$ is rather small, corresponding to bag models (the QCD sum rules estimate \cite{svz79,narisson89}  gives $\chi_0\approx 350$ MeV, leading to an almost pure NJL behavior). 
\item We have given above arguments to claim that the gluon condensate is a meaningful quantity to describe the chiral transition,  even if QCD does not scale.
It is however clear that our approach could have a very strong shortcoming: we use the scale anomaly at zero temperature. When there is no quark in the theory ($m_u=m_d=m_s\rightarrow\infty$), eq.~(\ref{gapchi2}) gives $\chi=\chi_0$ for any temperature or density\footnote{In the NJL model, the gluons have been eliminated to the benefit of an effective 4-quark interaction. In  our scaled model, some effects of the gluons have been included (scale anomaly). However we still do not have a gluon sector. It is then not astonishing that removing the quarks does not lead to a  gluon phenomenology: for example we do not have the right number of gluonic d.o.f.}.  However, the gluon condensate is expected to be flat\footnote{See e.g. eq. (3) of \cite{schafer96b} or lattice studies in the pure gluonic sector (e.g. figure 4 from \cite{karsch95}).}  below the transition, so that its temperature and density variations can be neglected, as a first approximation, in the chirally broken phase.
\end{itemize}

\section{The A-scaled NJL model at finite temperature and density}
\label{sectionchap2}

In the previous section, we have carefully described  the motivations leading to the A-scaled NJL model and we have commented on the drawbacks of the approach. This section gathers results  at finite temperature and density. We shall first focus on analytical results for the thermodynamics in section \ref{sectionthermodynamics}. We shall then show our numerical results concerning the condensates,  the thermodynamics and the critical surface $T_c(\mu_u,\mu_s)$ in section \ref{results}. We shall end the section with a study of the anomalous decay of the neutral pion into two photons. This will give us the opportunity to extend the model above the threshold.

\subsection{Thermodynamics}
\label{sectionthermodynamics}
\label{sectioncondensates}
It is believed that QCD has two phase transitions: a deconfinement transition corresponding to going from a hadronic gas to a quark-gluon plasma, and a transition leading to a phase where the chiral symmetry is restored. Lattice calculations \cite{laermann96,karsch95} suggest that these two transitions coincide. In a purely gluonic theory, it seems that the transition is of first order (for $N_c=3$). When quarks  are introduced, the order of the transition depends on the number of light flavors.  With $N_f=3$ massless flavors, QCD has a first order chiral transition, not connected to the pure gluonic one. When the mass of the quarks is varied, the two first order transitions are separated by a region in which there is a crossover (to keep things simple, we shall call it a second order transition, however). According to the type of calculations -- Wilson fermions \cite{iwasaki96b} or staggered (Kogut-Susskind) fermions \cite{brown90} -- QCD ($m_u\approx m_d\approx 10$ MeV, $m_s\approx 200$ MeV) appears to be in the first order region, or in the second order region, respectively. It is then not so clear which case occurs and we have the freedom to play with the parameters in such a way as to allow for both types of transitions. Note that the distinction between pure gauge deconfinement and light quark chiral phase transition is of prime importance: the energy scales are different. Pure gauge transitions occur at a temperature of about 260 MeV while, with two light flavors, the critical temperature is around 150 MeV \cite{laermann96,schafer96b,karsch95,detar95,lattice96}, or even as low as 140 MeV according to \cite{kogut91,brown93b}. \par

The deconfining transition is hard to study in  existing models. However,  chiral symmetry breaking and its restoration can be studied in effective models such as NJL. Since both transitions are seen to be coincident on the lattice, we shall take the point of view that the study of the chiral phase transition can shed  light on the deconfining transition. To simplify the discussion we shall call hadronic phase the phase where  chiral symmetry is spontaneously broken while we shall call quark-gluon plasma phase the phase where chiral symmetry is restored\footnote{Strictly speaking, only the quark condensate goes to zero in our model. However we include the gluons in this plasma phase when the gluon condensate reaches its minimum value.}. We shall also take the point of view that  gluon confinement is linked to the gluon condensate $\chi_s$, although this hypothesis is questionable \cite{roberts92b,brown93c}. There is a constraint between the quark and gluon condensates, which shows up into the form of three coupled equations for the three condensates.  We can then get, according to the strength of this constraint,  a second order transition (continuous transition from one phase to the other, see figure \ref{figuresu31b.eps}) or a first order transition (discontinuous passage from one phase to the other, see figure \ref{figuresu32a.eps}). 

The preliminary results given in this section have been discussed in \cite{vdb96,cugjamvdb96}. As already stated we work at the mean field level. The groups of Rostock \cite{schmidt95,blaschke95}, Heildelberg \cite{klevansky94b,klevansky94c,zhuang95} and Nikolov et al. \cite{ripka96} go beyond this approximation, studying the first $1/N_c$ corrections and showing that they are not negligible at low temperature and density (because pions are almost massless). However one can also take the point of view that an effective theory cannot be used above the mean field so that corrections should not be included. As a first approximation, we shall keep this argument.

\subsubsection{Pressure, energy density, entropy density, bag constant}
\label{sectionpressure}
In this section we want to understand the equilibrium properties of a strongly interacting matter, i.e.  determine the relations associated to the thermodynamics of  a hot and dense system. The system is modelized by an effective action of the NJL type (free massive constituent quarks) with a dilaton field included\footnote{For the thermodynamics of a  scaled linear sigma model, see \cite{ellispj98}.}, eq.~(\ref{modeleA}). In a grand-canonical system, the partition function is given by
\be
{\cal Z}=\exp(-\beta\bf\Omega),
\label{eqthermo1}
\ee
where {$\mathbf\Omega$} is the thermodynamical potential (or grand potential). We  exclusively consider a system in equilibrium: all the descriptions (micro-canonical, canonical, grand-canonical) are equivalent. However the grand-canonical description is the  easiest. In the canonical formalism, the basic quantity is the Helmholtz free energy and the independent variables are the temperature, the pressure and the densities (one for each chemical potential). Since phase transitions occur at a constant chemical potential, and not at a constant density, it has always to be checked if a lower energy solution, obtained by separating the system into subsystems,  exists \cite{diu89}. Working in the grand-canonical formalism, where the independent variables are the temperature, the pressure and the chemical potentials, there is a direct access to the solutions corresponding to the minimum of the thermodynamical potential. Eq.~(\ref{eqthermo1}) leads to the identification
\be
I_{eff}=\beta\bf\Omega.
\label{thermo2}
\ee
Since \cite{fetwal71,diu89}
\be
d{\bf\Omega}=-SdT-Pd\Omega-\rho_id{\mu_i}\ \ (i=u,d,s),
\label{thermopot}
\ee
we get
\be
P=-\left(
\f{\p{\bf\Omega}}{\p\Omega}
\right)_{T,\mu_i}.
\ee
This implies
\be
P=
\left.
\f{1}{\beta}\f{\p\ln({\cal Z})}{\p\Omega}
\right\}_{T,\mu_i}.
\ee
\par
Physically, the pressure is not an absolute quantity. We have to consider it with respect to a reference system which is chosen to be the (non-perturbative) vacuum, of pressure $P_0$. Defining $P-P_0=P'$ and replacing $P'$ by $P$, we then have
\be
P=
\left.
\f{1}{\beta}\f{\p\ln({\cal Z}/{\cal Z}_0)}{\p\Omega}
\right\}_{T,\mu_i}.
\label{thermo3}
\ee
\par

As explained for eq. (\ref{regA}), we can separate the action into a Fermi part   and a Dirac part. Subtracting the vacuum, the lowest order of the action is 
\beqn
\lefteqn{
I_{eff}^s(\varphi_a^s,\chi_s)\equiv
I_{eff}^s(M_u,M_s,\chi_s)=}\nonumber\\
&&I_{(\mu,\beta)}^s(M_u,M_s,\chi_s)
+I_{(0,\infty)}^s(M_u,M_s,\chi_s)-
I_{(0,\infty)}^s(M_u^0,M_s^0,\chi_0).
\label{actiontotale}
\eeqn
The first term is the Fermi part, which does not need  being regularized,
\begin{eqnarray}
\lefteqn{
{I}_{\left(\mu,\beta\right)}^s
\left({M}_{u},{M}_{s},{\chi}_{s}\right)=-\beta\Omega
\frac{{2N}_{c}}{{2\pi}^{2}\beta}\sum_{i=u,s}{a}_{i}
\int_{0}^{\infty}{k}^{2}dk} \nonumber\\
&&\mbox{}\times\bigg\{\ln\Big(1+\exp\left[-\beta\left({E}_{i}+{\mu}_{i}
\right)\right]\Big)+\ln\Big(1+\exp\left[-\beta
\left({E}_{i}-{\mu}_{i}\right)\right]\Big)\bigg\},
\label{eq8}
\end{eqnarray}
where\footnote{When  no  confusion is possible between the 4-momentum $k$ and the 3-momentum $\vec{k}$, we use the notation $k$ for $|\vec{k}|$.}
$E_i=\sqrt{k^2+M_i^2}$, and $a_u=2, a_s=1$,
while the Dirac part ($T=\mu=0$) has to be regularized and is given by
\begin{eqnarray}
\lefteqn
{I_{(0,\infty)}^s(M_u,M_s,\chi_s)-
I_{(0,\infty)}^s(M_u^0,M_s^0,\chi_0)
=-\beta\Omega\Bigg\{\f{2N_c}{32\pi^2}\sum_{i=u,s}a_i}\nonumber\\
&&\mbox{}\times\bigg[{\left({\Lambda{\chi}_{s}}\right)}^{4}\ln 
\f{{\left({\Lambda{\chi}_{s}}\right)}^{2}+{M_i}^2}{
{\left({\Lambda{\chi}_{0}}\right)}^{2}+{M_i^0}^2}-
{M_i}^{4}\ln \f{{\left({\Lambda{\chi}_{s}}\right)}^{2}+{M_i}^2}{
{M_i}^2}\nonumber\\
&&+\mbox{}
{M_i^0}^{4}\ln \f{{\left({\Lambda{\chi}_{0}}\right)}^{2}+{M_i^0}^2}{
{M_i^0}^2}\nonumber\\
&&\mbox{}-\Big(\f{1}{2}(\Lambda\chi_s)^4-\f{1}{2}(\Lambda\chi_0)^4
\Big)+\Big((M_i\Lambda\chi_s)^2-(M_i^0\Lambda\chi_0)^2\Big)\bigg]
\nonumber\\
&&\mbox{}+\bigg[(\chi_s^4-\chi_0^4)\Big(\f{b^2}{16}+\f{a^2}{4\chi_0^2}
(\sigma_0^{s2}+\sigma_8^{s2})\Big)\nonumber\\
&&\mbox{}-\sum_{i=u,s}a_i\Big(\f{a^2\chi_s^2}{4}{M_i}^2
(1-\f{m_i}{M_i})^2-\f{a^2\chi_0^2}{4}{M_i^0}^2
(1-\f{m_i}{M_i^0})^2\Big)\nonumber\\
&&\mbox{}-\f{b^2}{16}\chi_s^4\ln\left(
\f{\chi_s}{\chi_0}\right)^4\bigg]\Bigg\}.
\label{eqdirac}
\end{eqnarray}

With the action (\ref{actiontotale}),   eq.~(\ref{thermo3}) gives
\be
P=-\f{1}{\beta}\f{\p I_{eff}^s(M_u,M_s,\chi_s)}{\p\Omega},
\label{thermo2bisbis}
\ee
\par

which leads to
\begin{eqnarray} 
P&=&-{\frac{1}{\beta\Omega}}\left[{{I}_{\left({\mu,\beta}\right)}^{s}
\left({{M}_{u},{M}_{s},{\chi}_{s}}\right)+{I}_{\left({0,\infty}
\right)}^{s}\left({{M}_{u},{M}_{s},{\chi}_{s}}\right)-{I}_{\left({0,
\infty}\right)}^{s}\left({{M}_{u}^{0},{M}_{s}^{0},{\chi}_{0}}\right)}
\right].
\label{eqthermo17}
\end{eqnarray}
The physical meaning of this equation is that the grand potential is an extensive quantity:
\be
{\bf\Omega}=-P\Omega.
\label{thermo2bis}
\ee
\par
In section \ref{dsgap}, we have seen that, mathematically, the mean field approximation corresponds to finding the saddle-point solution of the equations~(\ref{gapsigma},\ref{gapchi}), i.e. to finding the minimum of the action. Eq.~(\ref{thermo2bis}) shows the physics attached to this condition: the system  chooses the phase where the pressure is a  maximum.

Quark densities can be evaluated from
~(\ref{thermopot}) and are given by (see (\ref{eqB3}) for the definition of the functions $n_{i\pm}$)
\be
\rho_i=-\left(
\f{\p{\bf\Omega}}{\p\mu_i}
\right)_{T,\Omega}=
-{\frac{1}{\beta\Omega}}\left({{\frac{{\partial
I}_{eff}^{s}}
{{\partial\mu}_{i}}}}\right)=-{\frac{{N}_{c}}{{\pi}^{2}}}
\int_{0}^{\infty}{k}^{2}\left({{n}_{i+}-{n}_{i-}}\right)dk.
\label{eqthermo15}
\ee
Since we work in the isospin limit ($m_u=m_d$) for a symmetric matter ($\mu_u=\mu_d$), it is clear that $\rho_u=\rho_d$.

The entropy density can also be evaluated from~(\ref{thermopot}) and is given by
\be
s\equiv\frac{S}{\Omega}=-{\frac{1}{\Omega}}
\left({1-\beta{\frac{\partial}{\partial\beta}}}\right){I}_{\left({\mu,
\beta}\right)}^{s}\left({{M}_{u},{M}_{s},{\chi}_{s}}\right).
\label{eqthermo19}
\ee
Only the Fermi part appears in this formula since this is the only one which depends upon temperature. When $T\rightarrow0$ the entropy density $s$ goes to zero,  in agreement with the third principle of thermodynamics.\par

Finally, the internal energy is given by \cite{fetwal71,diu89}
\be
E\equiv {\bf\Omega}+TS+\mu_i\rho_i=\left(
1+\beta\f{\p}{\p\beta}-\mu_i\f{\p}{\p\mu_i}
\right)
{\bf\Omega}=\left(
\f{\p}{\p\beta}-\f{1}{\beta}\mu_i\f{\p}{\p\mu_i}
\right)
I_{eff}^s(M_u,M_s,\chi_s).
\ee
This gives the energy density
\beqn
\epsilon\equiv{\frac{E}{\Omega}}
&=&{\frac{1}{\Omega}}\left({{\frac{\partial}{\partial\beta}}-
{\frac{1}{\beta}}{\mu}_{i}
{\frac{\partial}{{\partial\mu}_{i}}}}\right){I}_{\left({\mu,\beta}
\right)}^{s}\left({{M}_{u},{M}_{s},{\chi}_{s}}\right)\nonumber\\
&+&{\frac{1}{\beta\Omega}}\left[{{I}_{\left({0,\infty}\right)}^{s}
\left({{M}_{u},{M}_{s},{\chi}_{s}}\right)-{I}_{
\left({0,\infty}\right)}^{s}
\left({{M}_{u}^{0},{M}_{s}^{0},{\chi}_{0}}\right)}\right].
\label{eqthermo18}
\eeqn
\par

Like the pressure, the energy density is a relative quantity~: (\ref{eqthermo18}) gives the density energy of the system w.r.t. the vacuum energy.
\subsubsection{Bag constant $B$}
\label{sectionsac}

Following
\cite{brown93b,asakawa89}, we write
\beqn 
P={P}_{\boite{ideal gas}}-B\left({\beta,{\mu}_{u},{\mu}_{s}}\right),
\label{eqthermo20}\\
\epsilon={\varepsilon}_{\boite{ideal gas}}+B\left({\beta,{\mu}_{u},{\mu}_{s}}\right),
\label{eqthermo21}\\
Ts&=&{P}_{\boite{ideal gas}}+{\varepsilon}_{\boite{ideal gas}}-{\mu}_{i}{\rho}_{i},
\label{eqthermo22}
\eeqn
with
\beqn
{P}_{\boite{ideal gas}}&=&-{\frac{{I}_{\left({\mu,\beta}\right)}^{s}
\left({{M}_{u},{M}_{s},{\chi}_{s}}\right)}{\beta\Omega}}=
{\frac{{N}_{c}}{{3\pi}^{2}}}\sum\limits_{i=u,s}^{}{a}_{i}
\int_{0}^{\infty}{\frac{{k}^{4}}{{E}_{i}}}\left({{n}_{i+}+{n}_{i-}}
\right)dk,
\label{eqthermo23}\\
{\varepsilon}_{\boite{ideal gas}}&=&{\frac{1}{\Omega}}
\left({{\frac{\partial}{\partial\beta}}-{\frac{1}{\beta}}
{\mu}_{i}{\frac{\partial}{{\partial
\mu}_{i}}}}\right){I}_{\left({\mu,\beta}\right)}^{s}
\left({{M}_{u},{M}_{s},{\chi}_{s}}\right)\nonumber\\
&=&{\frac{{N}_{c}}{{\pi}^{2}}}\sum\limits_{i=u,s}{a}_{i}
\int_{0}^{\infty}{k}^{2}{E}_{i}\left({{n}_{i+}+{n}_{i-}}\right)dk,
\label{eqthermo24}
\eeqn
being quantities relative to a massive free quark system. The interaction measure, which is an indication of non-perturbative effects, is
\be
\varepsilon-3P=
4B+{\frac{{N}_{c}}{{\pi}^{2}}}\sum\limits_{i=u,s}{a}_{i}M_i^2\int_0^{\infty}
\f{k_i^2}{E_i}\left({{n}_{i+}+{n}_{i-}}\right)dk.
\label{eminus3P}
\ee

In eqs.~(\ref{eqthermo20}) and (\ref{eqthermo21}), we have defined a temperature and density dependent (through $M_u,M_s,\chi_s$) bag constant\footnote{In
\cite{brown93b,asakawa89}, there is no glueball: the bag constant is  purely chiral. Our bag constant is then a generalization of these references.}
$B$. It depends on the Dirac contribution (\ref{eqdirac}) to the pressure:
\be
B\left({\beta,{\mu}_{u},{\mu}_{s}}\right)={\frac{1}{\beta\Omega}}
\left\{{{I}_{\left({0,\infty}\right)}^{s}
\left({{M}_{u},{M}_{s},{\chi}_{s}}
\right)-{I}_{\left({0,\infty}\right)}^{s}
\left({{M}_{u}^{0},{M}_{s}^{0},{\chi}_{0}}\right)}\right\}.
\label{eqthermo25}
\ee
\par

 The definition~(\ref{eqthermo25})
is  different from
\cite{jamvdb95,ellis90b,klevansky94c,brown93} because two new effects are implemented: {\em i}) we take into account the explicit breaking of  chiral symmetry (even if
$m_u=0$, we have
$m_s\ne0$: the strange quark contribution can be non-negligible);  {\em ii)}  as already stated, eq.~(\ref{eqthermo25}) takes
into account the effects of the gluons. 
Moreover, (\ref{eqthermo25}) conceptually differs  from the definition
\cite{klevansky94c} where the bag constant is zero in the chirally restored phase.
Finally, it is different from the bag constant $B'$ introduced in
\cite{jamvdb94,jamvdb94b,jamrip92}. In these references, it is 
only obtained at zero temperature and density, through the definition of $B'$, being identical to the 
energy difference between the perturbative vacuum and  the true vacuum\footnote{One can also take the equivalent definition of considering it only through the glueball Lagrangian (decoupling between the glueball and the other fields).
 We have
$B=\f{1}{4}<\theta_{\mu\mu}>=\f{b^2}{16}\chi^4=\f{1}{16}m_{GL}^2\chi^2$.}.

In the perturbative vacuum, the chiral symmetry is restored and the gluon condensate vanishes. Then, according to
\cite{jamvdb94},
\be
B'={\frac{1}{\beta\Omega}}
\left\{{{I}_{\left({0,\infty}\right)}^{s}\left({0},{0},{0}\right)-
{I}_{\left({0,\infty}\right)}^{c}
\left({{M}_{u}^{0},{M}_{s}^{0},{\chi}_{0}}
\right)}\right\}=\frac{1}{16}{m}_{GL}^{2}{\chi}_{0}^{2}.
\label{eqthermo26}
\ee
In $SU(3)$, implementing
$m_s=0$ in the definition of the perturbative vacuum makes no sense. However, we shall see that the value of $B'^{1/4}$
is not so far from that of $B^{1/4}$~(\ref{eqthermo25}) (for $T>>$) so that the use of~(\ref{eqthermo26}) in
\cite{jamvdb94,jamvdb94b,jamrip92} is verified {\em a posteriori}. This remark also applies for the effect of $\chi_s$
which does not go down to zero.\par

Note that in \cite{brown93}, the authors study both the chiral symmetry restoration and the effects of the gluon condensate. They however define two bag constants, one associated to the restoration of  scale symmetry, the other to the restoration of  chiral symmetry.

\subsubsection{High temperature zero density limit ($T>T_c$)}
\label{sectionhightemplimit}
\label{sectionthese522}

In a phase where  chiral symmetry is restored -- in this section, we mean the phase where the constituent quark mass goes to the current quark mass
 ($M_i=m_i$, $i=u,s$) even if we are not in the chiral limit --  we have $m_s/T\lesssim 1$: a high temperature expansion in $m_s/T$ is possible
 \cite{cugjamvdb96,dolanjackiw74,kapusta89}. Calculations are lengthy and left to appendix \ref{Appendixhightemplimit}, taken from \cite{vdb96}. We work at zero density. Our results are a generalization of \cite{cugjamvdb96,dolanjackiw74,kapusta89} where only a limited number of terms in the $m_s/T$ expansion have been retained while we are able to give here the full expansion, involving only elementary functions. Note that to describe the results in section \ref{results},
the first four terms will be enough ($T>T_c$, with $T_c$ the critical temperature) so that we only keep them in the following.
For the pressure, we get
\be
{P}_{\boite{ideal gas}}\approx
T^4\left(
{\frac{7}{60}}{N}_{c}{\pi}^{2}
-{\frac{{N}_{c}}{12}}\f{{m}_{s}^{2}}{{T}^{2}}-
{\frac{{N}_{c}}{{8\pi}^{2}}}\f{{m}_{s}^{4}}{T^4}\ln{\frac{{m}_{s}}{\pi
T}}+{\frac{{N}_{c}}{{16\pi}^{2}}}\left({{\frac{3}{2}}-2\gamma}\right)
\f{{m}_{s}^{4}}{T^4}+\ ...
\right),
\label{eqthermo27}
\ee
where $\gamma$ is the Euler constant\footnote{In \cite{klevansky94c}, the massless free quark limit is unreachable, by construction: the $d^3k$ regularization introduces a cut-off $\Lambda$ for each Fermi or ideal gas quantity. These quantities, for example ${P}_{\boite{ideal gas}}$,
behave then at high temperature as $\Lambda/T$, decreasing to zero.
This drawback is not present in  this work, where we have chosen a $d^4k$ regularization for the vacuum while the Fermi part is not regularized.}.\par

Note that the case of finite chemical potentials is much more complex and, to our knowledge, has never been treated to all orders in the fermionic case. (In the bosonic case, the constraint $\mu_i<M_i$ (not present in the fermionic case) allows a high temperature expansion ($m_i/T$ and $\mu_i/T \lesssim 1$) to all orders\cite{weldon81,weldon82,weldon82bis}.)

Above  $T_c$, the bag constant $B$ is temperature independent and writes
\be
B={\frac{1}{\beta\Omega}}\left\{{{I}_{\left({0,\infty}\right)}^{s}
\left({0,{m}_{s},\chi_c}\right)-{I}_{\left({0,\infty}\right)}^{s}
\left({{M}_{u}^{0},{M}_{s}^{0},{\chi}_{0}}\right)}\right\},
\label{eqthermo28}
\ee
where $\chi_c$ is the  gluon condensate  above $T_c$. Note that,  for any set of parameters ($M_u^0,\chi_0$), we could not get $\chi_c=0$. In our model there is never a complete gluon deconfinement. This is related to the fact that gluons are only poorly incorporated in our formalism (we have no explicit temperature dependence  of the gluon condensate since the modeling of the gluon anomaly is through a temperature independent potential (\ref{eq3.21}); we also do not have the right number of gluonic d.o.f.).\par

The energy density is given by (\ref{eqthermo18})
\be
\varepsilon_{\boite{ideal gas}}=\f{1}{\Omega}\f{\p
I_{(\mu,\beta)}^s}{\p \beta},
\ee
i.e., using (\ref{thermo2}) and (\ref{thermo2bis}),
\be
\varepsilon_{\boite{ideal gas}}=-\f{\p}{\p\beta}(\beta P_{\boite{ideal gas}}).
\label{eqthermo1.30}
\ee

With (\ref{eqthermo27}), the high temperature, zero density, expansion gives (appendix \ref{Appendixhightemplimit})
\be
{\varepsilon}_{\boite{ideal gas}}\approx 
T^4\left(
{\frac{7}{20}}
{N}_{c}{\pi}^{2}-{\frac{{N}_{c}}{12}}\f{{m}_{s}^{2}}{{T}^{2}}+
{\frac{{N}_{c}}{{8\pi}^{2}}}\f{{m}_{s}^{4}}{T^4}\ln{\frac{{m}_{s}}{\pi
T}}+{\frac{{N}_{c}}{{16\pi}^{2}}}\left({2\gamma+{\frac{1}{2}}}\right)
\f{{m}_{s}^{4}}{T^4}+\ ...
\right).
\label{eqthermo37}
\ee
\par

Finally, (\ref{thermo2}) and (\ref{thermo2bis}) applied to~(\ref{eqthermo19}) lead to
\be
s=\left(
1-\beta\f{\p}{\p \beta}
\right)
\beta P_{\boite{ideal gas}},
\ee
or, equivalently, to
\be
s=-\beta^2\f{\p}{\p \beta}P_{\boite{ideal gas}}=\f{\p}{\p T} P_{\boite{ideal gas}}.
\label{eqthermo1.33}
\ee
The  high temperature, zero density, expansion of the entropy density is then
(appendix \ref{Appendixhightemplimit})
\be
Ts\approx
T^4\left({\frac{7}{15}}{N}_{c}{\pi}^{2}-
{\frac{{N}_{c}}{6}}\f{{m}_{s}^{2}}{T^2}+{\frac{{N}_{c}}{{8\pi}^{2}}}
\f{{m}_{s}^{4}}{T^4}+...
\right).
\label{eqthermo41}
\ee
\par

\espace

The results~(\ref{eqthermo27}), (\ref{eqthermo37})
and~(\ref{eqthermo41}) are used in section~\ref{results}.

\subsubsection{Low temperature zero density limit}
\label{sectionlowtemplimit}

To simplify the discussion, we  limit ourselves to the zero density case. If
 $M_u^0$ is of the order of 400 MeV, the low temperature expansion (see appendix \ref{Appendixhightemplimit}) is valid up to
$T\approx 100$ MeV.  Indeed, 
masses and condensates  are  not varying within this range of temperatures, see section \ref{results}, and the expansion parameters $\beta M_u$ and $\beta M_s$ are then large enough
-- we have at worst
$\beta M_u\approx 4$ -- to allow a stationary phase expansion of~(\ref{eqthermo23}):
\be
{P}_{\boite{ideal gas}}\approx{\frac{{4N}_{c}{\beta}^{-5/2}}{{\left({2\pi}
\right)}^{3/2}}}\sum\limits_{i=u,s}^{}{a}_{i}{M}_{i}^{3/2}{e}^{- {\beta
M}_{i}}\ +\ \mbox{corrections},
\label{eqthermo31}
\ee
where $a_u=2, a_s=1$.
To get the corrections\footnote{Corrections to  (\ref{eqthermo31}) are negligible only if $\beta M_i\gtrsim 40$. However their number is limited for  $\beta M_i$ as low as 4.
In appendix~\ref{Appendixhightemplimit}, we quantitatively discuss the importance of these corrections with respect to the value of $\beta M_i$.}, it is better to work with~(\ref{eqthermo23}) written in terms of $K_2$,
of which the asymptotic behavior is well known. The method is explained in appendix \ref{Appendixhightemplimit} which also contains the low temperature zero density expansion of the energy and entropy densities.

The results relative to this section are given in section \ref{resultsThermodynamics}. They necessitate the knowledge of the behavior of the condensates as a function of temperature. 

\subsection{Results: condensates and thermodynamical functions at finite temperature and density}
\label{results}

We  show the variation as a function of temperature and density of the quark and gluon condensates  in section \ref{resultsCondensates}. The  results concerning  the critical surface $T(\mu_u,\mu_s)$ are given in section \ref{sectioncriticalsurface}. We discuss the results relative to the thermodynamics  in section \ref{resultsThermodynamics}.

\subsubsection{Condensates}
\label{resultsCondensates}

The quark and gluon condensates are given by eqs. (\ref{eq13}--\ref{gapchi2}). The constituent quark masses $M_u$ and $M_s$ used in these equations come from the gap equations (\ref{gapup},\ref{gapstrange}). When $m_u=0$, $M_u=0$ is always a solution of (\ref{gapup}), so that it has to be checked if it corresponds to a greater pressure. We define a second order transition to be a transition for which the slope of the pressure as a function of the external parameters $T,\mu$ is continuous (this includes both true second order transitions and crossovers); otherwise it is said to be of the first order. We show results for four sets of parameters
\beqn
(i)\  & M_u^0= 450\ \mbox{MeV}, &\chi_0=350\ \mbox{MeV,}
\label{eq222a}\\ 
(ii)\  & M_u^0= 600\ \mbox{MeV},&\chi_0=125\ \mbox{MeV,}
\label{eq222b}\\
(iii)\  & M_u^0= 300\ \mbox{MeV},&\chi_0=80\ \mbox{MeV,}
\label{eq222c}\\ 
(iv)\  & M_u^0= 800\ \mbox{MeV},&\chi_0=90\ \mbox{MeV,}
\label{eq222d}
\eeqn
for which the model parameters have been obtained in table~\ref{table1}, section \ref{sectionparameters}. Different behaviors show up according to the chosen set and to the external parameters: we can have two first order transitions, coincident or not, or two second order transitions. We can also have a second order transition for one species of quark while the other experiences a first order transition. A discontinuous slope for the pressure means a discontinuous condensate: there is a mass gap. This is illustrated in figure \ref{figure1.eps}.

Since the pressure at point A is identical to the pressure at  point D, the transition takes place between these two points. Only the region from B to C is unstable: there is metastability between A and B and between C and D.\par

Note that looking graphically at the pressure to find the transition point is manageable only in the two degenerate flavor model at zero density or temperature. In that case, the strategy is the following. Given  $M_u$, the corresponding $\chi_s$ can be extracted from eq.~(\ref{gapchi2}) with the strange quark contribution removed (two-flavor case).
We can then use this couple ($M_u,\chi_s$) to extract the corresponding temperature from the gap equation, eq.~(\ref{gapup}). There is only one solution. (Should we have fixed the temperature, we should have to solve two coupled equations (\ref{gapup},\ref{gapchi2}) with the problem that, for a first order transition, several solutions are possible.) We  then get the ABCD curve of figure \ref{figure1.eps}(a). To get the transition point it is then enough to look at the pressure, figure \ref{figure1.eps}(b).

\begin{minipage}[t]{\textwidth}
\begin{figure}
\vbox{\begin{center}\psfig{file=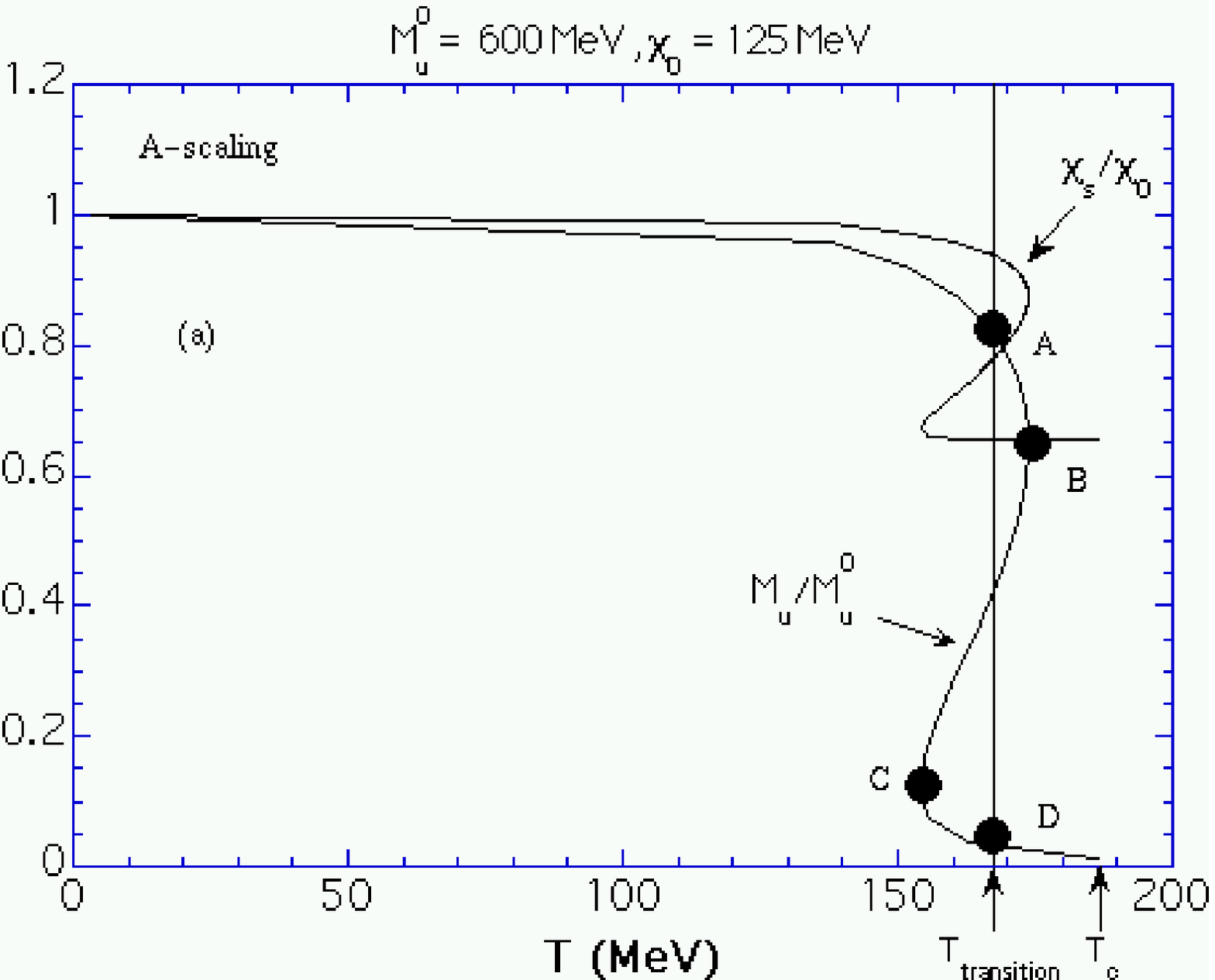,width=\widthepsxmgr}\end{center}}

\vspace{-0.5cm}

\vbox{\begin{center}\quad\psfig{file=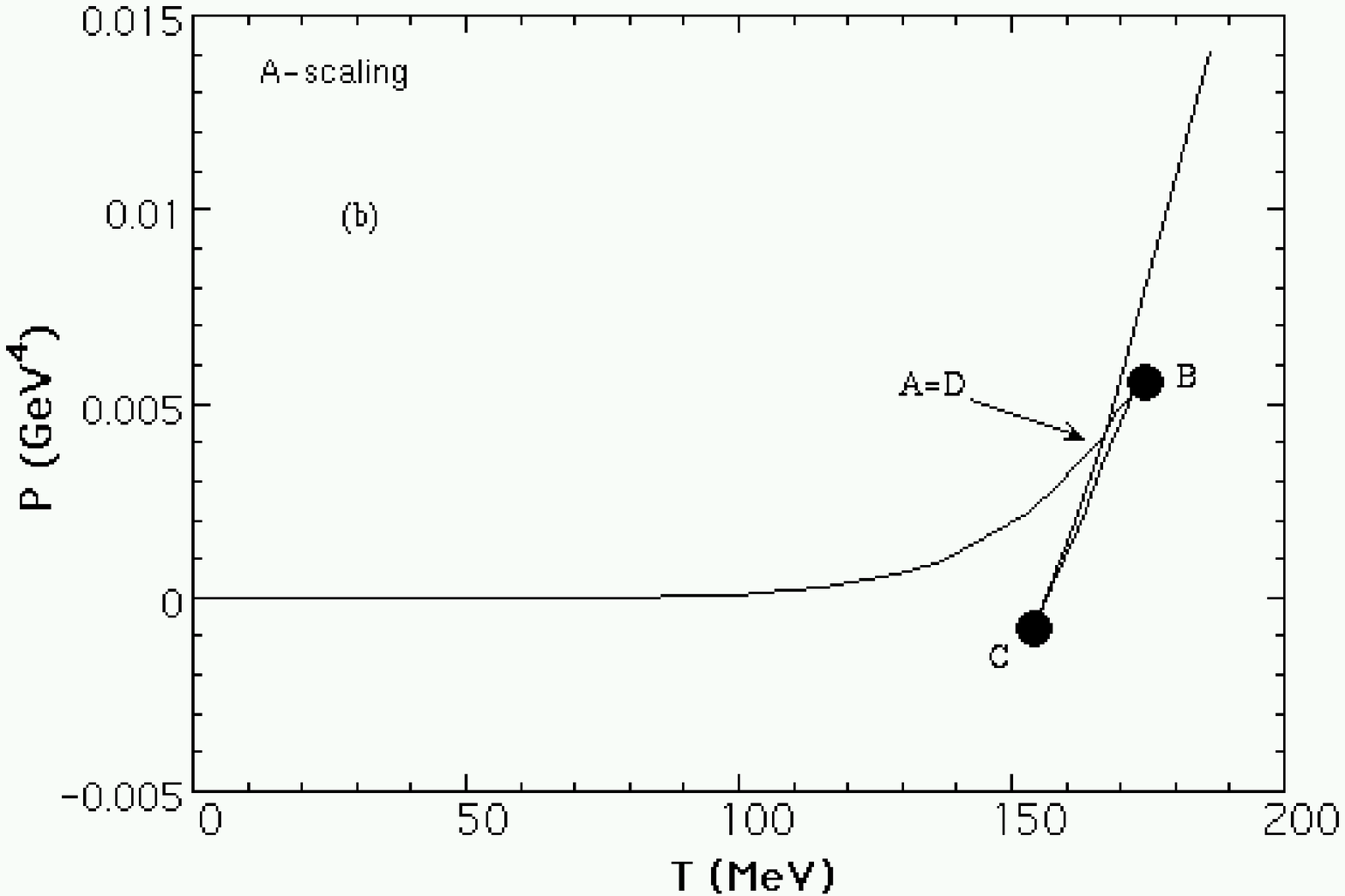,width=\widtheps}\end{center}}
\caption{\em Up constituent quark mass transition (with $m_u\ne0$).
\label{figure1.eps}}
\end{figure}
\end{minipage}

\vspace{0.7cm}

Of course, as soon as more variables ($\mu_u,\mu_s,M_s$) are introduced, this strategy is not anymore of interest. We have to solve  the full set of coupled equations (\ref{gapup},\ref{gapstrange},\ref{gapchi2}) as a function of chemical potentials and temperature. Because of possible first order transitions, several local extrema of the pressure show up so that it is necessary to use a numerical algorithm searching for global extrema. In our work \cite{cugjamvdb96}, we used the simulated annealing algorithm that we  adapted from \cite{simann}.\par

\begin{figure}[hbt]
\vbox{\begin{center}\psfig{file=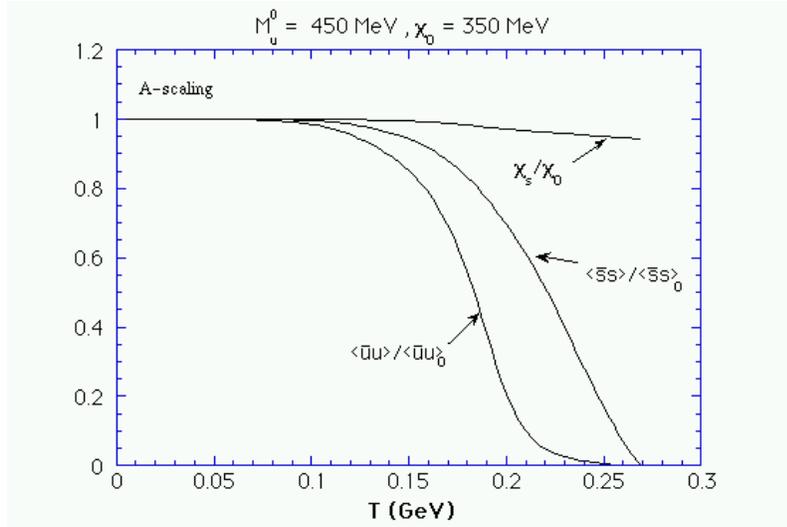,width=\widtheps}\end{center}}
\caption{\em Quark and gluon condensates as a function of temperature for the parameters $M_u^0=450 \mbox{ \rm MeV}, \chi_0=350$ {\rm MeV} ($\approx$ pure NJL model).
\label{figuresu31a.eps}}
\end{figure}

\begin{figure}[hbt]
\vbox{\begin{center}\psfig{file=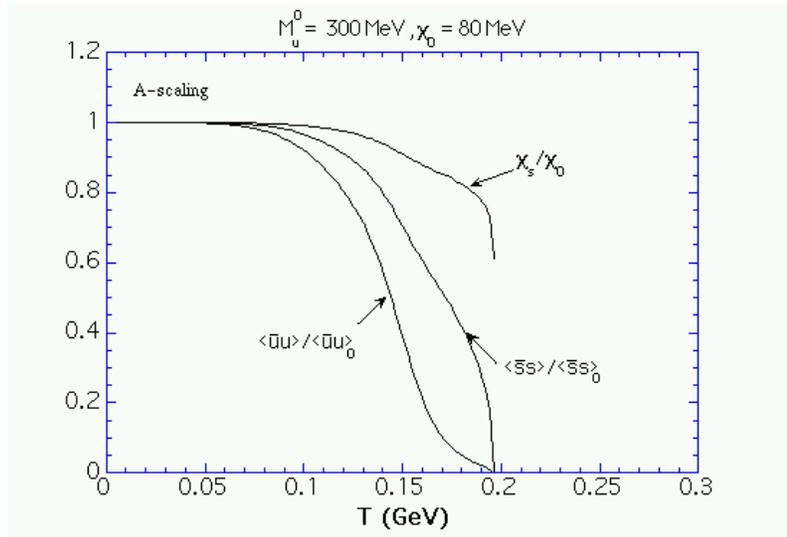,width=\widtheps}\end{center}}
\caption{\em Quark and gluon condensates as a function of temperature for the parameters $M_u^0=300 \mbox{ \rm MeV}, \chi_0=80$ {\rm MeV}.
\label{figuresu31b.eps}}
\end{figure}

Figures \ref{figuresu31a.eps} and \ref{figuresu31b.eps} are for the sets (\ref{eq222a}) and (\ref{eq222c}), respectively, and correspond to a second order phase transition\footnote{See figure \ref{figureZeit1.eps}, section \ref{sectionparameters}, for the order of the transition versus the parameters ($M_u,\chi_s$).} w.r.t. the temperature for vanishing chemical potentials.
 They are similar to results obtained in the two-flavor case \cite{jamvdb94,jamvdb94b}.
 The choice (\ref{eq222a})  almost corresponds to a pure NJL model, while the choice (\ref{eq222c}) leads to a lower critical temperature, in agreement with lattice results \cite{karsch95,lattice96,kogut91,brown93b}. \par

If we take the chiral limit analogue of figure \ref{figuresu31a.eps}, the critical temperature is found to be $T_c\approx 193$ MeV. The gluon condensate is almost flat, so that the quark and gluon condensates are almost uncoupled. The scaled model is then essentially a pure NJL. In figure \ref{figuresu31b.eps} this coupling is more important and the chiral limit gives $T_c\approx 140$ MeV.
 Working in the three-flavor version of the model, these two pictures show also the strange quark condensate, for which we can make two remarks:
\begin{itemize}
\item $<\bar{s}s>$ decreases slower than $<\bar{u}u>$, in agreement with \cite{blaizot91}. This can be traced back to the greater constituent strange quark mass compared to the up quark (see table~\ref{table1}) which, in turn, is a consequence of a greater current strange quark mass compared to the up one\footnote{Note that, in the two-flavor limit where $M_s,m_s\rightarrow\infty$, we have  $<\bar{s}s>/<\bar{s}s>_0=1$.}. Note however that our results show a  faster decrease of $<\bar{s}s>$  with temperature than in \cite{klevansky95}; 
\item because of the coupling between the condensates, the gluon condensate at the transition, $\chi_c$, is smaller than in the two-flavor case. For the set (\ref{eq222c}), we have $\chi_c(SU(2))\approx 0.8$ while figure \ref{figuresu31b.eps} shows $\chi_c(SU(3))\approx 0.6$.
\end{itemize}

\vspace{-0.5cm}

\vspace{-1.0cm}

\begin{figure}[hbt]
\vbox{\begin{center}\psfig{file=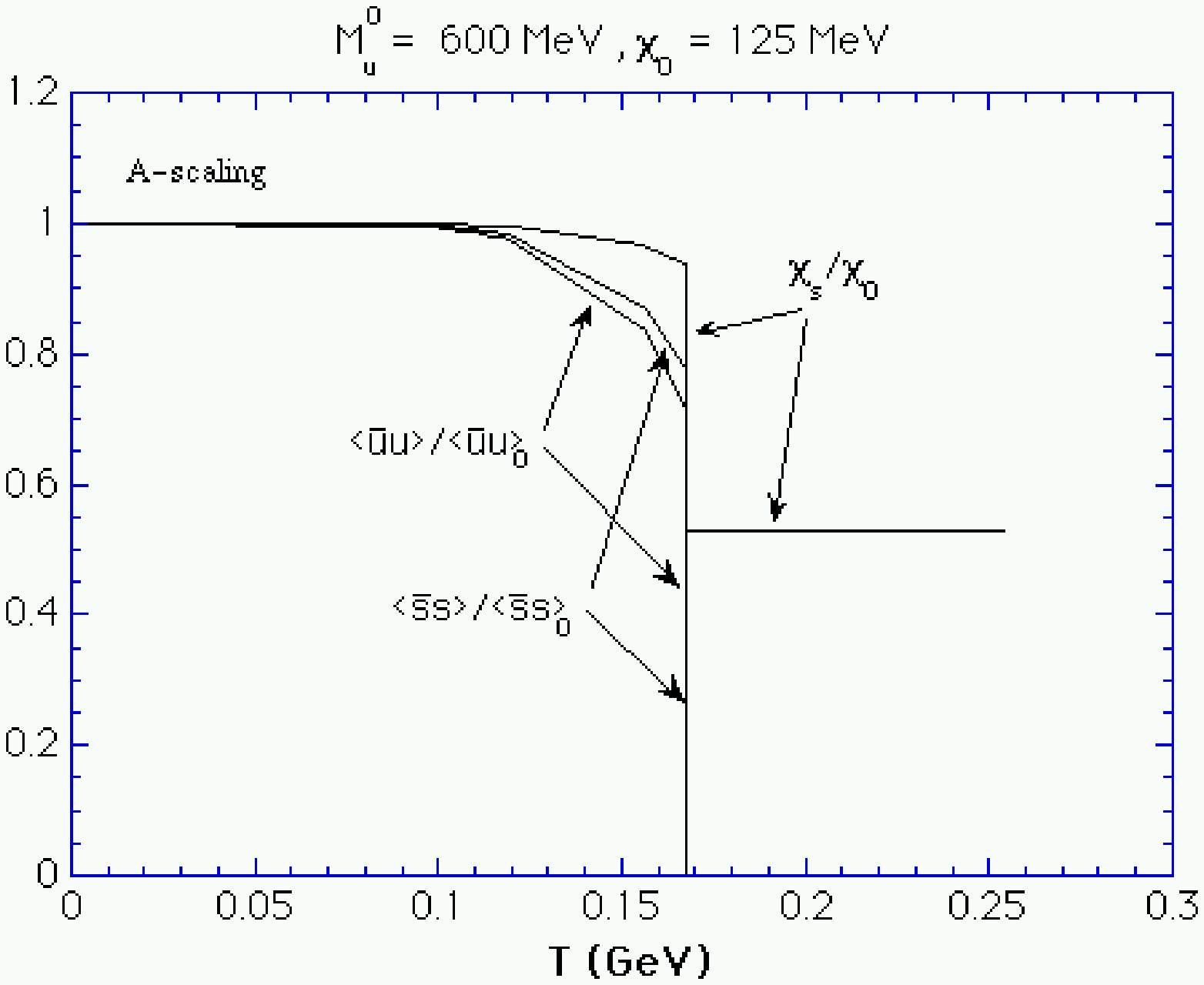,width=\widtheps}\end{center}}
\caption{\em Quark and gluon condensates as a function of temperature for the parameters $M_u^0=600 \mbox{ \rm MeV},\chi_0=125$ {\rm MeV}.\label{figuresu32a.eps}
}
\end{figure}

\begin{figure}[hbt]
\vbox{\begin{center}\psfig{file=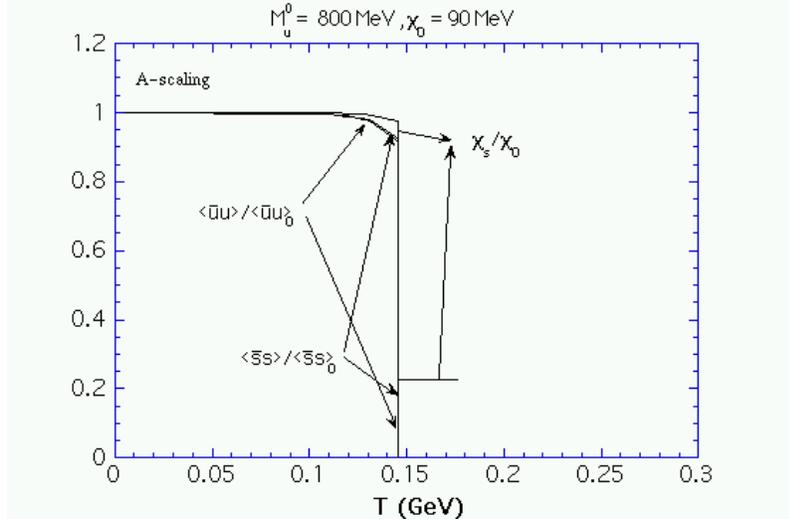,width=\widtheps}\end{center}}
\caption{\em Quark and gluon condensates as a function of temperature for the parameters $M_u^0=800 \mbox{ \rm MeV}, \chi_0=90$ {\rm MeV}.\label{figuresu32b.eps}
}
\end{figure}

Figures \ref{figuresu32a.eps} and \ref{figuresu32b.eps} are the analogues of figures \ref{figuresu31a.eps} and \ref{figuresu31b.eps} for the sets of parameters (\ref{eq222b}) and (\ref{eq222d}), respectively. These two sets allow to get a first order phase transition \cite{jamvdb94,jamrip92}. With the second one, however, the critical temperature is  lowered down to $T_c\approx140$ MeV, in agreement with lattice QCD \cite{karsch95,lattice96,brown93b}.\par

With these  two sets of parameters, the coupling between quark and gluon condensates is so strong that all the condensates undergo the transition together. Although these figures suggest that a first order transition is connected to a low value of the gluon condensate, both $M_u^0$ and $\chi_0$ really determine it, as shown in figure \ref{figureZeit1.eps} and confirmed in figure \ref{figuresu31b.eps}. 


\vspace{-1cm}

At zero density, the above analysis shows that:
\begin{itemize}
\item we can reproduce both first and second order phase transitions. A first order transition is typically an effect due to the gluon condensate which then does not show up in pure NJL models;
\item a low critical temperature, as low as 140 MeV,  can be reproduced. This is clearly related to the coupling between quark and gluon condensates. Pure NJL models are then unable to reproduce such a low temperature: they cannot go below $T_c\approx 190$ MeV. According to eq.~(\ref{eqtempcritique}), 
scaled models allow (for  second order transitions) a reduction of the critical temperature in the ratio $\chi_c/\chi_0$ compared to a pure NJL model;
\item the coupling between quark and gluon condensates is mainly driven by the value of the vacuum gluon condensate $\chi_0$. With large $\chi_0$, the coupling is weak while  the coupling becomes more and more important as we decrease $\chi_0$. The quark or gluon condensate can then be considered as the order parameter for the phase transition.
\end{itemize}

It should however not be forgotten that all the above analysis is performed without vector mesons. Without them, our results show that high values of the gluon condensate are needed to get a transition above the normal nuclear matter density $\rho_0$. This is illustrated for two flavors in figure \ref{figureZeit5a.eps}. \par

\vspace{-0.5cm}

\begin{figure}[hbt]
\vbox{\begin{center}\psfig{file=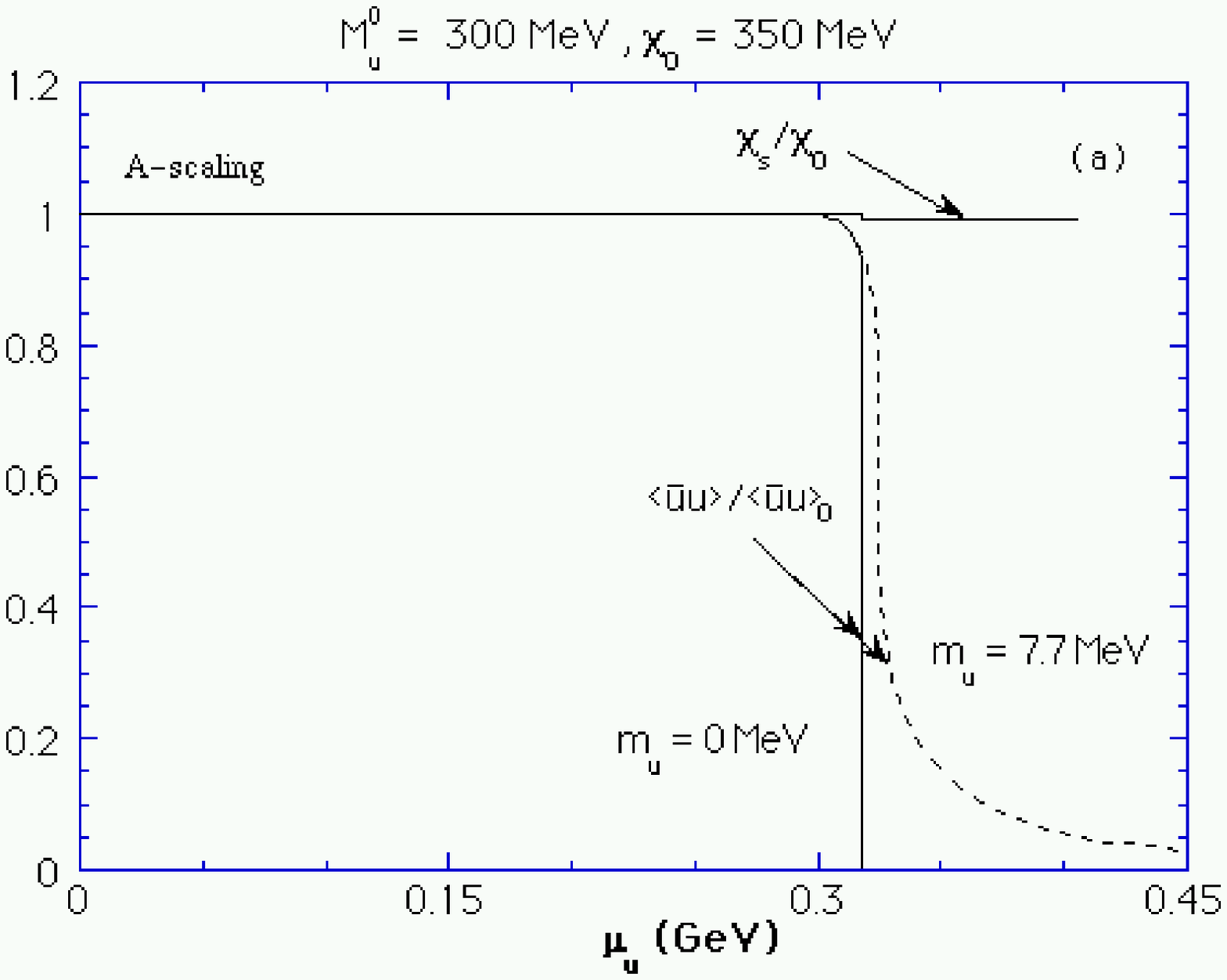,width=\widtheps}\end{center}}
\end{figure}

\vspace{-1.5cm}

\begin{figure}[hbt]
\vbox{\begin{center}\psfig{file=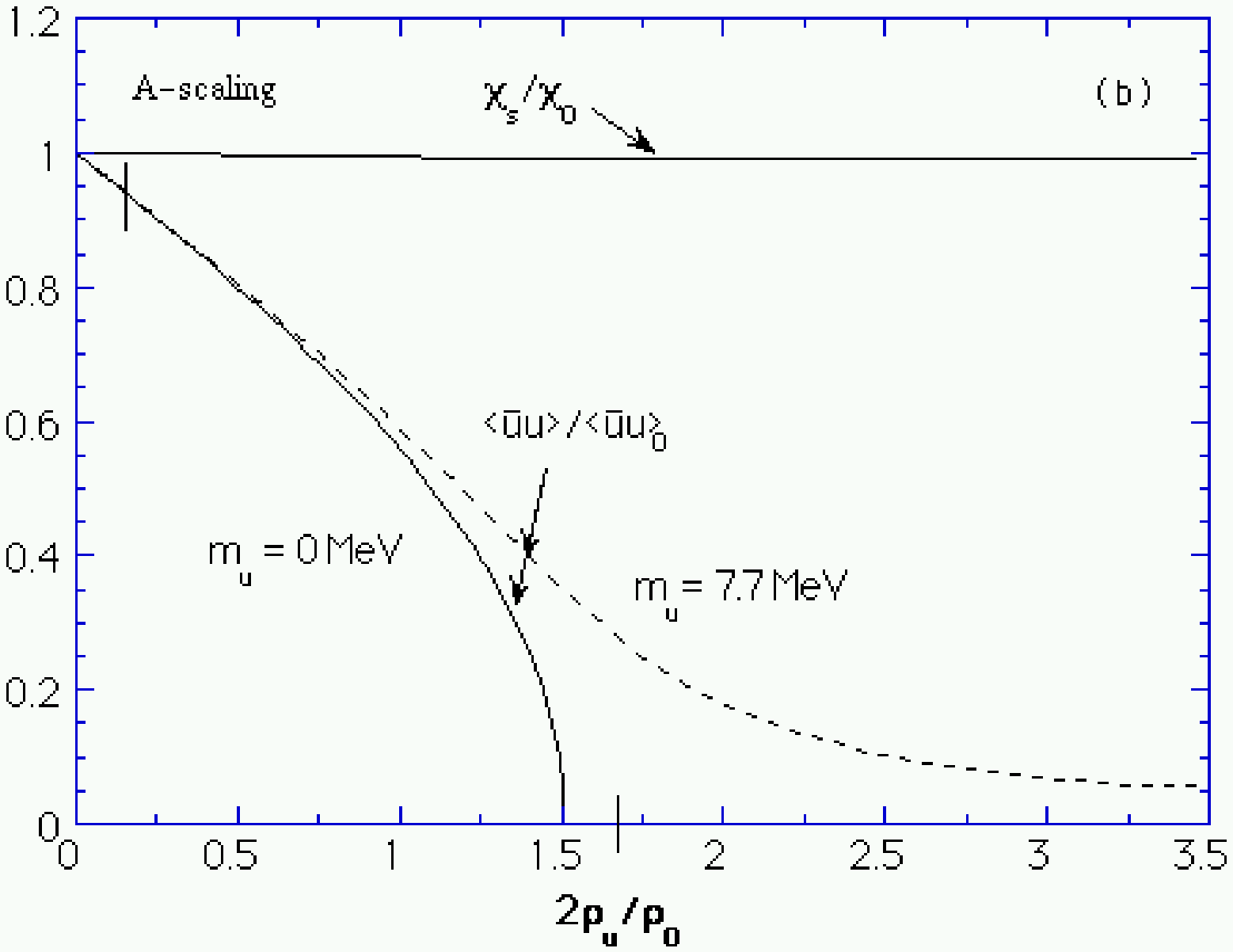,width=\widtheps}\end{center}}
\end{figure}

\vspace{-1.5cm}

\begin{figure}[hbt]
\vbox{\begin{center}\psfig{file=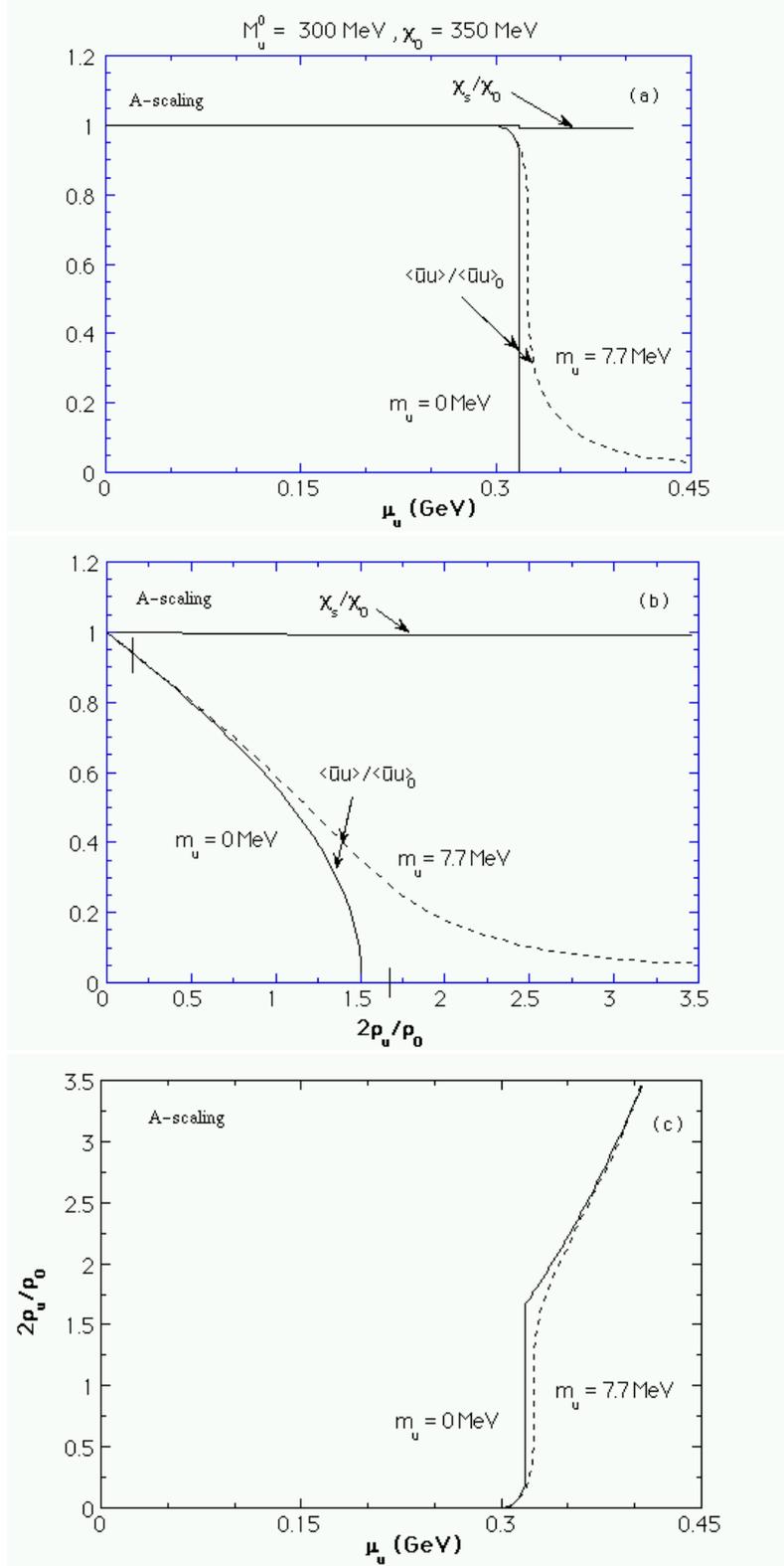,width=\widtheps}\end{center}}
\caption{\em Quark and gluon condensates for the parameters $M_u^0=300 \mbox{ \rm MeV},\chi_0=350$ {\rm MeV} for $m_u=0$ (plain line) and $m_u\ne0$ (dashed line) as a function of the chemical potential (a); as a function of the density (b); (c) shows  $\rho_u$ versus $\mu_u$.\label{figureZeit5a.eps}
}
\end{figure}


For $M_u^0=300$ MeV, the smallest required value of the vacuum gluon condensate, for a second order transition, is $\chi_0=350$ MeV. Even in this case the chiral limit leads to a first order transition. It is however clear from figure \ref{figureZeit5a.eps}(a)  that, although sharp, the transition is of  second order if $m_u$ is different from zero. Figure \ref{figureZeit5a.eps}(b) is also instructive is the sense that it shows clearly that looking at the density is misleading to have insights on the transition. The dashed curve is a second order transition while the plain curve is a first order transition. The transition occurs from below the normal nuclear density to above the normal nuclear density (the phase between the two vertical bars is unstable). The fact that a small current quark mass can stabilize the system has already been noticed in \cite{weise92b} in a pure NJL. Since a pure NJL does not lead to a first order transition at zero density finite temperature, this indicates that the temperature has a stabilizing effect on the transition.\par

Although the scaled models are not  able to reproduce a transition above the normal nuclear density for a small value of the gluon condensate (which leads to a low critical temperature), it  does not mean that they are inefficient. Indeed, it is well known that vector mesons make the vacuum stiffer against the restoration of chiral symmetry \cite{jamrip91,jamrip92bis}. Including these mesons will then correct what seems to be, at first sight, a drawback of the model.

\subsubsection{Critical surface $T_c(\mu_u,\mu_s)$}
\label{sectioncriticalsurface}

\vspace{-.5cm}

\begin{figure}[hbt]
\vbox{\begin{center}\psfig{file=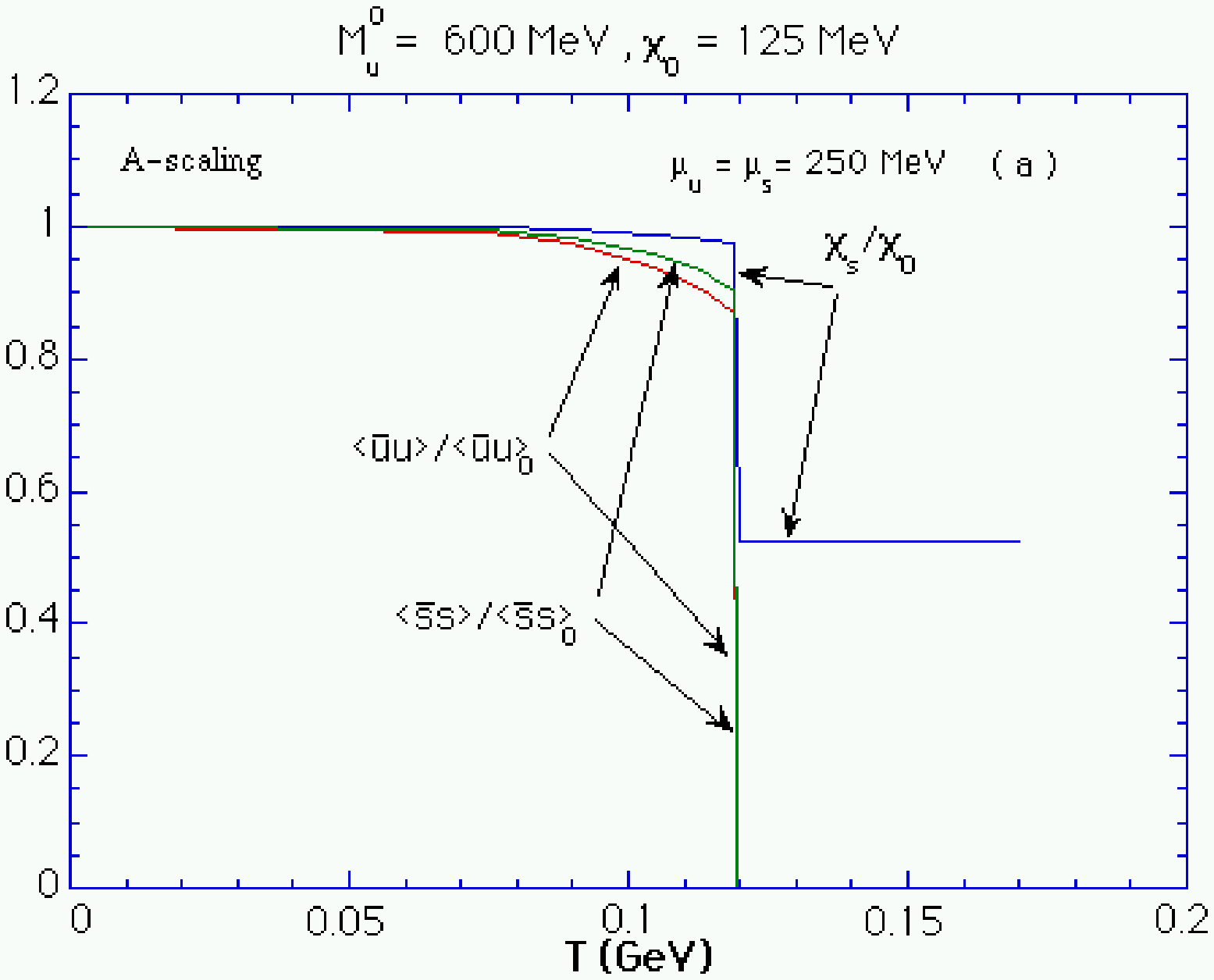,width=\widtheps}\end{center}}
\end{figure}

\vspace{-1.35cm}

\begin{figure}[hbt]
\vbox{\begin{center}\psfig{file=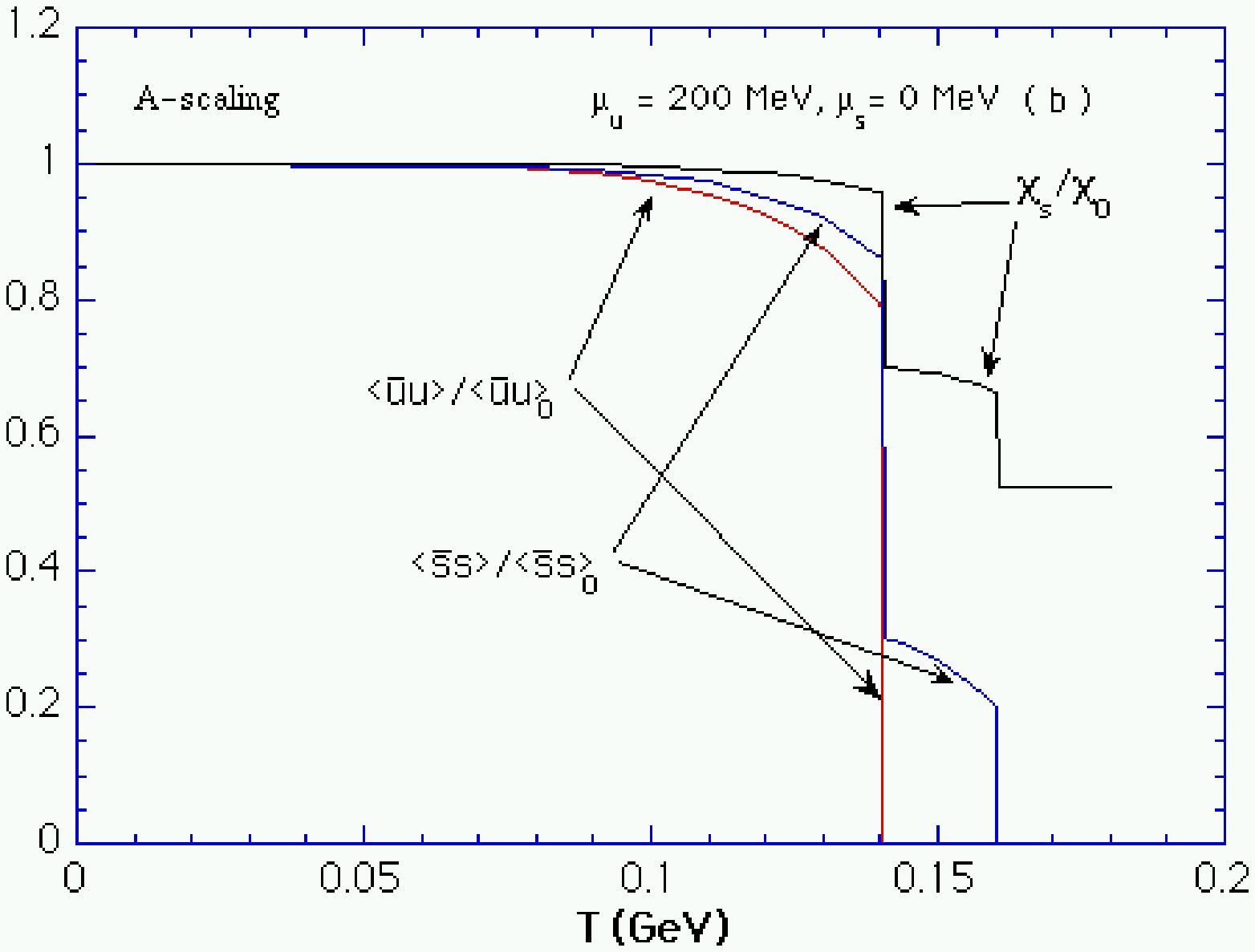,width=\widthepstwo}\end{center}}
\caption{\em Quark and gluon condensates as a function of temperature for the parameters $M_u^0=600 \mbox{ \rm MeV},\chi_0=125$ {\rm MeV} in the light quark chiral limit $m_u=m_d=0$ for $\mu_u=\mu_d=\mu_s=250$ {\rm MeV} (a); for $\mu_u=\mu_d=200, \mu_s=0$ {\rm MeV} (b).\label{figurethermo2a.eps}}
\end{figure}

In this section, we add to the results of the previous section the effect of the density through the chemical potentials  $\mu_u=\mu_d$ (symmetric matter) and $\mu_s$.
We base our discussion on the work \cite{cugjamvdb96} which is a systematic study of the chiral phase transition, for three flavors, as a function of temperature  and chemical potentials. To eliminate the long tail due to finite current quark masses, we work  in the chiral limit for the light quarks ($m_u=m_d=0$) while we keep $m_s\ne0$ which is the only source (apart from the chemical potentials) of  chiral symmetry breaking.  We also restrict ourselves to the sets of parameters (\ref{eq222b},\ref{eq222c}) which yield, at zero density, a low transition temperature of  respectively first or second order. \par

Adding chemical potentials to the temperature study is interesting because it shows new behaviors:  separated transitions for up and strange quarks, both of  first order or first and second order. This is illustrated in figures \ref{figurethermo2a.eps} and \ref{figurethermo4a.eps}.

Figure \ref{figurethermo2a.eps}(a) is quite similar to figure \ref{figuresu32a.eps}: both transitions are of  first order and  coincident. However, the addition of a chemical potential shows that the critical temperature is reduced, and eventually vanishes, if the density is high enough. This is illustrated in figure \ref{artisticview600.eps}. 
Figure \ref{figurethermo2a.eps}(b) gives a new behavior: the up and strange transitions are disconnected except through the gluon condensate $\chi$. In fact, when the up quarks make their transitions, eq.~(\ref{gapchi2}) for the $\chi$ field  shows a discontinuity which also implies that the strange quarks feel the up ones making the transition. This is what happens around $T\approx 140$ MeV. However this is not the true strange quark transition: at $T\approx 160$ MeV, the strange quarks perform their own first order transition.\par

\vspace{-0.5cm}

\begin{figure}[hbt]
\vbox{\begin{center}\psfig{file=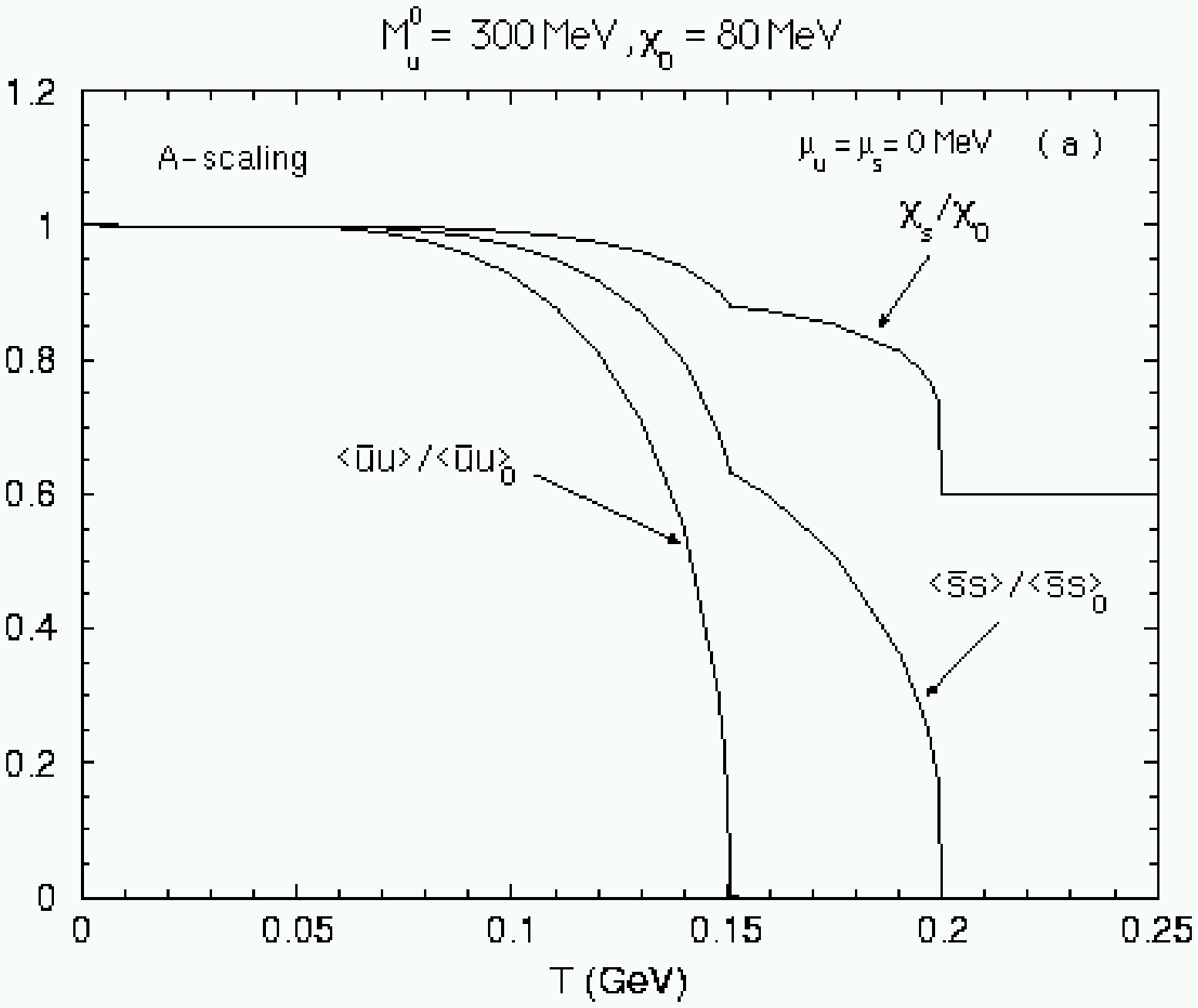,width=\widthepsxmgrtwo}\end{center}}
\end{figure}

\vspace{-1.75cm}

\begin{figure}[hbt]
\vbox{\begin{center}\psfig{file=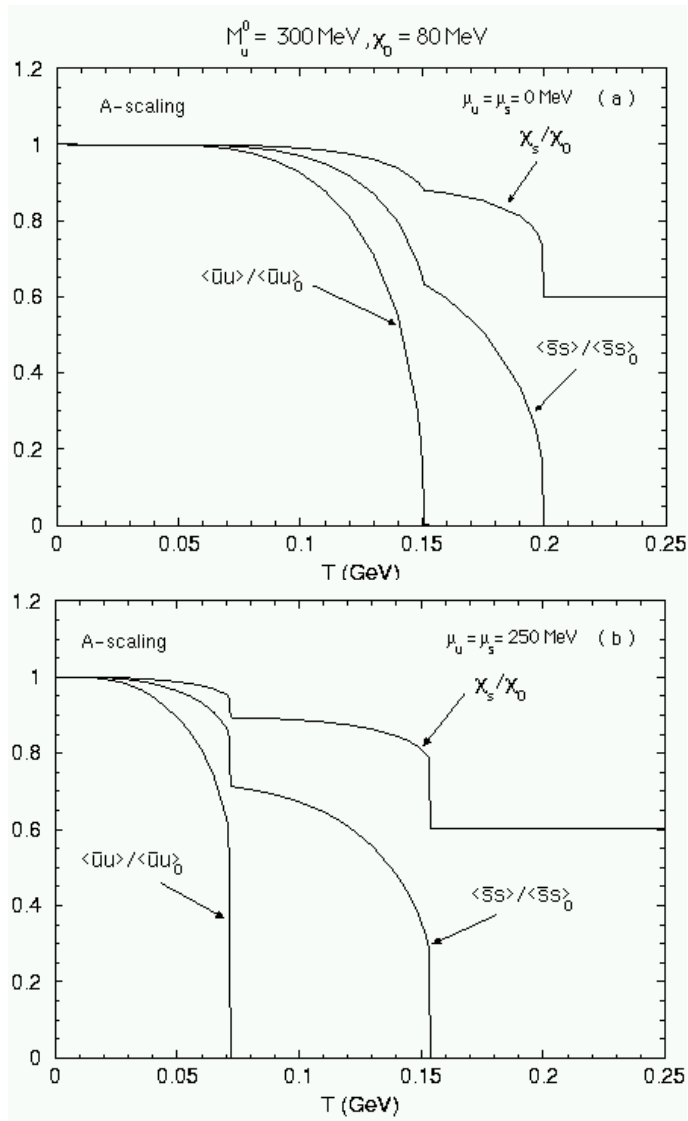,width=\widthepsxmgrtwo}\end{center}}
\caption{\em Quark and gluon condensates as a function of temperature for the parameters $M_u^0=300 \mbox{ \rm MeV},\chi_0=80$ {\rm MeV} in the light quark chiral limit $m_u=m_d=0$ for $\mu_u=\mu_d=\mu_s=0$ {\rm MeV}~(a); for $\mu_u=\mu_d=\mu_s=250$~{\rm MeV}~(b).\label{figurethermo4a.eps}}
\end{figure}

Figure \ref{figurethermo4a.eps}(a) is the analogue of figure \ref{figuresu31b.eps} for vanishing light quark masses. It shows a second order transition for the up condensate while the strange condensate feels a weak first order transition.

 It is interesting to notice that the tail in figure \ref{figuresu31b.eps} has completely disappeared, showing that the transition for the up quarks (for vanishing density) occurs at about 150 MeV (140 MeV, would we have only two flavors). It also shows that a small up current quark mass has a stabilizing effect for the strange transition.  This is traced back to the mixing of the up and strange quark condensates through the gluon condensate.

Figure \ref{figurethermo4a.eps}(b) is for the same set of parameters as figure 
\ref{figurethermo4a.eps}(a) but for nonvanishing chemical potentials. This explains why the transition temperatures are reduced. Note that we have two noncoincident first order phase transitions. Because the coupling between the condensates is smaller than in the case shown in figure \ref{figurethermo2a.eps}, the first transition of the strange quarks, connected to the transition of the up quarks, leads only to a small mass gap. Its true transition occurs around $T\approx 150$ MeV.\par
As we have seen, up and strange transitions can be completely disconnected. This is also illustrated in figures \ref{figurethermo1a.eps} and \ref{figurethermo5a.eps} for the sets of parameters (\ref{eq222b},\ref{eq222c}), respectively.
Open circles represent the phase transition $\mu_{sc}$ as a function of $\mu_{uc}$. For figure \ref{figurethermo1a.eps},  they correspond to a vanishing up quark condensate $<\bar{u}u>$ at a temperature  (figure \ref{figurethermo1a.eps}) $T_c=0$ MeV (a), $T_c=100$ MeV (b) and $T_c=161$ MeV (c). The transition associated with the strange quark condensate $<\bar{s}s>$ is given by the crosses. 

We also observe  on figure \ref{figurethermo1a.eps} that, at low temperature, up and strange transitions are independent as in a pure NJL without 't Hooft determinant, except in a small region of the plane $(\mu_{sc},\mu_{us})$ where the coupling with the gluon condensate is high enough to constrain the two condensates to vanish coincidently. This region is the only one to survive when the temperature is increased.
 At any  point $(T_c,\mu_{sc},\mu_{us})$, the transition is of the first order,
which is well illustrated on figure \ref{figurethermo2a.eps} in the case of a strong coupling between the condensates (figure \ref{figurethermo2a.eps}(a): $T_c\approx$ 120
MeV, $\mu_{uc}$ = $\mu_{sc}$  = 250 MeV) and in the case of a weaker coupling (figure \ref{figurethermo2a.eps}(b): $T_{uc}$ =  140
MeV, $T_{sc}$ = 160 MeV,  $\mu_{uc}$ = 200 MeV, $\mu_{sc}$  = 0
MeV).

\vspace{-0.3cm}

\begin{figure}
\vbox{\begin{center}\psfig{file=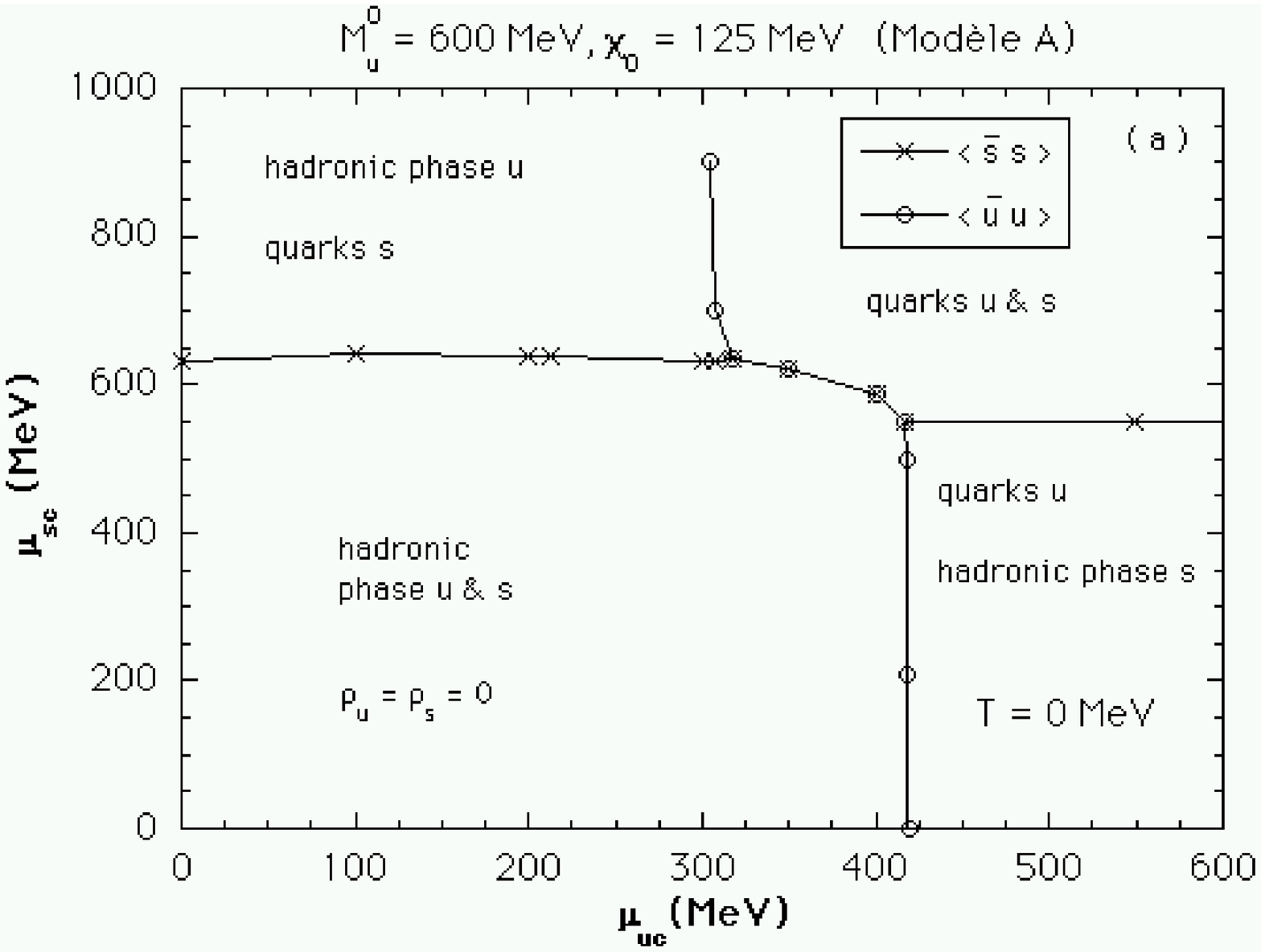,width=\widtheps}\end{center}}
\end{figure}

\vspace{-1.5cm}

\begin{figure}
\vbox{\begin{center}\psfig{file=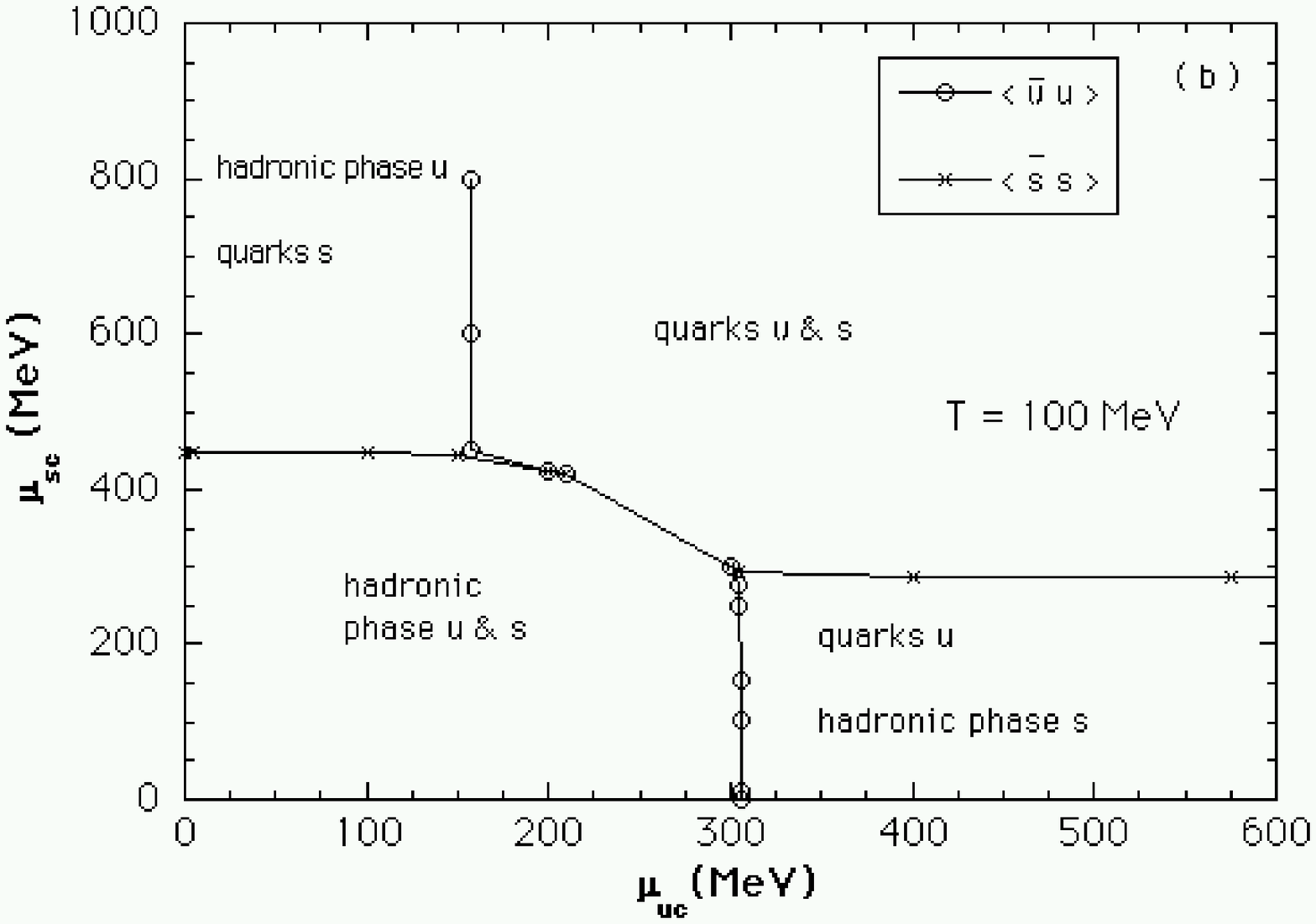,width=\widtheps}\end{center}}
\end{figure}

\vspace{-1.25cm}

\begin{figure}
\vbox{\begin{center}\psfig{file=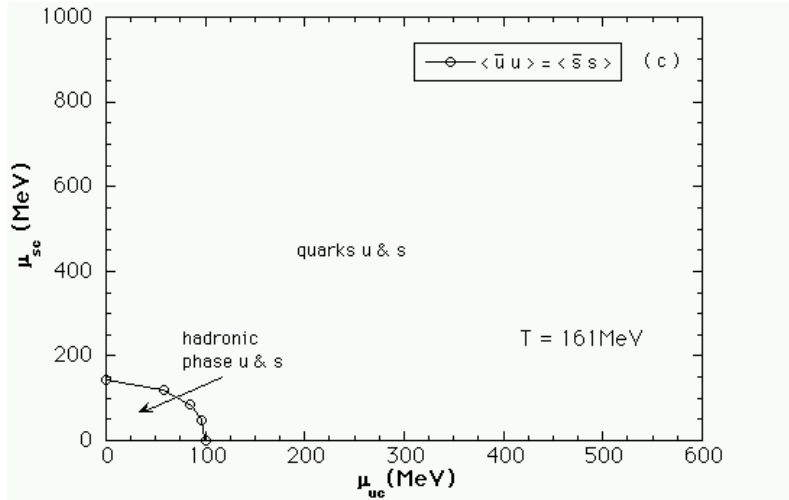,width=\widtheps}\end{center}}
\caption{\em Chiral phase transition for the parameters  $M_u^0=600 \mbox{ \rm MeV},\chi_0=125$ {\rm MeV} in the light quark chiral limit $m_u=m_d=0$. $\mu_{sc}$ as a function of $\mu_{uc}$ for $T=0$ {\rm MeV} (a); for $T=100$ {\rm MeV} (b); for $T=161$ {\rm MeV} (c).\label{figurethermo1a.eps}}
\end{figure}

\vspace{.7cm}

Note that, in our model, a discontinuity in the behavior of one of the quark condensates implies a discontinuity of the gluon condensate, in contradiction with the results\footnote{The authors of this reference also make the distinction between the two scales of deconfinement and restoration of  chiral symmetry, arguing that values of the MIT bag constant has more to do with chiral symmetry than with quark confinement (see also \cite{brown93b,brown93}).} of \cite{brown93c}. Note also that, above the transition, $\chi_s\ne0$: gluons are never completely deconfined.

Figure \ref{artisticview600.eps} sums up the preceding results in the form of an ``artistic'' view.


\begin{figure}[hbt]
\vbox{\begin{center}\psfig{file=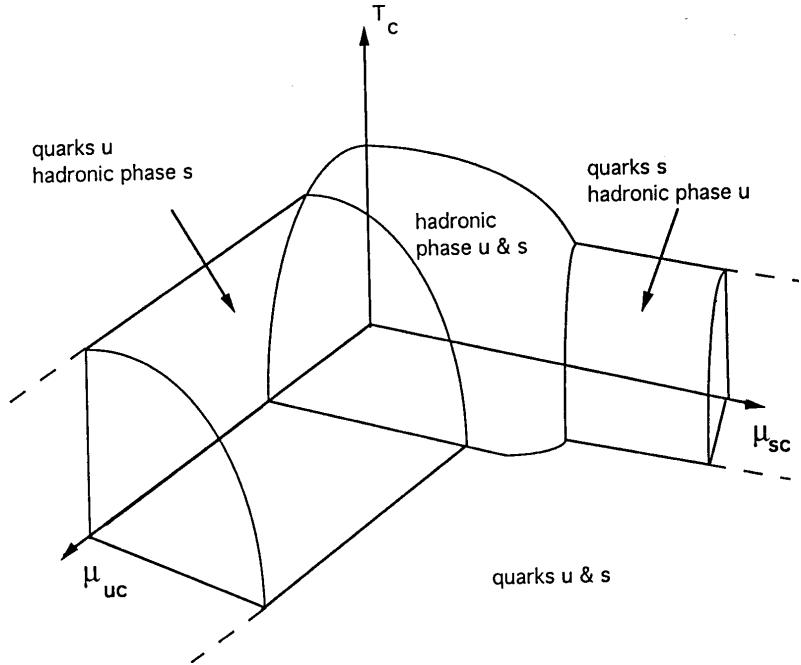,width=\widtheps}\end{center}}
\caption{\em Chiral phase transition summing up results from figure \ref{figurethermo1a.eps} for the set of parameters (\ref{eq222b}).\label{artisticview600.eps}}
\end{figure}

The space is divided into four regions: the internal region corresponds to quarks in the condensed phase or, in our terminology, in the hadronic phase; the external region is interpreted as the quark-gluon plasma phase; the last two regions, extending along the axis up to infinity, correspond to mixed phases with free up (strange) quarks in presence of strange (up) quarks in the hadronic phase. We notice an interesting prediction of the model: if non-strange matter (with a high enough  $up$ chemical potential) is heated, we observe the appearance of light free quarks in thermodynamical equilibrium with strange hadrons (of zero net strangeness, because $\mu_s=0$). These strange hadrons then transform into a free strange quark gas if the temperature is again increased.\par

As already mentioned, a drawback of the model shows up when densities are calculated. When the transition is of  first order, there is a density jump at the transition\footnote{Transitions occur at  constant chemical potential.}. The problem is that the jump occurs from before to after the normal nuclear density: in our terminology, the hadronic phase would be unstable. For example, at $\mu_{uc}=\mu_{dc}=\mu_{sc}=250$ MeV, we have $\rho_{u\ \boite{before}}/\rho_0$ = 0.362,
$\rho_{s\ \boite{before}}/\rho_0$ = 0.050 and
$\rho_{u\ \boite{after}}/\rho_0$ = 2.646,
$\rho_{s\ \boite{after}}/\rho_0$ = 0.978, where $\rho_0$ is the normal nuclear density of quarks (0.51/fm$^3$). To cure this defect,  one has to take a large value for the vacuum gluon condensate, while keeping  a small value for the constituent quark mass in the vacuum $M_u^0$, see figure \ref{figureZeit5a.eps}. This means that the model behaves as a pure NJL. In our model, scalar meson masses would then be badly reproduced (roughly speaking, $M_{meson}\approx 2M_u$) and the critical temperature would be large ($T_c\approx 200$ MeV), in contradiction with lattice calculations \cite{karsch95,lattice96,brown93b}. However, as  already indicated, one expects \cite{jamrip93,jamrip91} that the introduction of the vector mesons into the formalism will solve this problem, making the vacuum stiffer against the restoration of  chiral symmetry. This would then allow to get larger densities before the restoration takes place, while maintaining intact the general behavior of the phase transition, since it is mainly related to the chiral mesons and their coupling to the condensates.\par

The preceding results remain qualitatively unchanged when the set of parameters (\ref{eq222c}) is studied, except for the order of the transition: at low  density, the transition is of  second order (see figure \ref{figurethermo4a.eps}(a)). \par
Figures \ref{figurethermo5a.eps} and \ref{artisticview300.eps} are the equivalent of figures \ref{figurethermo1a.eps} and \ref{artisticview600.eps}, respectively.

 Figures \ref{figurethermo1a.eps} and \ref{figurethermo5a.eps} are qualitatively identical, except at high temperature where the quark condensates are always coupled, for any  choice of ($\mu_{uc},\mu_{sc}$), for the set (\ref{eq222b}). For the set (\ref{eq222c}), the up quarks can only exist in the deconfined phase with a restoration of chiral symmetry for strange quarks at $\mu_{sc}\approx$ cst. The results from figure \ref{figurethermo5a.eps} can be summed up in a 3-dimensional space, of which we give an ``artistic'' view in figure \ref{artisticview300.eps}. 

The results from this section form the basis to determine those  on thermodynamics, which are given in the next section.


\begin{figure}
\vbox{\begin{center}\psfig{file=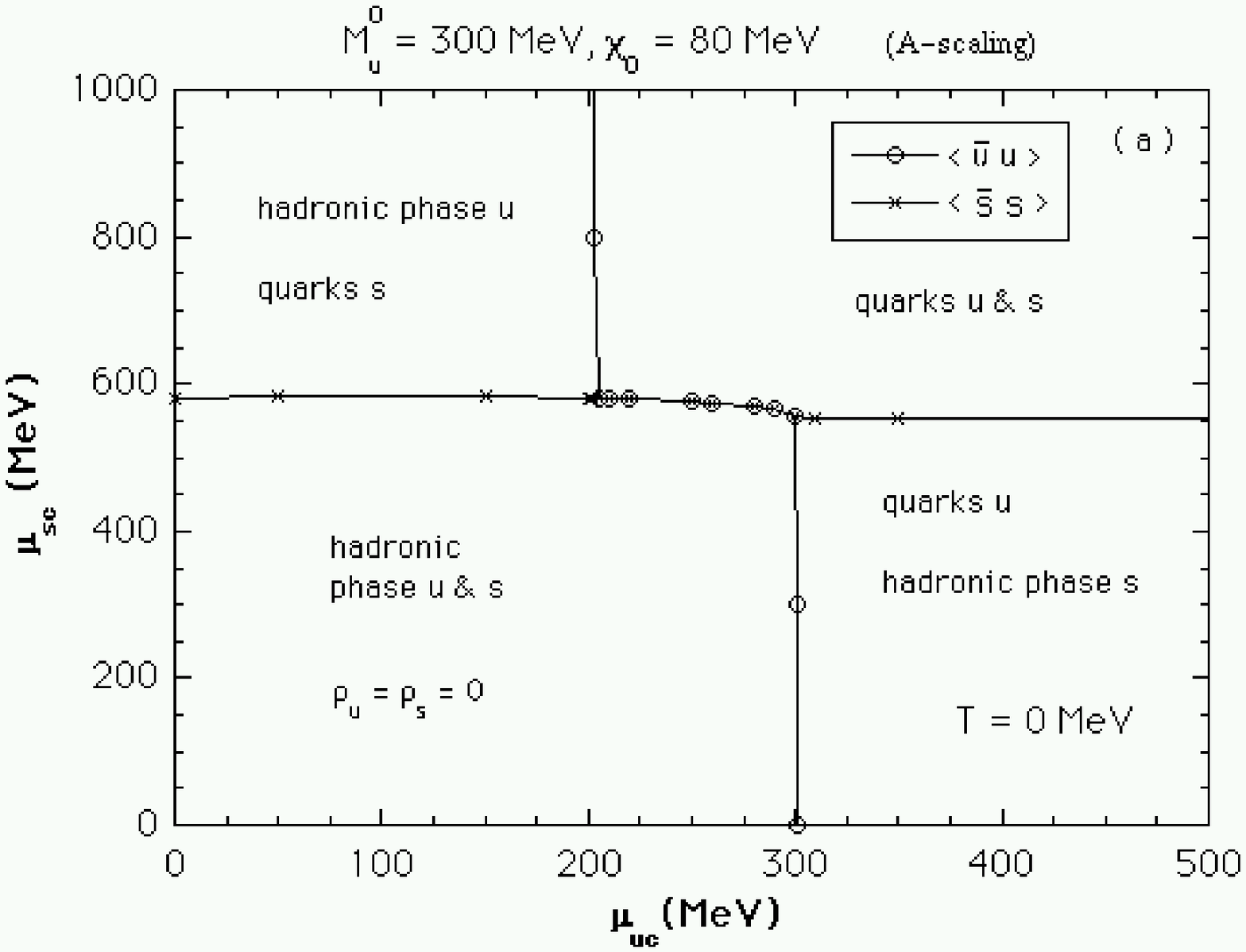,width=\widthepsthree}\end{center}}
\end{figure}

\vspace{-1.5cm}

\begin{figure}
\vbox{\begin{center}
\psfig{file=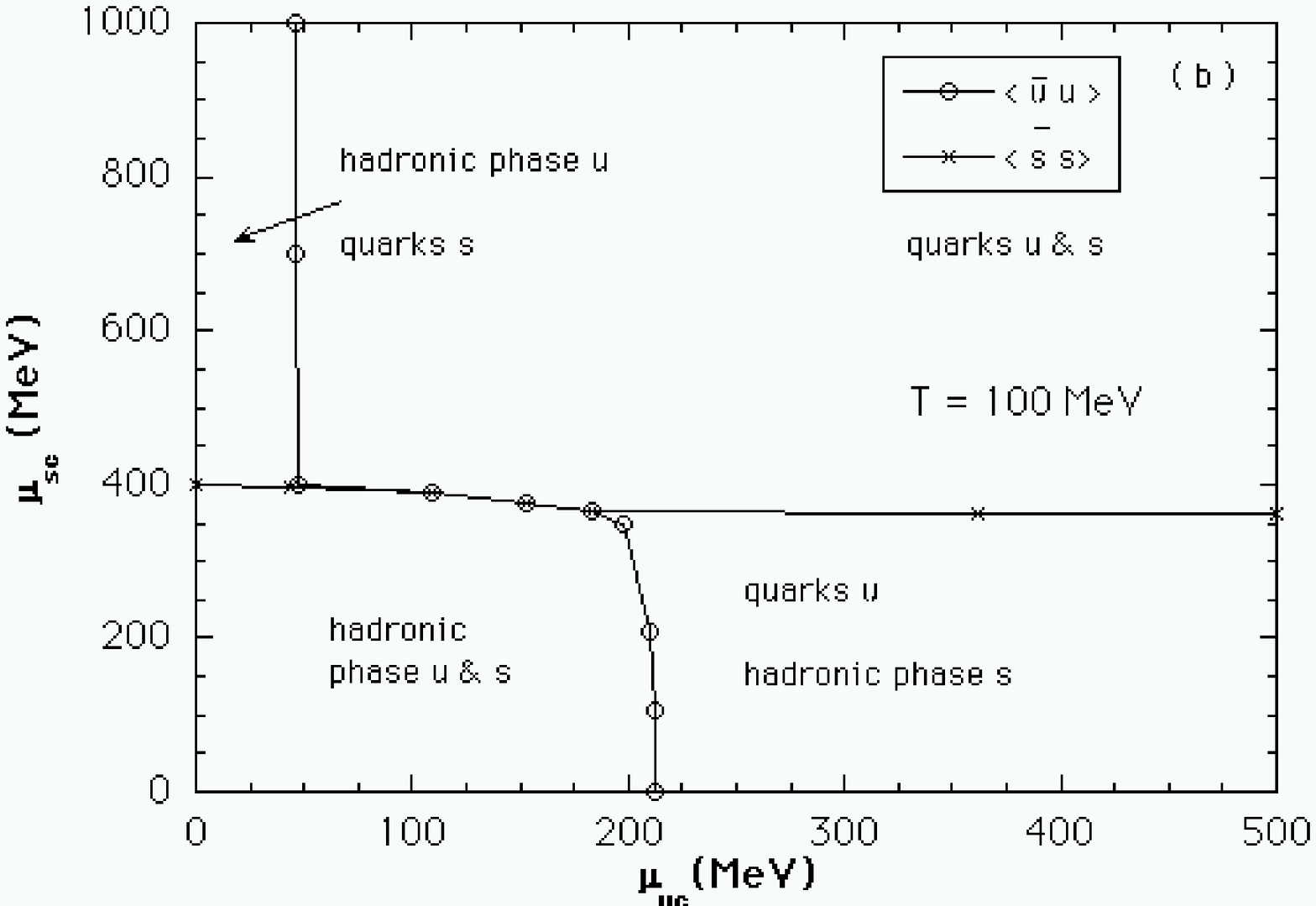,width=\widthepsfour}
\phantom{co}
\end{center}}
\end{figure}

\vspace{-1.5cm}

\begin{figure}
\vbox{\begin{center}\psfig{file=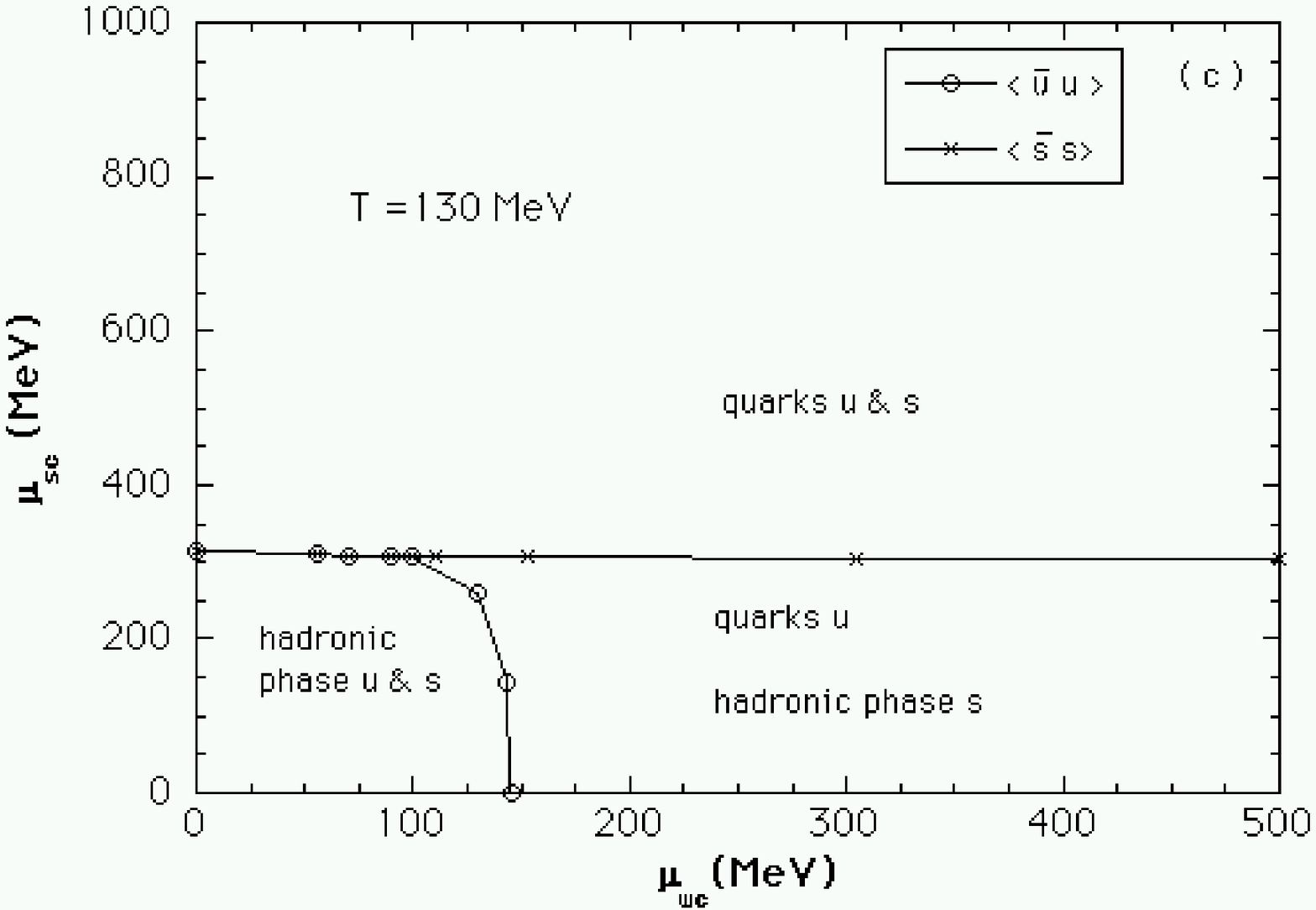,width=\widthepsthree}\end{center}}
\end{figure}

\vspace{-1.5cm}

\begin{figure}
\vbox{\begin{center}\psfig{file=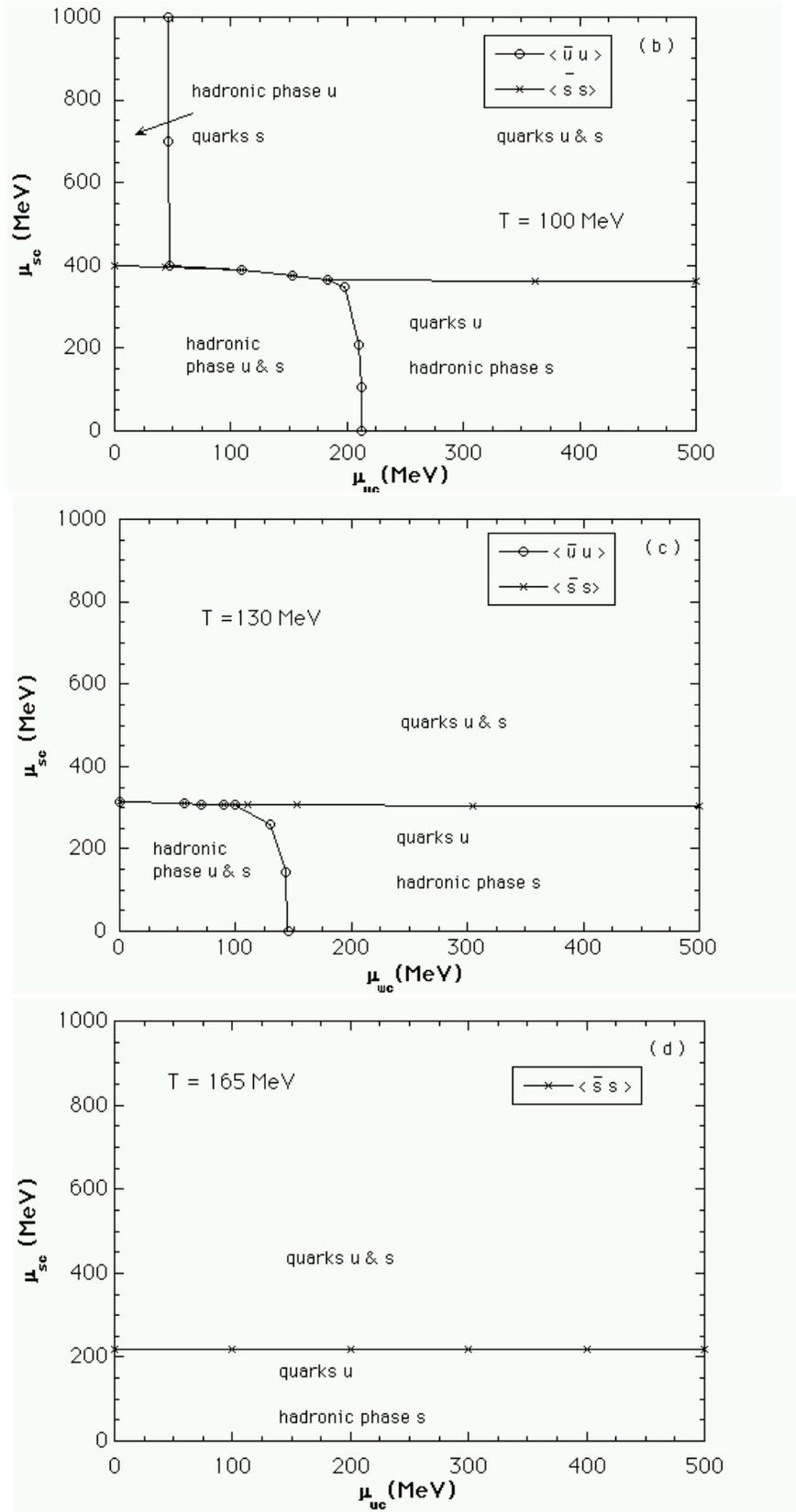,width=\widthepsthree}\end{center}}
\caption{\em Chiral phase transition for the parameters  $M_u^0=300 \mbox{ \rm MeV},\chi_0=80$ {\rm MeV} in the light quark chiral limit $m_u=m_d=0$. $\mu_{sc}$ as function of $\mu_{uc}$ for $T=0$ {\rm MeV} (a); for $T=100$ {\rm MeV} (b); for $T=130$ {\rm MeV} (c);  for $T=130$ {\rm MeV} (d).\label{figurethermo5a.eps}}
\end{figure}

\vspace{-1.0cm}

\begin{figure}[hbt]
\vbox{\begin{center}\psfig{file=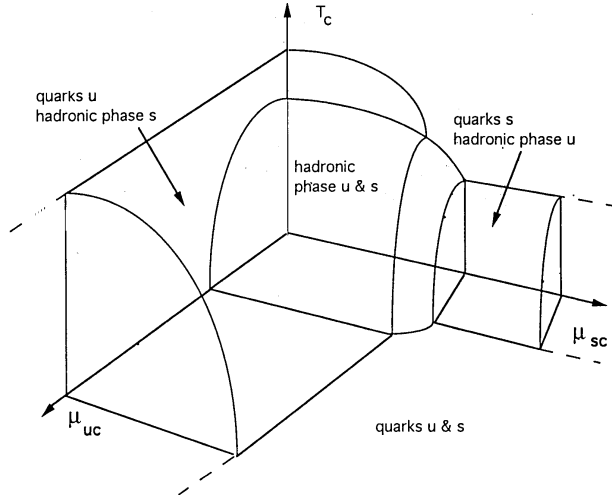,width=\widthepsxmgr}\end{center}}
\caption{\em Chiral phase transition summing up results from figure \ref{figurethermo5a.eps} for the set of parameters (\ref{eq222c}).\label{artisticview300.eps}}
\end{figure}

\subsubsection{Thermodynamics}
\label{resultsThermodynamics}

In this section, we present results relative to the pressure (equation of state), the energy density and the entropy density that we have adapted from \cite{vdb96,cugjamvdb96}. Because we would like to emphasize some points relative to fits, we  first take the somehow unusual way to present these quantities as a function of $T^4$ (pressure, energy density) or $T^3$ (entropy density). This allows us to separate curves corresponding to different parameters. Once these results will have been presented, we shall redraw some of our results in the usual way (pressure or energy density or $T$ times the entropy density, over $T^4$, as a function of $T$), allowing us to make a more direct comparison with the general shape of these quantities as obtained in lattice calculations.
Pressure, energy and entropy densities are obtained from 
(\ref{eqthermo17}), (\ref{eqthermo18}) and~(\ref{eqthermo19}),
respectively. The general behavior of these quantities can be understood (at vanishing density) from eqs. (\ref{eqthermo20}--\ref{eqthermo22}) with asymptotic behaviors given in appendix~\ref{Appendixhightemplimit} and summarized in 
 section~\ref{sectionpressure}. \par

\espace

\souligne{Pressure}

\espace

The behavior of the pressure is shown in figure \ref{figurethermo7a.eps}(a) for a vanishing density. 
As the strange quark mass $m_s$ is different from zero, the behavior of the pressure above $T_c$ is not the usual $T^4$ law: in addition to the bag constant 
(\ref{eqthermo28}) which is taken into account through the decomposition (\ref{eqthermo20}), the massive free gas part has the temperature expansion~(\ref{eqthermo27}).

\vspace{-0.75cm}

\begin{figure}[hbt]
\vbox{\begin{center}\psfig{file=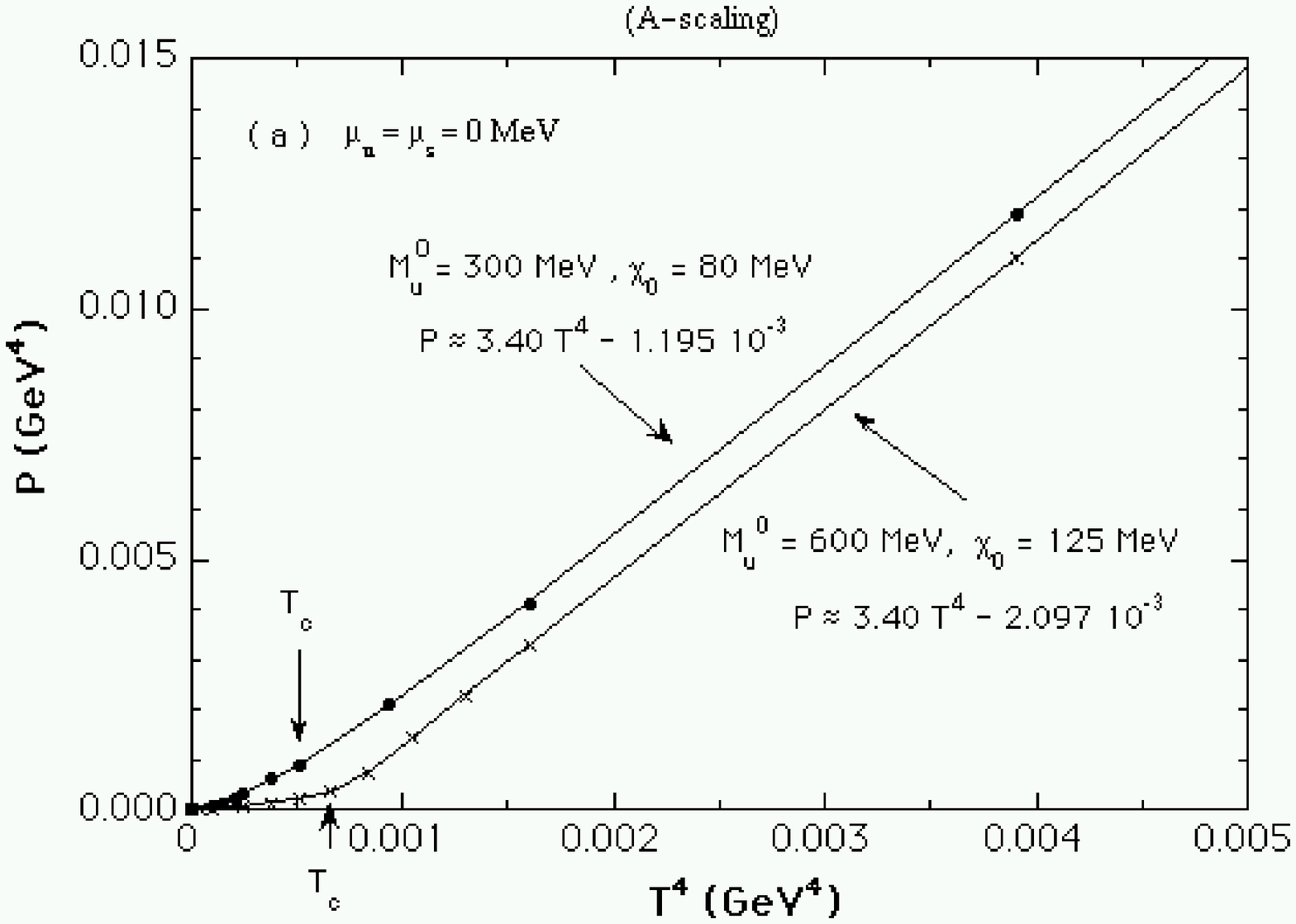,width=\widtheps}\end{center}}
\end{figure}

\vspace{-1.25cm}

\begin{figure}[hbt]
\vbox{\begin{center}\psfig{file=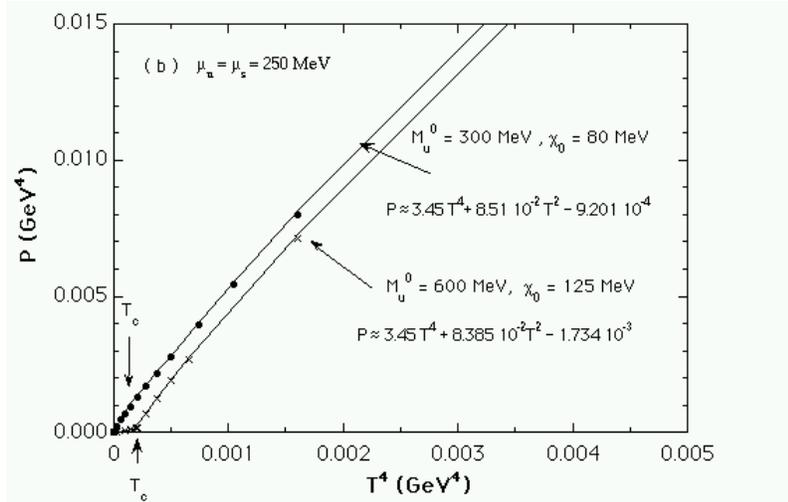,width=\widtheps}\end{center}}
\caption{\em Pressure as a function of $T^4$  for the set of parameters (\ref{eq222b},\ref{eq222c}) for $\mu_u=\mu_d=\mu_s=0$ {\rm MeV} (a); for $\mu_u=\mu_d=\mu_s=250$ {\rm MeV} (b).\label{figurethermo7a.eps}}
\end{figure}

Figure \ref{figurethermo7a.eps}(a) shows also the linear
fit in $T^4$ which is valid in the phase where the chiral symmetry is restored:
\be
P\approx 3.40\ T^4-2.097\ 10^{-3}
\label{eqthermo29}
\ee
for the set (\ref{eq222b}), and
\be
P\approx 3.40\ T^4-1.195\ 10^{-3}
\label{eqthermo30}
\ee
for the set (\ref{eq222c}). Note that the coefficient of $T^4$ is not
equal to the coefficient $7N_c\pi^2/60$ (=3.454) of the term  $T^4$ in the expansion (\ref{eqthermo27}). Because of the limited range of temperatures investigated in figure \ref{figurethermo7a.eps}, 
the corrective terms to the factor
$T^4$ in the expansion~(\ref{eqthermo27}) change the apparent slope, and the constant term in (\ref{eqthermo29}) cannot be identified
with the bag constant $B$. 
When a fit is realized with all the terms of the expansion~(\ref{eqthermo27})\footnote{By all the terms we mean the terms which are included in 
(\ref{eqthermo27}). Indeed the precision is sufficient and we do not have to take into account more terms defined in~(\ref{thermopression}).}, plus a
constant
$C$ to adjust, this one equals  $C^{1/4}\approx$ 207 MeV for
the set~(\ref{eq222b}) and $C^{1/4}\approx$ 179 MeV for the set
(\ref{eq222c}). These values are close to
$B^{1/4}\approx$ 209 MeV (set~(\ref{eq222b})) and
$B^{1/4}\approx$ 183 MeV (set~(\ref{eq222c})) obtained from a direct calculation of the exact bag constant (\ref{eqthermo28}).
This shows the consistency of our numerical results and the fast convergence of the expansion~(\ref{eqthermo27}), even if the
expanding parameter $m_s/T$ is not so small! Note also that the numerical values
of $B^{1/4}$ extracted from the exact bag constant (\ref{eqthermo28}) or extracted from the fits are not so far from $B'^{1/4}$ given by the approximate
equation (\ref{eqthermo26}). The latter is valid if we ignore the coupling between the quark and gluon condensates, and gives $B'^{1/4}$ = 201 MeV
(set~(\ref{eq222b})) and
$B'^{1/4}$ = 161 MeV (set~(\ref{eq222c})). 
This justifies {\em a posteriori} the introduction of the approximate bag constant (\ref{eqthermo26}) in the references
\cite{jamvdb94,jamvdb95,jamrip92}\footnote{It should however be stressed that it is better to work with the exact expression since an error in
$B^{1/4}$ is amplified when going to $B$.}.\par

This discussion shows that it could be quite dangerous to extract numerical values from fits: one could be tempted to identify the bag constant from the constant term in (\ref{eqthermo29}) and (\ref{eqthermo30}). The above analysis shows clearly that it would  be wrong.

Beyond the transition, the behavior of the pressure versus  temperature depends on the order of the transition. For the set (\ref{eq222b}), $\beta M_i$ ($i=u$, $s$) remains quite large, so that the $T$ behavior is described in a first
approximation by (\ref{eqthermo31}). 
However, a careful analysis shows that this expansion is not well suited for
$\beta M_s\le20$, or even for $\beta M_s\le40$, and that the expansion
 (\ref{termenegal2}) should be used instead. The
exponential behavior $e^{-\beta M_s}$ is however still correct.
For the set (\ref{eq222c}), the phase transition is of  second order
so that $M_u$ is 
progressively decreasing. This implies that the expansion (\ref{termenegal2})
 cannot be used over the whole range $T<T_c$ and, even more, that it is too crude to describe this behavior. We have however found that
the expansion (\ref{besselasymptpression}), with
 the sum limited to  $n=1$ and $n=2$, is well suited for temperatures below
 100 MeV.  \par

Note that for an analysis based on the 
 $1/N_c$ expansion
\cite{klevansky94b,klevansky94c,blaschke95c,schmidt95,zhuang95},
it has been shown  that the pions  give the largest contribution to the thermodynamical quantities at low temperature\footnote{This is in agreement with the results 
\cite{gasser87} based on  chiral perturbation theory
\cite{gasser84}.}. In the chiral limit where the pions are massless, their behavior is in 
$T^4$, which effectively shows they have a bigger contribution than (\ref{eqthermo31}) based on the high constituent quark mass. This is an example where the $1/N_c$ approach to the lowest order is not valid (see \cite{ripka97}).

The above analysis shows that an important ingredient is not included: indeed, the $T^4$ obtained for $P_{\boite{ideal gas}}$ does not contain the gluonic d.o.f., as indicated in section \ref{drawbacks}.
In the chiral limit, we should have
\be
P_{\boite{ideal gas}}=\f{\pi^2}{90}
\left\{
2N_g+\f{7}{8}N_cN_f4
\right\}T^4
\approx 5.2\ T^4,
\label{eqthermo32}
\ee
with $N_g$ = 8 if $N_c$ = 3.

Although the gluon condensate is in part due to these gluonic d.o.f., gluons do not contribute to the thermodynamics.
To take into account the thermodynamics of a purely gluonic system, we should add to the Lagrangian~(\ref{Lagchi}) a temperature dependent potential
$V_{\chi}(T)$. 
This has for example been noticed in 
\cite{ellis90b,weise94,weise92c}. The choice of this potential should be such that it  leads to the behavior (\ref{eqthermo32}).\par

Figure \ref{figurethermo7a.eps}(b) is the analogue of
figure~\ref{figurethermo7a.eps}(a) for the choice $\mu_u=\mu_s$ = 250 MeV.
These values correspond to a strong coupling between the condensates 
\mbox{$<\bar{u}u>$} and \mbox{$<\bar{s}s>$} for the
set~(\ref{eq222b}), and to a weak coupling 
for the set~(\ref{eq222c}), as shown in 
figures \ref{figurethermo2a.eps}(a) and \ref{figurethermo4a.eps}(b). With a chemical potential, there is no possible $T^4$ linear fit. This can be seen considering the chiral limit ($m_u=m_d=m_s=0$) of~(\ref{eqthermo20})
with~(\ref{eqthermo23}). This leads to
\be
P={\frac{7}{60}}{N}_{c}{\pi}^{2}{T}^{4}+{\frac{{N}_{c}}{2}}{\mu}^{2}
{T}^{2}+{\frac{{N}_{c}{\mu }^{4}}{{4\pi}^{2}}}-B.
\label{eqthermo36}
\ee
\par

Because $m_s\ne0$, eq. (\ref{eqthermo36}) has to be modified. Figure \ref{figurethermo7a.eps}(b)
shows that the  $T^4$ part is not modified compared to
(\ref{eqthermo36}), while the  $T^2$ part is slightly smaller. Once again, this  shows that one has to be very careful when making fits.
The coefficient of the $T^4$ term is not identical to the one of (\ref{eqthermo29}): the supplementary terms in (\ref{eqthermo27}), and terms coming from the chemical potential,
have a repercussion upon all the terms of the fit.\par

\espace

\souligne{Energy}

\espace

The behavior of the energy density versus temperature, as given by 
(\ref{eqthermo21}) and (\ref{eqthermo24}), is represented in
figure \ref{figurethermo8a.eps}. At high temperature and for vanishing density,
the whole expansion~(\ref{thermoenergie}) can be restricted to the  first four terms (\ref{eqthermo37}), which  perfectly describe
the curves above the chiral transition. For the sets (\ref{eq222b})
and (\ref{eq222c}), we get the linear $T^4$ fits
\be
\varepsilon\approx 10.269\ T^4+1.620\ 10^{-3}
\label{eqthermo38}
\ee
and
\be
\varepsilon\approx 10.271\ T^4+0.877\ 10^{-3},
\label{eqthermo39}
\ee
respectively.\par

It is useful to stress once more the difficulties one encounters to extract meaningful information form fits such as
(\ref{eqthermo38})  and (\ref{eqthermo39}), since the constant terms
in (\ref{eqthermo38})
and (\ref{eqthermo39}) are not the opposite of (\ref{eqthermo29})
 and (\ref{eqthermo30}), while the exact eqs.
(\ref{eqthermo20}) and (\ref{eqthermo21}) show clearly that they should. In fact, all the terms from the expansion (\ref{eqthermo37}) contribute to the determination of the coefficients of the fits in the restricted range of temperatures investigated.\par

The expansion (\ref{eqthermo37}) is perfectly adequate since 
a fit  from its different terms plus a 
constant  $C$ to be adjusted gives $C^{1/4}\approx$ 208 MeV for the set (\ref{eq222b}) and  $C^{1/4}\approx$ 182 MeV for the
set (\ref{eq222c}), in excellent agreement with the results obtained from the behavior of the pressure.\par

Figure \ref{figurethermo8a.eps} shows that the order of the transition and the nature of the coupling (strong or weak) between the quark condensates are very well visualized with the help of the energy density curves: there is a jump at  $T_c$ if the transition is of  first order, while there is a change of slope quite visible if it is a true\footnote{We mean a transition which is not a crossover, i.e. we are in the chiral limit.}  second order transition.

\begin{figure}[hbt]
\vbox{\begin{center}\psfig{file=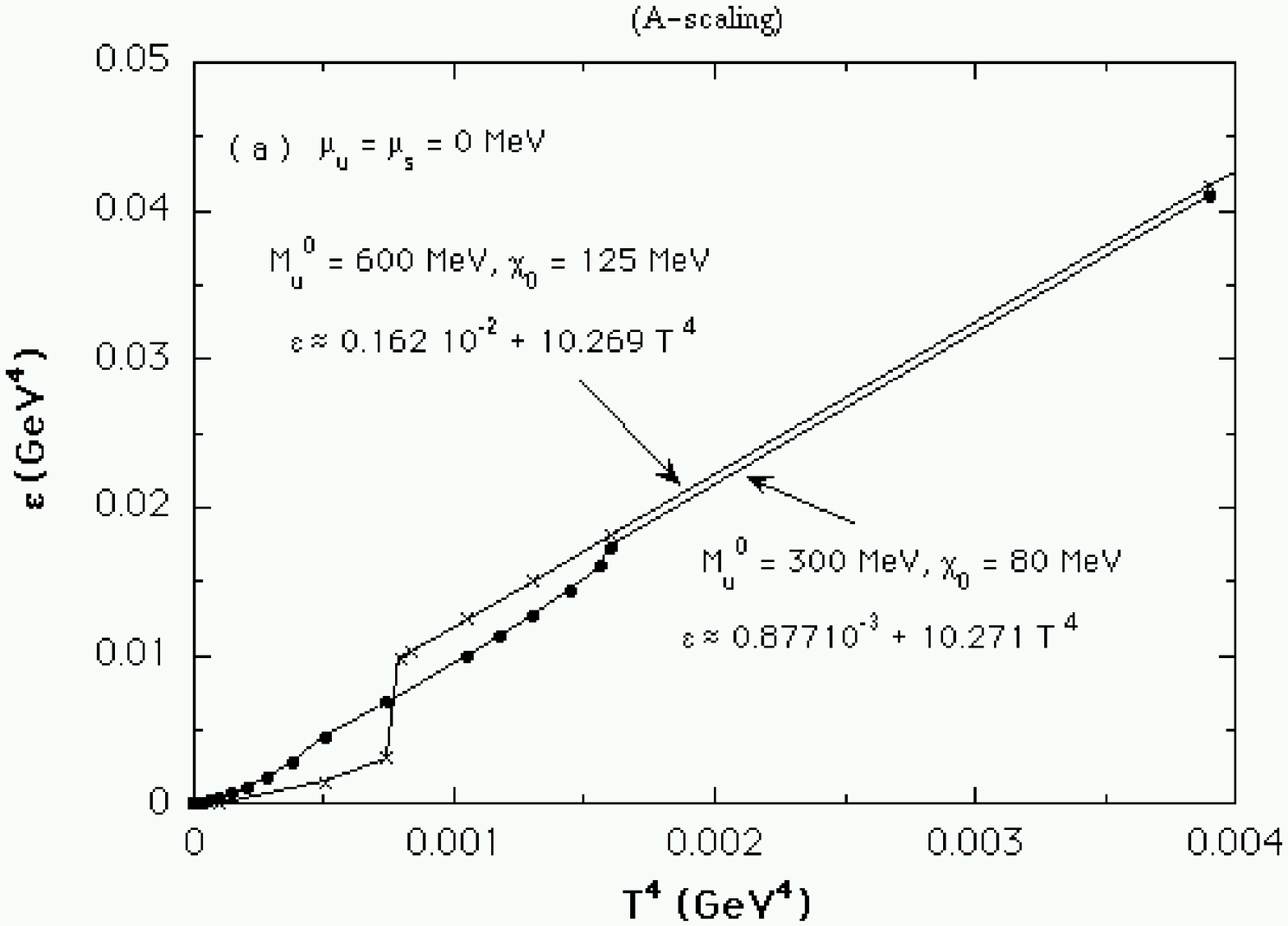,width=\widtheps}\end{center}}
\end{figure}

\vspace{-1.25cm}

\begin{figure}[hbt]
\vbox{\begin{center}\psfig{file=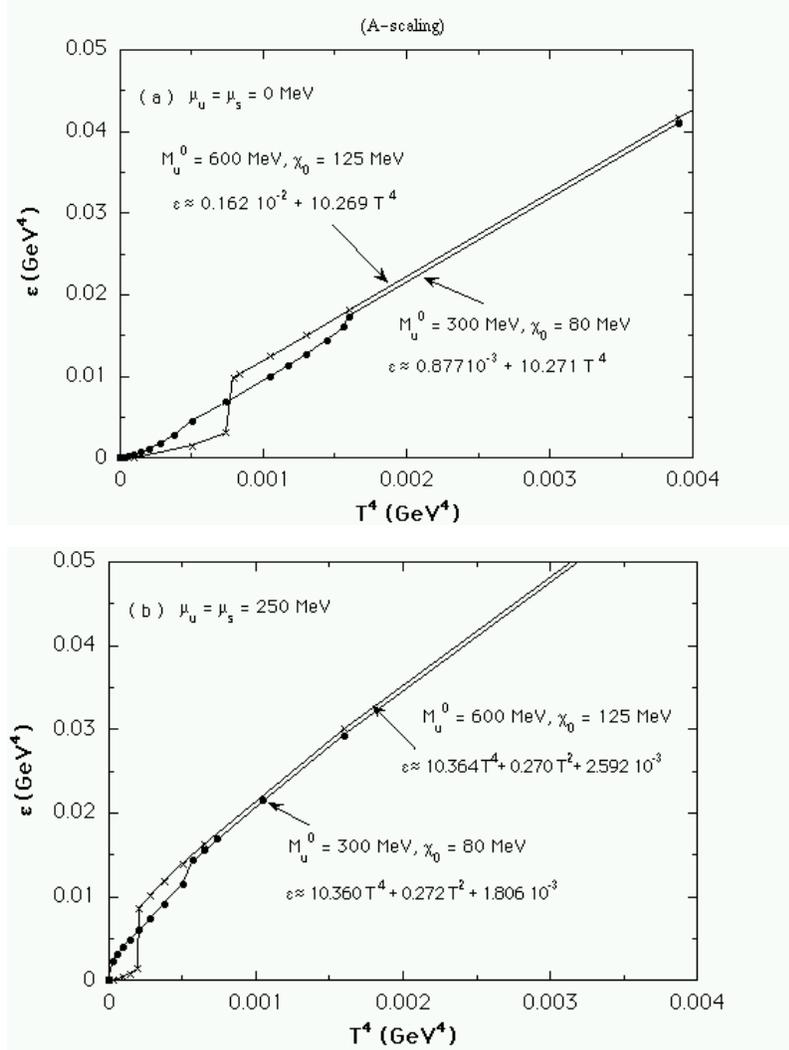,width=\widtheps}\end{center}}
\caption{\em Energy density as a function of $T^4$  for the set of parameters (\ref{eq222b},\ref{eq222c}) for $\mu_u=\mu_d=\mu_s=0$ {\rm MeV} (a); for $\mu_u=\mu_d=\mu_s=250$ {\rm MeV} (b).\label{figurethermo8a.eps}}
\end{figure}

\vspace{0.7cm}

This is the case for  the set (\ref{eq222c}), see  figure \ref{figurethermo4a.eps}(a):  the transition is of  second order for the up quarks while the strange quarks feel a first order transition.  There is then a change of the slope of $\varepsilon(T)$ at $T_{uc}$ and an energy jump at $T_{sc}$, see figure \ref{figurethermo8a.eps}(a).\par

The behavior below the transition can be understood from
eq. (\ref{besselasymptenergie}), with the same restrictions as in the case of the pressure, for the corresponding set of parameters.

Figure \ref{figurethermo8a.eps}(b) is the analogue of
figure \ref{figurethermo8a.eps}(a) for $\mu_u=\mu_s$ = 250 MeV. Chemical potentials introduce a 
$T^2$ dependence in the simpler case of chiral limit. We have, with $\mu\equiv\mu_u=\mu_s$,
\be
\varepsilon={\frac{7}{20}}{N}_{c}{\pi}^{2}{T}^{4}+{\frac{3}{2}}
{N}_{c}{\mu}^{2}{T}^{2}+{\frac{3}{4}}{N}_{c}{
\frac{{\mu}^{4}}{{\pi}^{2}}}+B.
\label{eqthermo40}
\ee
This equation comes from 
(\ref{eqthermo24}) or, more directly, from
(\ref{eqthermo18}) together with (\ref{thermo2})
and (\ref{thermo2bis}). Figure
\ref{figurethermo8a.eps}(b) shows also that having
$m_s$ different from zero  does not affect too much the second term of
(\ref{eqthermo40}), and introduces a constant supplementary term. All the remarks concerning the fits are also valid here.\par
 
\espace

\souligne{Entropy}

\espace

For a massless free quark gas, the entropy density behaves like
$T^3$. 
Since the strange current quark mass does not vanish, the high temperature expansion has correcting terms.
Eqs. (\ref{eqthermo22}), (\ref{eqthermo27}) and (\ref{eqthermo37}) give the first three terms of $Ts$ (temperature times the entropy density) in powers of $T^2$, leading to
eq. (\ref{eqthermo41}). The complete expansion is given by (\ref{thermoentropie}).

Figure \ref{figurethermo9a.eps}(a) shows that 
eq. (\ref{eqthermo41}) can be approached by a linear expression in
$T^3$ in the given range of temperatures:
\be
s\approx 13.868\ {T}^{3}-0.136\ {10}^{-1},
\label{eqthermo42}
\ee
and
\be
s\approx 13.839\ {T}^{3}-0.912\ {10}^{-2},
\label{eqthermo43}
\ee
for the set of parameters (\ref{eq222b}) and (\ref{eq222c}),
respectively.\par

 Although the numerical results seem to fit exactly the relation
(\ref{eqthermo19}), the $T^3$ coefficients 
 of (\ref{eqthermo42})  and (\ref{eqthermo43})
do not correspond to the combination of eqs. ~(\ref{eqthermo29},\ref{eqthermo30},\ref{eqthermo38},\ref{eqthermo39}).

\vspace{-0.5cm}

\begin{figure}[hbt]
\vbox{\begin{center}\psfig{file=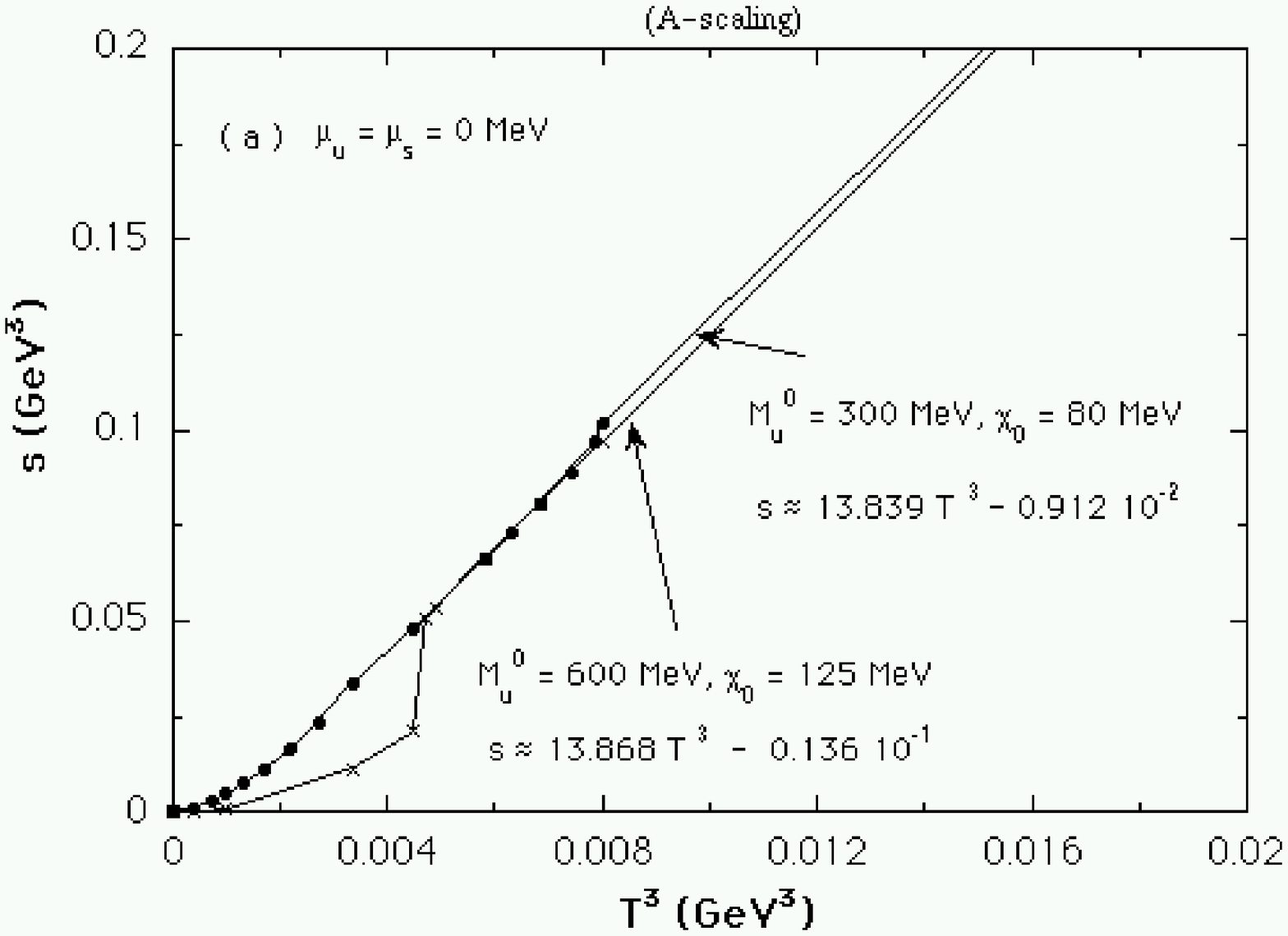,width=\widtheps}\end{center}}
\end{figure}

\vspace{-1.5cm}

\begin{figure}[hbt]
\vbox{\begin{center}\psfig{file=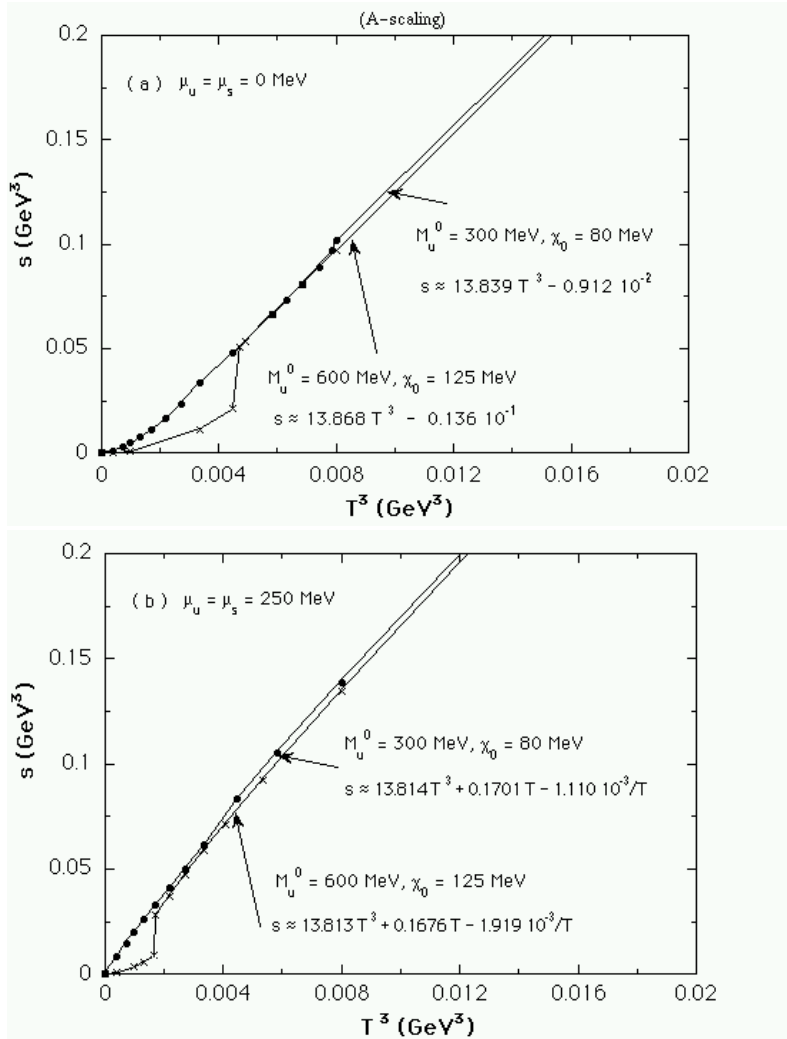,width=\widtheps}\end{center}}
\caption{\em Entropy density as a function of $T^4$  for the set of parameters (\ref{eq222b},\ref{eq222c}) for $\mu_u=\mu_d=\mu_s=0$ {\rm MeV} (a); for $\mu_u=\mu_d=\mu_s=250$ {\rm MeV} (b).\label{figurethermo9a.eps}}
\end{figure}

\vspace{0.7cm}

This remark confirms once more what we claimed in the  pressure and energy density case: it is dangerous to extract information from fits if we do not take care of possible corrections.

In our case, fits in $T^4$
(for pressure and energy) and in $T^3$ (for entropy) do not help
to get the bag constant or the number of excited d.o.f. The more complete forms (\ref{eqthermo27}),
(\ref{eqthermo37}) and (\ref{eqthermo41}) have to be considered.\par

The behavior of the entropy below the chiral transition can be understood using
(\ref{besselasymptentropie}) with the same warning as in the pressure case, for the corresponding set of parameters.\par

Figure \ref{figurethermo9a.eps}(b) shows the entropy behavior for nonvanishing chemical potentials.
Using eqs. (\ref{eqthermo22}), (\ref{eqthermo36})
and (\ref{eqthermo40}), and the density
(\ref{eqthermo15}) in the limit where all masses are vanishing
\be
\rho\equiv\sum_i\rho_i=
{\frac{{N}_{c}{\mu}^{3}}{{\pi}^{2}}}+{N}_{c}{\mu}{T}^{2},
\label{eqthermo44}
\ee
we get
\be
Ts={\frac{7}{15}}{N}_{c}{\pi}^{2}{T}^{4}+{N}_{c}{\mu}^{2}{T}^{2},
\label{eqthermo45}
\ee
which has to be modified in order to take into account the finite value of 
 $m_s$. It is worth  noticing that  $m_s$ only slightly affects the  $T^2$
term of (\ref{eqthermo45}). \par

\espace

As a conclusion to these results, we can mention that, above the chiral transition, all the d.o.f. of the model are excited.
However the vanishing mass limit of QCD is not reached for two reasons
\cite{weise92b}:
\begin{itemize}
\item the gluon d.o.f. are not included at high temperature.
A temperature and density dependent potential
$V_{\chi}(T,\mu)$ should be taken into account;
\item the strange quark mass is not negligible. We need going to very high temperatures in order that the lowest term in 
$T^4$  subsists in the expansion $m_s/T$.
\end{itemize}

We should also notice that the $1/N_c$ corrections in 
references
\cite{klevansky94b,klevansky94c,blaschke95c,schmidt95,zhuang95}  show that
the low temperature thermodynamics is driven by pion motion, pions being much lighter than the constituent quark mass. Since the quark loop contribution takes into account thermal excitations of quarks with mass $M\gtrsim300$ MeV (whose probability is reduced by the Boltzmann factors  $\exp(-\beta M)$), the low temperature thermal excitations are completely dominated by the almost massless pions. To obtain the effects of pions in our results, we should integrate over the meson fields in the path integral formalism, which is however beyond the scope of this paper.

Even with the above mentioned limitations, the scaled NJL models have important new features: some gluonic effects are included through the gluon condensate
$\chi$ which couples  the  up and strange quark condensates.
Thanks to this coupling, our model allows simultaneous transitions for the up and strange sectors (strong coupling), even though they tend to remain uncoupled for high chemical potentials.
The coupling then allows  first order transitions as a function of temperature while, within a pure NJL, they are always of the second order\footnote{The pure NJL model does, however, allow first order transitions w.r.t. density, e.g. \cite{klevansky92}.} (e.g. \cite{hatsuda94,weise90c}).\par

\subsubsection{Comparison with lattice QCD}
\label{comparisonlattice}

To make a comparison with lattice QCD\footnote{Because lattice QCD has only turned  recently towards finite density, see e.g. \cite{kogut97c}, we restrict ourselves to $\mu_u=\mu_s=0$.}, it can be advantageous to normalize $P$, $\varepsilon$, $Ts$ and the interaction measure $(\varepsilon-3P)$ to $T^4$. The interaction measure gives the non-perturbative contribution to the thermodynamics: it vanishes in the Stefan-Bolzmann limit. It is also interesting to plot $3P$ and $\varepsilon$ in the same picture to see how sharp is the increase of the corresponding thermodynamical function. The origin of the coefficient $3$ in front of $P$ compared to $\varepsilon$ comes from the coincidence of their respective asymptotic $T^4$ behavior (see the comparison between eq. (\ref{eqthermo27}) and eq. (\ref{eqthermo37})). In the same spirit, one can normalize the entropy density by a factor $3/4$ (see eq. (\ref{eqthermo41})). In this way, $3P$, $\varepsilon$ and $3sT/4$ have the same asymptotic value $7N_c\pi^2/20$, which is a direct consequence of the number of d.o.f. which enters the model. Note that the quantities we examine  are  relative to the quarks.  In our simplified model, the glueball only enters through the bag constant.

\begin{figure}[hbt]
\vbox{\begin{center}\psfig{file=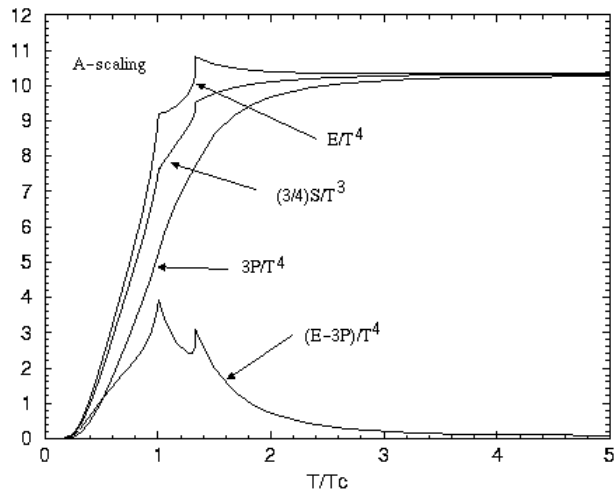,width=\widthepsxmgr}\end{center}}
\caption{\em Pressure, energy density, entropy density and interaction measure for the set of parameters $M_u^0=300 \mbox{ \rm MeV},\chi_0=80 \mbox{ \rm MeV}$.\label{latticePE.eps}}
\end{figure}

In figure \ref{latticePE.eps} we show both the pressure and the energy density of the A-scaling  NJL model versus $T/T_c$ ($T_c=150$ MeV) for the set of parameters (\ref{eq222c}) ($M_u^0=300 \mbox{ \rm MeV},\chi_0=80 \mbox{ \rm MeV}$), with $m_u=0$. Here, we have taken  the critical temperature corresponding to the chiral symmetry restoration connected to the up quarks. 

Several interesting points have to be mentioned. One expects from lattice studies, e.g. \cite{laermann96,karsch95,lattice96}, that the thermodynamical quantities are almost vanishing below $T_c$, then increasing. This increase is very sharp for $\varepsilon$ and $Ts$, while the pressure approaches the Stefan-Boltzmann limit very slowly. Lattice calculations show also that $\varepsilon/T^4$ has a peak\footnote{This is not the case for a pure gauge theory.} just above $T_c$, then approaching its asymptotic value from above. Finally, they also show that $\varepsilon-3P\ne0$ above $T_c$.
Our results, summarized in figure \ref{latticePE.eps}, show that the model is in qualitative agreement with lattice results. The quantitative difference can be understood in the following way: lattice calculations show a rapid variation of the entropy density in a narrow region of $T$ ($\approx 10$ MeV), which is traced back to the liberation of quarks and gluons. It seems then quite trivial to relate this fast increase to the confinement-deconfinement properties, which are not included in our model. This is clearly seen in the entropy density calculated with our model where the entropy is already increasing (although not as fast as near $T_c$) for $T$ as low as $0.2T_c$. Once this entropy curve is understood, the general behavior of $P$ and $\varepsilon$ can also be deduced, see e.g. \cite{hatsuda96}. It is  explicitly shown in that reference that, starting with a sharp entropy density, the energy density has a peak, and that the pressure increase above $T_c$ is low. In fact, would the entropy be approximated by a step, we should have the exact result
\be
\f{P(T)}{P_{SB}(T)}\sim1-\left(\f{T_c}{T}\right)^4,
\ee
which gives $P/P_{SB}=$ 50\% (90\%) for $T/T_c=$ 1.2 (1.8), independently of the details of the model. Since we are far from a step for the entropy, $P/T^4$ has an even weaker $T$ dependence. This is shown in the general model for the entropy  \cite{hatsuda96} and is confirmed by our particular model. On the same ground, it can also be shown that the interaction measure $(\varepsilon-3P)/T^4$, given in eq. (\ref{eminus3P}), has a peak above $T_c$.

We have seen that the general behavior can be understood from the analysis of \cite{hatsuda96} which, together with the lack of confinement of our model, explains the quantitative disagreement between lattice gauge calculations and scaled NJL ones. However, our figure shows  nice features not discussed extensively in the literature. If we concentrate on the energy density, it is clear that the peak has its slope broken in two places. These broken slope points coincide with the temperature where the chiral symmetry is restored. Since the current up quark mass is zero, the transition corresponding to the up quarks has no tail (see figure \ref{figurethermo4a.eps}(a)), leading to the first slope discontinuity while, because the transition  of the strange quarks is of  first order, there is in fact a jump in the energy density. Since this jump is small, it looks like a discontinuous slope. To get a nice peak, one then has  to consider only crossovers  (second order transition with nonvanishing current quark masses). Note also that a gap in energy only transforms into a change of slope for  the pressure, while a change of slope in the energy plot is almost invisible in the pressure. It is  evident that the energy density is the adequate quantity to be investigated in order to have insights on the order of the transition, and for extracting the critical temperature\footnote{These informations can of course be obtained from the reconstruction of the quark condensates.}.

\begin{figure}[hbt]
\vbox{\begin{center}\psfig{file=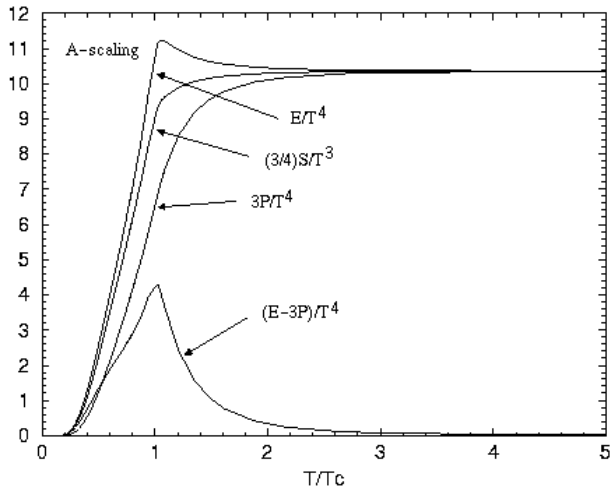,width=\widthepsxmgr}\end{center}}
\caption{\em Pressure, energy density, entropy density and interaction measure for the set of parameters $M_u^0=M_s^0=400 \mbox{ \rm MeV},\chi_0=350 \mbox{ \rm MeV}$.\label{latticePE2.eps}}
\end{figure}

Figure \ref{latticePE2.eps}  illustrates  that the broken peak of figure \ref{latticePE.eps} is due the combined effect of a second order phase transition for the up quarks (in the chiral limit $m_u=0$) and  a weak first order transition for the strange quarks. We have taken the set of parameters ($M_u^0=M_s^0=400 \mbox{ \rm MeV},\chi_0=350 \mbox{ \rm MeV}$). In that case, there is only one critical temperature, and the transition is of  second order (see figure \ref{figureZeit1.eps}), with the critical temperature given  precisely by eq. (\ref{eqtempcritique}).

All we have said for the energy density remains valid for the interaction measure (two peaks in figure \ref{latticePE.eps} which, with $m_u\ne0$ and a second order transition for the strange quarks, would lead to a single peak). This gives, in the limit of three degenerate flavors,  the interaction measure of figure \ref{latticePE2.eps}. Note that, in the chiral limit, the interaction measure just gives $4B/T^4$ (for $T\ge T_c$), $B$ being the bag constant, see eq. (\ref{eminus3P}).

\subsection{Anomalous decay $\pi_0\rightarrow\gamma\gamma$}
\label{sectionanomalouspiondecay}

Until now we did not pay too much attention to the fact that the model is not confining, allowing the unphysical decay $meson\rightarrow\bar{q}+q$ as soon as the meson is more massive than the sum of its constituent quarks. If the unphysical width is small compared to the meson mass, it is tempting to argue that the model is  still meaningful. We just need  extending it above threshold, performing calculations, and checking that the unphysical width is small.
Although there exist calculations which can be considered as exact\footnote{By exact, we mean calculations which take into account the full unphysical width, for any value of its magnitude.} above  threshold \cite{blin97}, we prefer to work in  the small width approximation. The calculations we give here are adapted from \cite{vdb96,jamvdb95b}. As an application we show, in the 2-flavor version of the A-scaling model, the variation as a function of temperature of the (physical) decay width $\pi_0\rightarrow\gamma\gamma$. This is interesting for two reasons: {\em i}) it is an anomalous process\footnote{It is here a QED anomaly which does not have to be introduced by hand as the strong axial anomaly or the scale anomaly. Here, the anomaly  shows up as a result of the introduction of the electromagnetic field.} and {\em ii}) we shall show that we expect a strong enhancement of the decay width near  the deconfining transition.\par

The calculation of the two-photon decay width of the neutral pion  $\Gamma_{\pi\rightarrow\gamma\gamma}$ requires the knowledge of the pion mass $\mpi$ and the pion coupling constant to the quarks $\gpiqq$. The lack of confinement  above the threshold $\mpi(T)>2M_u(T)$ implies imaginary parts for both the mass and the coupling constant. Because of the two different scales connected to QCD and QED ($\alpha_s$ and $\alpha$, respectively), the decay width $\Gamma_{\pi\rightarrow\gamma\gamma}$ has to be calculated non-perturbatively for the strong part, as it has been done previously, while the electromagnetic part can be treated perturbatively. To make things simple, we only consider a second order phase transition. Because we want a low critical temperature, we  shall keep the set (\ref{eq222c}): ($M_u^0=350 \mbox{ \rm MeV},\chi_0=80 \mbox{ \rm MeV}$). Taking a larger $\chi_0$ does  not change the conclusions, except that the transition occurs at a higher temperature (see figure \ref{figureZeit1.eps}).

\subsubsection{Definition of the model above the threshold $\pi\rightarrow \bar{q}q$}
\label{modeleseuil}

Although there is an exact study \cite{blin97}, we work in the small width approximation, in the same spirit as in
\cite{klevansky94}.

To remove any ambiguity about the sign of the width\footnote{We do not have any $i\epsilon$ as in Minkowski space.} $\Gamma_{\pi\rightarrow \bar{q}q}$, we make a continuation of the model over the whole real line ($q^2\in]-\infty,+\infty[$), before searching for poles in the Bethe-Salpeter equations (or, equivalently, in the definition of the pion mass as described in section \ref{bs}). Other ways can be found in the literature. For example,
\begin{itemize}
\item  the authors of \cite{blin90} 
define the 
mass and width above  threshold by looking at the weight position and the half-height of the imaginary part of the propagator, respectively. Below  threshold,  this imaginary part is reduced to a Dirac distribution whose position is identified with the pion mass. However this approach has been criticized in  
\cite{takizawa91} in the vector sector. In the more recent reference \cite{blin97}, the authors show why there is a discrepancy and give a way to cure  it;
\item  the authors of \cite{hatsuda94} simplify the problem, identifying the expression above  threshold with its Cauchy principal value. Although very simple, this method completely ignores  the process $\pi\rightarrow \bar{q}q$ which, while unphysical, is nevertheless present. 
\end{itemize}

The method we have developed in \cite{vdb96,jamvdb95b} is the following: let the inverse propagator be
$G^{-1}(q_0^2)$ (we work in the static limit $\vec{q}$ = 0), in Euclidean space.  The  pion propagator\footnote{For the kaon, the threshold is at
$q_0^2=-(M_u+M_s) ^2$.} is then defined in the range
$q_0^2\in ]-4M_u^2,+\infty[$. We extend it in the range
$]-\infty,-4M_u^2]$ by defining
\be
G^{-1}(q_0^2)\equiv\lim_{\varepsilon\rightarrow 0}G^{-1}(q_0^2-i
\varepsilon)={\cal A}(q_0^2)+i{\cal B}(q_0^2),
\label{prolongement}
\ee
where ${\cal A}(q_0^2)$ and ${\cal B}(q_0^2)$ are real functions of the real variable $q_0^2$. Below  threshold, 
${\cal B}(q_0^2)$ vanishes identically. With this definition, the pion propagator is defined for real $q_0^2$. The pion mass and its unphysical decay width are obtained from this expression, by searching for the zero of the propagator in the complex plane:

\be
G^{-1}(-(m_{\pi}-i\Gamma/2)^2)={\cal A}(-(m_{\pi}-i\Gamma/2)^2)
+i{\cal B}(-(m_{\pi}-i\Gamma/2)^2)=0.
\label{zeropropagateur1}
\ee

Since $q_0^2$ is now a complex variable, the quantites ${\cal A}$ and
${\cal B}$ are also complex. We write ${\cal A}$ = ${\cal A}_R+i{\cal
A}_I$ and
${\cal B}$ =
${\cal B}_R+i{\cal B}_I$, so that looking for the zero of~(\ref{zeropropagateur1}) implies
\beqn
{\cal A}_R(-(m_{\pi}-i\Gamma/2)^2)-{\cal
B}_I(-(m_{\pi}-i\Gamma/2)^2)&=&0,\\
{\cal A}_I(-(m_{\pi}-i\Gamma/2)^2)+{\cal
B}_R(-(m_{\pi}-i\Gamma/2)^2)&=&0.
\eeqn
\par

We apply this method to the inverse pion propagator (eq. (\ref{eqA37a}))
and we write  $\lim_{\varepsilon\rightarrow
0}Z_{\pi}^u(q_0^2-i\varepsilon)=$ 
$A(q_0^2)$ +
$iB(q_0^2)$.
The analogue of (\ref{zeropropagateur1}) is then
\beqn
&&-(m_{\pi}-i\Gamma/2)^2
\bigg(
A_R(-(m_{\pi}-i\Gamma/2)^2)-B_I(-(m_{\pi}-i\Gamma/2)^2)\nonumber\\
&&\ \mbox{}+i
\Big(A_I(-(m_{\pi}-i\Gamma/2)^2)+B_R(-(m_{\pi}-i\Gamma/2)^2)\Big)
\bigg)
+a^2\chi_s^2\f{m_u}{M_u}=0.
\eeqn
In the limit $\Gamma/2<<m_{\pi}$, we can neglect
$B_I$ and $A_I$ in front of $A_R$ and $B_R$, so that this equation can be rewritten in the form
\be
-(m_{\pi}-i\Gamma/2)^2
\left(
A(-m_{\pi}^2)
+iB(-m_{\pi}^2)
\right)
+a^2\chi_s^2\f{m_u}{M_u}=0.
\ee
We just need  solving\footnote{Our small width approximation is not exactly the same as in \cite{blin97}. They  are however equivalent by definition of a small width approximation.}
\be
-(m_{\pi}-i\Gamma/2)^2Z_{\pi}^u(-m_{\pi}^2-i0^+)+
a^2\chi_s^2\f{m_u}{M_u}=0.
\label{zeroetlargeur1}
\ee

This equation is valid in the chirally broken phase, where we made use of the gap equation (\ref{gapup}) to obtain (\ref{eqA37a}). The inverse propagator which is valid in both phases is given by
\be
P^{\pi \pi}=
\Big(
-8N_cg_{M_u,\beta,\mu}+Z_{\pi}^u(q)q^2+a^2\chi_s^2\Big).
\label{eqA36}
\ee
Below  threshold, we then have\footnote{In \cite{jamvdb95b}, we work in term of a polarization operator to facilitate the comparison with 
 Klevansky \cite{klevansky94}. We  keep here the same notation as in the previous sections.}
\be
-8N_cg_{M_u,\beta,\mu} -m_{\pi}^2Z_{\pi}^u(-m_{\pi}^2)+
{I}_{\pi \pi \ \rm scal}=0,
\label{chap8eq10}
\ee
while, above threshold, in the small width approximation
$\Gamma/2<<m_{\pi}$, we have to look for
\be
-8N_cg_{M_u,\beta,\mu}
-(m_{\pi}-i\Gamma/2)^2
Z_{\pi}^u(-m_{\pi}^2-i0^+)+{I}_{\pi \pi \ \rm scal}=0.
\label{zeroetlargeur2}
\ee
Eq. (\ref{zeroetlargeur2}) is  a system of two coupled equations for two unknowns
 $m_{\pi}$ and $\Gamma$. They are
however decoupled because the half-width is neglected in
$Z_{\pi}^u$. These equations are
\beqn
&&\Gamma=\f{1}{m_{\pi}}\f{B(-m_{\pi}^2)}{A^2(-m_{\pi}^2)+B^2(-m_{\pi}^2)}
(a^2\chi_s^2-8N_cg_{M_u,\beta,\mu}),\\
\label{largeurdupion}
&&m_{\pi}^4+m_{\pi}^2\f{A(-m_{\pi}^2)}{A^2(-m_{\pi}^2)+B^2(-m_{\pi}^2)}
(8N_cg_{M_u,\beta,\mu}-a^2\chi_s^2)\nonumber\\
&&\hspace{1cm}\mbox{}-\f{1}{4}
\f{B^2(-m_{\pi}^2)}{\Big(A^2(-m_{\pi}^2)+B^2(-m_{\pi}^2)\Big)^2}
(8N_cg_{M_u,\beta,\mu}-a^2\chi_s^2)^2=0.
\label{massedupion}
\eeqn
\par

\begin{figure}[hbt]
\vbox{\begin{center}\psfig{file=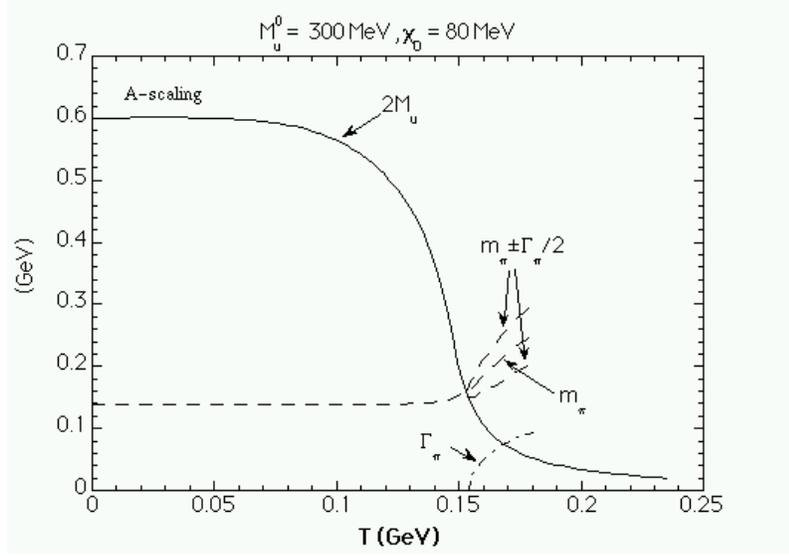,width=\widtheps}\end{center}}
\caption{\em Twice the constituent quark mass (plain line), the pion mass (dashed line) and the unphysical width $\Gamma_{\pi}$ (dot-dashed line) versus temperature.\label{figuredecay3b.eps}}
\end{figure}

In figure \ref{figuredecay3b.eps}, we show the pion mass and twice the 
constituent quark mass $M_u$ versus temperature.
An unphysical width appears as soon as the threshold $m_{\pi}=2M_u$ is reached.

The usual definition of the pion to quarks coupling constant $\gpiqq$  starts from the  polarization operator ($J^{\pi\pi}$
in the notation of \cite{jamvdb95b} or $\Pi$ in the one of
\mbox{Klevansky} \cite{klevansky94})
\be
\gpiqq=\left.\left(
\f{\p\Pi}{\p\omega^2}
\right)\!\right|_{\omega=m_{\pi}-i\Gamma/2}.
\ee

In our formalism, the coupling constant can  directly be expressed  from the pion propagator through the identification
\be
\f{1}{q^2Z_{\pi}^u(q)+(a^2\chi_s^2-8N_cg_{M_u,\beta,\mu})}=\f{\gpiqq^2}{q^2+m_{\pi}^2},
\ee
which is consistent with the identification of the on-shell coupling constant through the action. 

Indeed, with\footnote{Because $\vec{q}=0$, we have $f(q)\rightarrow
f(q_0,\vec{0})\rightarrow f(q_0^2)$  for any function  $f$. To simplify the notation, we write
$q_0^2\rightarrow q^2$.}
\be
G^{-1}_{\pi}(q^2)\equiv-8N_cg_{M_u,\beta,\mu}+Z_{\pi}^u(q^2)q^2
+a^2\chi_s^2,
\label{eqthese916}
\ee
a Taylor expansion around 
$q^2=-m_{\pi}^2$ (below  threshold) gives
\be
G^{-1}_{\pi}(q^2)\approx G^{-1}_{\pi}(-m_{\pi}^2)+\left.\f{\p
G^{-1}_{\pi}(q^2)}{\p
q^2}\!\right|_{q^2=-m_{\pi}^2}(q^2+m_{\pi}^2)+...
\ee

By definition of the pion mass,  $G^{-1}_{\pi}(-m_{\pi}^2)=0$, we  then have
\be
G^{-1}_{\pi}(q^2)\approx \left.\f{\p
\left(
q^2Z_{\pi}^u(q^2)
\right)
}{\p
q^2}\!\right|_{q^2=-m_{\pi}^2}(q^2+m_{\pi}^2)+...,
\label{pionnorm}
\ee
i.e.
\be
\gpiqq^{-2}(-m_{\pi}^2)=\left.\f{\p
\left(
q^2Z_{\pi}^u(q^2)
\right)
}{\p q^2}\!\right|_{q^2=-m_{\pi}^2},
\label{eqdecay21}
\label{eqthese919}
\ee
whose generalization above  threshold is obvious:
\be
\gpiqq^{-2}(-(m_{\pi}-i\Gamma/2)^2)=\left.\f{\p
\left(
q^2Z_{\pi}^u(q^2-i0^+)
\right)
}{\p q^2}\!\right|_{q^2=-(m_{\pi}-i\Gamma/2)^2}.
\label{eqdecay22}
\ee
This is nothing else than (small width approximation)
\be
\gpiqq^{-2}(\mbox{on-shell})=\f{1}{2}
\left[
Z_{\pi}^u(-m_{\pi}^2-i0^+)+Z_{\pi}^u(0)
\right]+\f{1}{2}(m_{\pi}-i\Gamma/2)^2J_3(-m_{\pi}^2-i0^+),
\label{eqdecay24}
\ee
where
\be
J_3(q^2)=4N_c\int\f{d^4k}{(2\pi)^4}\f{1}{(k^2+M_u^2)[(k-q)^2+M_u^2]^2}.
\label{fonctionJ3}
\ee

\begin{figure}[hbt]
\vbox{\begin{center}\psfig{file=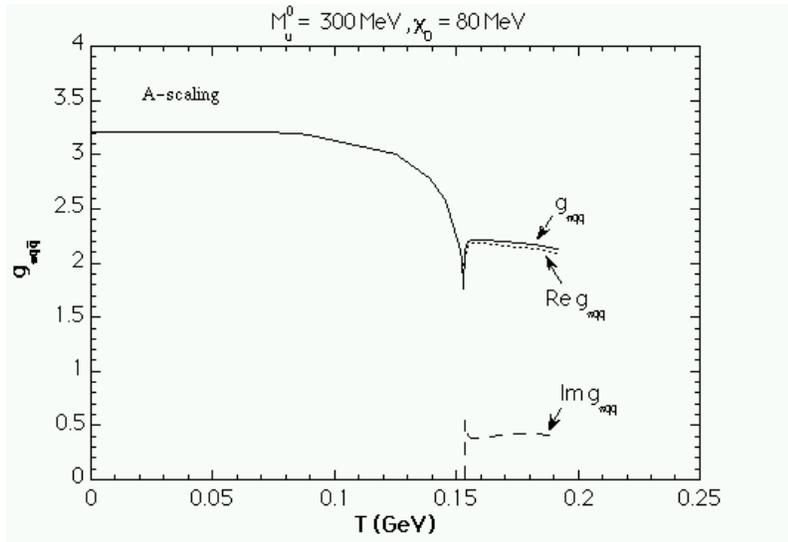,width=\widtheps}\end{center}}
\caption{\em The pion to quarks coupling constant $\gpiqq$ versus temperature (plain line), its real part (dotted line) and its imaginary part (dashed line).\label{figuredecay5b.eps}}
\end{figure}

Note that, with the approximation $\gpiqq(-m_{\pi}^2)\approx \gpiqq(0)=1/\sqrt{Z_{\pi}^u(0)}$ valid at low temperature, where pions are essentially massless, we have, with eq.~(\ref{fpiapprox})
\be
f_{\pi}^2g_{\pi q\bar{q}}^{2}(-m_{\pi}^2)\approx M_u^2,
\label{goldberger}
\ee
which is the {\em Goldberger-Treiman} relation.
With $M_u^0$ = 300 MeV and $f_{\pi}(T=0)$ = 93 MeV,
eq. (\ref{goldberger}) gives $\gpiqq(-m_{\pi}^2)\approx$ 3.23, in excellent agreement with an exact numerical calculation, as illustrated in figure \ref{figuredecay5b.eps}. 
This  figure  gives the behavior
of the real and imaginary parts of the on-shell coupling constant
$\gpiqq$ given by (\ref{eqdecay24}), together with its modulus\footnote{To simplify the terminology, we define from now on the coupling constant as being its modulus.}.  Near  threshold there is a  huge effect: the coupling constant goes to zero. This is explained by the fact that ${\rm Im}
J_3$, and then ${\rm Im} \ \gpiqq^{-2}$, are infinite at this point.

\subsubsection{Disintegration $\pi_0\rightarrow\gamma\gamma$}
\label{desintegration}

\begin{figure}[hbt]
\vbox{\begin{center}\psfig{file=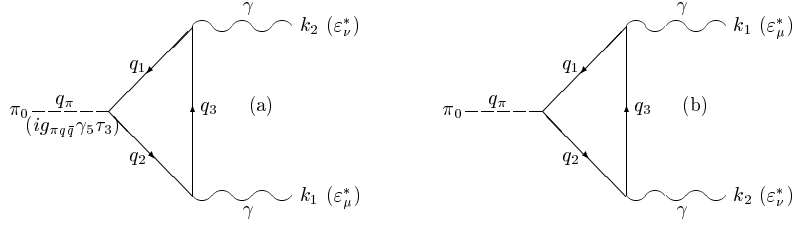,width=\widtheps}\end{center}}
\caption{\em Feynman diagrams for the decay width $\pi_0\rightarrow\gamma\gamma$: direct diagram (a); crossed diagram (b).\label{figuredecay7.eps}}
\end{figure}

The  process $\pi_0\rightarrow\gamma\gamma$ can only be explained by considering it as an anomalous process.
Because the gluons are not present in the model, the strong axial anomaly has to be included by hand. In that way, this quantum breaking of the axial symmetry can be treated at tree level.
However, the electromagnetic field  is included in the formalism. That means that we can have access to the full quantum effects without having to add a term by hand. Note that we have to go beyond the tree level approximation. The electromagnetic field can be studied as a perturbation ($\alpha\approx 1/137$).
From a formal point of view, the amplitude for the process $\pi_0\rightarrow\gamma\gamma$ can be obtained starting from the bosonized action (\ref{modeleA}). The term
$\Tr\ln$ is the fermion determinant and contains all the quantum effects of the fermions. The derivative
 $\dslash$ has to be replaced by the covariant one
$D\!\!\!\!/$ = $\dslash+i\hat{Q}A\!\!\!/\ $, where $\hat{Q}$ is the
charge quark matrix ${\rm diag}(e_u,e_d)$
= ${\rm diag}(\f{2}{3}e,-\f{1}{3}e)$. An expansion of the logarithm has to be performed until it includes the considered process\footnote{Electromagnetic corrections to a given order are expected to be small because $\alpha$ is small. The strong corrections, which should be included because the strong coupling constant $\alpha_s$ is not small, are eliminated on the basis of the $1/N_c$ ordering.}.
This is very tedious so that it is much more efficient to work with Feynman diagrams.
 Two of them have to be examined: the direct diagram (fig. \ref{figuredecay7.eps}(a)) and the crossed one
(fig.~\ref{figuredecay7.eps}(b)). From the action (\ref{modeleA}), it is clear that the pion vertex is\footnote{The pion to quarks coupling constant enters  the vertex because we have normalized the pion wave function in order to get a usual kinetic term for the pion propagator (see eq. (\ref{pionnorm})). This corresponds to a canonical normalization.}
$\gpiqq i\gd{5}\vec{\tau}_i$ ($i=1,...,3$). Vertices connected to photons are given by\footnote{$\varepsilon_{\mu}$ is the photon polarization.}
$\varepsilon_{\mu}^*(k_i)\hat{Q}\gd{\mu}$ ($i=1,2$).

 The flavor trace over the amplitude associated to figures \ref{figuredecay7.eps}(a) and
\ref{figuredecay7.eps}(b) is  
\begin{equation}
\tr\left\{
\left(
\begin{array}{cc}
\pi_0&\sqrt{2}\pi^+\\
\sqrt{2}\pi^-&-\pi_0
\end{array}
\right)
\left(
\begin{array}{cc}
e_u&0\\
0&e_d
\end{array}
\right)
\left(
\begin{array}{cc}
e_u&0\\
0&e_d
\end{array}
\right)
\right\}
=\pi_0 (e_u^2-e_d^2),
\end{equation}
which shows explicitly that only the neutral pion  has a  two-photon width (as it should from  electric charge conservation).\par

The color trace gives a factor $N_c$ (global gauge invariance of the model), so that only the Dirac trace and the loop contribution have to be determined. The transition amplitude for the process
$\pi_0\rightarrow\gamma\gamma$ associated to the direct diagram \ref{figuredecay7.eps}(a) gives then (in the Euclidean space
and in the isospin limit $m_u=m_d$)
\cite{klevansky94,jamvdb95b}:
\beqn
&&{\cal T}_{\pi_0\rightarrow\gamma\gamma}=N_c(e_u^2-e_d^2)g_{\pi
q\bar{q}}\varepsilon_{\mu}^*(k_1)
\varepsilon_{\nu}^*(k_2)\int\f{d^4q_3}{(2\pi)^4}\tr_{\rm D}
\Big\{
\gd{5}({q\!\!\!/}_2+M_u)\gd{\mu}({q\!\!\!/}_3
+M_u)\nonumber\\
&&\qquad\qquad\mbox{}\times\gd{\nu}({q\!\!\!/}_1 +M_u)
\Big\}
\f{1}{(q^2_1+M_u^2)(q_2^2+M_u^2)(q_3^2+M_u^2)}
\label{eqdecay35}
\eeqn
which has the general form, as shown by $\tr_D$,
\be
{\cal T}_{\pi_0\rightarrow\gamma\gamma}\propto
\epsilon^{\mu\nu\alpha\beta}\varepsilon_{\mu}^*(k_1)k_{1\nu}
\varepsilon_{\alpha}^*(k_2)k_{2\beta},
\label{eqdecay35bis}
\ee
which could have been guessed on the basis of Lorentz, parity and gauge invariance
\cite{donoghue92,bhaduri88}. The summation over photon polarizations in (\ref{eqdecay35}) leads to
\be
\left|
{\cal T}_{\pi_0\rightarrow\gamma\gamma}
\right|^2
=512\pi^2\alpha^2\f{m_{\pi}^4}{M_u^2}
\left|
g_{\pi q \bar{q}}M_u^2J_4(-m_{\pi^2}-i0^+)
\right|^2,
\label{eqdecay37}
\ee

where 
\be
J_4(q)=
\int\f{d^4q_3}{(2\pi)^4}
\f{1}{[(q_3-k_1)^2+M_u^2][(q_3+k_2)^2+M_u^2)(q_3^2+M_u^2)}.
\label{fonctionJ4}
\ee

Finally,
\beqn
\Gamma_{\pi_0\rightarrow\gamma\gamma}(T)&=&
\f{1}{2}\f{1}{2m_{\pi}}\f{1}{(2\pi)^2}\int\f{d^3k_1}{2|\vec{k}_1|}
\f{d^3k_2}{2|\vec{k}_2|}\left|
{\cal T}_{\pi_0\rightarrow\gamma\gamma}
\right|^2\delta(q_{\pi}-k_1-k_2)\nonumber\\
&=&\f{1}{32\pi m_{\pi}}\left|
{\cal T}_{\pi_0\rightarrow\gamma\gamma}
\right|^2,
\label{eqdecay38}
\eeqn
\goodbreak
where the factor $1/2$ takes into account the presence of two identical particles in the final state.\par

\begin{figure}[hbt]
\vbox{\begin{center}\psfig{file=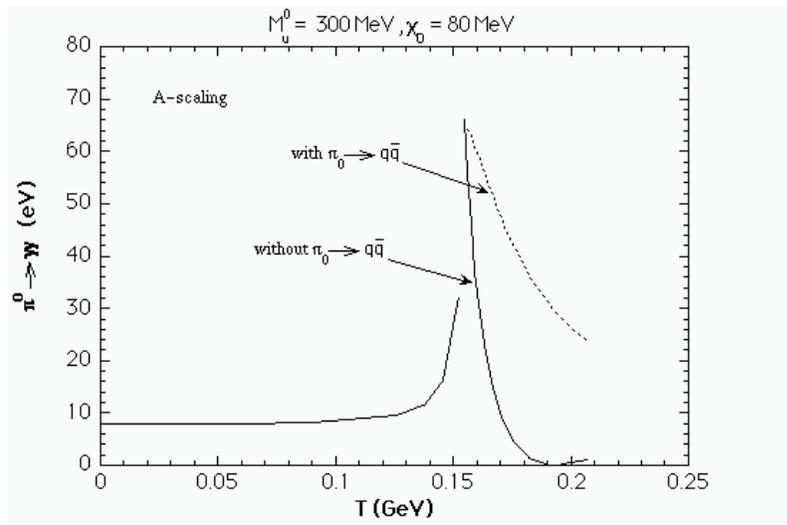,width=\widtheps}\end{center}}
\caption{\em Decay width $\Gamma_{\pi_0\rightarrow\gamma\gamma}$ with the unphysical process $\pi_0\rightarrow\bar{q}q$ included (dashed line) or not (plain line). \label{figuredecay9b.eps}}
\end{figure}

The temperature behavior of $\Gamma_{\pi_0\rightarrow\gamma\gamma}$
is given in figure \ref{figuredecay9b.eps}\footnote{Because of the behavior of $\gpiqq$, the width goes to zero at the threshold, then increases sharply. This is clearly an artifact of crossing the threshold, which would be removed provided we would smooth $\gpiqq$ around it. Because of this artifact, we do not draw this part of the width.}. The dashed line takes into account the unphysical process
$\pi_0\rightarrow \bar{q}q$. The plain line gives
$\Gamma_{\pi_0\rightarrow \gamma\gamma}$ where we artificially removed by hand the imaginary part of
$J_4$ and
$\gpiqq^{-2}$. The pion mass is  still given by (\ref{massedupion}). 

Note that a pure NJL would give the same figure except that the peak would be shifted to a larger temperature (around 190 MeV instead of 140 MeV). This is due to the fact that scaled NJL models incorporate the coupling of quarks to the gluon condensate. The zero temperature result is however not changed from a pure NJL because $\chi_0$ enters into the game only through the combination $\Lambda\chi_0$, which is fixed by the pion weak decay constant $\fpi$.\par

The numerical value of
$\Gamma_{\pi_0\rightarrow\gamma\gamma}$ is very sensitive to the actual pion mass, being proportional to
$m_{\pi}^3$.
A small pion mass error on $J_4(-m_{\pi}^2)$ and  
 $\gpiqq$ can also have some importance.
A great numerical precision of
$m_{\pi}$ is then required, and also for quantities which depend on it\footnote{This remark implies that we should take into account the mass splitting between  $\pi_0$ and
$\pi^{\pm}$ in the determination of
$\Gamma_{\pi_0\rightarrow\gamma\gamma}$. Table~\ref{tabledecay2}
has been obtained with $m_{\pi}$ = 139 MeV, as is usually done in the literature, e.g. \cite{klevansky94}. One expects a better agreement with experiment if the value $m_{\pi}$ = 135
MeV is chosen. We have verified this is actually the case. \addtocounter{footnote}{-1}
\refstepcounter{footnote}\label{notempion135}}.

At finite temperature, we get the table \ref{tabledecay2} where the experimental value is from \cite{pdb96}.

\begin{minipage}[c]{\textwidth}
\begin{table}
\caption{\em Decay width
$\Gamma_{\pi^0\rightarrow\gamma\gamma}$}
\begin{tabular}{cccccc}
&Exp. &NJL$_{\Lambda = 0.9\rm\ GeV}$ &NJL$_{\Lambda = \infty}$ &
Non-local &A-scaling \\ \hline
$\Gamma_{\pi^0\rightarrow\gamma\gamma}$ [eV] & 7.7$\pm$ 0.5
& 5.50 &8.75 & 7.22  &8.75
\end{tabular}
\label{tabledecay2}
\end{table}
\end{minipage}

We also compare our results with an instantaneous 3-dim\-ensional non-local NJL model\cite{schmidt95,blaschke94,blaschke95b}.\par

An important feature has to be noticed: each  value  given in table \ref{tabledecay2} is relative to a cut-off put to infinity in the convergent integrals, except for the third column. As one can see, this prescription is indispensable to reproduce 
the experimental result
(see also \cite{blin88}).
One can wonder about the prescription ``putting a cut-off only for diverging quantities''. Does the consistency of the model not require  to keep the cut-off everywhere? However, it is shown in the work
\cite{ripkaball94}  that a sharp cut-off is responsible for this fact. It then seems  reasonable to accept the prescription. One can also argue
 \cite{reinhardt95,reinhardt95b} that the diagrams of
figure \ref{figuredecay7.eps} are the only ones to contribute to the anomaly if the cut-off is infinite, as in case of a renormalizable theory. In  a
non-renormalizable one, the ratio
$M_u/\Lambda$ is finite and other diagrams have to be included (see appendix A of \cite{reinhardt95b}).

\vspace{-0.5cm}

If the cut-off is kept for each integral in the NJL model, these diagrams should then have to be taken into account. One can imagine that these diagrams would compensate for the discrepancy between the third and fourth column. However, it is  simpler  to only consider the diagrams of figure  (\ref{figuredecay7.eps}) and to take an infinite cut-off wherever  possible. The functions  $J_3$ and $J_4$ are evaluated in this way.

Figure (\ref{figuredecay9b.eps}) shows a sharp peak. It is present both in the scaled model and in  a pure NJL model 
\cite{klevansky94,jamvdb95b,hashimoto88}. At high temperature, it goes to zero. We have shown in  \cite{jamvdb95b} that the peak is sharper in the scaled model than in pure and  non-local NJL models\footnote{We have shown in the B-scaling NJL model \cite{jamvdb95b} that the two-photon decay width of the neutral pion  goes to zero at the transition. This is connected to the fact that, in this model,  the pion mass goes  continuously to zero below the chiral transition, and vanishes at the transition. The width
$\Gamma_{\pi_0\rightarrow\gamma\gamma}$ then vanishes  at
$T$ = $T_c$. However, this is true only in the two-flavor version of the B-scaling model, see \cite{jamvdb95}.}. It is located at the pion threshold temperature which is almost coincident with the critical temperature (it would be exactly the critical temperature in the chiral limit). The peak is mainly due to the behavior of 
$J_4$.

As a  conclusion to this section, we can write that:
\begin{itemize}
\item taking the cut-off to infinity whenever  possible  (table~\ref{tabledecay2}) gives similar results in pure NJL and A-scaling models  for  $\Gamma_{\pi_0\rightarrow\gamma\gamma}$ at zero temperature. There is a
14~\% discrepancy with the experimental result
\cite{pdb96}. The non-local NJL model of the Rostock group
\cite{schmidt95,blaschke94,blaschke95b} is in better agreement (6~\%). The footnote~(\ref{notempion135}) suggests, however, that the discrepancy between theory and experiment is connected to the pion mass we take in our calculations
($\mpi=m_{\pi^{\pm}}=139$ MeV). 
Indeed, with (\ref{goldberger},\ref{eqdecay37},\ref{eqdecay38}) and
$J_4(-m_{\pi}^2)$ at zero temperature and density
\be
J_4(-m_{\pi}^2)_{\beta=\infty,\mu=0}=
-\f{1}{32\pi^2M_u^2}
\left(
\f{2M_u}{m_{\pi}}\mbox{arcsin}
\f{m_{\pi}}{2M_u}
\right)^2,
\label{J4Tnulle}
\ee
we straightforwardly obtain
\be
\Gamma_{\pi_0\rightarrow\gamma\gamma}\approx
\f{\alpha^2m_{\pi}^3}{64\pi^3f_{\pi}^2}
\left(
\f{2M_u}{m_{\pi}}\mbox{arcsin}\f{m_{\pi}}{2M_u}
\right)^4.
\label{gammaapproche}
\ee
With the parameters from this section, eq.~{\ref{gammaapproche}} gives
$\Gamma_{\pi_0\rightarrow\gamma\gamma}$ = 8.65 eV, very close to the exact theoretical value (8.75 eV).
This shows that the
Goldberger-Treiman relation (\ref{goldberger}) is a very accurate
approximation. If we had chosen $m_{\pi}$ =
135 MeV, eq. (\ref{gammaapproche}) would have given
$\Gamma_{\pi_0\rightarrow\gamma\gamma}$ = 7.91 eV, which is now very close to the experimental value given in table~\ref{tabledecay2};
\item eq. (\ref{gammaapproche}) can still be more simplified. In the limit $\mpi/(2M_u)\rightarrow0$, we have
\be
\Gamma_{\pi_0\rightarrow\gamma\gamma}\approx
\f{\alpha^2m_{\pi}^3}{64\pi^3f_{\pi}^2},
\label{gammaapprochebis}
\ee
identical to the result quoted in \cite{itzykson85}. For $m_{\pi}$ = 135 MeV, this gives 7.64 eV;
\item in the A-scaling model (and in the non-local NJL model) \cite{jamvdb95b}, the decay width 
 $\Gamma_{\pi_0\rightarrow\gamma\gamma}$ has a sharp peak near the critical temperature. This is similar to a pure NJL model \cite{klevansky94,hashimoto88}, except that the peak of the latter is rejected to a higher temperature;
\item above  threshold, the unphysical process $\pi_0\rightarrow \bar{q}q$
is possible, since the models are non-confining. The pion acquires an unphysical width,
$\gpiqq$ and $J_4$ become complex quantities. Considering this process consistently in the model leads to a non-negligible influence on 
$\Gamma_{\pi_0\rightarrow\gamma\gamma}$ versus
temperature, as shown by the comparison between the plain and dashed curves in
figure \ref{figuredecay9b.eps}. Note that, taking this unphysical process into account, we still have a peak, in contradiction with \cite{klevansky94};
\item the constituent up quark mass  which enters  the calculation of  $\Gamma_{\pi_0\rightarrow\gamma\gamma}$ (either explicitly or implicitly through the pion mass) is extracted from the gap equation
(\ref{gapup}). In the case of a coupling with the
electromagnetic field, this equation has to be modified
\cite{lemmer89,vautherin89}. It is however shown in these references that it is justified to limit oneself to the usual gap equation for 
a result valid up to order $O(\alpha)$;
\item it should also be mentioned that our results are at complete variance with those of Pisarski et al. \cite{pisarski96,pisarski97}, who obtain a continuous decrease of the decay width with the temperature.
\end{itemize}

\section{Mixing between scalar isoscalar mesons and the glueball}
\label{sectionmixing}

In the previous section, we focused on the conventional pseudoscalar channel, plus the effect of a scalar glueball. Because of its scalar nature, it can be interesting to consider this channel. The scalar glueball is an isoscalar. It can then couple to similar d.o.f. In our three flavor models, this implies a coupling between three particles: the glueball and two $\bar{q}q$-like mesons. In our formalism or in GCM type  models, these mesons are associated to the canonical vertex $\one$ in Dirac space. This is probably a very crude approximation for the scalar channel. However, as a first approximation, it can give insights into the way of treating these mesons.\par

The interest in scalar mesons  mainly lies in the search for the scalar glueball, a particle which should show up in QCD because of the color charge of the gluons, and which is seen on the lattice (see below). In our three-flavor model, the scalars form a set of ten particles (singlet, octet plus glueball). The identification of the scalar nonet is anything but clear, both experimentally and theoretically. Does it have to include the $a_0(1450),f_0(1300),f_0(1590),K_0^*(1430)$ as suggested by Montanet \cite{montanet95}, leaving the $a_0(980)$ and $f_0(980)$ as $K\bar{K}$ molecules and the $f_0(1500)$ as a solid candidate for the glueball? Or should the nonet include the $f_0(980)$, as proposed by Palano \cite{palano95} (in that work, this $f_0(980)$ is also said to be mainly a $\bar{s}s$ state)? Törnqvist \cite{tornqvist96}  presents the mesons $f_0(980)$ and $f_0(1300)$ as two manifestations of the same $\bar{s}s$ state. Lindenbaum and Longacre \cite{lindenbaum92} claim that the $0^{++}$ data can be reproduced with the four mesons $f_0(980),f_0(1300),f_0(1400)$ and $f_0(1710)$. In the latter case, it is also assumed that the $\theta(1710)\equiv f_0(1710)$ has a spin 0.\par

Theoretically, the situation is also very obscure. For example, Münz, Klempt et al. \cite{munz95,munz96} calculate masses and decay properties of the scalar nonet in a relativistic model with  an instanton-induced interaction and a linear confinement. They give support to identify the nonet with $a_0(1450),K_0^*(1430), f_0(980)$ and $f_0(1500)$. (There is no glueball in their model.) Lattice QCD, in the UKQCD Collaboration, predicts a scalar glueball of mass around 1500 MeV \cite{bali93}. This  favors its identification with the $f_0(1500)$ \cite{amsler95,amsler95b}.  However, the valence (quenched) approximation of Sexton et al. \cite{sexton95} gives a glueball with a mass of approximately 1710 MeV, suggesting its identification with the $f_J(1710)$, provided its total angular momentum is $J=0$.  The coupling with $\bar{q}q$ states would then increase the glueball mass of about 60 MeV. The fact that the glueball disperses over three resonances is also shown in the K-matrix fit formalism of Anisovich et al. \cite{anisovich96,anisovich97b} which, however, favors a glueball at 1500 MeV, as also shown  in  \cite{anisovich95}.

This scalar resonance $f_0(1500)$ has been discovered at LEAR by the Crystal Barrel Collaboration \cite{anisovich94,amsler94} and also seen\footnote{In the Mark III data on  the radiative decay  $J/\psi\rightarrow4\pi+\gamma$, a sharp peak is seen in the $4\pi$ spectrum, having the quantum numbers $0^{++}$ and a mass and width compatible with $f_0(1500)$ MeV \cite{bugg95}.} by this group in the decay $f_0(1500)\rightarrow K\bar{K}$ MeV \cite{CBC96}. Note that this requires a mixing between the glueball and the neighboring $\bar{q}q$ states, as shown in \cite{close97}. As already mentioned, Sexton et al. \cite{sexton95} have also noted a mixing between the glueball and the $\bar{q}q$ states. In their work, and in subsequent researches from one of the authors \cite{weingarten97,weingarten97b}, the glueball is the $f_J(1710)$. Its mixing with $f_0(1370)$ and $f_0(1500)$ in turn requires  that $J=0$ \cite{close97}.

Although we have varied the parameters of the scaled NJL model in order to examine both the $f_0(1500)$ and the $f_0(1710)$ as possible glueball states in the study \cite{jamvdb97}, we shall mainly focus on \cite{jamvdb98} in the following, and restrict ourselves to the $f_0(1500)$ as the scalar glueball.

We concentrate on the minimal list of light scalars given in \cite{pdb96}. It includes the $f_0(400-1200)$, $f_0(980)$, $f_0(1370)$, $f_0(1500)$, $a_0(980)$, $a_0(1450)$ and $K_0^*(1430)$. We also consider the heavier $f_0(1710)$ as a $\bar{s}s$ candidate. We give in section \ref{bs} and appendix \ref{PandGamma} the masses of the various scalars within our model. It is clear that, without coupling with the glueball, the $a_0$ and $f_{0,\bar{u}u}$ are degenerate and, in the chiral limit, of mass $m=2M_u^0$. The $f_{0,\bar{s}s}$ has then a mass $m=2M_s^0$ and $m_{K_0^*}=M_u^0+M_s^0$. To make contact with our model, the $f_0(400-1200)$ has to be rejected without explanation: there is no room left for it in our formalism. One has also to choose between the sets ($f_0(980)$,$a_0(980)$) and ($f_0(1370)$,$a_0(1450)$). (Because of the meson masses in the chiral limit as given above, it is not possible to have, for example, $f_0(980)$ with $a_0(1450)$.) To compare our work  with \cite{close97}, we choose to keep the set ($f_0(1370)$,$a_0(1450)$), noticing that the $f_0(980)$ and the $a_0(980)$ can be interpreted as {$K\bar{K}$} molecules\footnote{For the alternative choice, see \cite{jamvdb97}.}.

In section \ref{sectionhybrids}, 
we describe how the scalars are treated in the scaled NJL model.
In section \ref{sectioncoupling},
 we study the mixing angles and meson to quarks coupling constants. Finally, results concerning the two-photon decay width of the scalar isoscalars  is given in section \ref{sectionscalardecay}.

\subsection{Hybrids: from unphysical pure $\bar{n}n,\bar{s}s$ and glueball fields to physical particles}
\label{sectionhybrids}

In appendix \ref{PandGamma} we give the general structure of the matrix propagator of both pseudoscalar $P$ and scalar\footnote{In a  3-flavor NJL model with a 't Hooft determinant, scalars have already been investigated by Hatsuda and Kunihiro \cite{hatsuda94,hatsuda91} and by Bajc et al. \cite{blin94}.} $\G$ particles. In the following, we are interested in the latter. The cases of the $a_0$ and the $K_0^*$ are trivial since  these particles are unmixed. However, there are three coupled states in the isoscalar sector: two members of the nonet (the singlet and the eighth part of the octet) and the glueball\footnote{We are working in the isospin limit, which explains why all the components of $a_0$ and  $K_0^*$ are unmixed and why the third component of the octet does not contribute to the isoscalars.}. Going from mixed  to physical states can be treated just like an eigenvalue (e.v.)/eigenvector (E.V.) problem and is discussed in some detail  in appendix \ref{appendixmixing}. This is generally a non-trivial problem because of the relativistic nature of the theory: a relativistic theory does not allow one to describe observable states as superpositions of other states with different masses. Mathematically, this is traced back to the energy dependence of the $\G$ matrix elements. Obtaining the physical masses is however  simple: one has just to look for the zeroes of the $\G$ e.v. \cite{jamvdb97,jamvdb98}. The only difficulty consists in attaching the right e.v. to the right particle. This can be done by varying (slowly enough to follow the evolution of the e.v.) the coupling to the glueball from zero to its actual value. 
We can also use the analytical solutions for the e.v. Since we have a 
 $3\times3$ problem, we have just to look for the solutions of a cubic equation\footnote{The cubic equation $x^3+ax^2+bx+c=0$ has the three {\em real} solutions
\beqn
x_1&=&-2\sqrt{Q}\cos(\f{\theta}{3})-\f{a}{3},\label{cubic1}\\
x_2&=&-2\sqrt{Q}\cos(\f{\theta+2\pi}{3})-\f{a}{3},\label{cubic2}\\
x_3&=&-2\sqrt{Q}\cos(\f{\theta-2\pi}{3})-\f{a}{3},\label{cubic3}\\
\eeqn
with $\theta=\arccos(R/\sqrt{Q^3})$ and $Q=(a^2-3b)/9,R=(2a^3-9ab+27c)/54$, provided that $R^2<Q^3$ and $R,Q$ are real. One has then to find a one-to-one correspondence between each  of these solutions and the physical states. This is performed by comparing these solutions to the known mass of the mesons when the coupling with the glueball is removed.
}. Once we have identified to which e.v. corresponds each state (this is done by putting to zero the coupling to the glueball), we can switch on this coupling without having to  care about the e.v. ordering\footnote{In the purely numerical case, this is not possible because the e.v. are not ordered once for all.}.
We proceed as follows: for a vanishing coupling to the glueball (matrix elements (\ref{eqA73a},\ref{eqA74a}) put to zero), the glueball mass can be taken from eq. (\ref{eqA75a}). We can then identify from (\ref{cubic1}--\ref{cubic3}) the e.v. to attach to it. The remaining $2\times2$ system (eqs. (\ref{eqA64a}--\ref{eqA72a})) can be exactly diagonalized, giving for one of the $f_0$ the same mass as the $a_0$, eq.~(\ref{sigma1a3}). The other $f_0$ has the same mass expression, except for $M_s^0$ instead of $M_u^0$. We can then also attach one of the e.v. (\ref{cubic1}--\ref{cubic3}) to it.
Since we know which particle $f_0(1370), f_0(1500), f_0(1710)$  is attached to each e.v. in (\ref{cubic1}--\ref{cubic3}),   we are then able to study quite easily the mass evolution with respect to the parameters $(M_u^0,\chi_0$). Because the meson to quarks coupling constants  can be written as an eigenvector (E.V.) problem (see next section), the identification of the e.v. to their corresponding state is also of prime importance to  determine these couplings.

In the preceding sections, we used a glueball mass of  1300 MeV in order to find the $b^2$ parameter in a sigma model limit of the scaled NJL model.
This was done to obtain $b^2$ without having to think about the scalar sector. It is clear from this section that this is only an approximation. In fact, as this study shows, one has to impose the value of the glueball mass. Once its e.v. has been identified, one can look at the value\footnote{With the approximate method of the previous sections, we get $b=4.69$ for ($M_u^0=450 \mbox{ \rm MeV},\chi_0=350 \mbox{ \rm MeV}$), see   table \ref{table1}. In the correct procedure developed here, we have $b=5.02$. Should we have taken a glueball mass of 1500 MeV in the preceding sections, as done here, we would have got $b=5.16$. This shows the consistency of all the  performed approximations. Note that, although playing almost no role on the thermodynamics and on the variation of the condensates as a function of temperature and density,  this difference in $b^2$ is  important for the scalar sector and explains why we have to take the exact expression in this section.} of the glueball mass without coupling. Since it has to coincide with the mass extracted from eq. (\ref{eqA75a}), we can then deduce $b^2$.

Another subtlety shows up in the scalar sector: scalars are always unbound, lying above the unphysical threshold $f_0\rightarrow\bar{q}+q$. The technique of section \ref{modeleseuil}  has to be applied. Note that this makes sense only if this unphysical width is small. In our studies, this is always the case except for the $f_0(1710)$. However, it can be argued (see below) that meaningful quantities, such as two-photon decay widths, can still be extracted.

In the following, we want to calculate the two-photon decay widths of the scalar isoscalars. Close, Farrar and Li \cite{close97} have used general ideas to estimate the relative strength of these decay widths. They consider two schemes for the mixing between $f_0(1370), f_0(1500), f_0(1710)$. Firstly, they consider the case where the $f_0(1500)$ has a large glue content while $f_0(1710)$ is mainly a $\bar{s}s$ excitation. This scheme allows to understand the $f_0(1500)\rightarrow K\bar{K}$ data \cite{CBC96}. The second scheme consists in considering that the glueball lies above the scalar nonet, as suggested in \cite{sexton95,weingarten97,weingarten97b}:  $f_{0}(1710)$ can then be the glueball while $f_{0}(1500)$ is mainly a $\overline{s}s$ excitation. Within these two schemes, Close et al. \cite{close97}
have
estimated the relative strength of the $2\gamma$ widths for the three $f_0$
states:
\beqn
f_0(1370):f_0(1500):f_0(1710)&\approx&12:1:3 \hspace{1.4cm} (\mbox{scheme 1}),\label{scheme1}\\
f_0(1370):f_0(1500):f_0(1710)&\approx&13:0.2:3 \hspace{1.1cm} (\mbox{scheme 2}).\label{scheme2}
\eeqn
For any mixing, the decay width of $f_{0}(1500)$ is always the
smallest. This suggests that the experimental estimation of the $2\gamma$ widths  might  be a
good test of the general idea of $\overline{q}q$  and glue mixing.

We shall restrict ourselves to scheme 1. Note that, although derived using very general assumptions, the ratios (\ref{scheme1},\ref{scheme2}) are obtained assuming $M_u^0=M_s^0$ ($SU(3)$ symmetry) and that loop phase-space effects can be ignored (see eq.~(6.2) of \cite{close97}). We show in the following that these effects are in fact crucial, at least for the crude model we are using.

\begin{figure}[hbt]
\vbox{\begin{center}\psfig{file=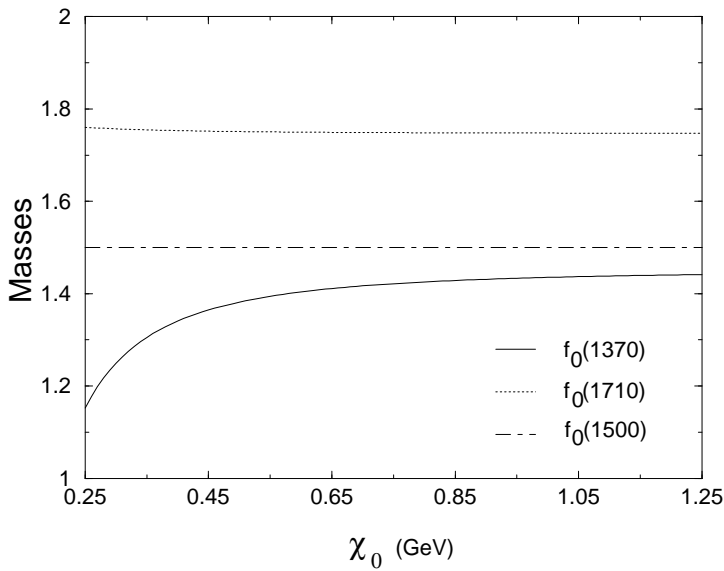,width=\widthepsxmgrthree}\end{center}}
\caption{\em $\chi_0$ behavior of the mass of the three states $f_0(1370)$,
$f_0(1500)$ and $f_0(1710)$.}
\label{figurescalarmass.eps}
\end{figure}

Considering that the three scalar isoscalar mesons we want to investigate are $f_0(1370), f_0(1500), f_0(1710)$ leaves little choice for the value of $M_u^0$: without coupling with the glueball, and in the chiral limit, we would get $m_{f_0(1370)}=2M_u^0$. The state $a_0$ has no coupling with other particles and its mass is exactly $2M_u^0$ in the chiral limit. This leads us to take $a_0(1450)$ as a member of the scalar nonet, rejecting $a_0(980)$ (and ($f_0(980)$) as $K\bar{K}$ molecules. We then take $M_u^0=725$ MeV. The mass of the isoscalars is dependent on the choice of $\chi_0$. However, we  immediately get $m_{a_0(1450)}=1450$ MeV and $m_{K_0^*(1430)}=1600$ MeV, the latter being quite poorly reproduced. We shall not focus on this problem here. We show in figure \ref{figurescalarmass.eps} the $\chi_0$ behavior of the mass of the three $f_0$ states.

For large values of $\chi_0$, the states take the mass
they would have without coupling ($m_{f_{0}(1370)}= 2M_u^0=1450$ MeV
, $m_{f_{0}(1710)}=2M_s^0=1728$ MeV in the chiral limit). The mass of
$f_{0}(1710)$ is  quite
 stable, while an acceptable value for the mass of $f_{0}(1370)$ limits the
value of $\chi_0$ to a domain between 300 and 450 MeV.

\subsection{Mixing angles and coupling constants as an eigenvector problem}
\label{sectioncoupling}

\begin{figure}[hbt]
\vbox{\begin{center}\psfig{file=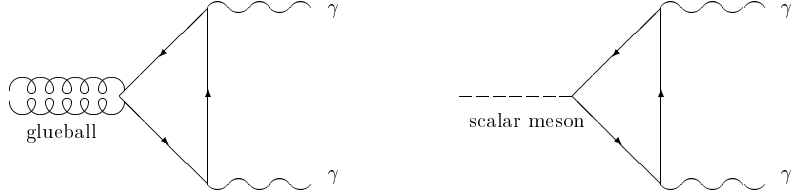,width=\widtheps}\end{center}}
\caption{\em Two-photon decay of the scalars.}
\label{scalar2gamma.eps}
\end{figure}

We explain in appendix \ref{appendixmixing} how to go from mixed states to physical states. The derivation shows how to obtain the masses (they are the zeros of the e.v.) and how to define the physical states.  This enables to get the meson to quarks coupling constants. Similarly to what has been performed in the $\pi_0\rightarrow\gamma\gamma$ case, the decay of scalar mesons into two photons  is represented as a triangle diagram. Since the glueball field is not pure  (it mixes with quarks states), two  types of diagrams have to be investigated, as shown in figure \ref{scalar2gamma.eps}.

In the limit of a large $\chi_0$, the glueball decouples from the nonet. This is then the vacuum gluon condensate which fixes the mixing\footnote{This is also the only parameter at our disposal since we have fixed $M_u^0$ to reproduce the $a_0(1450)$.}: the smaller $\chi_0$, the larger the mixing. This is confirmed in the mass plot, figure \ref{figurescalarmass.eps}. This plot shows that, in the range of $\chi_0$ investigated to reproduce the meson masses, the $f_0(1710)$ remains mainly a $\bar{s}s$ excitation, while the strength of the coupling to the $u$ and $s$ quark has about the same order of magnitude for $f_0(1370)$ and $f_0(1500)$. This has of course to be checked, looking at the meson to quarks coupling constants. This is the subject of this section.\par

 In our model, the three scalar states have a glue content. As mentioned, the $f_0(1500)$ is the one which yields a pure glueball if no mixing. This state has the largest glue content even if that of $f_0(1370)$ can become important. We then work with scheme 1, eq. (\ref{scheme1}).

A mixing between the three glueball and scalar fields modifies the transition amplitude. Without coupling, the $f_0(1370)$, for example, can decay into two photons only through the production of a $\bar{u}u$ pair. As soon as a mixing with the other states is allowed, the production of a $\bar{s}s$ pair has to be considered. The coupling constants can be obtained from appendix \ref{appendixmixing}. We recall here the main result\footnote{Note that in ref. \cite{jamvdb98} we use a prediagonalization for the system $f_0(1370)$, $f_0(1710)$ which leads directly to the physical states if no coupling with the glueball. In that case, eq. (\ref{getoilemoins2}) can be used.}, 

\begin{minipage}[t]{\textwidth}
\[
\hfill \ginvdeux_*(\qq=-\micarre)_{ii}=\sum_{k,l}\ginvdeux_{kl}(\qq=-\micarre)V_*(-\micarre)_{ki}V_*(-\micarre)_{li},
\hfill (\mbox{\ref{getoilemoins3}})
\]
\end{minipage}

with $(\ginvdeux_*)_{i,j\ne i}=0$. $V_*$ is the E.V. matrix which allows going from mixed states to physical states and $g$ is the mixed scalar to quarks coupling constant matrix.
Eq.~(\ref{getoilemoins3}) is  the basis to obtain the physical coupling constants to quarks. Indeed, the functional (\ref{Znjl}) contains the term $(\bar{q}\varphi^a\Gamma^a)q$, where $\varphi^a$ denotes the mixed states. For the scalar isoscalars, this gives $\bar{q}\sigma_a\ld{a}/2q$. Using the relation going from the mixed states to the physical ones, 

\begin{minipage}[t]{\textwidth}
\[
\hfill \Sigma=\ginv_*(-\micarre)\Vinv_*(-\micarre)\s,
\hfill (\mbox{\ref{newfields}})
\]
\end{minipage}

it is clear that the meson to quarks coupling constants can be identified from\footnote{The conventions of appendix \ref{appendixmixing} are used, where we denote by $A(\qq=-\micarre)$ a matrix $A$ of which each element is evaluated at the same $\qq$, in opposition to $A(-\micarre)$ which indicates a matrix whose  first column is evaluated at $-m_1^2$, the second at $-m_2^2$, ..., the $n$th at $-m_n^2$.}
\be
\bar{q}\left(
V_*(-m_i^2)g_*(-m_i^2)\Sigma
\right)_a\f{\ld{a}}{2} q,
\ee
which implies that the meson to quarks coupling constants  are (no summation over $i$)
\beqn
g_i^{*\bar{u}u}=g_i^{*\bar{d}d}&=&\left[
(\lambda_0)_{11}{V}_{*1i}(-m_i^2)+(\lambda_8)_{11}{V*}_{2i}(-m_i^2)
\right]g_*(-m_i^2)_i,
\label{eq27}
\\
g_i^{*\bar{s}s}&=&\left[
(\lambda_0)_{33}{V_*}_{1i}(-m_i^2)+(\lambda_8)_{33}{V_*}_{2i}(-m_i^2)\right]
g_*(-m_i^2)_i,\label{eq28}
\eeqn
where $g_*(-m_i^2)_i\equiv g_*(-m_i^2)_{ii}$ and $i=1\equiv f_0(1370)$, $i=2\equiv f_0(1710)$, $i=3\equiv f_0(1500)$.

\begin{figure}[hbt]
\vbox{\begin{center}\psfig{file=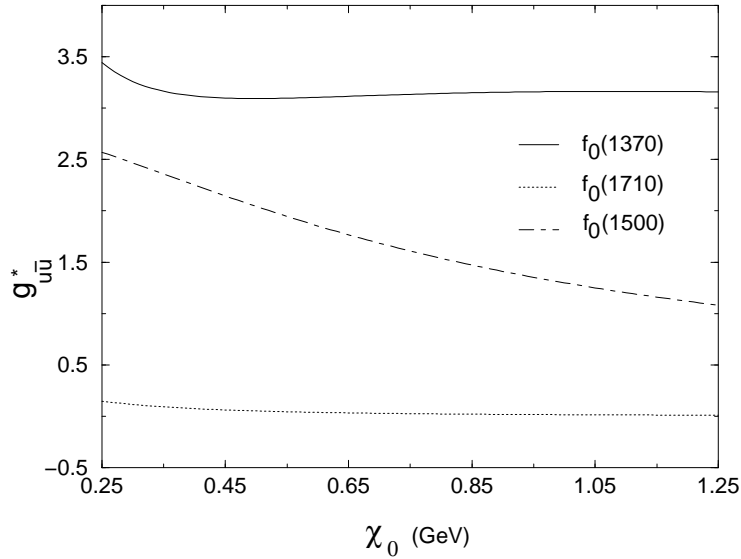,width=\widthepsxmgrthree}\end{center}}
\caption{\em $\chi_0$ behavior of the coupling constant to up quarks,
 ${g}^{\bar{u}u}$, of the three states $f_0(1370)$,
$f_0(1500)$ and $f_0(1710)$.}
\label{guu.eps}
\end{figure}

\begin{figure}[hbt]
\vbox{\begin{center}\psfig{file=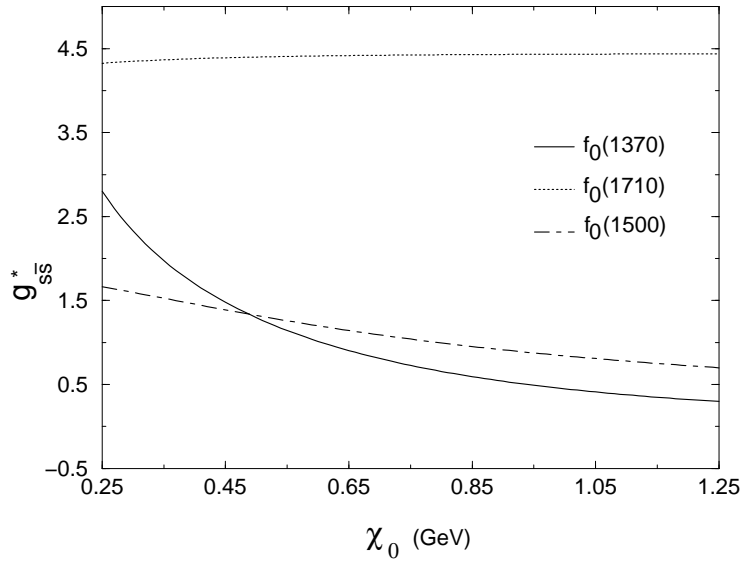,width=\widthepsxmgrthree}\end{center}}
\caption{\em $\chi_0$ behavior of the coupling constant to the strange quarks,
 ${g}^{\bar{s}s}$, of the three states $f_0(1370)$,
$f_0(1500)$ and $f_0(1710)$.}
\label{gss.eps}
\end{figure}

This expression clearly shows  that, as soon as the E.V. matrix $V_*$ and the meson masses are obtained, the coupling constants can be algebraically constructed. They are plotted, as a function of $\chi_0$, in figures \ref{guu.eps} and \ref{gss.eps}, for $g_i^{*\bar{u}u}$ and $g_i^{*\bar{s}s}$, respectively.\par

\vspace{-1cm}   

We see from picture \ref{guu.eps} that the coupling constant of $f_0(1710)$ to the up quarks is small, vanishing quickly as $\chi_0$ is increased. This shows that this meson can be considered as an almost  pure $\bar{s}s$ excitation\footnote{This confirms what was only intuitive in figure \ref{figurescalarmass.eps}.}. Note also that, because this coupling constant is small, the processes involving up quark diagrams for this meson will be suppressed. This is interesting since the threshold coming from $2M_u^0\equiv 1450$ MeV $< 1710$ MeV implies that the unphysical decay width into free quarks could be important, so that {\em i}) the small width approximation would not be correct; {\em ii}) unphysical effects  would radically change physical quantities. Because $g_i^{*\bar{u}u}(-m_{f_0(1710)}^2)$ is small, it is however clear that these unphysical processes play no role.

This picture also shows that, although  figure \ref{figurescalarmass.eps} indicates that the masses attain almost their asymptotic value  at large $\chi_0$ (no coupling of the quark states with the glueball state), the quark content   of the glueball is not negligible.
This is seen  from the nonvanishing of $g_i^{*\bar{u}u}(-m_{f_0(1500)}^2)$ and in  picture \ref{gss.eps}.
The latter also shows  that the $f_0(1370)$ has still a $\bar{s}s$ component for $\chi_0$ as large as 1250 MeV. 

The conclusion we can draw from these plots is that the glueball to quarks coupling constant is a slow decreasing function of the vacuum gluon condensate $\chi_0$, showing that the glueball has  a non-negligible quark content. This is particularly true for values of $\chi_0$ in the regime giving ``good'' masses for the scalar mesons ($\chi_0\in[300,450] MeV)$. Moreover, for this set of parameters, the $f_0(1370)$ has also an important strange quark content.\par

Having the coupling constants, it is now easy to determine the two-photon decay width.

\subsection{Two-photon decay of scalar isoscalars}
\label{sectionscalardecay}

\begin{figure}[hbt]
\vbox{\begin{center}\psfig{file=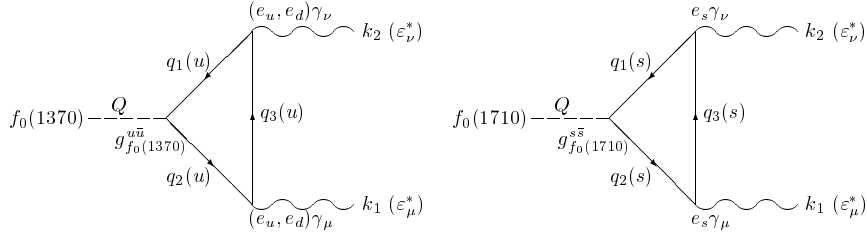,width=\widthepsthree}\end{center}}
\caption{\em Direct Feynman  diagrams for $2\gamma$ decay of $f_0(1370)$ (left)
and  of $f_0(1710)$ (right).}
\label{scalartriangle.eps}
\end{figure}

With the meson masses and the coupling constants to quarks, we are in a position to calculate the two-photon decay of scalar isoscalars. Without coupling with the glueball, we would just have to evaluate diagrams similar to 
figure \ref{figuredecay7.eps}. They are given in figure \ref{scalartriangle.eps}, where we omit the crossed diagram.

Since, in the scaled NJL model, the mesons $f_0(1370)$ and $f_0(1710)$ are not anymore pure $\bar{u}u$ or $\bar{s}s$ excitations, these diagrams have to be supplemented by similar ones with a permutation of the quark flavor. We then arrive to the following amplitudes:
\beqn
{\cal
T}_{{f}_{0}(1370)}&=&{N}_{c}({e}_{u}^{2}+{e}_{d}^{2}){g}_{{f}_
{0}(1370)}^{*u\overline{u}
}{\varepsilon }_{\mu }^{*}({k}_{1}){\varepsilon }_{\nu
}^{*}({k}_{2})U+{N}_{c}{e}_{s}^{2}{g}_{{f}_{0}(1370)}^{*s\overline{s}
}{\varepsilon
}_{\mu }^{*}({k}_{1}){\varepsilon }_{\nu }^{*}({k}_{2})S,
\label{eqscalar10}\\
{\cal
T}_{{f}_{0}(1710)}&=&{N}_{c}{e}_{s}^{2}{g}_{{f}_{0}(1710)}^{*s\overline{s}
}{\varepsilon }_{\mu }^{*}({k}_{1}){\varepsilon }_{\nu
}^{*}({k}_{2})S+{N}_{c}({e}_{u}^{2}+{e}_{d}^{2}){g}_{{f}_{0}(1710)}^{*u\overline{u}
}{\varepsilon }_{\mu }^{*}({k}_{1}){\varepsilon }_{\nu
}^{*}({k}_{2})U,
\label{eqscalar11}
\eeqn 
for the $f_0(1370)$ and $f_0(1710)$ states while, having some $\bar{q}q$ excitations, the glueball can also decay to two photons:
\beqn
{\cal
T}_{{f}_{0}(1500)}&=&{N}_{c}({e}_{u}^{2}+{e}_{d}^{2}){g}_{{f}_{0}(1500)
}^{*u\overline{u}\
}{\varepsilon }_{\mu }^{*}({k}_{1}){\varepsilon }_{\nu
}^{*}({k}_{2})U+{N}_{c}{e}_{s}^{2}{g}_{{f}_{0}(1500)}^{*s\overline{s}
}{\varepsilon }_{\mu }^{*}({k}_{1}){\varepsilon }_{\nu
}^{*}({k}_{2})S.
\label{eqscalar12}
\eeqn

In eqs. (\ref{eqscalar10}--\ref{eqscalar12}), we have written 
\beqn
(U,S)&=&\int{d^4q_3 \over (2\pi {)}^{
4}}\tr_D\left\{{({q\!\!\!/}_2+{M}_{u,s}^0){\gamma
}_{\mu}({q\!\!\!/}_3+{M}_{u,s}^0){\gamma
}_{\nu}({q\!\!\!/}_1+{M}_{u,s}^0)}\right\}\nonumber\\
&&\hspace{4cm}\mbox{}\times{1 \over
({q}_{1}^{2}+({{M}_{u,s}^0})^{2})({q}_{2}^{2}+({{M}_{u,s}^0})^{2})
({q}_{3}^{2}+({{M}_{u,s}^0})^{2})},
\label{eqscalar7}
\eeqn
which is similar to the trace quantity appearing in eq.~(\ref{eqdecay35}) except for the $\gamma_5$ which is due to the pseudoscalar nature of the pion and which leads to the $\epsilon^{\mu\nu\alpha\beta}$ structure of eq.~(\ref{eqdecay35bis}). In the scalar case, the amplitude still leads to the function $J_4$ of eq.~(\ref{fonctionJ4}) that we write  $J_u$ and $J_s$ according to the flavor which enters the equations. We finally get, after a summation over the photon polarizations (and for an incoming meson at rest)
\be \Gamma_i={1 \over 2}{1 \over 2{m}_{i}}{1 \over (2\pi
{)}^{2}}\int\int{{d}^{3}{k}_{1} \over
2\left|{{\vec{k}}_{1}}\right|}{{d}^{3}{k}_{2} \over
2\left|{{\vec{k}}_{2}}\right|}{\left|{{\cal T}_{i}}\right|}^{2}
\delta(Q-{k}_{1}-{k}_{2})
={1 \over 32\pi }{1 \over {m}_{i}}
{\left|{{\cal T}_{i}}\right|}^{2}
\label{eqdecay5}
\ee
with 
\be
\left|{\cal T}_i\right|^2=\f{12800}{9}\pi^2\alpha^2
\bigg|(m_i^2-4{M_u^0}^2)
M_u^0 g_i^{*\bar{u}u}J_u
(-m_i^2)+\f{1}{5}(m_i^2-4{M_s^0}^2)M_s^0 g_i^{*\bar{s}s}J_s
(-m_i^2)\bigg|^2,
\label{eqdecay13}
\ee
where $i=f_0(1370),  f_0(1710), f_0(1500)$.

\begin{figure}[hbt]
\vbox{\begin{center}\psfig{file=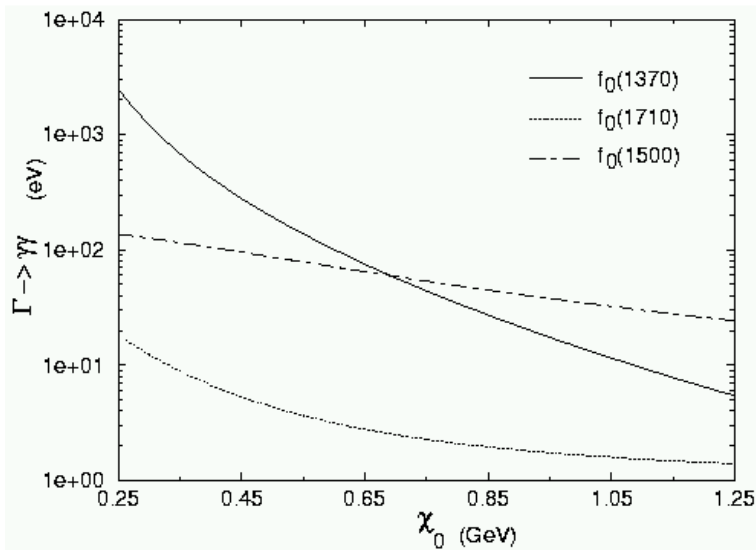,width=\widthepsxmgrthree}\end{center}}
\caption{\em $\chi_0$ behavior of the  $2\gamma$ decay width of the three states
$f_0(1370)$, $f_0(1500)$ and $f_0(1710)$.}
\label{scalarwidth.eps}
\end{figure}

With these equations, we are now able to plot the two-photon decay width of the scalar mesons, see figure \ref{scalarwidth.eps}. 

The three widths decrease with increasing $\chi_0$, as it should since, in the limit $\chi_0\rightarrow\infty$, one of the corresponding coupling constants vanishes. $\Gamma_{f_0(1710)}$ is always the smallest:  the first term of the r.h.s. of (\ref{eqdecay13}) is small due to the small
value of ${g}_{{f}_{0}(1710)}^{*\bar{u}u}$ (i.e. this meson is mainly a $\bar{s}s$ excitation), while the second one nearly
vanishes due to the fact that $m_{f_0(1710)} \approx 2M_s^0$. Note that, in the chiral limit, and for a vanishing mixing between the scalar isoscalars, the model predicts that the decay widths  vanish exactly.
For the $\bar{q}q$ states, this can be traced back to the quark loop which gives 
  a phase-space like factor\footnote{Note that this phase-space factor can be negative. In that case, the coupling constant is also negative. The products $g_i^{*\bar{u}u}\times (m_i^2-4{M_u^0}^2)$ and $g_i^{*\bar{s}s}\times (m_i^2-4{M_s^0}^2)$ are however always positive: the triangle diagrams with up quarks and strange quarks always add. Because of this change of  sign of the coupling constant, we have plotted their modulus in figures \ref{guu.eps} and \ref{gss.eps}.} ($m_i^2-4({M_{u,s}^0})^2$).
For the glueball state, it is clear that the absence of coupling  implies
 $g_i^*=0$ (no quark content), so that the glueball has  also no decay width.
This vanishing of the decay widths of $f_0(1370)$ and $f_0(1710)$ in the pure NJL model (no mixing) in the chiral limit is a consequence of the chiral symmetry in the scalar sector. This has already been shown by Bajc et al.  \cite{blin94}. It means that, even if the meson to quarks coupling contant is large, the decay width into two photons can be small. We can conclude that this decay width may not be  a good observable to claim whether a meson is glue rich or not.\par

Figure \ref{scalarwidth.eps} also shows  that the meson and glueball widths can vary from one to three order of magnitude and that their relative magnitude is  $\chi_0$ dependent (see fig. 7 of \cite{jamvdb98}).\par

\espace

Our results are in complete disagreement with those of Close et al. \cite{close97}, summarized in eq. (\ref{scheme1}) for  scheme 1. The relative strengths are not reproduced, even qualitatively. In fact, we have, for $\chi_0=350$ MeV,
\be
f_0(1370):f_0(1500):f_0(1710)= 5.5:1:0.08.
\ee

Although being a glue rich particle, the $f_0(1500)$ has a two-photon decay width  much larger than that of $f_0(1710)$, which is a quark rich particle. It arises from  the fact that the latter has a mass almost corresponding to the chiral limit without coupling ($m_{f_0(1710)}\approx 2M_s^0$), with small coupling to the up quarks ($g_i^{*\bar{u}u}\approx 0$), while the decay width of the glueball is not suppressed by the factors ($m_{f_0(1500)}^2-4{M_u^0}^2, m_{f_0(1500)}^2-4{M_s^0}^2$). Moreover, as can be seen on figures \ref{guu.eps} and \ref{gss.eps}, the glueball to quarks coupling constant is not so small.

One can wonder where is the origin of the discrepancy with the  results  \cite{close97}. Close et al. give very general arguments, while our results are probably dependent on the crudeness of the model. However, they show  clearly that chiral symmetry effects ($m_i^2-4({M_{u,s}^0})^2$) are important and that mass effects should also be considered (see eq. (6.2) of \cite{close97} where the authors do not consider the $u-s$ mass difference in the meson mixing; in the present model, these effects are considered).
Should we have  worked with the same hypothesis as in \cite{close97} ($M_u^0=M_s^0$ in the mixing matrix and in the chiral symmetry reduction factors\footnote{Note that because the $a_0(1450)$ is not mixed to other channels, its mass is always very close to the chiral limit value $2M_u^0$. Due to the reduction factor, this implies that its width into $2\gamma$ is very small, being exactly zero in the chiral limit.} ($m_i^2-4({M_{u,s}^0})^2$), we would have obtained
\be
f_0(1370):f_0(1500):f_0(1710)= 10.7:1:5.7,
\ee
in surprisingly good agreement with eq. (\ref{scheme1}). It seems  that, working with the same hypothesis as \cite{close97}, our model would be able to reproduce the same order of magnitude for the decay ratios. Because our results are very sensitive to the difference $M_u^0\ne M_s^0$, our model could then suggest  that the $M_u^0=M_s^0$ ($SU(3)$ symmetry) hypothesis is a very drastic one and has to be rejected. We have however to keep in mind that this might be an artifact of the crudeness of our  model.\par

Finally, it is worth noticing that the two-photon decay width  of $f_0(1370)$ is always too small compared to the one given in \cite{pdb96} ($5.4\pm2.3$ keV). Even restricting to the lower bound of the experimental value, one needs considerering low gluon condensates. This  gives rise to a too small mass for $f_0(1370)$ (of the order of 1.2 GeV).

\section{Conclusions}

We have reviewed the 3-flavor A-scaled Nambu--Jona-Lasinio model both in the vacuum and at finite temperature and density. In section \ref{introduction}, we have mainly described the features of the model and its connections to QCD. Starting from a symmetry-anomaly point of view, we have used Fierz identities (section \ref{fierztransfo}) to rewrite the local 4-point color-color interaction into the NJL form. For the sake of completeness, we have derived Fierz identities for a general covariant gauge, as well as for diquark modes. Although not needed in the present study, these generalized identities (compared to a Feynman-like gauge) are important to make contact with Global Color Models or Schwinger-Dyson equations, or any model which incorporates a covariant gauge. In particular, the Landau gauge is included in our Fierz identities. To reproduce the strong axial and scale anomalies, we have motivated the introduction of a new parameter $(\xi)$ and of a dilaton field $(\chi)$, respectively (section \ref{3flavors}). After having bosonized the model, we have derived the gap equations, an equation for the dilaton field and the meson propagators.

Apart from the drawbacks that we have indicated throughout the text, we have also included a specific section to discuss this important subject.

In section \ref{sectionchap2}, we have applied the model to study the thermodynamics of the 3-flavor A-scaled NJL model. We have investigated the quark condensates, the pressure and the energy and entropy densities. 
For vanishing densities, we have given both a low and a high temperature expansion. We have carefully defined a bag constant and have made warnings about the way to extract it from fits.
We have also included a section to compare the NJL behavior with lattice QCD (section \ref{comparisonlattice}), as well as an application of the finite temperature formalism to the $\pi_0\rightarrow\gamma\gamma$ anomalous process (section \ref{sectionanomalouspiondecay}).

Finally, we have included a section on scalar mesons. The study, limited to zero temperature and densities, is interesting due to the coupling between the scalar isoscalar members of the $\bar{q}q$ nonet with the glueball. In order to obtain the physical states from a 3-particle coupled system, which is far from trivial because of the relativistic nature of the theory, we have indicated a way to extract masses and meson to quarks coupling constants. The procedure we have developed consists in studying the energy-dependent mass matrix. As an application, we have determined the two photon decay width of the scalar isoscalars. Although the model has to be taken with caution, especially in the scalar sector where $\bar{q}q$ mesons are unbound due to an unphysical threshold coming from the non-confining nature of the NJL type of models, we have shown that these decays might be not a good choice for  the determination of the glue content of scalar particles. 

\begin{acknowledgments}
Most of this review is based on common works with M. Jaminon. I wish to thank her for friendly and critical discussions. 

I wish also to thank   D. Blaschke and Y. Kalinovsky for physics inputs as well as for having facilitated my participation to the ``Deconfinement at Finite Temperature and  Density'' Workshop, Dubna (Russia), October 1997. I thank L. Kalinovskaya for  having  made it running smoothly, and the Heisenberg-Landau program of the BMBW for having made it possible.

I would like to acknowledge the help of people who clarified the material included here.  Most of all, I am indebted to M. Bawin, J.-R. Cudell, J. Cugnon, M. Mathot and F. Stancu.

I am very grateful  to M.-Cl. Pontus for her technical assistance during the typesetting process, and to V. and V. Makoveev for logistic support in Dubna.

This work has been completed with the support of the Institut Interuniversitaire des Sciences Nucléaires de Belgique.
\end{acknowledgments}

\appendix
\section{Conventions}
\label{conventions}

\begin{itemize}
\item The units $\hbar=c=k_B=1$, where $\hbar$ is the Planck constant normalized to $2\pi$, $c$ the speed of light and
$k_B$ the Boltzmann's constant, are used throughout this review.
\item In Minkowski metric, the metric tensor is
\begin{displaymath}
g_{\mu\nu}=g^{\mu\nu}=
\left(
\begin{array}{cccc}
1&0&0&0\\
0&-1&0&0\\
0&0&-1&0\\
0&0&0&-1
\end{array}
\right),
\end{displaymath}
so that for $x^{\mu}=(t,x,y,z)$, we get
$x_{\mu}=g_{\mu\nu}x^{\nu}=(t,-x,-y,-z)$ and
$x^2=x_{\mu}x^{\mu}=t^2-(x^2+y^2+z^2)$.
\item In Euclidean metric, we define
$x_{\mu}=x^{\mu}=(it,x,y,z)$, which implies $x^2= x_{\mu}x_{\mu}=
x^{\mu}x^{\mu}=-t^2+x^2+y^2+z^2$.
\item Dirac matrices are taken from \cite{itzykson85}.
In this reference, $\gd{5}=i\gu{0}\gu{1}\gu{2}\gu{3}$. We define
$\gu{\mu}=(\gu{0},\vec{\gamma})$ and we work in the Dirac representation, which means
$\gu{0}=\beta\equiv{\rm diag}(1,1,-1,-1)$. In Euclidean metric, we
define $\gd{0}=i\beta$, $\gd{5}$ being unchanged (i.e.  defined
through $\beta$ instead of
$\gd{0}$). The relation $\{\gu{\mu},\gu{\nu}\}=2g^{\mu\nu}$ from Minkowski metric becomes $\{\gu{\mu},\gu{\nu}\}=-2\delta^{\mu\nu}$.
\item When not explicitly indicated, summation over repeated indices is used.
\item The totally  antisymmetrical tensor $\epsilon^{\mu\nu\alpha\beta}$ is defined by the relation:
\begin{displaymath}
\epsilon^{\mu\nu\alpha\beta}=\left\{\begin{array}{cl}
+1&\{\mu,\nu,\alpha,\beta\}\ \mbox{even premutation of \{0,1,2,3\}},\\
-1& \mbox{odd permutation},\\
0& \mbox{otherwise.}
\end{array}\right. 
\end{displaymath}
Indices are lowered with the help of the metric tensor. Note that
$\epsilon_{0123}=-1$.
\end{itemize} 

\section{Color quark current-current interaction}
\label{fierzappendix}

We start from the definition (\ref{gluonprop})  of the gluon propagator, for which the interaction term of the GCM model (\ref{LGCM}) is $\bar{q}(x)(\lambda_a/2)\gamma^{\mu}q(x)D_{\mu\nu}(x,y)\bar{q}(y)(\lambda_a/2)\gu{\nu}q(y)$. 
We note that the spin part
involves only the vector part of the $\Gamma$ matrices (sse section \ref{fierztransfodirac}): $\Gu{\alpha}=\gamma^{\mu}, \Gd{\alpha'}=\gamma_{\nu}$.

Making use of eq. (\ref{xongamma}), one can project 
$\gamma^{\mu}\Gu{\beta}$ and $\gamma_{\nu}\Gd{\beta}$
onto the basis $\Gamma$. The calculation can be performed using\footnote{When using  eq. (\ref{xongamma}), we simplify the summation by noting that:
\[
\frac{1}{4}\sum_{T=1}^6\Gd{T}\tr(X\Gu{T})=
\frac{1}{4}\sum_{\mu}\sum_{\nu>\mu}\sigma_{\mu\nu}\tr(X\sigma^{\mu\nu})
=\frac{1}{4}\frac{1}{2}\sum_{\mu,\nu}\sigma_{\mu\nu}\tr(X\sigma^{\mu\nu}).
\]
\label{eqtensor}}
 (together with the vanishing of the trace of an odd number of $\gamma$-matrices)
\beqn
\tr(\gamma^5\gamma^{\mu})&=&0,\\
\tr(\gamma^{\mu}\gamma^{\nu})&=&4g^{\mu\nu},\\
\tr(\sigma^{\mu\nu})&=&0,\\
\tr(\gamma^5\gamma^{\mu}\gamma^{\nu})&=&0,\\
\tr(\gamma^{\mu}\gamma^{\nu}\gamma^{\rho}\gamma^{\sigma})&=&
4(g^{\mu\nu}g^{\rho\sigma}-g^{\mu\rho}g^{\nu\sigma}+g^{\mu\sigma}g^{\nu\rho}),\\
\tr(\gamma^5\gamma^{\mu}\gamma^{\nu}\gamma^{\rho}\gamma^{\sigma})&=&
-4i\epsilon^{\mu\nu\rho\sigma},
\eeqn
which allows one to obtain the following decomposition (for $\gamma^{\mu}$ on $\Gu{\alpha}$):
\beqn
\gamma^{\mu}\one&=&\gamma^{\mu},\\
\gamma^{\mu}\gamma^{\mu'}&=&g^{\mu\mu'}\one-i\sigma^{\mu\mu'},
\label{gammaugammauprime}\\
\gamma^{\mu}\gamma^5\gamma^{\mu'}&=&-g^{\mu\mu'}\gamma^5
-\frac{1}{2}\sigma^{\mu''\nu''}{\epsilon_{\mu''\nu''}}^{\mu\mu'},\\
\gamma^{\mu}\sigma^{\mu'\nu'}&=&i(\gamma^{\nu'}g^{\mu\mu'}-\gamma^{\mu'}
g^{\mu\nu'})-\gamma^5\gamma^{\mu''}{\epsilon_{\mu''}}^{\mu\mu'\nu'},
\label{gammasigma}\\
\gamma^{\mu}i\gamma^5&=&-i\gamma^5\gamma^{\mu}.
\eeqn
With these equations, we are now able to evaluate (we leave aside the color and flavor d.o.f., reintroducing them when appropriate)
\be
\bar{q}(x)\gamma^{\mu}q(x)D_{\mu\nu}(x,y)\bar{q}(y)\gamma^{\nu}q(y)
=-\frac{1}{4}\bar{q}(x)\gamma^{\mu}\Gu{\beta}q(y)D_{\mu\nu}(x,y)
\bar{q}(y)\gamma^{\nu}\Gd{\beta}q(x),
\ee
where the tensor summation is evaluated as described in the footnote (\ref{eqtensor}).
We are then left with the evaluation of 
\barray
\hspace{-1.25cm}
-\frac{1}{4}D_{\mu\nu}(x,y)\Bigg\{
\Big(\qbar(x)\gu{\mu}q(y)\Big) \Big(\qbar(y)\gu{\nu}q(x)\Big)+
\Big(\qbar(x)\gu{\mu}\gu{5}q(y)\Big)\Big(\qbar(y)\gu{\nu}\gu{5}q(x)\Big)\\
\mbox{}+\Big(\qbar(x)\gu{\mu}\gu{\mu'}q(y)\Big)\Big(\qbar(y)\gu{\nu}\gd{\mu'}q(x)\Big)-\Big(\qbar(x)\gu{\mu}\gu{5}\gu{\mu'}q(y)\Big)\Big(\qbar(y)\gu{\nu}\gu{5}\gd{\mu'}q(x)\Big)\nonumber\\
\mbox{}+\frac{1}{2}\Big(\qbar(x)\gu{\mu}\sigma^{\mu'\nu'}q(y)\Big)\Big(\qbar(y)\gu{\nu}\sigma_{\mu'\nu'}q(x)\Big)
\Bigg\}.
\label{fierz}
\earray
The first two terms are already on the $\Gamma$ basis. The evaluation of the last three terms can be performed using eqs.~(\ref{gammaugammauprime}--\ref{gammasigma}). Using the property $D_{\mu\nu}(x,y)=D_{\nu\mu}(y,x)$, one can then show that
\barray
\hspace{-1.25cm}
-\frac{1}{4}D_{\mu\nu}(x,y)\Big(\qbar(x)\gu{\mu}\gu{\mu'}q(y)\Big)\Big(\qbar(y)\gu{\nu}\gd{\mu'}q(x)\Big)=\\
\hspace{0cm}-\frac{1}{4}D_{\mu\nu}(x,y)
\Bigg\{
g^{\mu\nu}\Big(\qbar(x)q(y)\Big)\Big(\qbar(y)q(x)\Big)-
\Big(\qbar(x)\sigma^{\mu\mu'}q(y)\Big)
\Big(\qbar(y){\sigma^{\nu}}_{\mu'}q(x)\Big)\\
+2i\Big(\qbar(x)q(y)\Big)\Big(\qbar(y)\sigma^{\mu\nu}q(x)\Big)
\Bigg\},
\earray
\barray
\hspace{0cm}
-\frac{1}{4}D_{\mu\nu}(x,y)\Big(-\qbar(x)\gu{\mu}\gu{5}\gu{\mu'}q(y)
\Big)\Big(\qbar(y)\gu{\nu}\gu{5}\gd{\mu'}q(x)\Big)=\\
\hspace{1cm}-\frac{1}{4}D_{\mu\nu}(x,y)
\Bigg\{
-g^{\mu\nu}\Big(\qbar(x)\gu{5}q(y)\Big)\Big(\qbar(y)\gu{5}q(x)\Big)-
\Big(\qbar(x)\sigma^{\mu'\nu'}q(y)\Big)
\Big(\qbar(y)\gu{5}q(x)\Big){\epsilon_{\mu'\nu'}}^{\mu\nu}\\
\hspace{1cm}-\frac{1}{4}\Big(\qbar(x)\sigma^{\mu''\nu''}q(y)\Big)
\Big(\qbar(y)\sigma^{\mu'''\nu'''}q(x)\Big)
{\epsilon_{\mu''\nu''}}^{\mu\mu'}{{\epsilon_{\mu'''\nu'''}}^{\nu}}_{\mu'}
\Bigg\},
\label{epsilonepsilon1sameindice}
\earray
and
\barray
\hspace{-0.25cm}
-\frac{1}{4}D_{\mu\nu}(x,y)
\frac{1}{2}\Big(\qbar(x)\gu{\mu}\sigma^{\mu'\nu'}q(y)\Big)\Big(\qbar(y)\gu{\nu}\sigma_{\mu'\nu'}q(x)\Big)=\\
\hspace{0.5cm}-\frac{1}{4}D_{\mu\nu}(x,y)
\Bigg\{
-g^{\mu\nu}\Big(\qbar(x)\gu{\mu'}q(y)\Big)\Big(\qbar(y)\gd{\mu'}q(x)\Big)+
\Big(\qbar(x)\gu{\nu}q(y)\Big)
\Big(\qbar(y)\gu{\mu}q(x)\Big)\\
\hspace{0.5cm}+2i\Big(\qbar(x)\gu{\mu'}q(y)\Big)
\Big(\qbar(y)\gu{5}\gu{\nu'}q(x)\Big){{\epsilon_{\nu'}}^{\mu\nu}}_{\mu'}
+\frac{1}{2}\Big(\qbar(x)\gu{5}\gu{\mu''}q(y)\Big)
\Big(\qbar(y)\gu{5}\gu{\mu'''}q(x)\Big)
{\epsilon_{\mu''}}^{\mu\mu'\nu'}{{\epsilon_{\mu'''}}^{\nu}}_{\mu'\nu'}
\Bigg\}.
\label{epsilonepsilon2sameindice}
\earray
With $\epsilon^{\mu\nu\rho\sigma}{\epsilon_{\mu}}^{\nu'\rho'\sigma'}=
-\mbox{det}(g^{\alpha\alpha'}),\ (\alpha ;\alpha')=(\mu,\rho,\sigma;\mu',\rho',\sigma')$, eq.~(\ref{epsilonepsilon1sameindice}) gives
\barray
\hspace{-1cm}
-\frac{1}{4}D_{\mu\nu}(x,y)\Big(-\qbar(x)\gu{\mu}\gu{5}\gu{\mu'}q(y)
\Big)\Big(\qbar(y)\gu{\nu}\gu{5}\gd{\mu'}q(x)\Big)=\\
\hspace{0cm}-\frac{1}{4}D_{\mu\nu}(x,y)
\Bigg\{
-g^{\mu\nu}\Big(\qbar(x)\gu{5}q(y)\Big)\Big(\qbar(y)\gu{5}q(x)\Big)-
\Big(\qbar(x)\sigma^{\mu'\nu'}q(y)\Big)
\Big(\qbar(y)\gu{5}q(x)\Big){\epsilon_{\mu'\nu'}}^{\mu\nu}\\
+\frac{1}{2}\Big(\qbar(x)\sigma_{\mu'\nu'}q(y)\Big)
\Big(\qbar(y)\sigma^{\mu'\nu'}q(x)\Big)g^{\mu\nu}
-\Big(\qbar(x){\sigma^{\nu}}_{\mu'}q(y)\Big)
\Big(\qbar(x)\sigma^{\mu\mu'}q(y)\Big)
\Bigg\},
\earray
whereas
$\epsilon^{\mu\nu\rho\sigma}{\epsilon_{\mu\nu}}^{\rho'\sigma'}=
-2(g^{\rho\rho'}g^{\sigma\sigma'}-g^{\rho\sigma'}g^{\rho'\sigma})$
applied to
 eq.~(\ref{epsilonepsilon2sameindice}) gives 
\barray
\hspace{-1.5cm}
-\frac{1}{4}D_{\mu\nu}(x,y)
\frac{1}{2}\Big(\qbar(x)\gu{\mu}\sigma^{\mu'\nu'}q(y)\Big)\Big(\qbar(y)\gu{\nu}\sigma_{\mu'\nu'}q(x)\Big)=\\
-\frac{1}{4}D_{\mu\nu}(x,y)
\Bigg\{
-g^{\mu\nu}\Big(\qbar(x)\gu{\mu'}q(y)\Big)\Big(\qbar(y)\gd{\mu'}q(x)\Big)+
\Big(\qbar(x)\gu{\nu}q(y)\Big)
\Big(\qbar(y)\gu{\mu}q(x)\Big)\\
-\Big(\qbar(x)\gu{5}\gu{\mu'}q(y)\Big)
\Big(\qbar(y)\gu{5}\gd{\mu'}q(x)\Big)g^{\mu\nu}
+\Big(\qbar(x)\gu{5}\gu{\nu}q(y)\Big)
\Big(\qbar(y)\gu{5}\gd{\mu}q(x)\Big)\\
+2i\Big(\qbar(x)\gu{\mu'}q(y)\Big)
\Big(\qbar(y)\gu{5}\gu{\nu'}q(x)\Big){{\epsilon_{\nu'}}^{\mu\nu}}_{\mu'}
\Bigg\}.
\earray
Gathering these results gives the final answer for an arbitrary propagator $D_{\mu\nu}(x,y)$ within a current-current interation\footnote{
To write down this Fierz-transformed action, we used the property
$D_{\mu\nu}(x,y)=D_{\nu\mu}(y,x)$. Without using this property, one can find the Fierz transformation of the product $(\gu{\mu})_{rs}(\gu{\nu})_{tu}$, which is the generalization of the usual form 
\begin{eqnarray}
(\gd{\nu})_{rs}(\gu{\nu})_{tu}=(\one)_{ru}(\one)_{ts}+(i\gd{5})_{ru}
(i\gd{5})_{ts}+(i\f{\gd{\mu}}{\sqrt{2}})_{ru}
(i\f{\gd{\mu}}{\sqrt{2}})_{ts}+(i\f{\gd{\mu}\gd{5}}{\sqrt{2}})_{ru}
(i\f{\gd{\mu}\gd{5}}{\sqrt{2}})_{ts},
\label{fierzdirac}
\nonumber
\end{eqnarray}
and is given by $(\gu{\mu})_{rs}(\gu{\nu})_{tu}=(\gu{\mu}\Gu{\beta})_{ru}(\gu{\nu}\Gd{\beta})_{ts}$.
This leads to
\barraystar
\hspace{-1.5cm}
4(\gu{\mu})_{rs}(\gu{\nu})_{tu}=
g^{\mu\nu}\delta_{ru}\delta_{ts}+g^{\mu\nu}(i\gu{5})_{ru}(i\gu{5})_{ts}
+(\gu{\mu})_{ru}(\gu{\nu})_{ts}+(\gu{\nu})_{ru}(\gu{\mu})_{ts}
-g^{\mu\nu}(\gu{\mu'})_{ru}(\gd{\mu'})_{ts}\\
+(\gu{5}\gu{\mu})_{ru}(\gu{5}\gu{\nu})_{ts}+
(\gu{5}\gu{\nu})_{ru}(\gu{5}\gu{\mu})_{ts}
-g^{\mu\nu}(\gu{5}\gu{\mu'})_{ru}(\gu{5}\gd{\mu'})_{ts}\\
+\frac{1}{2}g^{\mu\nu}(\sigma^{\mu'\nu'})_{ru}(\sigma_{\mu'\nu'})_{ts}
-(\sigma^{\mu\mu'})_{ru}({\sigma^{\nu}}_{\mu'})_{ts}
-({\sigma^{\nu}}_{\mu'})_{ru}(\sigma^{\mu\mu'})_{ts}\\
+i\Big[
\delta_{ru}(\sigma^{\mu\nu})_{ts}-(\sigma^{\mu\nu})_{ru}\delta_{ts}
+\frac{1}{2}(\sigma^{\mu'\nu'})_{ru}(i\gu{5})_{ts}{\epsilon_{\mu'\nu'}}^{\mu\nu}
-\frac{1}{2}(i\gu{5})_{ru}(\sigma^{\mu'\nu'})_{ts}{\epsilon_{\mu'\nu'}}^{\mu\nu}\\
+(\gu{\nu'})_{ru}(\gu{5}\gu{\mu'})_{ts}{\epsilon_{\mu'\nu'}}^{\mu\nu}
-(\gu{5}\gu{\mu'})_{ru}(\gu{\nu'})_{ts}{\epsilon_{\mu'\nu'}}^{\mu\nu}
\Big].
\earraystar
For diquarks, we start from  $(\gd{\mu})_{rs}(\gu{\nu})_{tu}=(\gd{\mu})_{rs}((\gu{\nu})^T)_{ut} = 
(\gd{\mu})_{rs}(C\gu{\nu}C)_{ut}=C_{ua}C_{bt}
(\gu{\mu})_{rs}(\gu{\nu})_{ab}$, $C=i\gu{2}\beta$ being the charge conjugation  matrix. This allows to write, using the previous Fierz transformation, 
\barraystar
\hspace{0cm}
4(\gu{\mu})_{rs}(\gu{\nu})_{tu}=
g^{\mu\nu}C_{rt}C_{us}+g^{\mu\nu}(i\gu{5}C)_{rt}(Ci\gu{5})_{us}
+(\gu{\mu}C)_{rt}(C\gu{\nu})_{us}+(\gu{\nu}C)_{rt}(C\gu{\mu})_{us}
-g^{\mu\nu}(\gu{\mu'}C)_{rt}(C\gd{\mu'})_{us}\\
\hspace{1.5cm}+(\gu{5}\gu{\mu}C)_{rt}(C\gu{5}\gu{\nu})_{us}+
(\gu{5}\gu{\nu}C)_{rt}(C\gu{5}\gu{\mu})_{us}
-g^{\mu\nu}(\gu{5}\gu{\mu'}C)_{rt}(C\gu{5}\gd{\mu'})_{us}\\
\hspace{1.5cm}+\frac{1}{2}g^{\mu\nu}(\sigma^{\mu'\nu'}C)_{rt}(C\sigma_{\mu'\nu'})_{us}
-(\sigma^{\mu\mu'}C)_{rt}(C{\sigma^{\nu}}_{\mu'})_{us}
-({\sigma^{\nu}}_{\mu'}C)_{rt}(C\sigma^{\mu\mu'})_{us}\\
\hspace{1.5cm}+i\Big[
C_{rt}(C\sigma^{\mu\nu})_{us}-(\sigma^{\mu\nu}C)_{rt}C_{us}
+\frac{1}{2}(\sigma^{\mu'\nu'}C)_{rt}(Ci\gu{5})_{us}{\epsilon_{\mu'\nu'}}^{\mu\nu}
-\frac{1}{2}(i\gu{5}C)_{rt}(C\sigma^{\mu'\nu'})_{us}{\epsilon_{\mu'\nu'}}^{\mu\nu}\\
\hspace{1.5cm}+(\gu{\nu'}C)_{rt}(C\gu{5}\gu{\mu'})_{us}{\epsilon_{\mu'\nu'}}^{\mu\nu}
-(\gu{5}\gu{\mu'}C)_{rt}(C\gu{\nu'})_{us}{\epsilon_{\mu'\nu'}}^{\mu\nu}
\Big].
\earraystar
%
}:
\barray
\hspace{-0.75cm}
\bar{q}(x)\gamma^{\mu}q(x)D_{\mu\nu}(x,y)\bar{q}(y)\gamma^{\nu}q(y)
=-\frac{1}{4}D_{\mu\nu}(x,y)\\
\hspace{-0.25cm}\times\Bigg\{
g^{\mu\nu}\Big(\qbar(x)q(y)\Big)\Big(\qbar(y)q(x)\Big)
+g^{\mu\nu}\Big(\qbar(x)i\gu{5}q(y)\Big)\Big(\qbar(y)i\gu{5}q(x)\Big)\\
+\Big(\qbar(x)\gu{\mu}q(y)\Big) \Big(\qbar(y)\gu{\nu}q(x)\Big)
+\Big(\qbar(x)\gu{\nu}q(y)\Big) \Big(\qbar(y)\gu{\mu}q(x)\Big)\\
-g^{\mu\nu}
\Big(\qbar(x)\gu{\mu'}q(y)\Big) \Big(\qbar(y)\gd{\mu'}q(x)\Big)
+
\Big(\qbar(x)\gu{5}\gu{\mu}q(y)\Big)
\Big(\qbar(y)\gu{5}\gu{\nu}q(x)\Big)\\
+
\Big(\qbar(x)\gu{5}\gu{\nu}q(y)\Big)
\Big(\qbar(y)\gu{5}\gu{\mu}q(x)\Big)
-g^{\mu\nu}
\Big(\qbar(x)\gu{5}\gu{\mu'}q(y)\Big)
\Big(\qbar(y)\gu{5}\gd{\mu'}q(x)\Big)\\
+2i\Big(\qbar(x)q(y)\Big)\Big(\qbar(y)\sigma^{\mu\nu}q(x)\Big)
+\frac{1}{2}\Big(\qbar(x)\sigma_{\mu'\nu'}q(y)\Big)
\Big(\qbar(y)\sigma^{\mu'\nu'}q(x)\Big)g^{\mu\nu}\\
-\Big(\qbar(x){\sigma^{\nu}}_{\mu'}q(y)\Big)
\Big(\qbar(y)\sigma^{\mu\mu'}q(x)\Big)
-\Big(\qbar(x)\sigma^{\mu\mu'}q(y)\Big)
\Big(\qbar(y){\sigma^{\nu}}_{\mu'}q(x)\Big)\\
+2i\Big(\qbar(x)\gu{\mu'}q(y)\Big)
\Big(\qbar(y)\gu{5}\gu{\nu'}q(x)\Big){{\epsilon_{\nu'}}^{\mu\nu}}_{\mu'}
+i\Big(\qbar(x)\sigma^{\mu'\nu'}q(y)\Big)
\Big(\qbar(y)i\gu{5}q(x)\Big){\epsilon_{\mu'\nu'}}^{\mu\nu}
\Bigg\}.
\earray
%
Using the invariance of the propagator under $\mu\leftrightarrow\nu$, we are left with
\barray
\hspace{-1cm}
\bar{q}(x)\gamma^{\mu}q(x)D_{\mu\nu}(x,y)\bar{q}(y)\gamma^{\nu}q(y)
=-\frac{1}{4}D_{\mu\nu}(x,y)\\
\hspace{-0.25cm}\times\Bigg\{
g^{\mu\nu}\Big(\qbar(x)q(y)\Big)\Big(\qbar(y)q(x)\Big)
+g^{\mu\nu}\Big(\qbar(x)i\gu{5}q(y)\Big)\Big(\qbar(y)i\gu{5}q(x)\Big)\\
+2\Big(\qbar(x)\gu{\mu}q(y)\Big) \Big(\qbar(y)\gu{\nu}q(x)\Big)
-g^{\mu\nu}
\Big(\qbar(x)\gu{\mu'}q(y)\Big) \Big(\qbar(y)\gd{\mu'}q(x)\Big)\\
+2
\Big(\qbar(x)\gu{5}\gu{\mu}q(y)\Big)
\Big(\qbar(y)\gu{5}\gu{\nu}q(x)\Big)
-g^{\mu\nu}
\Big(\qbar(x)\gu{5}\gu{\mu'}q(y)\Big)
\Big(\qbar(y)\gu{5}\gd{\mu'}q(x)\Big)\\
+\frac{1}{2}g^{\mu\nu}\Big(\qbar(x)\sigma_{\mu'\nu'}q(y)\Big)
\Big(\qbar(y)\sigma^{\mu'\nu'}q(x)\Big)
-2\Big(\qbar(x){\sigma^{\nu}}_{\mu'}q(y)\Big)
\Big(\qbar(y)\sigma^{\mu\mu'}q(x)\Big)
\Bigg\}.
\earray
With the Fierz transformations for color and flavor (see section \ref{fierztransfocolor} and \ref{fierzflavor}), we have -- taking then into account the appearance of diquarks
  --
\beqn
\lefteqn{
\hspace{-1cm}
\bar{q}(x)\f{\lambda_a}{2}_c\gamma^{\mu}q(x)D_{\mu\nu}(x,y)
\bar{q}(y)\f{\lambda_a}{2}_cq(y)=-\frac{1}{4}D_{\mu\nu}(x,y)\frac{1}{3}}
\nonumber\\
&&\times\Bigg\{
\bigg[
g^{\mu\nu}\sum_{a,e}\Big(\qbar(x)G^eK'^aq(y)\Big)
\Big(\qbar(y)G^eK'_aq(x)\Big)\nonumber\\
&&\mbox{}-2\Big(\qbar(x)G^ei\gu{\mu}q(y)\Big) \Big(\qbar(y)G^ei\gu{\nu}q(x)\Big)
-2\Big(\qbar(x)G^ei\gu{5}\gu{\mu}q(y)\Big)
\Big(\qbar(y)G^ei\gu{5}\gu{\nu}q(x)\Big)\nonumber\\
&&\mbox{}-4\Big(\qbar(x)G^e\frac{{\sigma^{\nu}}_{\mu'}}{\sqrt{2}}q(y)\Big)
\Big(\qbar(y)G^e\frac{\sigma^{\mu\mu'}}{\sqrt{2}}q(x)\Big)
\bigg]
\nonumber\\
&&\mbox{}+\bigg[
g^{\mu\nu}\sum_{a,e}\Big(\qbar(x)G^e_{A,S}\frac{i\epsilon^{\rho}}{\sqrt{2}}K'^aC\qbar^T(y)\Big)
\Big(q^T(y)CG^e_{A,S}\frac{i\epsilon^{\rho}}{\sqrt{2}}K'_aq(x)\Big)\nonumber\\
&&\mbox{}-2\Big(\qbar(x)G^e_Si\gu{\mu}\frac{i\epsilon^{\rho}}{\sqrt{2}}C\qbar^T(y)\Big) \Big(q^T(y)CG^e_S\frac{i\epsilon^{\rho}}{\sqrt{2}}i\gu{\nu}q(x)\Big)\nonumber\\
&&-2\Big(\qbar(x)G^e_A\frac{i\epsilon^{\rho}}{\sqrt{2}}i\gu{5}\gu{\mu}C\qbar^T(y)\Big)
\Big(q^T(y)CG^e_A\frac{i\epsilon^{\rho}}{\sqrt{2}}i\gu{5}\gu{\nu}q(x)\Big)\nonumber\\
&&\mbox{}-4\Big(\qbar(x)G^e_S\frac{i\epsilon^{\rho}}{\sqrt{2}}\frac{{\sigma^{\nu}}_{\mu'}}{\sqrt{2}}C\qbar^T(y)\Big)
\Big(q^T(y)CG^e_S\frac{i\epsilon^{\rho}}{\sqrt{2}}\frac{\sigma^{\mu\mu'}}{\sqrt{2}}q(x)\Big)
\bigg]
\Bigg\},\\
\nonumber
\eeqn
where we have defined the shortened notation\footnote{The factor $1/\sqrt{2}$ for the tensor part comes from the convention defined in the footnote (\ref{eqtensor}).} $K'^{a}=(1,i\gu{5},i\gu{\mu'},i\gu{5}\gu{\mu'},\sigma^{\mu'\nu'}/\sqrt{2})$. Note that a color triplet diquark state is antisymmetric in the exchange of colors; then -- this is the Pauli principle -- it must be antisymmetric under the exchange of the other coordinates (both spatial and internal). This implies that $G^e_A$ has to be associated to $(1,i\gu{5},i\gu{5}\gu{\mu'})$ while $G^e_S$ has the Dirac counterpart $K'^{a}=(i\gu{\mu'},\sigma^{\mu'\nu'}/\sqrt{2})$.
Mathematically, these relations can also be shown to hold in the path integral formalism, where the quark fields are  anticommuting variables. This implies that (see  eq.~(\ref{interqq}) for the explanation of this term) $\bar{q}^TC\Gu{\alpha}q=-\bar{q}^T(C\Gu{\alpha})^Tq$ which, in turn, implies 
$C\Gu{\alpha}=-(C\Gu{\alpha})^T$. Because the color part is antisymmetric, it is clear that the expected relation shows up: $(CK'^aG^e_{A,S})^T=(CK'^aG^e_{A,S})$.

\section{Propagator matrices}
\label{PandGamma}

The propagator matrices indicated in eqs. (\ref{actionscalaire},\ref{eq2135})
are defined below for symmetric matter ($\mu_u=\mu_d\ne\mu_s$) in the isospin limit ($m_u=m_d\ne m_s,\Delta\mu\equiv \mu_u-\mu_s$).

\vspace{0.5cm}

\souligne{$P^{\pi\pi}$}
\be
P^{\pi\pi}=\Big(
{q}^{2}{Z}_{\pi}(\beta
,\mu)+{a}^{2}{\chi_s^2}\frac{m_u}{M_u}
\Big),
\label{eqA37a}
\ee

\goodbreak

\souligne{$P^{KK}$}
\be
P^{KK}=
\bigg[
\frac{a^2\chi_s^2}{2}
\left(
\frac{m_u}{M_u}+
\frac{m_s}{M_s}
\right)
+
\left[
(q_0\pm i\Delta\mu)^2+\vec{q}^2
+(M_u-M_s)^2
\right]
Z_{K^{\pm}}(\beta,\mu )
\bigg],
\label{eqA45a}
\ee

\souligne{$P^{00}$}
\be
{P}^{00}=\frac{2}{3}
\left(
Z_{\pi}^u+\f{1}{2}
Z_{\pi}^s
\right)
{q}^{2}+a^2\chi_s^2
\left[
\f{2}{3}
\left(
\f{m_u}{M_u}+\f{m_s}{2M_s}
\right)
+\xi
\right],
\label{eqA51a}
\ee

\souligne{$P^{88}$}
\be
{P}^{88}=\frac{1}{3}
\left(Z_{\pi}^u+2Z_{\pi}^{s}
\right)
{q}^{2}+\f{1}{3}a^2\chi_s^2
\left(
\frac{m_u}{M_u}+
\frac{2m_s}{M_s}
\right),
\label{eqA56a}
\ee

\souligne{$P^{08}=P^{80}$}
\be
{P}^{08}=\frac{\sqrt{2}}{3}
\left(
Z_{\pi}^u-Z_{\pi}^s
\right)
{q}^{2}+\frac{\sqrt{2}}{3}a^2\chi_s^2
\left(
\frac{{m}_{u}}{{M}_{u}}-
\frac{{m}_{s}}{{M}_{s}}
\right),
\label{eqA60a}
\ee

\souligne{$\Gamma^{a_0a_0}$}
\be
\Gamma^{a_0a_0}=
\left[
(q^2+4M_u^2)Z_{\pi}(\beta,\mu)+a^2\chi_s^2\f{m_u}{M_u}
\right],
\label{sigma1a3}
\ee

\souligne{$\Gamma^{K_0^*K_0^*}$}
\be
\Gamma^{K_0^*K_0^*}=
\bigg[
\f{a^2\chi_s^2}{2}\left(
\f{m_u}{M_u}+\f{m_s}{M_s}
\right)+
\left[
(q_0\pm i\Delta\mu)^2+\vec{q}^2+\left(M_u+M_s\right)^2
\right]Z_{K^{\pm}}(\beta,\mu)
\bigg],
\label{sigma4a7}
\ee

\souligne{$\Gamma^{00}$}

\be
{\Gamma}^{00}=\frac{2}{3}
\left(
Z_{\pi}^{u}+\frac{{Z}_{\pi}^{s}}{2}
\right)
{q}^{2}+\frac{8}{3}
\left(
Z_{\pi}^{u}{M}_{u}^{2}+\frac{1}{2}Z_{\pi}^{s}
{M}_{s}^{2}
\right)
+\frac{2}{3}a^2\chi_s^2
\left(
\frac{{m}_{u}}{{M}_{u}}+\frac{{m}_{s}}{{2M}_{s}}
\right),
\label{eqA64a}
\ee

\souligne{$\Gamma^{88}$}
\be
{\Gamma}^{88}=\frac{1}{3}
\left(Z_{\pi}^{u}+2Z_{\pi}^{s}
\right)
{q}^{2}+\frac{4}{3}
\left(
Z_{\pi}^{u}{M}_{u}^{2}+2Z_{\pi}^{s}{M}_{s}^{2}
\right)
+\frac{1}{3}a^2\chi_s^2
\left(
\frac{{m}_{u}}{{M}_{u}}
+\frac{{2m}_{s}}{{M}_{s}}
\right),
\label{eqA68a}
\ee

\souligne{$\Gamma^{08}=\Gamma^{80}$}
\be
{\Gamma}^{08}=\frac{\sqrt{2}}{3}
\left(
Z_{\pi}^{u}-{Z}_{\pi}^{s}
\right)
{q}^{2}+
\frac{4\sqrt{2}}{3}
\left(
Z_{\pi}^{u}{M}_{u}^{2}-Z_{\pi}^{s}{M}_{s}^{2}
\right)
+\f{\sqrt{2}}{3}a^2\chi_s^2
\left(
\f{m_u}{M_u}-\f{m_s}{M_s}
\right),
\label{eqA72a}
\ee

\souligne{$\Gamma^{\chi\chi}$}
\beqn
{\Gamma}^{\chi \chi}&=&
{q}^{2}+{b}^{2}\chi_s^2-
\frac{{N}_{c}}{2{\pi}^{2}}{\Lambda}^{6}\chi_s^4
\left(
\frac{2}{\Lambda^2\chi_s^{2}+{M}_{u}^{2}}+
\frac{1}{\Lambda^2\chi_s^{2}+{M}_{s}^{2}}
\right)
\nonumber\\
&&\qquad\qquad
\mbox{}-2{a}^{2}
\Big[
(M_u-m_u)^2+\f{1}{2}(M_s-m_s)^2
\Big],
\label{eqA75a}
\eeqn

\souligne{$\Gamma^{0\chi}=\Gamma^{\chi0}$}
\be
{\Gamma}^{0\chi}=2a^2\chi_s\sigma_{0}^{s}-
\frac{N_c}{{\pi}^{2}}\Lambda^4\chi_s^3
\left(
\frac{2M_{u}/\sqrt{6}}{
(\Lambda\chi_s)^2+{M}_{u}^{2}
}
+\frac{M_{s}/\sqrt{6}}{
(\Lambda
\chi_s)^2+{M}_{s}^{2}}
\right),
\label{eqA73a}
\ee

\souligne{$\Gamma^{8\chi}=\Gamma^{\chi8}$}
\be
{\Gamma}^{8\chi}=2a^2\chi_s\sigma_{8}^{s}-
\frac{N_c}{{\pi}^{2}}\Lambda^4\chi_s^3
\left(
\frac{M_{u}/\sqrt{3}}{
(\Lambda\chi_s)^2+{M}_{u}^{2}
}
-\frac{M_{s}/\sqrt{3}}{
(\Lambda
\chi_s)^2+{M}_{s}^{2}}
\right).
\label{eqA74a}
\ee

\section{Thermodynamics}
\label{Appendixhightemplimit}
\label{appendicechapitre4}

\subsection{Pressure}

For a vanishing chemical potential, eq.~(\ref{eqthermo23}) reads
\be
{P}_{\boite{ideal gas}}=2{\frac{{N}_{c}}{{3\pi}^{2}}}\left\{
{2\int_{0}^{\infty}
{\frac{{k}^{4}}{E_u}\f{1}{1+{e}^{\beta E_u}}}dk+\int_{0}^{\infty}
{\frac{{k}^{4}}{{E}_{s}}}{\frac{1}{1+{e}^{\beta{E}_{s}}}}dk}
\right\}.
\label{eqthermoA2}
\ee
\par

Only one of these terms has to be analyzed. 
We define
\be
{P}_{\boite{ideal gas}}(s)=
{\frac{2{N}_{c}}{{3\pi}^{2}}}M_s^4\lim
\limits_{\varepsilon\rightarrow 0}^{}\int_{1}^{\infty}
{\frac{{\left({{y}^{2}-1}\right)}^{3/2}}{1+{e}^{\varepsilon}
{e}^{M_s\beta y}}}dy,
\label{eqthermoA4}
\ee
where the infinitesimal quantity $\varepsilon$ will allow to regularize the summation that will be encountered in the following.
Expanding, we get
\be
{P}_{\boite{ideal gas}}(s)={\frac{{2N}_{c}}{{3\pi}^{2}}}M_s^4
\lim\limits_{\varepsilon\rightarrow 0}^{}\sum\limits_{n=1}^{\infty}
{\left({-1}\right)}^{n+1}{e}^{-n\varepsilon}\int_{1}^{\infty}
{\left({{y}^{2}-1}\right)}^{3/2}{e}^{-nM_s\beta y}dy,
\label{eqthermoA5bis}
\ee
which can be expressed in terms of the modified Bessel function of order two \cite{rizik}
\be
{P}_{\boite{ideal gas}}(s)={\frac{{2N}_{c}}{{\pi}^{2}}}M_s^4\lim
\limits_{\varepsilon
\rightarrow 0}^{}\sum\limits_{n=1}^{\infty} {\left({-1}\right)}^{n+1}
{e}^{-n\varepsilon}{\frac{{K}_{2}\left({nM_s\beta}\right)}
{{n}^{2}M_s^{2}{\beta}^{2}}}.
\label{eqthermoA5}
\ee

This series  has also been investigated in \cite{jamvdb94}.
Because of the fast decrease due to the
$1/n^2$ factor and because of the asymptotic behavior of $K_2$, it can be numerically more advantageous to use~(\ref{eqthermoA5}) than~(\ref{eqthermoA5bis}).\par
\subsubsection{High temperature zero density expansion}

The converging factor $\varepsilon$ in
(\ref{eqthermoA5bis}) and~(\ref{eqthermoA5}) is necessary
to obtain nondiverging quantities in the high temperature expansion. In such an expansion, only the first two terms are finite.
Following \cite{rizik},
\be
{K}_{2}\left({z}\right)=2{z}^{-2}\left({1-{\frac{{z}^{2}}{4}}}
\right)
+{\frac{1}{8}}{z}^{2}\sum\limits_{k=0}^{\infty}{\frac{\psi
\left({k+1}\right)+\psi\left({k+3}\right)}{k!\left({2+k}\right)!}}
{\left({{\frac{{z}^{2}}{4}}}\right)}^{k}-\ln{\frac{z}{2}}
{I}_{2}\left({z}\right),
\label{eqthermoA6}
\ee
where $\psi$ is the Digamma function (defined as $d\ln\Gamma(z)/dz$ with
$\Gamma(z)$ the Euler Gamma function) \cite{rizik},
\be
\psi(n+1)=-\gamma+\sum_{k=1}^n\f{1}{k},\ \ \mbox{with}\ \ 
\psi(1)=-\gamma,
\label{eqpsi}
\ee
and where $I_2(z)$
is the other modified Bessel function of order two \cite{rizik}
\be
{I}_{2}\left({z}\right)={\frac{{z}^{2}}{4}}\sum\limits_{k=0}^{\infty}
{\frac{1}{k!\left({k+2}\right)!}}
{\left({{\frac{{z}^{2}}{4}}}\right)}^{k}.
\label{eqthermoA7}
\ee
In (\ref{eqpsi}), $\gamma$ is the Euler constant.\par

The logarithm $\ln(z)$ in $K_2(z)$ implies a
singularity at the origin. There is no Taylor expansion around it. This problem, and the way to circumvent it through the converging factor, has been established in \cite{dolanjackiw74} for a fermionic gas (only the first few terms of the expansion are given) and in
\cite{weldon81,weldon82,weldon82bis} for a bosonic case where a full expansion, valid also at non zero density, has been given.
\par

Combining (\ref{eqthermoA6}) and (\ref{eqthermoA7}), we obtain
\be
\left(
\f{2}{z^2}
\right)
\left(
1-\f{z^2}{4}
\right)
+
\left(
\f{z^2}{4}
\right)
\sum_{k=0}^{\infty}\f{1}{k!(k+2)!}
\left(
\f{z^2}{4}
\right)^k
\left[
\f{1}{2}
\left(
\psi(k+1)+\psi(k+3)
\right)
-\ln
\left(
\f{z}{2}
\right)
\right].
\label{thermobessel1}
\ee
This leads to
\beqn
P_{\boite{ideal gas}}(s)&=&\f{2N_c}{\pi^2}M_s^4\lim\limits_{\varepsilon\rightarrow 0}
\bigg\{
\sum_{n=1}^{\infty}(-1)^{n+1}\f{1}{(nM_s\beta)^2}
\left(
\f{2}{(nM_s\beta)^2}-\f{1}{2}
\right)\nonumber\\
&&\hspace{-2cm}\mbox{}+\f{1}{4}\sum_{n=1}^{\infty}\sum_{k=0}^{\infty}
(-1)^{n+1}e^{-n\varepsilon}\f{1}{k!(k+2)!}
\left(
\f{(nM_s\beta)^2}{4}
\right)^k
\left[
\f{1}{2}
\left(
\psi(k+1)+\psi(k+3)
\right)
-\ln
\left(
\f{(nM_s\beta)}{2}
\right)
\right]
\bigg\}.
\label{thermobessel2}
\eeqn

The first two terms are easy to determine and coincide with the two non-diverging terms from the Taylor expansion around
$M_s=0$. They give
\be
\f{2N_c}{\pi^2}M_s^4
\left\{
\f{2}{M_s^4\beta^4}\sum_{n=1}^{\infty}(-1)^{n+1}\f{1}{n^4}
-\f{1}{2M_s^4\beta^2}\sum_{n=1}^{\infty}(-1)^{n+1}\f{1}{n^2}
\right\}.
\ee
They are tabulated in \cite{rizik} and can be expressed through the use of the Riemann zeta function\footnote{For ${\rm Re}(z)\le0$ which we shall need in the following, a converging factor
$e^{-n\varepsilon}$ is needed.}
\beqn
\zeta(z)&=&\f{1}{1-2^{1-z}}\sum_{n=1}^{\infty}\f{(-1)^{n+1}}{n^z}\ ,\
\ \ \ {\rm Re}(z)>0\\
\zeta(-2m)&=&0, \mbox{ $m$=1,2,...},\label{zetamoins2m}\\
\zeta(2m)&=&\frac{2^{2m-1}\pi^{2m}|B_{2m}|}{(2m!)}, \mbox{ with $m$=1,2,... and $B$ the Bernoulli numbers},
\eeqn
which leads to
\be
\f{7}{180}N_c\pi^2T^4-\f{N_c}{12}M_s^2T^2.
\label{thermobessel3}
\ee

The contribution of the term $k=0$
is rewritten in the form
\be
\f{2N_c}{4\pi^2}M_s^4
\Bigg\{
\lim\limits_{\varepsilon \rightarrow 0}\f{1}{2}
\bigg[
\sum_{n=1}^{\infty}
(-1)^{n+1}e^{-n\varepsilon}
\Big(
\f{1}{2}(\psi(1)+\psi(3))-
\ln(
\f{M_s\beta}{2})
\Big)-\sum_{n=1}^{\infty}
(-1)^{n+1}e^{-n\varepsilon}\ln n
\bigg]
\Bigg\}.
\label{thermobessel4}
\ee
Using
\be
\lim\limits_{\varepsilon \rightarrow 0}
\sum_{n=1}^{\infty}
(-1)^{n+1}e^{-n\varepsilon}n^{z}=(1-2^{1+z})\zeta(-z),
\label{thermobessel5}
\ee
we have
\be
\f{d}{dz}\sum_{n=1}^{\infty}
(-1)^{n+1}e^{-n\varepsilon}n^{z}=\sum_{n=1}^{\infty}
(-1)^{n+1}e^{-n\varepsilon}n^{z}\ln n\equiv
\f{d}{dz}
\left(
(1-2^{1+z})\zeta(-z)
\right),
\label{thermobessel6}
\ee
which leads to
\be
\sum_{n=1}^{\infty}
(-1)^{n+1}e^{-n\varepsilon}n^{z}\ln n
=-(\ln 2) 2^{1+z}\zeta(-z)-(1-2^{1+z})\zeta'(-z),
\label{thermobessel7}
\ee
where $\left.\zeta'(-z)\equiv \f{d}{dz}\zeta(z)\right\}_{-z}$.

$\zeta(0)$ is obtained from
\be
-\sum_{n=1}^{\infty}
(-1)^{n+1}e^{-n\varepsilon}=-\f{e^{-n\varepsilon}}{1+e^{-n\varepsilon}}
=-\f{1}{2}\ \mbox{when $\varepsilon\rightarrow 0$},
\label{zeta0}
\ee
while $\zeta'(0)$ is given in \cite{rizik} 
\be
\zeta'(0)=-\f{1}{2}\ln 2\pi.
\label{zetaprime0}
\ee
\par

With (\ref{zeta0}) and (\ref{zetaprime0}), (\ref{thermobessel4}) gives
\be
\f{2N_c}{4\pi^2}M_s^4
\f{1}{2}
\Bigg\{
\f{1}{2}\cdot \f{1}{2}
\left(
\psi(1)+\psi(3)-2\ln(\f{M_s\beta}{2})
\right)
-\f{1}{2}\ln
\left(
\f{2}{\pi}
\right)
\Bigg\}.
\label{thermobessel8}
\ee
\par

We still need the $k>0$ terms of the expansion. With~(\ref{thermobessel5}) and~(\ref{thermobessel7}), they are
\beqn
&&\f{2N_c}{4\pi^2}M_s^4\lim\limits_{\varepsilon\rightarrow 0}
\Bigg\{
\sum_{k=1}^{\infty}
\f{(M_s\beta)^{2k}}{k!(k+2)!4^k}
\bigg[
\f{1}{2}
\left(
\psi(k+1)+\psi(k+3)
-2\ln
(\f{M_s\beta}{2})
\right)(1-2^{1+2k})\zeta(-2k)
\nonumber\\
&&\ \qquad\qquad \mbox{}+(\ln 2)2^{1+2k}\zeta(-2k)
+(1-2^{1+2k})\zeta'(-2k)
\bigg]
\Bigg\}.
\label{thermobessel9}
\eeqn
\par

Eq. (\ref{eqpsi}) implies
\beqn
\psi(1)&=&-\gamma,\nonumber\\
\psi(3)&=&-\gamma + \f{3}{2},
\eeqn
so that (\ref{zetamoins2m},\ref{thermobessel3},\ref{thermobessel8},\ref{thermobessel9}) lead to
\beqn
&&P_{\boite{ideal gas}}(s)=
\f{7}{180}N_c\pi^2T^4-\f{N_c}{12}M_s^2T^2\nonumber\\
&&\qquad\mbox{}
+\f{N_c}{16\pi^2}M_s^4
\left(
-2\gamma+\f{3}{2}
\right)
-\f{N_c}{8\pi^2}M_s^4\ln
\left(
\f{M_s\beta}{\pi}
\right)\nonumber\\
&&\qquad\mbox{}
+\f{N_c}{2\pi^2}M_s^4
\sum_{k=1}^{\infty}
\f{(M_s\beta)^{2k}}{k!(k+2)!4^k}
(1-2^{1+2k})\zeta'(-2k).
\label{thermobessel10}
\eeqn
\par

It is clear that the converging factor has regularized the summation\footnote{Summations of this kind are  called Euler sums \cite{bender}.} of the expansion. Note from~(\ref{thermobessel10}) that the separation into a logarithmic term and a constant one is arbitrary: one can always write
$\ln (ab\beta)=\ln (a\beta)+\ln b$ ($a$ and $b$ being dimensionless constants)
and put $\ln b$ into the constant term.
This has some importance for the interpretation of the high temperature results.

To write (\ref{thermobessel10}) into a form involving only elementary functions, we still need to know
$\zeta'(-2k) (k\ge1)$, i.e.
\be
-\f{1}{(1-2^{1+2k})}\sum_{n=1}^{\infty}
(-1)^{n+1}e^{-n\varepsilon}n^{2k}\ln n,
\label{shanks}
\ee
because of (\ref{thermobessel7}) and (\ref{zetamoins2m}). Using \cite{rizik}
\be
2^{1-z}\Gamma(z)\zeta(z)\cos 
\left(
\f{\pi z}{2}
\right)
=\pi^z\zeta(1-z)
\ee
and 
\be
\Gamma(z)\Gamma(1-z)=\f{\pi}{\sin(\pi z)},
\ee
we have
\be
2^{1-z}\f{\pi}{\sin(\pi z)}
\zeta(z)\cos 
\left(
\f{\pi z}{2}
\right)
=\pi^z\Gamma(1-z)\zeta(1-z),
\ee
so that 
\be
\lim\limits_{z\rightarrow -2k}
\f{\pi}{\sin(\pi z)}
\zeta(z)
=(-1)^{k}2^{-2k-1}\pi^{-2k}\Gamma(1+2k)\zeta(1+2k).
\ee
\par

With
\be
\lim\limits_{z\rightarrow -2k}
\f{\pi}{\sin(\pi z)}
\zeta(z)
=\f{0}{0}=\f{\pi\zeta'(-2k)}{\pi\cos(-2\pi k)}=\zeta'(-2k),
\ee
we finally obtain
\be
\zeta'(-2k)=\f{1}{2}(-1)^{k}(2\pi)^{-2k}\Gamma(1+2k)\zeta(1+2k).
\label{zetaimpaire}
\ee
Eq.~(\ref{thermobessel10}) is then rewritten into the form
\beqn
\lefteqn{
P_{\boite{ideal gas}}(s)=
\f{7}{180}N_c\pi^2T^4-\f{N_c}{12}M_s^2T^2
}
\nonumber\\
&&\mbox{}
+\f{N_c}{16\pi^2}M_s^4
\left(
-2\gamma+\f{3}{2}
\right)
-\f{N_c}{8\pi^2}M_s^4\ln
\left(
\f{M_s\beta}{\pi}
\right)\nonumber\\
&&\mbox{}
+\f{N_c}{2\pi^2}M_s^4
\sum_{k=1}^{\infty}
\f{(M_s\beta)^{2k}}{k!(k+2)!4^k}
(1-2^{1+2k})\f{1}{2}(-1)^{k}(2\pi)^{-2k}\Gamma(1+2k)\zeta(1+2k),
\nonumber\\
&&
\label{thermopression}
\eeqn
\par
which only necessitates the evaluation of known functions.

\subsubsection{Low temperature zero density expansion}

We can search for a low temperature expansion,  
$\beta\rightarrow\infty$, starting from~(\ref{eqthermoA2})
or~(\ref{eqthermoA5}). The last one is better suited because of the well known asymptotic expansion of $K_2$ \cite{stegun}
\be
K_i(z)\approx\sqrt{\f{\pi}{2z}}e^{-z}
\bigg\{
1+\f{4i^2-1}{8z}+\f{(4i^2-1)(4i^2-9)}{2!(8z)^2}
+\f{(4i^2-1)(4i^2-9)(4i^2-25)}{3!(8z)^3}+...
\bigg\}.
\label{besselasympt}
\ee

Combining  (\ref{eqthermoA5}) and (\ref{besselasympt}), we have
\beqn
&&{P}_{\boite{ideal gas}}(s)\approx{\frac{{2N}_{c}}{{\pi}^{2}}}M_s^4
\sum\limits_{n=1}^{\infty}
\f{(-1)^{n+1}}{n^2M_s^2\beta^2}
\sqrt{\f{\pi}{2nM_s\beta}}e^{-nM_s\beta}\nonumber\\
&&\qquad \mbox{}\times
\bigg\{
1+\f{15}{8nM_s\beta}+\f{15\cdot 7}{2!(8nM_s\beta)^2}
+\f{15\cdot 7 \cdot \mbox{}(-9)}{3!(8nM_s\beta)^3}+...
\bigg\}.
\label{besselasymptpression}
\eeqn
\par

When $\beta$ is large, we are in the chirally broken phase where the quark masses are the constituent masses. Since the expanding parameter is $\beta M_i$, $i=u,s$, the approximation~(\ref{besselasymptpression}) becomes better as
$\beta M_i$ is increased. In section~\ref{results}, it is shown that
the mass variation is low for $T\lesssim 100$ MeV. For a constituent quark mass of about
400 MeV (at $T=0$)
$\beta M_i$ is, at least, 4. The expansion~(\ref{besselasymptpression}) 
is then perfectly justified. In that case, the first term $n=1$ is enough and the pressure is
\beqn
&&{P}_{\boite{ideal gas}}(s)\approx\frac{4N_c\beta^{-5/2}}{(2\pi)^{3/2}}M_s^{3/2}
e^{-M_s\beta}\nonumber\\
&&\qquad \mbox{}\times
\bigg\{
1+\f{15}{8M_s\beta}+\f{15\cdot 7}{2!(8M_s\beta)^2}
+\f{15\cdot 7 \cdot \mbox{}(-9)}{3!(8M_s\beta)^3}+...
\bigg\}.
\label{termenegal2}
\eeqn

We have checked that for the set of parameters $(M_u^0=300,\chi_0=80)$ MeV (see section \ref{results})  this expansion is not well suited. In that case, the second term $n=2$ in~(\ref{besselasymptpression}),  as well as the three corrections to ``1'' for both $n=1$ and $n=2$,  are
necessary to reproduce results valid up to 100 MeV.

\subsubsection{Finite density, zero temperature}

For finite density at vanishing temperature, 
eq. (\ref{eqthermo23}) can be  exactly integrated.
We have $n_{i+}$ = 0,  $n_{i-}$ = 
$\theta(\mu_i-E_i)$ and $\mu_i=\sqrt{k_{F_i}^2+M_i^2}$, where
$k_{F_i}$ is the Fermi momentum of the $i$th flavor, so that
\be
P_{\boite{ideal gas}}(s)=\f{N_c}{3\pi^2}\int_0^{k_{F_s}}dk\ \f{k^4}{E_s},
=\f{N_c}{3\pi^2}
\bigg\{
k_{F_s}^3\mu_s-3
\Big[
\f{k_{F_s}}{4}\mu_s^3-\f{M_s^2}{8}k_{F_s}\mu_s-\f{M_s^4}{8}
\ln 
\left(
\f{k_{F_s}+\mu_s}{M_s}
\right)
\Big]
\bigg\}.
\label{pressionT0}
\ee
\par

\subsection{Energy density}

\subsubsection{High temperature zero density expansion}

Combining eqs.~(\ref{eqthermo1.30},\ref{thermopression}) immediatly gives
\beqn
\lefteqn{
\varepsilon_{\boite{ideal gas}}(s)=
\f{7}{60}N_c\pi^2T^4-\f{N_c}{12}M_s^2T^2
}
\nonumber\\
&&\mbox{}
+\f{N_c}{16\pi^2}M_s^4
\left(
2\gamma+\f{1}{2}
\right)
+\f{N_c}{8\pi^2}M_s^4\ln
\left(
\f{M_s\beta}{\pi}
\right)\nonumber\\
&&\mbox{}
-\f{N_c}{2\pi^2}M_s^4
\sum_{k=1}^{\infty}
\f{(M_s\beta)^{2k}}{k!(k+2)!4^k}(2k+1)
(1-2^{1+2k})\f{1}{2}(-1)^{k}(2\pi)^{-2k}\Gamma(1+2k)\zeta(1+2k).
\label{thermoenergie}
\eeqn
\par

\subsubsection{Low temperature zero density expansion}

Eq. (\ref{eqthermo1.30}) is not well-suited because it would imply  taking the derivative of a truncated series (see eq.~(\ref{besselasymptpression})). It is better to search for the expansion of the exact solution in terms of the modified Bessel function obtained in\cite{cugjamvdb96}
\be
\varepsilon_{\boite{ideal gas}}(s)=3P_{\boite{ideal gas}}(s)+\f{2N_c}{\pi^2}
M_s^4\sum_{n=1}^{\infty}(-1)^{n+1}\f{K_1(nM_s\beta)}{nM_s\beta}.
\label{bassetempenergie}
\ee
Its low temperature asymptotic expansion is obtained
from eq.~(\ref{besselasymptpression}) for the first term and from
eq.~(\ref{besselasympt}) with $i=1$
\cite{stegun}
for the second one. We then have
\beqn
&&{\varepsilon}_{\boite{ideal gas}}(s)\approx 3{\frac{{2N}_{c}}{{\pi}^{2}}}M_s^4
\sum\limits_{n=1}^{\infty}
\f{(-1)^{n+1}}{n^2M_s^2\beta^2}
\sqrt{\f{\pi}{2nM_s\beta}}e^{-nM_s\beta}\nonumber\\
&&\qquad \mbox{}\times
\bigg\{
1+\f{15}{8nM_s\beta}+\f{15\cdot 7}{2!(8nM_s\beta)^2}
+\f{15\cdot 7 \cdot \mbox{}(-9)}{3!(8nM_s\beta)^3}+...
\bigg\}
\nonumber\\
&&\quad\mbox{}+\f{2N_c}{\pi^2}
M_s^4\sum_{n=1}^{\infty}(-1)^{n+1}\f{1}{nM_s\beta}
\sqrt{\f{\pi}{2nM_s\beta}}e^{-nM_s\beta}
\nonumber\\
&&\qquad\mbox{}\times
\bigg\{
1+\f{3}{8nM_s\beta}+\f{3\cdot \mbox{}(-5)}{2!(8nM_s\beta)^2}
+\f{3\cdot\mbox{}(-5)\cdot\mbox{}(-21)}{3!(8nM_s\beta)^3}+...
\bigg\}.
\label{besselasymptenergie}
\eeqn
\par

Once again we can limit ourselves to $n=1$ or $n=1,2$, depending upon the chosen set of parameters $(M_u^0,\chi_0)$. However the first two or three corrections to ``1'' are necessary.

\subsubsection{Finite density, zero temperature}

As for the pressure, eq.~(\ref{eqthermo24}) can be  exactly integrated. It is however more judicious to use the vanishing nature of the entropy at $T=0$ in order to get
(eq.~(\ref{eqthermo22})),
\be
\varepsilon_{\boite{ideal gas}}=-P_{\boite{ideal gas}}+\mu_i\rho_i,
\label{tnulle1}
\ee
where $\rho_i$ is given by eq.~(\ref{eqthermo15}), i.e.
\be
\lim\limits_{T\rightarrow 0}\rho_i=\f{N_c}{3\pi^2}k_{F_i}^3.
\label{tnulle2}
\ee
\par

Using (\ref{pressionT0}), (\ref{tnulle1}) and (\ref{tnulle2}), we obtain
\be
\varepsilon_{\boite{ideal gas}}(s)=\f{N_c}{\pi^2}
\Big[
\f{k_{F_s}}{4}\mu_s^3-\f{M_s^2}{8}k_{F_s}\mu_s-\f{M_s^4}{8}
\ln 
\left(
\f{k_{F_s}+\mu_s}{M_s}
\right)
\Big].
\ee

\subsection{Entropy density}
\subsubsection{High temperature zero density expansion}

Using eqs.~(\ref{eqthermo1.33},\ref{thermopression}) gives
\beqn
\lefteqn{
s(s)=
\f{7}{45}N_c\pi^2T^3-\f{N_c}{6}M_s^2T
+\f{N_c}{8\pi^2}M_s^4\beta
}
\nonumber\\
&&\mbox{}
-\f{N_c}{2\pi^2}M_s^4\beta
\sum_{k=1}^{\infty}
\f{(M_s\beta)^{2k}}{k!(k+2)!4^k}2k
(1-2^{1+2k})\f{1}{2}(-1)^{k}(2\pi)^{-2k}\Gamma(1+2k)\zeta(1+2k).
\label{thermoentropie}
\eeqn
\par
\subsubsection{Low temperature zero density expansion}

We can obtain this expansion starting from eq.~(\ref{eqthermo22}) with
$\mu_i=0$:
\be
s(s)=\beta(P_{\boite{ideal gas}}(s)+\varepsilon_{\boite{ideal gas}}(s)),
\ee
so that, using eqs.~(\ref{besselasymptpression},\ref{besselasymptenergie}), we have
\beqn
&&s(s)\approx 4\beta{\frac{{2N}_{c}}{{\pi}^{2}}}M_s^4
\sum\limits_{n=1}^{\infty}
\f{(-1)^{n+1}}{n^2M_s^2\beta^2}
\sqrt{\f{\pi}{2nM_s\beta}}e^{-nM_s\beta}\nonumber\\
&&\qquad \mbox{}\times
\bigg\{
1+\f{15}{8nM_s\beta}+\f{15\cdot 7}{2!(8nM_s\beta)^2}
+\f{15\cdot 7 \cdot \mbox{}(-9)}{3!(8nM_s\beta)^3}+...
\bigg\}
\nonumber\\
&&\quad\mbox{}+\f{2N_c}{\pi^2}
M_s^4\beta\sum_{n=1}^{\infty}(-1)^{n+1}\f{1}{nM_s\beta}
\sqrt{\f{\pi}{2nM_s\beta}}e^{-nM_s\beta}
\nonumber\\
&&\qquad\mbox{}\times
\bigg\{
1+\f{3}{8nM_s\beta}+\f{3\cdot \mbox{}(-5)}{2!(8nM_s\beta)^2}
+\f{3\cdot\mbox{}(-5)\cdot\mbox{}(-21)}{3!(8nM_s\beta)^3}+...
\bigg\}.
\label{besselasymptentropie}
\eeqn
\par

Once again we can limit ourselves to $n=1$ or $n=1,2$, depending upon the chosen set of parameters $(M_u^0,\chi_0)$. However the first two or three corrections to ``1'' are necessary.

\subsubsection{Finite density, zero temperature}

It is clear that
\be
s(s)=0,
\ee
in agreement with the third principle of thermodynamics.

\section{Masses and on-shell coupling constants for mixed particles}
\label{appendixmixing}
In this section, we define the masses and on-shell coupling constants for mixed particles. In   appendix \ref{PandGamma}, the scalar isoscalars form a set of three coupled particles, whose inverse propagator is given by the $3\times3$ subset of the matrix $\G$. Searching for the physical particles means that this matrix has to be diagonalized. In the following, we show how these physical states can be obtained. This is a non-trivial procedure because of the energy dependence of the normalization functions $Z$. The following demonstration  is valid for a set of $n$ coupled particles.\par

In order to shorten the notation, we use the following convention throughout this appendix: we denote by $A(\qq=-\micarre)$ a matrix $A$ of which each element is evaluated at the same $\qq$ in opposition to $A(-\micarre)$, which indicates a matrix whose  first column is evaluated at $-m_1^2$, the second at $-m_2^2$, ..., the $n$th at $-m_n^2$.

\subsection{Generalities}

Let $\G$ be the $n\times n$ meson matrix propagator 
\be
\G_{ii}=A_i\qq+B_i, \G_{ji}=\G_{ij}.
\label{coeffAi}
\ee
It is better to first consider $A_i,B_i,\G_{ij}(j\ne i)$ as being independent of $\qq$. In this case, an exact diagonalization (valid for each $\qq$) can be performed. We proceed in the following way:
\be
\st\G\s=\st\ginv\G'\ginv\s,
\label{gammagammaprime}
\ee
where ($\s$ is the vector denoting the coupled fields)
\be
\G'_{ii}=\qq+\f{B_i}{A_i},\G'_{ij}=\f{\G_{ij}}{\sqrt{A_iA_j}}\mbox{ and }
\ginv=\ \diag(\sqrt{A_i},...,\sqrt{A_n}).
\ee
Writing $\G'=\qq\one+\G''$, which defines $\G''$, we deduce
\be
\det(\G'-\lambda'\one)=\det(\G''-(\lambda'-\qq)\one)\equiv\det(\G''-\lambda''\one).
\ee
Since $\G''$ is $\qq$-independent, so are the eigenvalues (e.v.) $\lambda''$. Writing $\ld{i}''=\micarre$, it is easy to see that the e.v. of $\G'$ are
\be
\ld{i}'=\qq+\micarre.
\ee
Moreover, it is clear that the eigenvector (E.V.) matrix associated with $\G'$ is identical to that of $\G''$. This E.V. matrix is denoted by $V$. It is $\qq$-independent (it can then  be evaluated exactly) and is orthogonal: $\Vinv=V^T$.  
In short:
\be
\st\G\s=\st\ginv V\diag(\qq+m_1^2,...,\qq+m_n^2)\Vinv\ginv\s, 
\ee
i.e. 
\be
\st\G\s=\Sigt\diag(\qq+m_1^2,...,\qq+m_n^2)\Sig,
\label{indirect}
\ee
with
\be
\Sig=\Vinv\ginv\s,
\label{defSigma}
\ee
the vector denoting the physical fields.\par

On the other hand, a direct diagonalization implies 
\be
\st\G\s=\st V_*(\qq)\diag(\ld{1},...,\ld{n})\Vinv_*(\qq)\s.
\label{direct}
\ee
At this level, everything is exact (because this is the same  $\qq$ for each E.V.), and we have the property $\Vinv_*(\qq)=V^T_*(\qq)$.

Writing 
\be
\Lambda(\qq)\equiv\Vinv_*(\qq)\G V_*(\qq)=\diag(\ld{1},...,\ld{n}),
\label{defvaleurpropre}
\ee
 and using 
\be
\f{\p}{\p\qq}(\Vinv_*(\qq)V_*(\qq))=0,
\ee
we have
\beqn
\ginvdeux_*(\qq)\equiv
\f{\p\Lambda}{\p\qq}&=&[\f{\p\Vinv_*(\qq)}{\p\qq}V_*(\qq),\Lambda(\qq)]
+\Vinv_*(\qq)\f{\p\G}{\p\qq}V_*(\qq)\nonumber\\
&=& [\f{\p\Vinv_*(\qq)}{\p\qq}V_*(\qq),\Lambda(\qq)]
+\Vinv_*(\qq)\ginvdeux V_*(\qq).
\label{gmoinsdeux}
\eeqn
The physical masses are clearly obtained from the zeroes of the e.v.
Since the trace of a commutator is identically zero for a finite matrix, we get the following relation between the exact quantities\footnote{We call ``exact'' the quantities which are all evaluated at the same $\qq$; approximations only play a role when quantities must be evaluated at their respective mass shell.} 
\be
\tr\ginvdeux_*(\qq)=\tr\ginvdeux.
\label{trace}
\ee

Another exact relation comes from
\be
\st V_*(\qq)\Lambda(\qq)\Vinv_*(\qq)\s=\st\ginv V\diag(\qq+m_1^2,...,\qq+m_n^2)\Vinv\ginv\s,
\ee
which implies
\be
\det(V_*(\qq)\Lambda(\qq)\Vinv_*(\qq))=
\det(\ginv V\diag(\qq+m_1^2,...,\qq+m_n^2)\Vinv\ginv),
\ee
i.e.
\be
\det(\Lambda(\qq))=\det(\ginvdeux V\diag(\qq+m_1^2,...,\qq+m_n^2)\Vinv)=\det(\ginvdeux)\det(\diag(\qq+m_1^2,...,\qq+m_n^2)),
\ee
i.e.
\be
\prod_i\ld{i}(\qq)=\left(\prod_i (\qq+\micarre)\right)\left(\prod_i A_i\right).
\label{produit}
\ee
Let us once more emphasize  that this relation is only valid as far as  the e.v. are all evaluated at the same $\qq$.

\subsection{First order Taylor expansion, as a function of $\qq$, around the respective zeroes of the eigenvalues} 

If each diagonal component of  eq.~(\ref{gmoinsdeux}) is evaluated at the zero of the corresponding e.v., the commutator gives a zero contribution. The latter can be established from 
\be
\ginvdeux_*(\qq=-\micarre)=\left.[\f{\p\Vinv_*(\qq)}{\p\qq}V_*(\qq),\Lambda(\qq)]\right|_{\qq=-\micarre}
+\Vinv_*(\qq=-\micarre)\ginvdeux V_*(\qq=-\micarre).
\ee
Since each quantity is here evaluated at the same $\qq$, the inverse E.V. matrix is identically equal to its transposed. Taking out the  $j,j$ component on both sides of the equality, we obtain (summation over  $k$ and $l$, no summation over $i$)
\beqn
\ginvdeux_*(\qq=-\micarre)_{jj}&=&\left.\left(\f{\p\Vinv_*(\qq)}{\p\qq}V_*(\qq)\right)_{jk}\Lambda(\qq)_{kj}\right|_{\qq=-\micarre}
-\left.\Lambda(\qq)_{jk} 
\left(\f{\p\Vinv_*(\qq)}{\p\qq}V_*(\qq)\right)_{kj}\right|_{\qq=-\micarre}
\nonumber\\
&&\hspace{3cm}\mbox{}+V_*^T(\qq=-\micarre)_{jk}\ginvdeux_{kl} V_*(\qq=-\micarre)_{lj}.
\label{complet}
\eeqn
Since $\Lambda(\qq)$ is diagonal, it is clear that the contribution of the commutator vanishes. Anyway, if we take $j=i$, this commutator gives also trivially zero since it implies the evaluation of $\lambda_i(\qq=-\micarre)$ which is zero by definition. Now, let us effectively take $j=i$. By definition of the transposed of a matrix, we then have the relation 
\be
\ginvdeux_*(\qq=-\micarre)_{ii}=\sum_k\ginvdeux_{kk}\left(V_*(\qq=-\micarre)_{ki}\right)^2,
\label{tronque}
\ee
which shows that the $i$th E.V. plays a role only for the component $i,i$ of $\ginvdeux_*(\qq=-\micarre)$. Then, we can generalize (\ref{tronque}) to
\be
\ginvdeux_*(-\micarre)_{ii}=\left(V_*^T(-\micarre)\ginvdeux V_*(-\micarre)\right)_{ii},
\label{correctii}
\ee
where,  according to the conventions of this appendix, each column of $V$ and $\ginvdeux_*$ is now evaluated at its respective zero.
Although we have demonstrated eq.~(\ref{correctii})  for diagonal components, it can be  extented to
\be
V_*^T(-\micarre)=\ginvdeux_*\Vinv_*\gdeux.
\label{correct}
\ee
It  includes the previous one for the diagonal components of $\ginvdeux_*$. But eq.~(\ref{correct}) can also be shown to imply the vanishing of the off-diagonal elements. This is seen as soon as we can prove that
\be
\Vinv_*(-\micarre)=g_*(-\micarre)\Vinv\ginv
\label{relDD}
\ee
is true, with being $\ginvdeux_*$ the diagonal matrix taken from the definition $\diag(\p\lambda_i(\qq)/\p\qq)$. 
The relation (\ref{relDD}) is suggested by a comparison between eqs.~(\ref{indirect}) and (\ref{direct}). Let us notice that it is the inverse  $\Vinv_*(-\micarre)$ which must be considered, not $V^T(-\micarre$) (E.V. evaluated at different $\qq$).

In order to prove eq.~(\ref{relDD}), we start from the E.V. definition (\ref{defvaleurpropre}):  
\be
\G(\qq) V_*(\qq)=V_*(\qq)\Lambda(\qq).
\ee
Taking the component $i,j$, evaluated at $\qq=-\mjcarre$, we obtain
\be
\G(\qq=-\mjcarre)_{ik} V_*(\qq=-\mjcarre)_{kj}=V_*(\qq=-\mjcarre)_{ik}\Lambda(\qq=-\mjcarre)_{kj},
\ee
i.e.,  (no summation over $j$)
\be
\G(\qq=-\mjcarre)_{ik} V_*(\qq=-\mjcarre)_{kj}=
V_*(\qq=-\mjcarre)_{ij}\lambda(-\mjcarre)_j.
\label{lastref}
\ee
It is clear that  the only E.V. which plays a role is associated with the $j$th e.v. Moreover, by definition of $\mjcarre$, the r.h.s. of eq.~(\ref{lastref}) is identically zero. We then have:
\be
\G(\qq=-\mjcarre)_{ik} V_*(-\mjcarre)_{kj}=0.
\label{relindices}
\ee
It is useful to stress that eq.~(\ref{relindices})  has to be verified component by component since the $\qq$ used for $\G$ is the one of the corresponding E.V. in $V_*$. If we denote by  $d_*^{(j)}$ the E.V. associated with the $j$th e.v., eq.~(\ref{relindices}) is then identical to
\be
\G(\qq=-\mjcarre)_{ik} d_{*,k}^{(j)}=0\leftrightarrow
\G(\qq=-\mjcarre)d_*^{(j)}=0.
\ee
The demonstration of eq.~(\ref{relDD}) consists in starting from the relation (\ref{relindices}) and in showing that (\ref{relDD}) is a possible solution thereof. It is then sufficient to show that the   normalization is correct for the uniqueness of the solution. (Note : the last relation does not imply the existence of $n$ orthogonal E.V. associated with  \ $\qq=-\mjcarre$. There is indeed only one E.V. associated with the e.v. $\lambda(-\mjcarre)=0$.)

If we take the inverse of (\ref{relDD}), i.e.
\be
V_*(-\mjcarre)=gV\ginv_*,
\label{inverse}
\ee
 and if we insert it on the l.h.s of  (\ref{relindices}), we have 
\be
\G(\qq=-\mjcarre)_{ik}(gV\ginv_*)_{kj}=\G(\qq=-\mjcarre)_{ik}(gV)_{kl}\ginv_{*,lj}=(\G(\qq=-\mjcarre)gV)_{il}\ginv_{*,lj},
\ee
i.e. (see eq.~(\ref{gammagammaprime}))
\be
\G(\qq=-\mjcarre)_{ik}(gV\ginv_*)_{kj}=(\ginv\G'(-\mjcarre)V)_{il}\ginv_{*,lj}.
\label{GgD}
\ee
By definition of $V$, we have
\be
\G'V=V\diag(\qq+m_1^2,...,\qq+m_n^2),
\ee
so that (there is no summation over $l$)
\be
(\ginv\G'V)_{il}=(\ginv V)_{ik}\diag(\qq+m_1^2,...,\qq+m_n^2)_{kl}\equiv
(\ginv V)_{il}(\qq+m_l^2).
\label{groumph}
\ee
Then, inserting (\ref{groumph}) in (\ref{GgD}), we get
\be
\G(\qq=-\mjcarre)_{ik}(gV\ginv_*)_{kj}=\sum_l (\ginv V)_{il}(-\mjcarre+m_l^2)\ginv_{*,lj}
\ee
and then, because $\ginv_*$ is diagonal,
\be
\G(\qq=-\mjcarre)_{ik}(gV\ginv_*)_{kj}=(\ginv V)_{ij}(-\mjcarre+\mjcarre)\ginv_{*,jj}\equiv 0.
\ee
We have thus shown that (\ref{inverse}) is a  solution of (\ref{relindices}). In fact,    (\ref{inverse}) is exact since the normalization is correct (because, from (\ref{inverse}), we deduce
(\ref{correct}) and then  (\ref{correctii}) which was shown to be true).

As a  conclusion to this appendix, we note that with the E.V. matrix $V_*(-\micarre)$, the physical fields are given by 
\be
\Sigma=\Vinv\ginv\s=\ginv_*(-\micarre)\Vinv_*(-\micarre)\s.
\label{newfields}
\ee
This definition will be kept even if the coefficients $A_i$, see eq.~(\ref{coeffAi}), are $\qq$-dependent (this is for example the case of the NJL model). Nevertheless, the previous relations are not all valid since the matrix $\ginv$ becomes also $\qq$-dependent. However, we can conclude 
from (\ref{complet}) that the equivalent of (\ref{tronque}) is:
\be
\ginvdeux_*(\qq=-\micarre)_{ii}=\sum_k\ginvdeux_{kk}(\qq=-\micarre)\left(V_*(\qq=-\micarre)_{ki}\right)^2=\sum_k\ginvdeux_{kk}(\qq=-\micarre)\left(V_*(-\micarre)_{ki}\right)^2,
\label{getoilemoins2}
\ee
as far as the coefficients $B_i$ and the elements $\G_{ij},i\ne j$ are $q^2$-independent. In case they are $q^2$-dependent, it is obvious from   (\ref{complet}) that
\beqn
\ginvdeux_*(\qq=-\micarre)_{ii}&=&V_*^T(\qq=-\micarre)_{ik}\ginvdeux_{kl}(\qq=-\micarre) V_*(\qq=-\micarre)_{li}\nonumber\\
&=&\sum_{k,l}\ginvdeux_{kl}(\qq=-\micarre)V_*(-\micarre)_{ki}V_*(-\micarre)_{li}
\label{getoilemoins3}
.
\eeqn

\newpage
\addcontentsline{toc}{section}{Bibliography}
\bibliography{echaya}
\bibliographystyle{osa}

\newpage

\listoftables
\listoffigures

\end{document}